\documentclass[letterpaper,11pt]{article}
\pdfoutput=1

\usepackage{jheppub} 

\usepackage[utf8]{inputenc}
\usepackage{color}
\usepackage{tikz}
\usetikzlibrary{shapes.geometric}
\usepackage{hepunits}
\usepackage{floatrow}
\usepackage{subfig}
\usepackage{units}

\newcommand{\Ref}[1]{ref.~\cite{#1}}
\newcommand{\Refs}[1]{refs.~\cite{#1}}
\newcommand{\Fig}[1]{fig.~\ref{#1}}
\newcommand{\Figs}[2]{figs.~\ref{#1} and \ref{#2}}

\newcommand{\Sec}[1]{sec.~\ref{#1}}

\newcommand{\App}[1]{app.~\ref{#1}}

\newcommand{\Eq}[1]{eq.~\eqref{#1}}
\newcommand{\Eqs}[2]{eqs.~\eqref{#1} and \eqref{#2}}

\newcommand{\be}{\begin{equation}}
\newcommand{\ee}{\end{equation}}

\newcommand{\SUB}{S}

\newcommand{\al}{\alpha}
\newcommand{\GFF}{\mathcal{F}}

\preprint{\vbox{\hbox{MIT-CTP 4887}
\hbox{NIKHEF 2017-002}
\hbox{CERN-TH-2017-064}}}

\title{Generalized Fragmentation Functions for \\ Fractal Jet Observables}

\author[a]{Benjamin T.~Elder,}
\author[b]{Massimiliano Procura,}
\author[a]{Jesse Thaler,}
\author[c,d]{\\ Wouter J.~Waalewijn,}
\author[a]{and Kevin Zhou}

\affiliation[a]{Center for Theoretical Physics, Massachusetts Institute of Technology, Cambridge, MA 02139, USA}
\affiliation[b]{Theoretical Physics Department, CERN, Geneva, Switzerland}
\affiliation[c]{Institute for Theoretical Physics Amsterdam and Delta Institute for Theoretical Physics, University of Amsterdam, Science Park 904, 1098 XH Amsterdam, The Netherlands}
\affiliation[d]{Nikhef, Theory Group, Science Park 105, 1098 XG, Amsterdam, The Netherlands}

\emailAdd{belder13@mit.edu}
\emailAdd{massimiliano.procura@cern.ch}
\emailAdd{jthaler@mit.edu}
\emailAdd{w.j.waalewijn@uva.nl}
\emailAdd{knzhou@mit.edu}

\abstract{
We introduce a broad class of fractal jet observables that recursively probe the collective properties of hadrons produced in jet fragmentation.
To describe these collinear-unsafe observables, we generalize the formalism of fragmentation functions, which are important objects in QCD for calculating cross sections involving identified final-state hadrons.
Fragmentation functions are fundamentally nonperturbative, but have a calculable renormalization group evolution.
Unlike ordinary fragmentation functions, generalized fragmentation functions exhibit nonlinear evolution, since fractal observables involve correlated subsets of hadrons within a jet.
Some special cases of generalized fragmentation functions are reviewed, including jet charge and track functions.
We then consider fractal jet observables that are based on hierarchical clustering trees, where the nonlinear evolution equations also exhibit tree-like structure at leading order.
We develop a numeric code for performing this evolution and study its phenomenological implications.
As an application, we present examples of fractal jet observables that are useful in discriminating quark jets from gluon jets.
}

\begin{document} 
\maketitle

\section{Introduction}
\label{sec:Introduction}

Fragmentation functions (FFs) have a long history in QCD for calculating cross sections for collinear-unsafe observables.  
Ordinary FFs are process-independent nonperturbative objects that describe the flow of momentum from a fragmenting quark or gluon into an identified final-state hadron 
\cite{Georgi:1977mg,Mueller:1978xu,Ellis:1978ty,Curci:1980uw,Altarelli:1981ax,Collins:1981uk,Collins:1981uw}.
Since the momentum of a single hadron is not collinear safe, cross sections for single-hadron observables have singularities beginning at $\mathcal{O}(\alpha_s)$.
These collinear singularities are absorbed by the FFs order by order in $\alpha_s$.
From this singularity structure, one can derive the renormalization group (RG) evolution for FFs, leading to the well-known DGLAP equations \cite{Lipatov:1974qm,Gribov:1972ri,Altarelli:1977zs,Dokshitzer:1977sg}.
This evolution is linear, since FFs depend only on the momentum of a single hadron in the final state.

In this paper, we present a formalism for generalized fragmentation functions (GFFs), which describe the flow of momentum  from a fragmenting quark or gluon into \emph{subsets} of final-state hadrons.
Because GFFs depend on correlations between final-state hadrons, their evolution equations are nonlinear and therefore more complicated than in the ordinary FF case.
Motivated by the structure of the DGLAP equations, we define \emph{fractal jet observables} where the evolution, albeit nonlinear, takes a special recursive form that is well-suited to numerical evaluation.\footnote{This should not be confused with ``extended fractal observables'' recently introduced in \Ref{Davighi:2017hok}, which are based on determining the fractal dimension of a jet.}

\begin{figure}[t]
\centering
\begin{tikzpicture}
\draw[ultra thick] (2.0,0) -- (5.0,0)
node[draw=none,fill=none,midway,sloped,anchor=center,above]{$\vec{p},x=\hat{x}(z,x_1,x_2)$};
\draw[ultra thick] (5.0,0) -- (7,1)
node[draw=none,fill=none,midway,sloped,anchor=center,above] {$\vec{p}_1$, $x_1$};
\draw[ultra thick] (5.0,0) -- (7,-1)
node[draw=none,fill=none,midway,sloped,anchor=center,below] {$\vec{p}_2$, $x_2$};
\end{tikzpicture}
\caption{Fractal jet observables are defined recursively on binary clustering trees. In each recursive step, the value $x$ for the mother is expressed in terms of the momentum fraction $z$ and the value $x_1$ and $x_2$ of the observable for the daughters.}
\label{fig:splitting}
\end{figure}
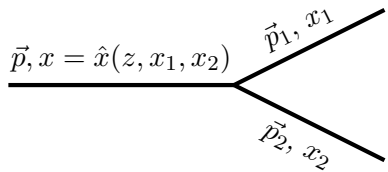

Specifically, we focus on observables defined using hierarchical binary clustering trees that mimic the leading-order tree-like structure of the evolution equations.
A fractal jet observable $x$ can then be defined recursively according to \Fig{fig:splitting} as
\be
\label{eq:recurse}
x = \hat{x}(z,x_1,x_2),
\ee
where $x_1$ and $x_2$ are the values of the observable on the branches of a $1 \to 2$ clustering tree, and $z$ is the momentum sharing between branches, defined by 
\be \label{eq:z_def}
z  \equiv \frac{E_1}{E_1+E_2}
\ee
with $E_i$ the energy of branch $i$.\footnote{While it would be more accurate to call \Eq{eq:z_def} the ``energy fraction'', we use momentum fraction since that is more common in the fragmentation function literature.}  With these definitions, the leading-order evolution equation of the corresponding GFF  takes the simplified form
\be
\label{eq:evolution}
\mu \frac{\text{d}}{\text{d}\mu} \GFF_i (x,\mu) = \frac{1}{2}\sum_{j,k} \int \text{d}z\, \text{d}x_1\, \text{d}x_2\, \frac{\alpha_s(\mu)}{\pi} P_{i\rightarrow jk}(z)\,\GFF_j (x_1,\mu)\, \GFF_k (x_2,\mu)\, \delta[x-\hat{x}(z,x_1,x_2)],
\ee
where $\GFF_i (x,\mu)$ is the GFF for parton $i = \{u, \bar{u}, d, \dots, g\}$, $P_{i\rightarrow jk}(z)$ is the $1 \to 2$ QCD splitting function, and $\mu$ is the $\overline{\rm MS}$ renormalization scale.
This evolution equation has the same structure as a $1 \to 2$ parton shower, which is sufficiently straightforward to implement numerically.  Although we mostly restrict ourselves to lowest order in perturbation theory, our framework allows for the systematic inclusion of higher-order corrections, in contrast to the semi-classical parton shower approach.

The class of fractal jet observables described by \Eq{eq:recurse} is surprisingly rich, allowing for many collinear-unsafe observables to be calculated with the help of GFFs.
For example, \Eq{eq:evolution} describes the evolution of weighted energy fractions,
\be
x = \sum_{a \in {\rm jet}} w_a \, z_a^\kappa, \qquad z_a \equiv \frac{E_a}{E_{\rm jet}},
\ee
where $w_a$ is a weight factor that depends on non-kinematic quantum numbers such as charge or flavor, $\kappa > 0$ is an energy weighting exponent, and the sum extends over all jet constituents.
These observables are defined by associative recursion relations, such that their value is independent of the choice of clustering tree.
Examples of weighted energy fractions include weighted jet charge \cite{FIELD19781}, whose nonlinear evolution was first studied in \Ref{Waalewijn:2012sv}; track functions which characterize the fraction of a jet's momentum carried by charged particles \cite{Chang:2013rca,Chang:2013iba}; and the observable $p_T^D$ used by the CMS experiment for quark/gluon discrimination~\cite{Pandolfi:1480598,Chatrchyan:2012sn}, whose nonlinear evolution was first studied in \Ref{Larkoski:2014pca}.
While we focus on the case of $e^+e^-$ collisions with jets of energy $E_{\rm jet}$, our formalism easily adapts to hadronic collisions with jets of transverse momentum $p_{T}^{\rm jet}$.

In addition to performing a more general analysis of weighted energy fractions, we also present examples of fractal observables with non-associative recursion relations. 
These quantities depend on the details of the clustering tree used to implement \Eq{eq:recurse}, providing a complementary probe of jet fragmentation.
In particular, while \Eq{eq:recurse} does not involve any explicit angular separation scales, the clustering tree does introduce an implicit angular dependence.
Remarkably, the details of the clustering do not affect the leading-order RG evolution in \Eq{eq:evolution} considered in this paper, beyond the requirement that particles are appropriately clustered in the collinear limit. 
An example of a non-associative fractal observable is given by node-based energy products,
\be
\label{eq:intro_nodes}
x = \sum_{\rm nodes} \left(4 z_L z_R  \right)^{\kappa/2},
\ee
where the observable depends on the momentum fractions carried by the left and right branches at each node in the clustering tree.
We also study observables defined entirely in terms of \Eq{eq:recurse}, with no obvious simplification.
This sensitivity to the tree structure allows non-associative observables to probe parton fragmentation from a different perspective  than previously-studied jet observables.
As one application, we consider the discrimination between quark- and gluon-initiated jets~(see e.g.~\cite{Gallicchio:2011xq,Gallicchio:2012ez,Larkoski:2013eya,Larkoski:2014pca,Bhattacherjee:2015psa,Badger:2016bpw,FerreiradeLima:2016gcz,Komiske:2016rsd,Gras:2017jty} for recent studies). We find that fractal observables are effective for this purpose, in some cases yielding improved quark/gluon separation power compared to weighted energy fractions.

For clustering trees obtained from the Cambridge/Aachen (C/A) algorithm~\cite{Dokshitzer:1997in,Wobisch:1998wt}, the depth in the tree is directly related to the angular separation scale between subjets.
This opens up the possibility of modifying the recursion relation $\hat x$ in \Eq{eq:evolution} to be a function of angular scale.
For example, starting from a jet of radius $R$, one can introduce a subjet radius parameter $R_{\rm sub} \ll R$ such that evolution equation takes a different form below and above $R_{\rm sub}$.
A particularly simple case is if the weighted energy fraction with $\kappa=1$ is measured on the branches below $R_{\rm sub}$, since this effectively amounts to defining fractal observables in terms of subjets of radius $R_{\rm sub}$.
In this case, the initial conditions for the GFF leading-order evolution is simply given by $\GFF_i(x,\mu_{\rm sub}) = \delta(1-x)$ at the initial scale $\mu_{\rm sub} = E_{\rm jet} R_{\rm sub} \gg \Lambda_{\rm QCD}$, such that no nonperturbative input is needed. 
By evolving the GFFs to $\mu = E_{\rm jet} R$, we achieve the resummation of leading logarithms of $R_{\rm sub}/R$.  Related evolution techniques have been used to resum logarithms of the jet radius $R$ in inclusive jet cross sections \cite{Dasgupta:2014yra,Kang:2016mcy,Dai:2016hzf}.

The formalism of GFFs is reminiscent of other multi-hadron FFs in the literature.  This includes dihadron fragmentation functions which describe the momentum fraction carried by pairs of final-state hadrons~\cite{Sukhatme:1980vs,Vendramin:1981te}, and fracture functions which correlate the properties of one initial-state and one final-state hadron \cite{Trentadue:1993ka,Graudenz:1994dq}.  In all of these cases, the RG evolution equations are nonlinear.  The key difference here is that fractal jet observables are not based on a fixed number of hadrons, but rather allow for arbitrary hadron multiplicities. Depending on the observable, this may require that all hadrons can be consistently labeled by non-kinematic quantum numbers (e.g.~charge).  As discussed in \Ref{Waalewijn:2012sv} for the case of weighted jet charge, the $n$-th moment of GFFs can sometimes be related to moments of $n$-hadron FFs.  At the level of the full distribution, though, GFFs are distinct from multi-hadron FFs, and thereby probe complementary aspects of jet fragmentation.

The rest of this paper is organized as follows.
In \Sec{sec:formalism}, we review the theoretical underpinnings of ordinary parton fragmentation and explain how to extend the formalism to generalized fragmentation and fractal observables.
We then construct generic fractal jet observables using clustering trees in \Sec{sec:fractal}.
In \Sec{sec:WEF}, we treat the case of weighted energy fractions, exploring their RG evolution for a range of parameters.
We introduce two new sets of non-associative fractal observables in \Sec{sec:NA}---node products and full-tree observables---and motivate their application in quark/gluon discrimination in \Sec{sec:quarkgluon}.
We briefly explain how our formalism also applies to fractal observables based on subjets rather than hadrons in \Sec{sec:angular}. 
We conclude in \Sec{sec:conclusion}, leaving calculational details and a description of the numerical RG implementation to the appendices.

\section{Formalism}\label{sec:formalism}

To motivate the definition of fractal jet observables, it is instructive to first review the formalism of standard fragmentation and then generalize it to arbitrary collinear-unsafe observables.  We give a general definition of fractal jet observables at the end of this section, which serves as a preamble to the explicit constructions in \Sec{sec:fractal}.

\subsection{Review of Standard Fragmentation}

Ordinary FFs, denoted by $D_i^h(x,\mu)$, are nonperturbative objects that describe the number density of hadrons of type $h$ carrying momentum fraction $x$ among the particles resulting from the fragmentation of a parton of type $i$.
They are the final-state counterpart to parton distribution functions (PDFs).
For any parton flavor $i$, they satisfy the momentum conservation sum rule
\begin{equation}\label{eq:FFsumrule}
\sum_h \int_0^1 \text{d}x\, x \, D_i^h(x,\mu) = 1\,.
\end{equation}
At leading order, the FFs are independent of the factorization scheme (see e.g.~\cite{Furmanski:1981cw}).

The field-theoretic definition of the bare unpolarized quark FF is given by \cite{Collins:1981uk,Collins:1981uw}
\begin{align}
\label{eq:FFoperator}
D_i^h(x,\mu) = \frac{1}{x} \int \text{d}^2 p_h^\perp \! \int \frac{\text{d}y^+\text{d}^2y_\perp}{2 (2 \pi)^3} e^{i p^-y^+} \sum_X \frac{1}{2N_C}  \mbox{Tr}\bigg[ \frac{\gamma^-}{2} \langle 0 \vert \psi_i(y^+,0,y_\perp) \vert hX \rangle \langle hX \vert \overline{\psi}_i(0)\vert 0\rangle \bigg],
\end{align}
where we are working in a frame with quark transverse momentum $\vec{p}_\perp = 0$ and using the gauge choice $A^- = 0$. 
The jet-like state $\vert hX \rangle$ contains an identified hadron $h$ of momentum $p_h$ with $p_h^- \equiv x p^-$, and $X$ refers to all other hadrons in that state.
The factor $1/(2N_C)$, where $N_C=3$ is the number of colors, accounts for averaging over the color and spin of the quark field $\psi$ of flavor $i$.
Here and in the rest of the paper, we adopt the following convention for decomposing a four-vector $w^\mu$ in light-cone coordinates: 
\begin{align}
w^\mu = w^- \frac{n^\mu}{2} + w^+ \frac{\bar n^\mu}{2} + w_\perp^\mu, \qquad
w^- = \bar n \cdot w, \qquad
w^+ = n \cdot w,
\end{align}
where $n^\mu$ is a light-like vector along the direction of the energetic parton, and $\bar n$ is defined such that $n^2 = \bar n^2 = 0$ and $n \cdot \bar n = 2$.
Thus at leading order, $p^- = 2E_{\rm jet}$.
Gauge invariance requires adding eikonal Wilson lines in \Eq{eq:FFoperator} (see e.g.~\cite{Collins:1989gx}), which we suppress here for notational convenience.
An analogous definition applies for the gluon FF. 

In the context of $e^+e^-$ annihilation, FFs are crucial ingredients in the factorization formula for the semi-inclusive cross section at leading power in $\Lambda_{\rm QCD}/\sqrt{s}$,
\begin{equation}
\label{eq:eecs}
\frac{1}{\sigma^{(0)}} \frac{\text{d}\sigma}{\text{d}x} (e^+e^-\rightarrow hX) = \sum_i \int_x^1 \frac{\text{d}z}{z} \, C_i(z,s,\mu) \, D_i^h(x/z,\mu),
\end{equation}
where $x = 2 E_h/\sqrt{s} \le 1$ is the hadron energy fraction, $\sigma^{(0)}$ is the tree-level cross section and $X$ represents all other final state particles in the process.\footnote{In the literature (see e.g.~\cite{Olive:2016xmw}), the cross section
$1/\sigma^{(0)}\, \text{d}\sigma/\text{d}x (e^+e^-\rightarrow hX) = F^h(x,\mu)$
is sometimes referred to as the total FF, in which case $D_i^h(x,\mu)$ is called the parton FF.}
The coefficients $C_i(z,s,\mu)$ are process-dependent perturbative functions that encode the physics of the hard subprocess. 
The FFs $D_i^h(x,\mu)$ are universal, process-independent functions, which appear (with appropriate PDF convolutions) in related channels such as $ep\rightarrow hX$ or $pp\rightarrow hX$.
Since the coefficients $C_i$ contain logarithms of $s/\mu^2$, in order to avoid terms that could spoil perturbative convergence in \Eq{eq:eecs}, the renormalization scale $\mu$ should be chosen close to $\sqrt{s}$.

While computing the FFs themselves requires nonperturbative information about the hadronic matrix elements in \Eq{eq:FFoperator}, their scale dependence is perturbatively calculable.
This allows us to, for example, take FFs extracted from fits to experimental data at one scale and evolve them to another perturbative scale.
The RG evolution of FFs is described by the DGLAP equations \cite{Lipatov:1974qm,Gribov:1972ri,Altarelli:1977zs,Dokshitzer:1977sg},
\begin{equation}
\label{eq:DGLAP}
\mu \frac{\text{d}}{\text{d}\mu} D_i^h(x,\mu) = \sum_j \int_x^1 \frac{\text{d}z}{z} \frac{\alpha_s(\mu)}{\pi} P_{ji}(z) D_j^h(x/z,\mu).
\end{equation}
Here, the splitting kernels $P_{ji}(z)$ can be calculated in perturbation theory, 
\begin{equation}
P_{ji} (z) = P_{ji}^{(0)}(z) + \frac{\alpha_s}{2\pi} P^{(1)}_{ji}(z) + \ldots,
\end{equation}
and are at lowest order the same as the splitting kernels for PDF evolution. The next-order splitting function $P^{(1)}_{ji}$ arises from $1\to 3$ splittings as well as loop corrections to $1 \to 2$ splittings.

In order to motivate the transition to generalized fragmentation, it is convenient to rewrite the lowest-order splitting function explicitly as a $1 \to 2$ process:
\be
\label{eq:Parrowdef}
P_{ji}^{(0)}(z) \equiv P_{i\rightarrow jk}(z),
\ee
where the parton $j$ carries momentum fraction $z$, e.g.\ $P_{g\rightarrow gg}(z)$ or $P_{q\rightarrow qg}(z) = P_{q\rightarrow gq}(1-z)$.  
With this notation, we can rewrite the leading-order DGLAP equation in a suggestive form\footnote{Because the splitting functions are divergent as $z\rightarrow 1$ and as $z\rightarrow 0$, plus-function regulators are required at both endpoints when integrating over the entire range $0 \leq z \leq 1$.}
\begin{align}
\label{eq:linear_as_double_integral}
\mu \frac{\text{d}}{\text{d}\mu} D_i^h(x,\mu) &= \frac{1}{2} \sum_{j,k} \int_0^1 \text{d}x_1\, \int_0^1 \text{d}x_2\,  \int_0^1 \text{d}z \, \frac{\alpha_s(\mu)}{\pi} \, P_{i\rightarrow jk}(z) \nonumber \\ & \quad
\times \Bigl( D_j^h\left(x_1,\mu\right) \, \delta[ x-zx_1 ] + D_k^h\left(x_2,\mu\right)\, \delta[ x-(1-z)x_2 ] \Bigr).
\end{align}
Though we have written \Eq{eq:linear_as_double_integral} as an integral over both $x_1$ and $x_2$, corresponding to the two final state branches from the $i \to jk$ splitting, the FFs only require information about one single final-state hadron in each term, so the evolution simplifies to the linear form in \Eq{eq:DGLAP}.
This will no longer be the case with generalized fragmentation, which depends on correlations between the final-state hadrons.

\subsection{Introducing Generalized Fragmentation}
\label{sec:generalFF}

We now extend the FF formalism to handle the distribution of quantities $x$ carried by {\it a subset} $\SUB$ of collinear particles, where $x$ can be more general than the simple momentum fraction and $\SUB$ is defined by non-kinematic quantum numbers.
For example, we will consider observables defined on all particles within a jet, but also on charged particles only.
For a given observable $x$, there is a GFF for each parton species $i$, which we denote by $\GFF_i(x,\mu)$. 
At lowest order in $\alpha_s$, the GFF is the probability density for the particles in $\SUB$ to yield a value of the observable $x$ from jets initiated by a parton of type $i$.
The GFF automatically includes information about hadronization fluctuations.
Being a probability density, the GFFs are normalized to unity for each parton type,
\begin{equation}
\label{eq:GFFsumrule}
\int \text{d}x\, \GFF_i (x,\mu) = 1.
\end{equation}

For any collinear-unsafe (but soft-safe) observable $x$, we can give an operator definition for GFFs analogous to that for fragmentation functions.
A (bare) quark GFF for the gauge choice $A^-=0$ is defined as
\begin{align}
\label{eq:GFFopQ}
\GFF_i(x,\mu) &=   \int \text{d}y^+ \text{d}^2y_\perp e^{ip^-y^+/2} \frac{1}{2N_C}\sum_{\SUB X} \delta[x-\tilde{x}(p^-,\SUB)] 
\nonumber \\ & \quad
\times 
\mbox{Tr}\bigg[\frac{\gamma^-}{2}\langle 0\vert \psi_i(y^+,0,y_\perp) \vert \SUB X \rangle \langle \SUB X \vert \overline{\psi}_i(0) \vert 0\rangle\bigg],
\end{align}
to be compared with \Eq{eq:FFoperator}. Here, $\vert \SUB X \rangle$ is the asymptotic final state divided into the measured subset $\SUB$ and unmeasured subset $X$, and $\tilde{x}(p^-,\SUB)$ is the functional form of the quantity being observed, which can depend on the overall jet momentum and any information from $\SUB$.
We stress that, in contrast to the standard FFs, a GFF involves a sum over polarizations and a phase-space integration over all detected particles in $\SUB$; if the measured set $\SUB$ consists of a single hadron, then \Eq{eq:GFFopQ} reduces to \Eq{eq:FFoperator} for a quark FF.
The definition for gluon-initiated jets is
\begin{align} \label{eq:GFFopG}
\GFF_g(x,\mu) &= -\frac{1}{(d-2)(N_C^2-1) p^-} \int \text{d}y^+ \text{d}^2y_\perp e^{ip^-y^+/2} \sum_{\SUB X} \delta[x-\tilde{x}(p^-,\SUB)]  \nonumber \\ & \quad
\times\langle 0\vert G^{-,a}_{\lambda}(y^+,0,y_\perp)\vert \SUB X \rangle \langle \SUB X \vert G^{-,a,\lambda}(0)\vert 0\rangle,
\end{align}
where $G^{-,a}_{\lambda} = \overline{n}^\mu G_{\mu\lambda}^a$ is the gluon field strength tensor for generator $T^a$, the factor of $1/(d-2)$ comes from averaging over the gluon polarizations in $d$ space-time dimensions, and the factor of $1/(N_C^2-1)$ comes from averaging over the color of the gluon. 

The definitions in \Eqs{eq:GFFopQ}{eq:GFFopG} extend the ones introduced in \Ref{Chang:2013rca} for track functions.
In the track function case, $x$ is the momentum fraction carried by the charged particles in the final states, irrespective of their individual properties or multiplicities.   
As mentioned in the introduction, GFFs are reminiscent of multi-hadron FFs~\cite{Sukhatme:1980vs,Vendramin:1981te}, with the key difference that multi-hadron FFs describe a fixed number of identified final-state hadrons (i.e.~two in the case of dihadron FFs), whereas GFFs allow for a variable number of final-state hadrons in the subset $\SUB$.

With these GFFs in hand, we can calculate the cross section differential in the fractal observable $x$ for an inclusive jet sample with radius parameter $ R \ll 1$. Letting $z_J$ be the fraction of the center-of-mass energy carried by the measured jet ($z_J \equiv 2 E_{\rm jet}/E_{\rm cm}$), we have
\begin{align}
\label{eq:cross_section}
& \frac{1}{\sigma^{(0)}}\frac{\text{d}\sigma}{\text{d}z_J\, \text{d}x}(e^+ e^- \to {\rm jet} + X) =  \sum_i \int\! \frac{\text{d} y'}{y'}\, C_i(z_J/y',E_{\rm cm},\mu) 
 \\ & \quad
\times \Big\{ \delta(1-y')\, \GFF_i(x,\mu) 
+  \sum_j \mathcal{J}^{(1)}_{i\to j} (y', E_{\rm jet} R,\mu)\, \GFF_j(x,\mu) 
 \nonumber \\ & \quad
+ \delta(1\!-\!y')\, \frac{1}{2}\sum_{j,k} \int \text{d}z\, \text{d}x_1\, \text{d}x_2\, \mathcal{J}^{(1)}_{i\rightarrow jk}(z, E_{\rm jet} R,\mu)\, \GFF_j (x_1,\mu)\, \GFF_k (x_2,\mu)\, \delta[x\!-\!\hat{x}(z,x_1,x_2)]
\nonumber \\ & \quad
+ \frac{1}{2}\sum_{j,k} \int \text{d}z\, \text{d}x_1\, \text{d}x_2\, \mathcal{J}^{(2)}_{i\rightarrow jk}(y',z, E_{\rm jet} R,\mu)\, \GFF_j (x_1,\mu)\, \GFF_k (x_2,\mu)\, \delta[x-\hat{x}(z,x_1,x_2)]
+ \dots \Big\},
\nonumber 
\end{align}
where the ellipsis includes further terms at next-to-next-to leading order and $\sigma^{(0)}$ denotes the tree-level cross section. 
There is a similar version of \Eq{eq:cross_section} for $pp$ and $ep$ collisions with the inclusion of PDFs, where the jet rapidity would appear in the $C_i$ coefficients.
As in \Eq{eq:eecs}, the effects of the hard interaction producing a parton $i$ 
are encoded in the coefficients $C_i$, which can be expanded perturbatively and depend on $z_J$ and $E_{\rm cm}$.  At leading order, the jet only consists of parton $i$, thus $C_i^{(0)}(z_J) = \delta(1-z_J)$ and the dependence on the fractal observable $x$ arising from parton production and hadronization is described simply by $\GFF_i$.  For most of the paper, we restrict ourselves to leading order, though we stress that \Eq{eq:cross_section} provides the tools to interface our GFF formalism with fixed-order calculations and to extract GFFs beyond leading order. 

At next-to-leading order in \Eq{eq:cross_section}, the parton $i$ can undergo a perturbative splitting into partons $j$ and $k$. If only $j$ is inside the jet then $z_J<1$, as described by the perturbative coefficient $\mathcal{J}_{i \to j}^{(1)}$ that can be derived from ref.~\cite{Kang:2016mcy}, and the $x$-dependence is described by $\GFF_j$. If both partons belong to the jet then again $z_J=1$, but the observable $x$ now follows from combining the values $x_1$ and $x_2$ of the GFFs for partons $j$ and $k$ with the momentum fraction $z$ of the perturbative splitting described by the $\mathcal{J}_{i \to jk}^{(1)}$ from ref.~\cite{Waalewijn:2012sv}. 
At next-to-next-to-leading order, there are even more contributions, including one with three partons in the jet involving $\mathcal{J}_{i \to jk\ell}^{(2)}$. In \Eq{eq:cross_section}, we displayed only the term with two partons belonging to the jet, since it is the first term that directly correlates $z_J$ and $z$. The natural scale of the coefficients $\mathcal{J}_{i \to j}, \mathcal{J}_{i \to jk}, \dots,$ is the typical jet invariant mass $E_{\rm jet} R$, so we conclude that the GFFs should be evaluated at
$\mu \simeq E_{\rm jet} R$ to minimize the effect of higher-order corrections.  If $R\gtrsim 1$, then $C_i$ and $\mathcal{J}$ can be combined, and the natural scale to evaluate the GFF would be $\mu \simeq E_{\rm jet}$.

It is important to note that \Eq{eq:cross_section} really combines two different formalisms.  The first is the formalism for GFFs discussed initially in \Refs{Waalewijn:2012sv,Chang:2013rca} for track-based observables and further developed here.  The second is the formalism for fragmentation in inclusive jet production of \Refs{Dai:2016hzf,Kang:2016ehg}, which builds upon work on fragmentation in exclusive jet samples~\Refs{Procura:2009vm,Jain:2011xz,Liu:2010ng,Jain:2011iu}.  Both of these formalisms are needed to perform higher-order jet calculations, though at leading order, the GFF formalism alone suffices.  For the interested reader, we provide all details of the matching for $e^+e^- \to {\rm jet}+ X$ at next-to-leading order in \App{app:matching}.  As in \Refs{Waalewijn:2012sv,Chang:2013rca}, we expect that the absorption of collinear divergences by GFFs can be carried out order-by-order in $\alpha_s$ due to the universality of the collinear limits in QCD.

\subsection{Introducing Fractal Observables}
\label{sec:fractal_preamble}

The above generalized fragmentation formalism works for any collinear-unsafe (but soft-safe) observable.
The RG evolution for a generic $\GFF_i(x,\mu)$, however, can be very complicated.
In order to deal with numerically tractable evolution equations, we focus on observables whose RG evolution simplifies to a nonlinear version of \Eq{eq:linear_as_double_integral}.
Specifically, we want to find the most general form of the function $\tilde{x}(p^-,\SUB)$ in \Eqs{eq:GFFopQ}{eq:GFFopG} such that the RG evolution of $\GFF_i(x,\mu)$ depends only on itself and other GFFs for the same observable, and does not mix with other functions.
An example of an observable that involves GFF mixing is given in \App{app:sums_wef}, where the evolution equation is considerably more complicated than considered below.

We define fractal observables as those whose GFFs obey the (leading-order) RG equation in \Eq{eq:evolution}, repeated here for convenience: 
\begin{equation}
\label{eq:evolution_repeated}
\mu \frac{\text{d}}{\text{d}\mu} \GFF_i (x,\mu) = \frac{1}{2}\sum_{j,k} \int \text{d}z\, \text{d}x_1\, \text{d}x_2\, \frac{\alpha_s(\mu)}{\pi} P_{i\rightarrow jk}(z)\, \GFF_j (x_1,\mu)\, \GFF_k (x_2,\mu)\, \delta[x-\hat{x}(z,x_1,x_2)],
\end{equation}
where $\hat{x}(z,x_1,x_2)$ is a function related to $\tilde{x}(p^-,\SUB)$, which now depends on the momentum $p$ only through the momentum sharing $z$.  
As advertised, the evolution of $\GFF_i (x,\mu)$ depends only on GFFs for the same observable $x$, and no other nonperturbative functions.
We leave a detailed discussion of higher-order evolution to future work, and focus primarily on the leading-order evolution here.
As a consistency check, the $\delta$ function in \Eq{eq:evolution_repeated} ensures that the RG evolution automatically preserves the GFF normalization,
\be
\mu \frac{\text{d}}{\text{d}\mu} \int \text{d}x\, \GFF_i (x,\mu) = \frac{1}{2}\sum_{j,k} \int\text{d}z\, \frac{\alpha_s(\mu)}{\pi} P_{i\rightarrow jk}(z) \int \text{d}x_1\, \GFF_j (x_1,\mu) \int \text{d}x_2\, \GFF_k (x_2,\mu) = 0,
\ee
where we used the fact that $\sum_{j,k} \int \text{d}z \, P_{i\rightarrow jk}(z) = 0$.

As a simple example of a fractal observable, consider the momentum fraction $x$ carried by a subset $\SUB$ of hadrons of a common type.
This case has already been studied in the context of track functions \cite{Chang:2013rca,Chang:2013iba}, where $\SUB$ corresponded to charged particles.
Treating the states $\vert \SUB X \rangle$ in \Eqs{eq:GFFopQ}{eq:GFFopG} partonically, the next-to-leading-order bare GFF in dimensional regularization with $d=4-2 \epsilon$ satisfies 
\begin{align}
\label{eq:GFFNLO}
\GFF_i^{(1)} (x) &=
 \frac{1}{2}\sum_{j,k} \int \text{d}z\, \frac{\alpha_s(\mu)}{2\pi} \left(\frac{1}{\epsilon_{\rm UV}} - \frac{1}{\epsilon_{\rm IR}}\right) P_{i\rightarrow jk}(z) \nonumber \\ & \quad
\times \int \text{d}x_1\, \text{d}x_2\, \GFF_j^{(0)}(x_1,\mu)\, \GFF_k^{(0)}(x_2,\mu)\, \delta[x-\hat{x}(z,x_1,x_2)].
\end{align}
Here, the function $\hat{x}(z,x_1,x_2)$ is the form of $\tilde{x}(p^-,\SUB)$ written in terms of two subjets,
\be
\hat{x}(z,x_1,x_2) = z \, x_1 + (1-z) \, x_2,
\ee
where $x_1$ and $x_2$ are the momentum fractions carried by particles belonging to subjets 1 and 2 within $\SUB$, and $z$ is the momentum fraction carried by subjet 1, as defined in \Eq{eq:z_def}.
Renormalizing the UV divergences in \Eq{eq:GFFNLO} in the $\overline{\rm MS}$ scheme leads directly to the RG equation in \Eq{eq:evolution_repeated}.
Thus, the momentum fraction $x$ carried by the final-state subset $\SUB$ is indeed a fractal observable.

In the above analysis, we implicitly assumed massless partons, since otherwise the parton mass $m$ would regulate the $1/\epsilon_{\rm IR}$ divergence.  As long as $m \ll E_{\rm jet} R$, it is consistent to take the $m \to 0$ limit, which resums the large logarithms of $E_{\rm jet} R/m$ in the cross section for the fractal observable.  At the scale $\mu = m$, one has to match the GFF evolution onto the appropriate heavy-quark description.

\section{Fractal Observables via Clustering Trees}
\label{sec:fractal}

\begin{figure}[t]
\centering
\begin{tikzpicture}
\node[circle,minimum size=1cm,draw] at (4,3) (x) {$x$};
\node[circle,minimum size=1cm,draw] at (0.95,0.3) (x12) {$x_{12}$};
\node[circle,minimum size=1cm,draw] at (7.05,0.3) (x34) {$x_{34}$};
\draw[->,ultra thick] (x12) -- (x)
node[draw=none,fill=none,midway,sloped,anchor=center,above] {$p_1+p_2$};
\draw[->, ultra thick] (x34) -- (x)
node[draw=none,fill=none,midway,sloped,anchor=center,above] {$p_3+p_4$};
\node[diamond,draw,inner sep=3pt,aspect=0.75] at (-0.7,-1.5) (w1) {$w_1$};
\node[diamond,draw,inner sep=3pt,aspect=0.75] at (2.6,-1.5) (w2) {$w_2$};
\node[diamond,draw,inner sep=3pt,aspect=0.75] at (5.4,-1.5) (w3) {$w_3$};
\node[diamond,draw,inner sep=3pt,aspect=0.75] at (8.7,-1.5) (w4) {$w_4$};
\draw[->,ultra thick] (w1) -- (x12) node[draw=none,fill=none,midway,sloped,anchor=center,above] {$p_1$};
\draw[->,ultra thick] (w2) -- (x12) node[draw=none,fill=none,midway,sloped,anchor=center,above] {$p_2$};
\draw[->,ultra thick] (w3) -- (x34) node[draw=none,fill=none,midway,sloped,anchor=center,above] {$p_3$};
\draw[->,ultra thick] (w4) -- (x34) node[draw=none,fill=none,midway,sloped,anchor=center,above] {$p_4$};
\end{tikzpicture}
\caption{Tree structure for fractal observables.  Each leaf node has a starting weight $w_a$.  Each edge has a momentum value $p_i$, which is used to calculate the  momentum fraction $z$ of the splitting at each non-leaf node.  The observable values at the non-leaf nodes are given by the $\hat{x}(z,x_1,x_2)$ recursion relation.  The final value of the observable measured on the tree as a whole is the value obtained at the root node.}
\label{fig:treeDiagram} 
\end{figure}
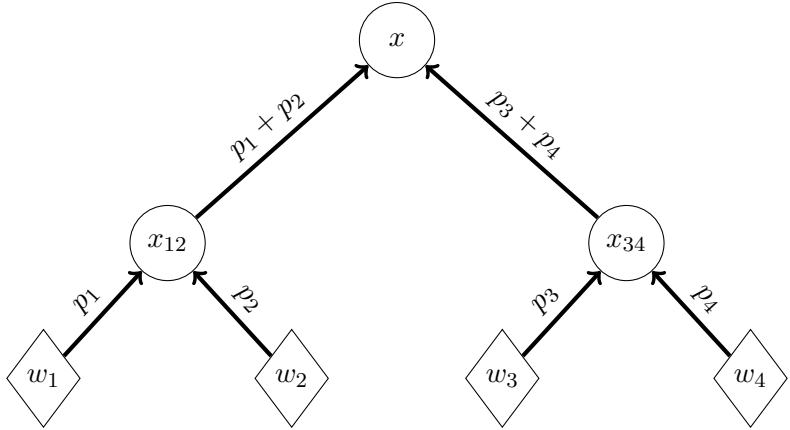

We now present a straightforward way to build a broad class of fractal observables that have the desired RG evolution in \Eq{eq:evolution_repeated}.
The idea is to use recursive clustering trees that mimic the structure of the leading-order RG evolution equations.
Our construction is based on the following three ingredients, as shown in \Fig{fig:treeDiagram}:
\begin{enumerate}
\item Weights $w_a$ for each final-state hadron; 
\item An IRC-safe binary clustering tree;
\item The recursion relation $\hat{x}(z,x_1,x_2)$.
\end{enumerate}
By implementing the function $\hat{x}$ directly on recursive clustering trees, the resulting observable is guaranteed to have fractal structure.

\subsection{Construction}

For this discussion, we start with a collection of hadrons from an identified jet, found using a suitable jet algorithm, e.g.~anti-$k_t$ \cite{Cacciari:2008gp} in the studies below.
As the initial boundary condition for the observable, each final-state hadron within the jet is assigned a weight $w_a$ (possibly zero) based on some non-kinematic quantum number associated with that hadron.
This weight controls how much each type of hadron contributes to the value of the jet observable.
For example, to construct an observable that only depends on the charged particles in the jet, all charged particles would be given weight 1 and all neutral particles weight 0.
It is crucial that $w_a$ is independent of the energy and direction of the hadron, otherwise the NLO  GFF would not take the form in \Eq{eq:GFFNLO}.

These final-state hadrons are then used as inputs to an IRC-safe binary clustering tree, which is in general different from any clustering algorithm used to determine the identified jet. 
For our studies, we use the generalized-$k_t$ family of jet clustering algorithms \cite{Cacciari:2008gp}, which are designed to follow the leading-order structure of the parton shower.
In the context of $e^+e^-$ collisions, these algorithms have the pairwise clustering metric 
\begin{align}
\label{eq:metric}
d_{ij} = \min[E_{i}^{2p},E_{j}^{2p}] \, \Omega_{ij}^2,
\end{align}
where the exponent $p$ parametrizes the tree-dependence of the observable, with $p = \{-1,0,1\}$ corresponding to the $\{\text{anti-$k_t$~\cite{Cacciari:2008gp}}, \text{C/A~\cite{Dokshitzer:1997in,Wobisch:1998wt}}, \text{$k_t$~\cite{Catani:1993hr,Ellis:1993tq}}\}$ clustering algorithms, and $\Omega_{ij}^2$ is a measure of the angular separation between two constituent's momenta scaled by the jet radius parameter $R$.\footnote{Since we start with the constituents of an identified jet, all of the particles are (re)clustered into a single tree.  For this reason, the single-particle distance measure and the jet radius parameter $R$ in the (re)clustering algorithm are irrelevant.}
For any value of $p$, generalized-$k_t$ provides a pairwise clustering structure that directly mimics \Eq{eq:evolution_repeated}.  For $pp$ collisions, one insteads use a form of \Eq{eq:metric} based on transverse momenta $p_T$ and distance $\Delta R_{ij}$ in azimuthal angle and rapidity.

From this clustering tree, one can determine the observable $x$ by applying the recursion relation $\hat{x}(z,x_1,x_2)$ at each stage of the clustering.
Specifically, the value of $x$ at each node depends on the momentum fraction $z$ given by the $2 \to 1$ merging kinematics as well as on the $x_1$ and $x_2$ values determined from the corresponding daughter nodes (which might be the initial weights $w_a$).
When all nodes are contained in a single connected tree, the root node represents the entire jet, and the root value of $x$ determines the final observable.

Even though the clustering tree is IRC safe, the resulting fractal observable $x$ is generally collinear unsafe.  These collinear divergences are absorbed into the GFFs, and are in fact responsible for the evolution in \Eq{eq:evolution}.

\subsection{Requirements}
\label{subsec:requirements}

There are a few fundamental limitations on the choice of $\hat{x}(z,x_1,x_2)$ dictated by the fact that this same function will appear in \Eq{eq:evolution_repeated}. 
First, the recursion relation must be symmetric under the exchange $z\leftrightarrow 1-z$, $x_1\leftrightarrow x_2$, since the assignment of these labels is unphysical.\footnote{In the case of jets with heavy flavor, one could use heavy-flavor tags to define asymmetric recursion relations (see e.g.~\cite{Ilten:2017rbd}).  We do not give a separate treatment of heavy-flavor GFFs in this work, and instead assume to always work in the $m_{b,c} \ll E_{\rm jet} R$ limit.} 
Second, the recursion relation has to be IR safe, since the GFF formalism only regulates collinear (and not soft) divergences.
In order that an emission with $z \to 0$ does not change the observable, IR safety translates into the conditions
\begin{equation}
\label{eq:IRsafety}
\lim_{z\rightarrow 1} \hat{x}(z,x_1,x_2) = x_1, \qquad \lim_{z\rightarrow 0} \hat{x}(z,x_1,x_2) = x_2,
\end{equation}
such that an arbitrarily soft branch in the clustering tree has no impact on the values of $x$.
Third, the recursion relation has to have unambiguous limits.
As a counterexample, $\hat{x}(z,x_1,x_2) = x_1^z x_2^{1-z}$ satisfies \Eq{eq:IRsafety} when $x_1$ and $x_2$ are non-zero, but not when they vanish.
Apart from these limitations, any choice of $\hat{x}(z,x_1,x_2)$ (along with starting weights and a clustering tree) defines a fractal observable.

The tree traversal prescription, along with the requirement in \Eq{eq:IRsafety}, helps ensure IR safety to all $\alpha_s$ orders.  As a counterexample, consider the sum over all tree nodes of some function $f(z)$ which vanishes as $z\rightarrow 0$ or $z\rightarrow 1$.  In that case, the resulting observable would receive no contribution from a single infinitely soft splitting, but subsequent finite $z$ splittings that followed the soft one would not be suppressed, violating IR safety.   By contrast, \Eq{eq:IRsafety} requires the contribution from an entire soft branch to be suppressed, as desired.

In this paper, we mainly focus on recursion relations that do not depend explicitly on the opening angle $\theta$ between branches in the clustering tree.
In \Sec{sec:angular}, we do discuss how the recursion relation gets modified if a threshold value for $\theta$ is introduced (i.e.~$\theta_{\rm thr}=R_{\rm sub} \ll R$).
Of course, fractal observables depend indirectly on angular information through the structure of the clustering tree, but as discussed below, the leading-order evolution equations do not depend on the clustering algorithm.
When explicit $\theta$-dependence is included in the $\hat{x}$ function, this sometimes results in a fully IRC-safe observable, requiring a different type of evolution equation that is beyond the scope of the present work (see e.g.~\cite{HarvardInProgress}).

\subsection{Evolution Equations}

The generalized-$k_t$ clustering tree has an obvious mapping to a parton branching tree, such that at order $\alpha_s$, the RG evolution is given precisely by \Eq{eq:evolution_repeated}, with the flavor of the GFF matching the flavor of the jet's initiating parton.
More formally, as discussed in \Sec{sec:fractal_preamble}, the NLO calculation of the bare GFF shows that the same recursion relation $\hat{x}(z,x_1,x_2)$ appears in \Eq{eq:GFFNLO}, as desired.

In fact, to order $\alpha_s$, the evolution in \Eq{eq:evolution_repeated} is insensitive to the clustering tree, as long as it is IRC safe, even if the fractal observable itself depends on the clustering order. We explicitly test this surprising feature in \Sec{sec:NA}.
Note that if the clustering tree is not collinear safe, in the sense that particles with collinear momenta are not clustered with each other first, then the collinear divergences in the GFF will not cancel against the collinear divergences in the hard matching coefficients of \Eq{eq:cross_section}.
If the clustering tree is not IR safe, then the observable $x$ is not IR safe, and the GFF formalism does not apply.

We stress that the evolution in \Eq{eq:evolution_repeated} is only valid to lowest order in $\alpha_s$.
At higher orders in $\alpha_s$, the evolution of fractal observables is more complicated, but, as discussed more in the paragraph below, still satisfies the property that the evolution of $\GFF_i (x,\mu)$ depends only on GFFs of the same observable. 
Schematically, this can be written as
\be
\label{eq:higherorder}
\mu \frac{\text{d}}{\text{d}\mu} \GFF_i = \frac{\alpha_s}{\pi} P_{i\rightarrow jk} \otimes \GFF_j  \otimes \GFF_k + \left(\frac{\alpha_s}{\pi}\right)^2 P_{i\rightarrow jk \ell} \otimes \GFF_j  \otimes \GFF_k \otimes \GFF_\ell + \ldots,
\ee
where $\otimes$ represents a convolution.
This equation includes $1\rightarrow n$ splittings at order $\al_s^{n-1}$.
There is no longer a one-to-one correspondence between pairwise clustering trees and GFF evolution trees, and one has to explicitly carry out the calculation in \Eq{eq:GFFNLO} to higher orders to determine the evolution.
In particular, there will be different clusterings of the $1 \to n$ splitting into a binary tree when integrating over phase space, which depend on the choice of clustering algorithm.
Because our specific realization of fractal observables in this section is based on recursive clustering trees, this guarantees that \Eq{eq:higherorder} depends only on GFFs of the same type as $\GFF_i$ at all perturbative orders.

To justify the structure of \Eq{eq:higherorder} in a bit more detail, it is instructive to take a closer look at the $1/\epsilon_{\rm UV}$ poles of $\GFF_i$.
As usual, the anomalous dimension of the GFFs is determined by the single $1/\epsilon_{\rm UV}$ poles.
At order $\al_s$, we get $(1/\epsilon_{\rm UV}) P_{i\to jk}$, as shown in \Eq{eq:GFFNLO}.
At order $\al_s^2$, the $1 \to 3$ splitting factorizes into a sequence of two $1\to 2$ splittings when the angles of the splittings are strongly ordered.
This leads to a term like $(1/\epsilon_{\rm UV}^2) P_{i \to jk} \otimes P_{j \to \ell m}$ which does not contribute to the GFF's anomalous dimension.
However, it does justify attaching $\GFF_j$ and $\GFF_k$ to the external splittings in \Eq{eq:evolution_repeated}, as it corresponds to the cross term between a one-loop renormalization factor and one-loop $\GFF_j$ (and tree-level $\GFF_k$).
Away from the strongly-ordered limit, the $1 \to 3$ splitting does have a genuine $1/\epsilon_{\rm UV}$ divergence, contributing to the second term in \Eq{eq:higherorder}.
The precise structure of this term depends on how the clustering algorithm maps the three partons to a binary tree.
The justification for attaching GFFs to each of the three external partons follows again by considering higher-order corrections with some strong ordering.
For example, consider a $1 \to 5$ splitting that is strongly ordered such that it factorizes in a $1 \to 3$ splitting, in which two partons undergo $1 \to 2$ splittings.
Such a term would have a $1/\epsilon_{\rm UV}^3$ divergence, corresponding to the cross term of the renormalization factor for the $1 \to 3$ splitting term at order $\al_s^2$ with two one-loop $\GFF$'s and one tree-level $\GFF$.
Finally, the $1/\epsilon_{\rm UV}$ from the one-loop virtual contribution to the $1 \to 2$ splitting gives a higher-order correction to the first term in \Eq{eq:evolution_repeated}.
For the remainder of this paper, we focus on the leading-order evolution, leaving an analysis at higher orders to future work.

\section{Weighted Energy Fractions}
\label{sec:WEF}

The procedure outlined in \Sec{sec:fractal} is very general, but for special choices of $\hat{x}(z,x_1,x_2)$, the definition of a fractal observable can simplify greatly.
In this section, we consider the recursion relation
\begin{equation}
\label{eq:wefrecur}
\hat{x}(z,x_1,x_2) = x_1 \, z^\kappa  + x_2 \, (1-z)^\kappa,
\end{equation}
where $\kappa > 0$ is an energy exponent.
As we will see, for any choice of pairwise clustering tree, the resulting observable simplifies to a sum over the hadrons in a jet, 
\begin{equation}
\label{eq:wefext}
x = \sum_{a \in {\rm jet}} w_a \, z_a^\kappa, \qquad z_a \equiv \frac{E_a}{E_{\rm jet}},
\end{equation}
where $\kappa$ is the same as in \Eq{eq:wefrecur}, and $w_a$ is the hadron weight factor.  We call these observables weighted energy fractions.

Several examples of weighted energy fractions have already been studied in the literature.
The weighted jet charge is defined for any $\kappa > 0$ and weights given by the electric charges of final-state hadrons \cite{FIELD19781,Krohn:2012fg,Waalewijn:2012sv}.
This quantity has, for example, been used in forward-backward asymmetry measurements at $e^+e^-$ experiments~\cite{Stuart:1989db,Decamp:1991se}, as well as to infer the charge of quarks~\cite{Berge:1979qg,Buskulic:1992sq,Abazov:2006vd}. Recently, the scale dependence of the average jet charge was observed in $pp \to$ dijets~\cite{Aad:2015cua}.
Track fractions correspond to the case of $\kappa = 1$, where charged particles are given weight 1 and neutral particles given weight 0 \cite{Chang:2013rca,Chang:2013iba}.
Jet $p_T^D$ is a weighted energy fraction with $\kappa = 2$ and all particles given weight 1~\cite{Chatrchyan:2012sn,Pandolfi:1480598}.
Weighted energy fractions with arbitrary $\kappa > 0$ and $w_a=1$ for all particles were studied in \Ref{Larkoski:2014pca} for applications to quark/gluon discrimination.

\subsection{Associativity}

\begin{figure}[t]
\centering
\begin{tikzpicture}[thick,scale=1.5, cross/.style={path picture={ 
  \draw[black]
(path picture bounding box.south east) -- (path picture bounding box.north west) (path picture bounding box.south west) -- (path picture bounding box.north east);
}}]
\draw[ultra thick] (0,0) -- (0.5,0);
\node at (1.5,0.8) (1) {1};
\node at (1.5,0) (2) {2};
\node at (1.5,-0.8) (3) {3};
\draw[ultra thick] (0.5,0) -- (1);
\draw[ultra thick] (0.5,0) -- (1,-0.4);
\draw[ultra thick] (1,-0.4) -- (2);
\draw[ultra thick] (1,-0.4) -- (3);
\node at (0.5,-1) {(A)};

\node at (4,0.8) (1) {1};
\node at (4,0) (2) {2};
\node at (4,-0.8) (3) {3};
\draw[ultra thick] (2.5,0) -- (2);
\draw[ultra thick] (3,0) -- (3.4,0.4);
\draw[ultra thick] (3.4,0.4) -- (3.5,0.2);
\draw[ultra thick] (3.4,0.4) -- (1);
\draw[ultra thick] (3.7,-0.2) -- (3);
\node at (3,-1) {(B)};

\draw[ultra thick] (5,0) -- (5.5,0);
\node at (6.5,0.8) (1) {1};
\node at (6.5,0) (2) {2};
\node at (6.5,-0.8) (3) {3};
\draw[ultra thick] (5.5,0) -- (3);
\draw[ultra thick] (5.5,0) -- (6,0.4);
\draw[ultra thick] (6,0.4) -- (1);
\draw[ultra thick] (6,0.4) -- (2);
\node at (5.5,-1) {(C)};
\end{tikzpicture}
\caption{The three binary trees which could be constructed by clustering three particles.  For associative observables studied in \Sec{sec:WEF}, the order of the clustering does not affect the final observable.  The ordering of the clustering will matter for the non-associative observables studied in \Sec{sec:NA}.}
\label{fig:trees}
\end{figure}
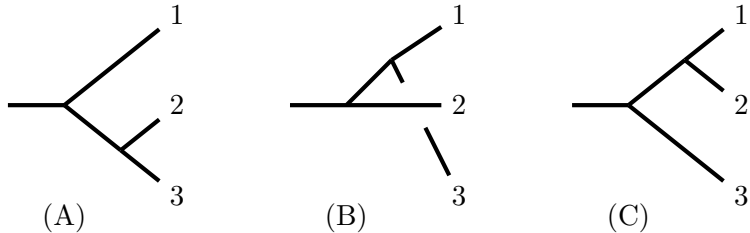

Weighted energy fractions have an associative recursion relation, meaning that the order of the clustering tree does not affect the final observable.
To see this, consider the case of just three particles with weights $\{w_1,w_2,w_3\}$ and respective momentum fractions $\{z_1,z_2,z_3\}$.
As shown in \Fig{fig:trees}, there are three clustering trees that can be built using only $1\rightarrow 2$ splittings, labeled as A, B, and C.\footnote{Of course, for a specific choice of kinematics, not all of these trees will be possible from generalized-$k_t$ clustering, particularly in the collinear limit.}
The corresponding observables are
\begin{align}
x_{\rm A} &= \hat{x}\left(z_1,w_1, \hat{x}\Bigl(\frac{z_2}{z_2 + z_3}, w_2, w_3\Bigr) \right), \nonumber\\ 
x_{\rm B} &= \hat{x}\left(z_2,w_2, \hat{x}\Bigl(\frac{z_3}{z_3 + z_1}, w_3, w_1\Bigr) \right),\nonumber\\
x_{\rm C} &= \hat{x}\left(z_3,w_3, \hat{x}\Bigl(\frac{z_1}{z_1 + z_2}, w_1, w_2\Bigr) \right).\label{eq:ABCtrees}
\end{align}
Using \Eq{eq:wefrecur} and the fact that $z_1 + z_2 + z_3 = 1$, it is straightforward to prove that
\be
x_{\rm A} = x_{\rm B} = x_{\rm C} = w_1 \, z_1^\kappa + w_2 \, z_2^\kappa + w_3 \, z_3^\kappa,
\ee
owing to the fact that the recursion relation has homogenous scaling with $z$.  This argument generalizes to an arbitrary numbers of particles, so the weighted energy fractions are indeed independent of the clustering tree.\footnote{Remember that this tree is one obtained from \emph{reclustering} the particles in the jet.  The value of a jet observable of course depends on the choice of initial jet algorithm, which may itself be a clustering algorithm.}

Of course, there are other observables that have non-associative recursion relations, where the observable does not simplify to a sum over final-state hadrons and the full tree traversal is necessary.  We explore some non-associative observables in \Sec{sec:NA}.

\subsection{Extraction of GFFs}
\label{sec:wefextraction}

In general, to extract GFFs, one has to numerically match the cross section in \Eq{eq:cross_section} using perturbatively calculated values for the coefficients $C_i$, $\mathcal{J}_{i \to j}$, $\mathcal{J}_{i \to jk}$, \dots.
For the parton shower studies in this paper, we limit ourselves to leading order where $C_i^{(0)}(z_J) = \delta(1-z_J)$, and we use parton-shower truth information to assign the parton label $i$.
To generate pure samples of quark- and gluon-initiated jets, we use the $e^+e^- \to \gamma/Z^* \to q \bar{q}$ and $e^+e^- \to H^* \to g g$ processes in \textsc{Pythia} 8.215~\cite{Sjostrand:2014zea}, switching off initial-state radiation.
We find jets using \textsc{FastJet} 3.2.0~\cite{Cacciari:2011ma}, with the $ee$-generalized $k_t$ algorithm with $p=-1$ (i.e.~the $e^+e^-$ version of anti-$k_t$ \cite{Cacciari:2008gp}) and then determine the various weighted energy fractions on the hardest jet in the event.
At leading order, the normalized probability distributions for the weighted energy fractions directly give the corresponding GFF $\GFF_i (x,\mu)$.  

As discussed in \Sec{sec:generalFF}, for jets of a given energy $E_{\rm jet}$ and radius $R$, the characteristic scale for GFFs is expected to be
\be
\mu = E_{\rm jet} R,
\ee
which is roughly the scale of the hardest possible splitting in the jet.
By varying $E_{\rm jet}$ and $R$ but keeping $\mu$ fixed, we can estimate part of the uncertainty in the extraction of the GFFs.
In addition, we assess the uncertainty from using different parton shower models.
Here, since our primary interest is in the perturbative uncertainty in different shower evolution equations, we test the native \textsc{Pythia} parton shower along with the \textsc{Vincia} 2.0.01~\cite{Giele:2007di} and \textsc{Dire} 0.900~\cite{Hoche:2015sya} parton shower plugins.
A further source of uncertainty would be given by the hadronization model, which enters the boundary conditions used for GFF evolution.
This is not included in our present study, since we decided to interface all of the showers above with the Lund string model.
In the context of an experimental analysis, one would also have statistical and systematic uncertainties from the extraction of GFFs from data.

For each observable $x$, there are 11 GFFs, corresponding to 5 quark flavors $\{u,d,s,c,b\}$, 5 anti-quark flavors, and the gluon.
To avoid a proliferation of curves, it is convenient to define singlet (denoted by $\langle \rm Quark \rangle$ in the figures below) and non-singlet combinations for the quark GFFs, respectively,
\begin{align}
\label{eq:colorsinglet}
\mathcal{S}(x,\mu) &= \frac{1}{2n_f}\sum_{i \in \{u,\overline{u},d,\ldots \overline{b}\}} \GFF_{i}(x,\mu),\nonumber \\
\mathcal{N}_{ij}(x,\mu) &= \GFF_{i}(x,\mu) - \GFF_j(x,\mu).
\end{align}
For the observables we study, the anti-quark GFFs are either identical to the quark GFFs or simply involve the replacement $x \to -x$, due to charge conjugation symmetry.
We start by showing numerical results for the gluon GFF and the quark-singlet combination, postponing a discussion of the non-singlet case to \Sec{sec:wefmoments}.

\begin{figure}
\subfloat[]{
		\includegraphics[width=0.45\textwidth,trim=0 0 0 0,clip]{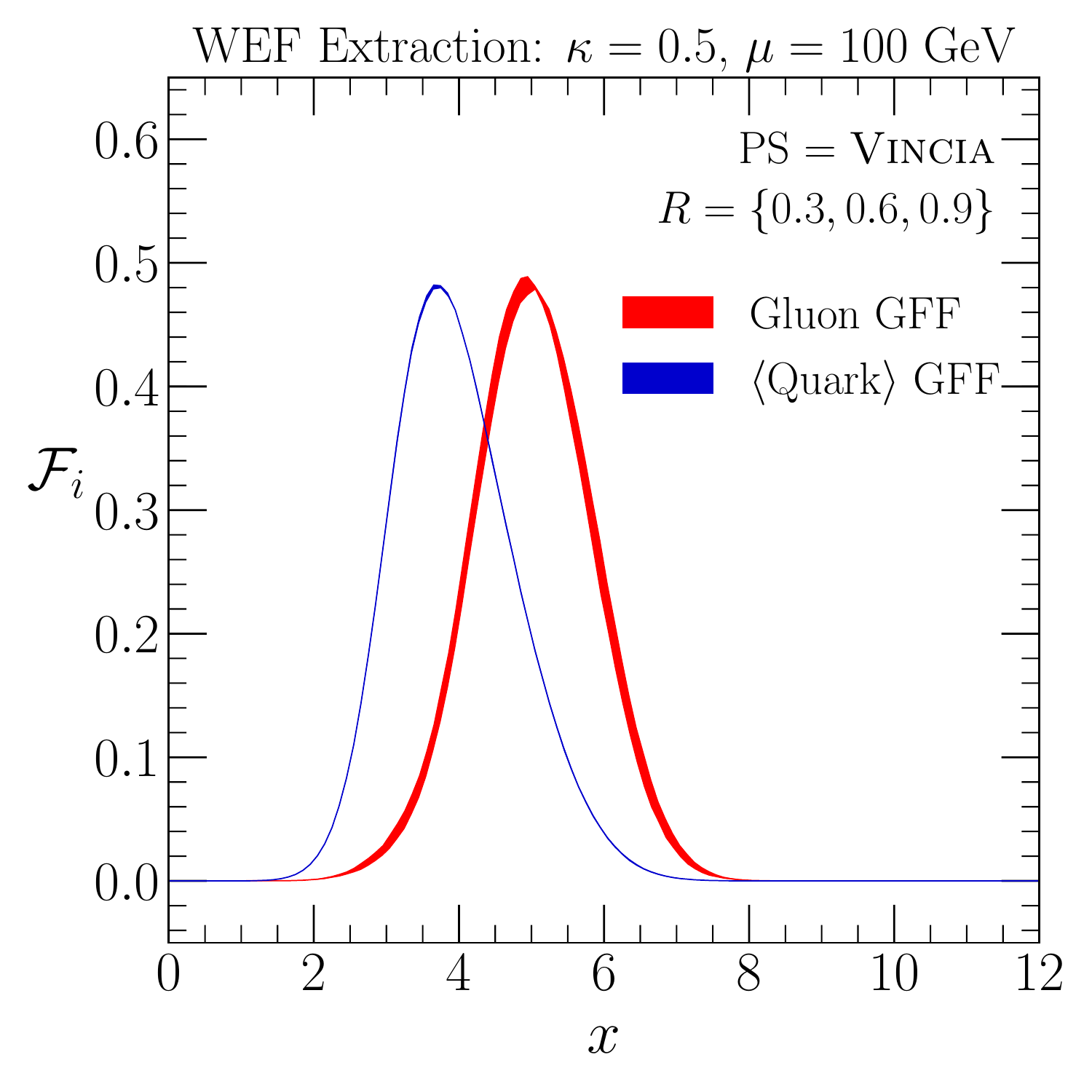}
		 \label{fig:extract-all-a}

}
\subfloat[]{
		\includegraphics[width=0.45\textwidth,trim=0 0 0 0,clip]{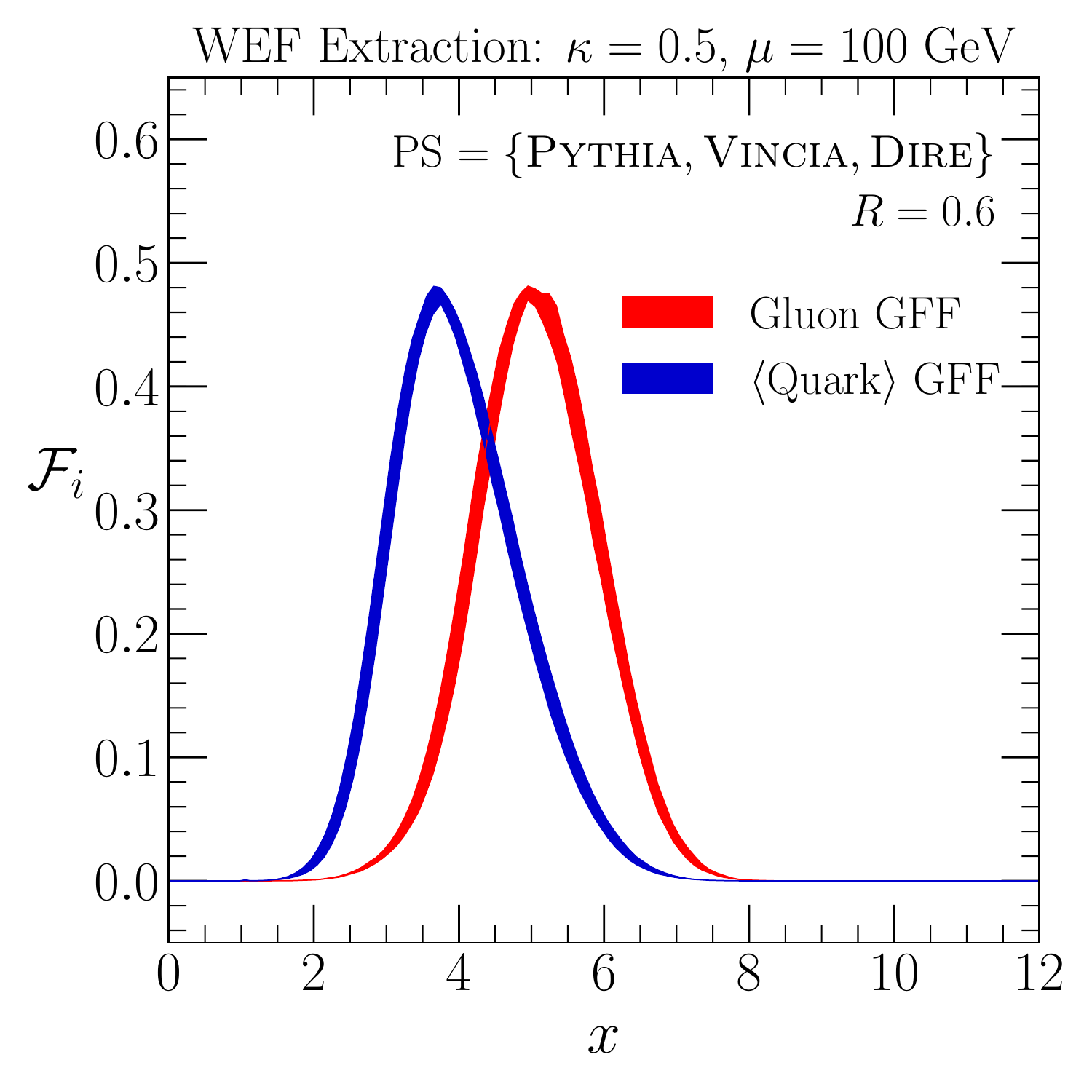}
		 \label{fig:extract-all-b}
}

\subfloat[]{
		\includegraphics[width=0.45\textwidth,trim=0 0 0 0,clip]{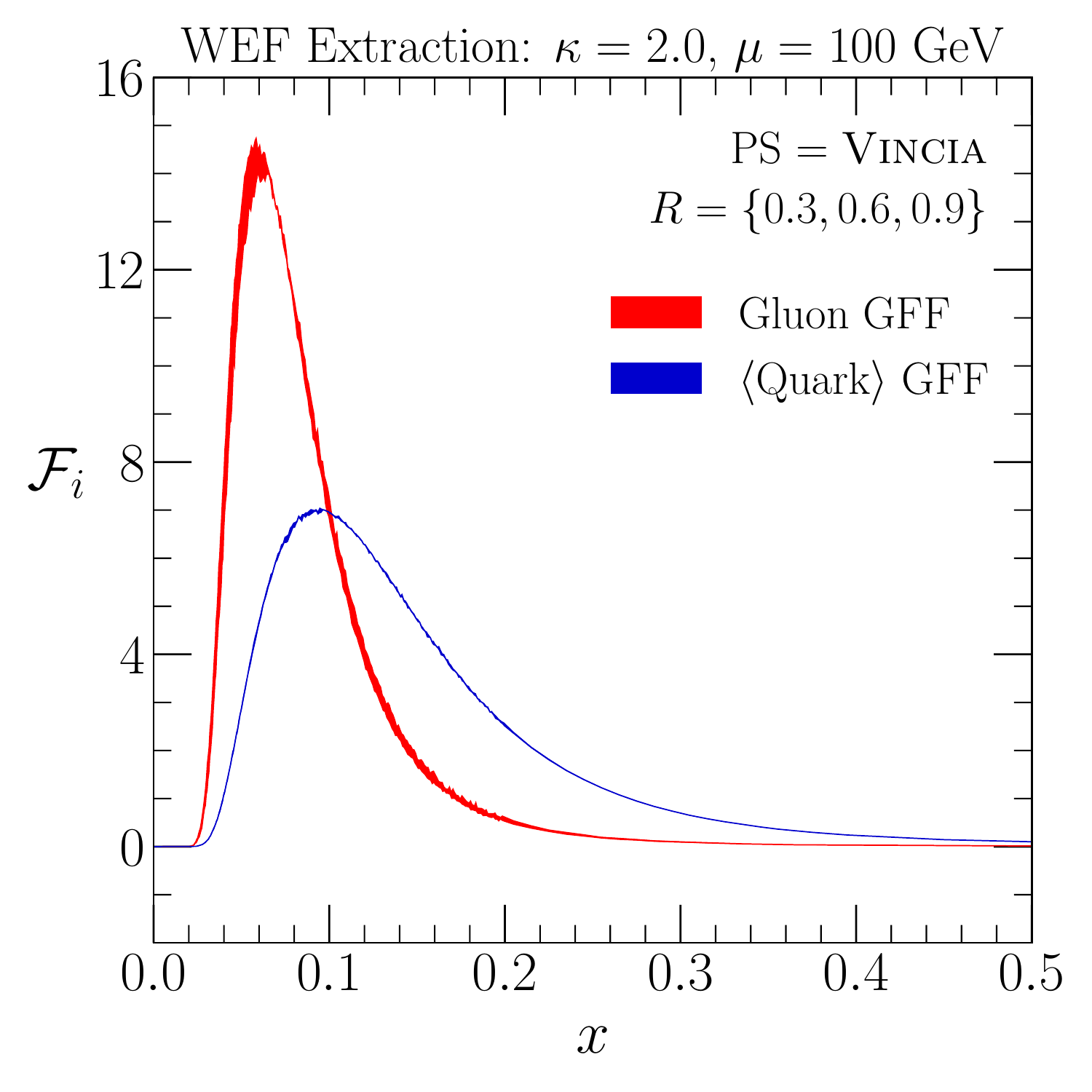}
		 \label{fig:extract-all-c}
}
\subfloat[]{
		\includegraphics[width=0.45\textwidth,trim=0 0 0 0,clip]{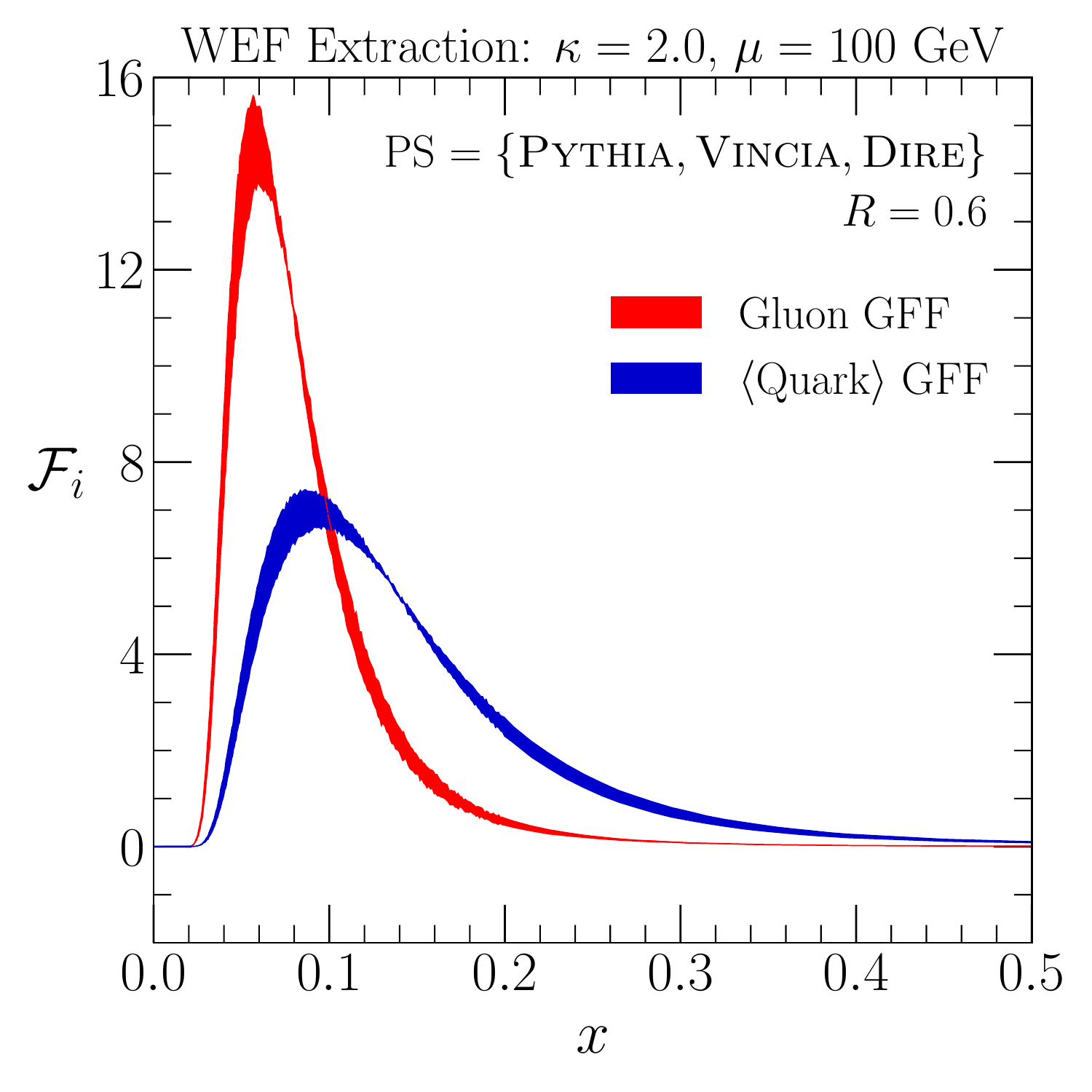}
		 \label{fig:extract-all-d}
}
	\caption{Gluon and quark-singlet GFFs for weighted energy fractions with (top) $\kappa = 0.5$ and (bottom) $\kappa=2$, with all particles given starting weight 1. These distributions were extracted at the scale $\mu = 100$ GeV.  The left column shows results from the \textsc{Vincia} parton shower, with uncertainty bands from varying $R=\{0.3,0.6,0.9\}$ while keeping $\mu$ fixed. The right column shows the fixed jet radius $R=0.6$, with uncertainty bands from testing three different parton showers: \textsc{Pythia}, \textsc{Vincia}, and \textsc{Dire}.  In this and subsequent figures, $\langle \text{Quark} \rangle$ always refers to the quark-singlet combination $\mathcal{S}(x,\mu)$ defined in \Eq{eq:colorsinglet}.}
	\label{fig:extractall} 
\end{figure}

In \Fig{fig:extractall}, we show the extracted gluon and quark-singlet GFFs at $\mu = E_{\rm jet} R = 100~\GeV$ for the weighted energy fractions with $w_a=1$, comparing $\kappa = 0.5$ and $\kappa = 2$. 
Since gluon jets have roughly a factor of $C_A/C_F$ larger hadron multiplicity than quark jets, the mean of the gluon GFF is roughly a factor of $(C_A/C_F)^{1-\kappa}$ higher than the mean of the quark-singlet GFF.
In the left column, we show the impact of changing the jet radius $R = \{0.3,0.6,0.9\}$, leaving $\mu$ fixed. 
The envelope from changing $R$ is very small, indicating that $\mu = E_{\rm jet} R$ is an appropriate definition for the RG scale.
In the right column, we show the impact of switching between the  \textsc{Pythia}, \textsc{Vincia}, and \textsc{Dire} parton shower models.
The envelope is larger, but still reasonably narrow, giving us confidence in the extraction of the GFFs, at least as far as changing the perturbative shower model is concerned.
Though not shown here, we checked that the GFFs for the $\kappa \to 1$ and $\kappa \to \infty$ limits behave sensibly as well (see \Sec{sec:limits} below).

\subsection{Evolution of GFFs}
\label{sec:wefevolution}

We now use these extracted GFFs as boundary conditions for the RG evolution in \Eq{eq:evolution_repeated}.
In \App{app:implementation}, we describe in detail the numeric implementation of the evolution.
Formally, the evolution equations work equally well running up or down in $\mu$, but in practice downward evolution is numerically unstable, as discussed further in \App{app:numerical_stability}.
As a proof of principle for our RG evolution code, we show upward evolution from $\mu=100~\GeV$ to $\mu = 4~\TeV$, comparing our RG evolution in \Eq{eq:evolution_repeated} to that obtained from parton showers.

\begin{figure}
\subfloat[]{
		\includegraphics[width=0.32\textwidth,trim=0 0 0 0,clip]{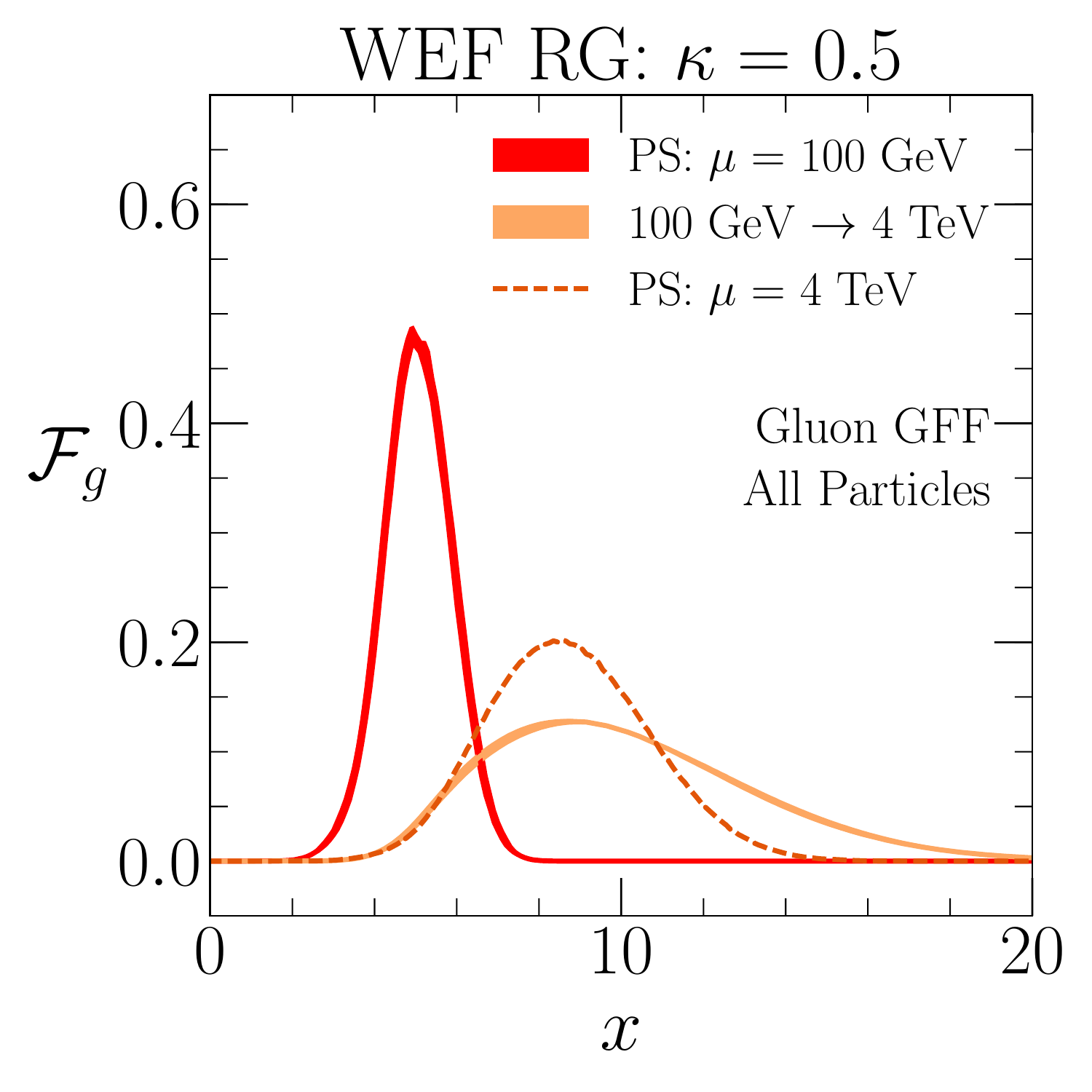}
		\label{fig:wef_gluona}
}
\subfloat[]{
		\includegraphics[width=0.32\textwidth,trim=0 0 0 0,clip]{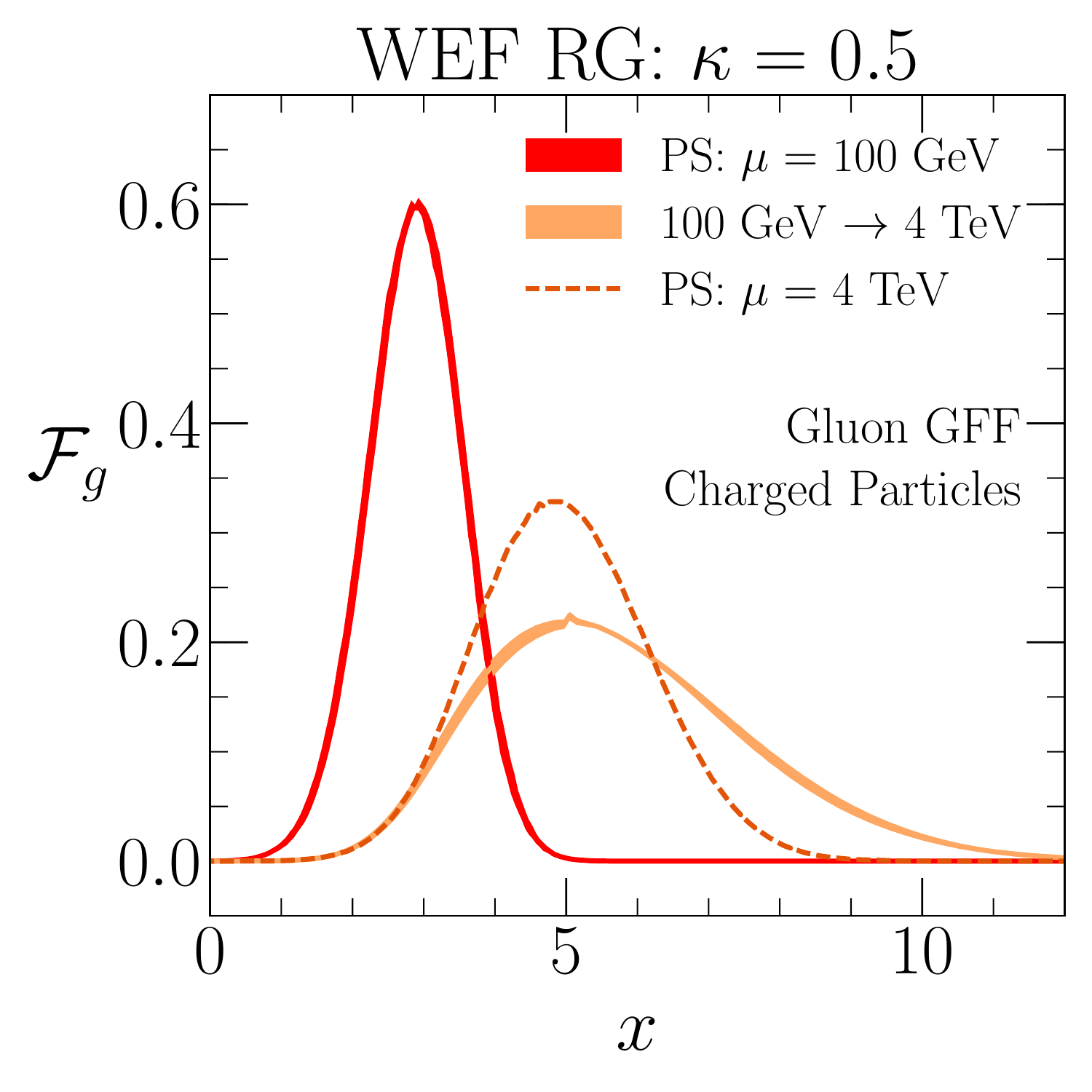}
		\label{fig:wef_gluonb}
}
\subfloat[]{
		\includegraphics[width=0.32\textwidth,trim=0 0 0 0,clip]{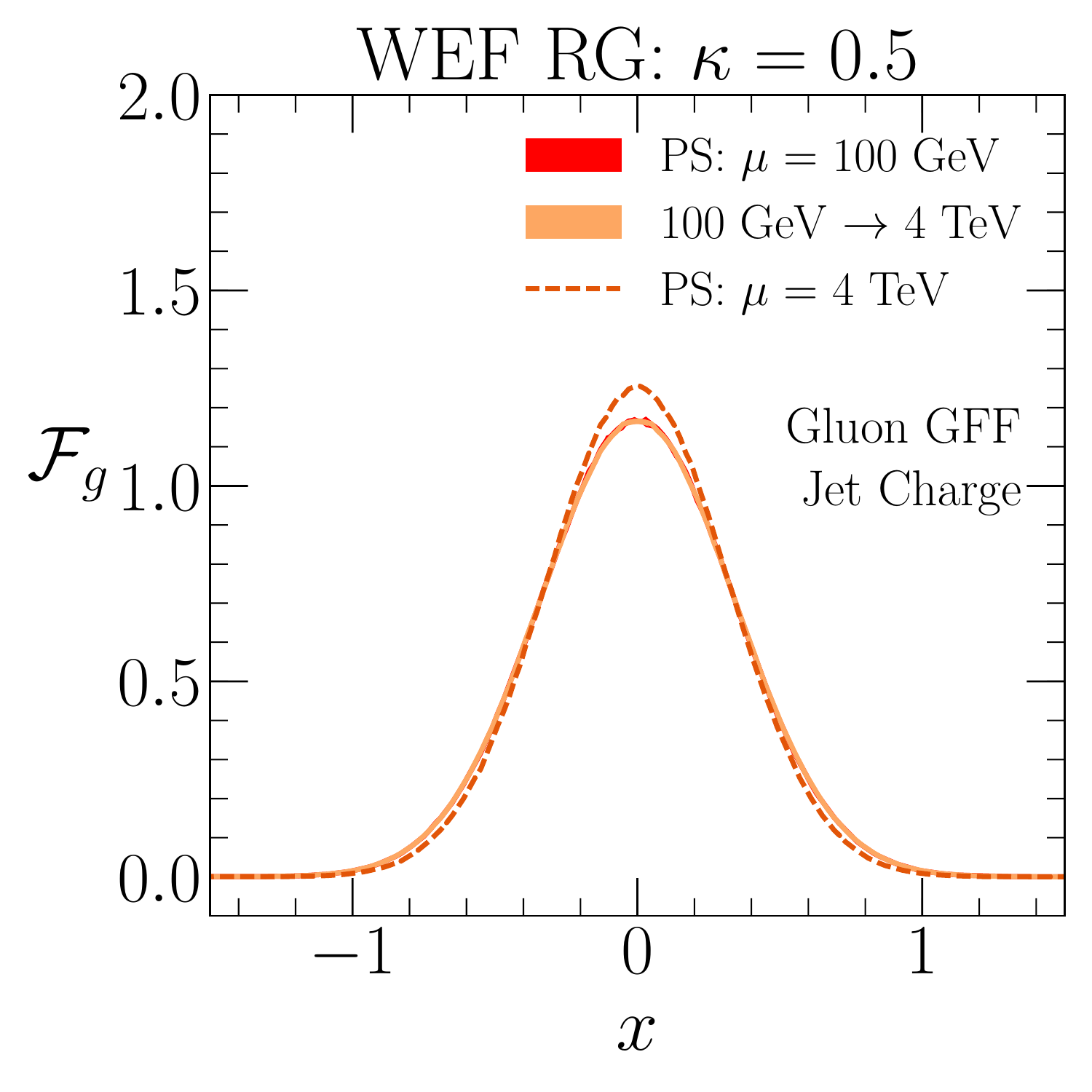}
		\label{fig:wef_gluonc}
}

\subfloat[]{
		\includegraphics[width=0.32\textwidth,trim=0 0 0 0,clip]{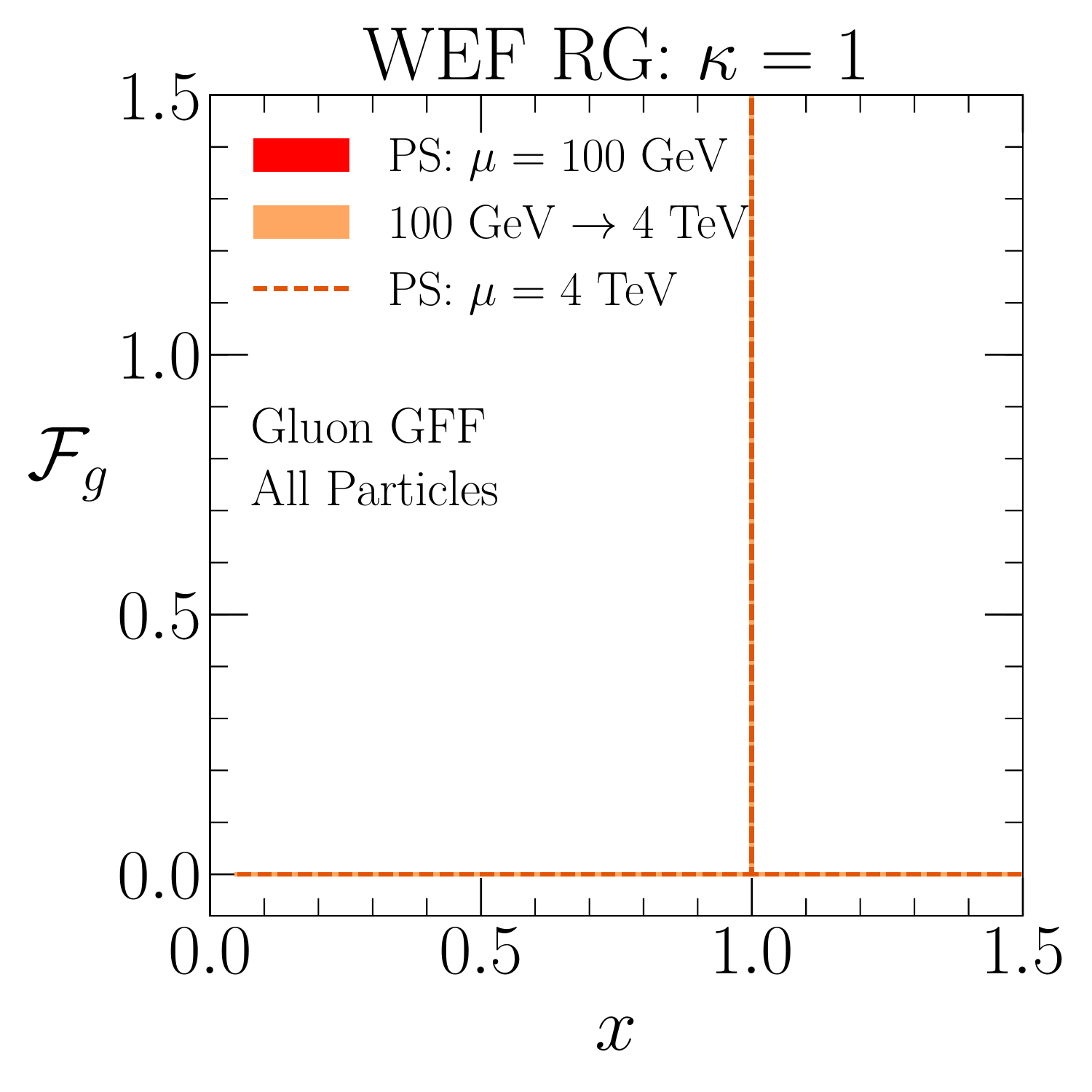}
		\label{fig:wef_gluond}
}
\subfloat[]{
		\includegraphics[width=0.32\textwidth,trim=0 0 0 0,clip]{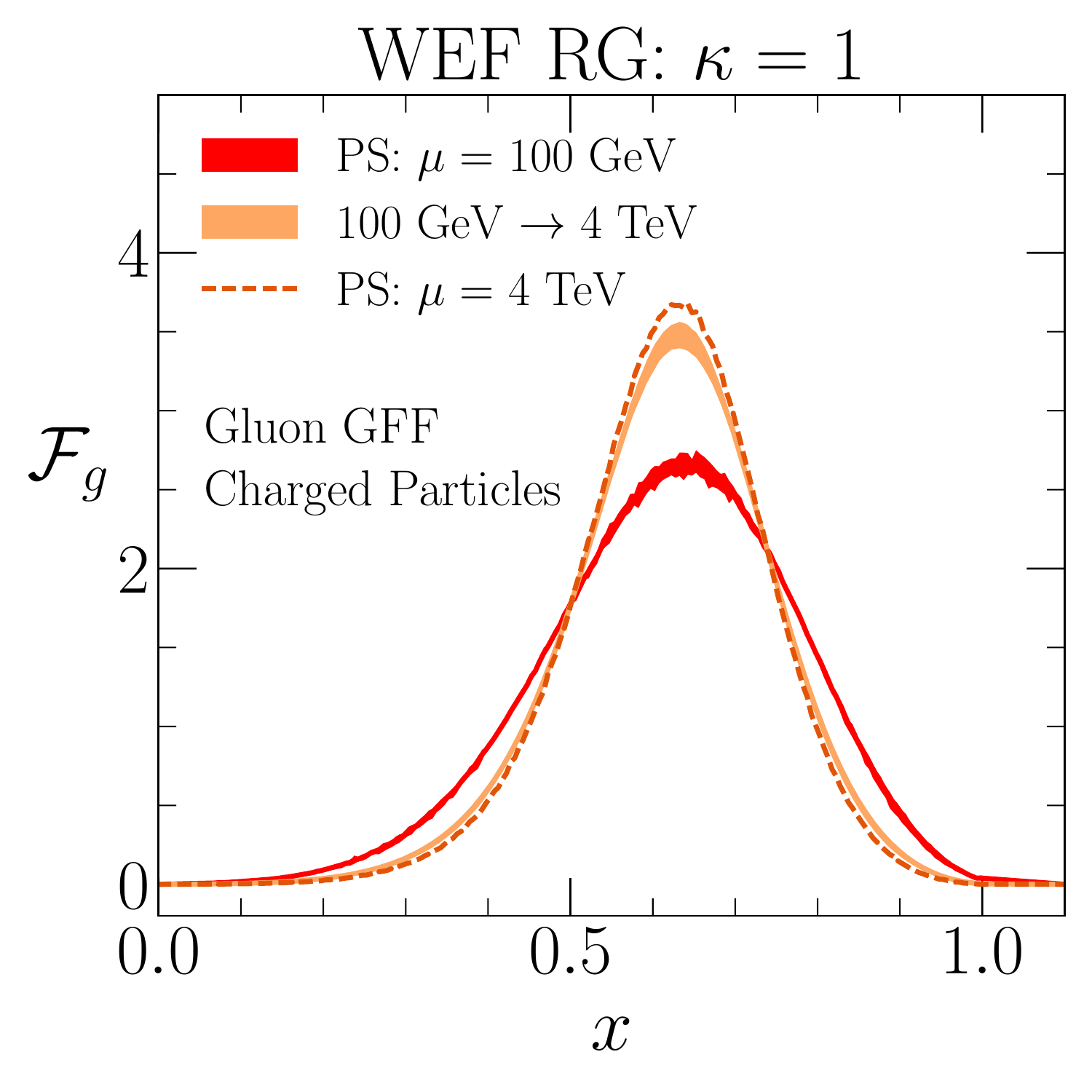}
		\label{fig:wef_gluone}
}
\subfloat[]{
		\includegraphics[width=0.32\textwidth,trim=0 0 0 0,clip]{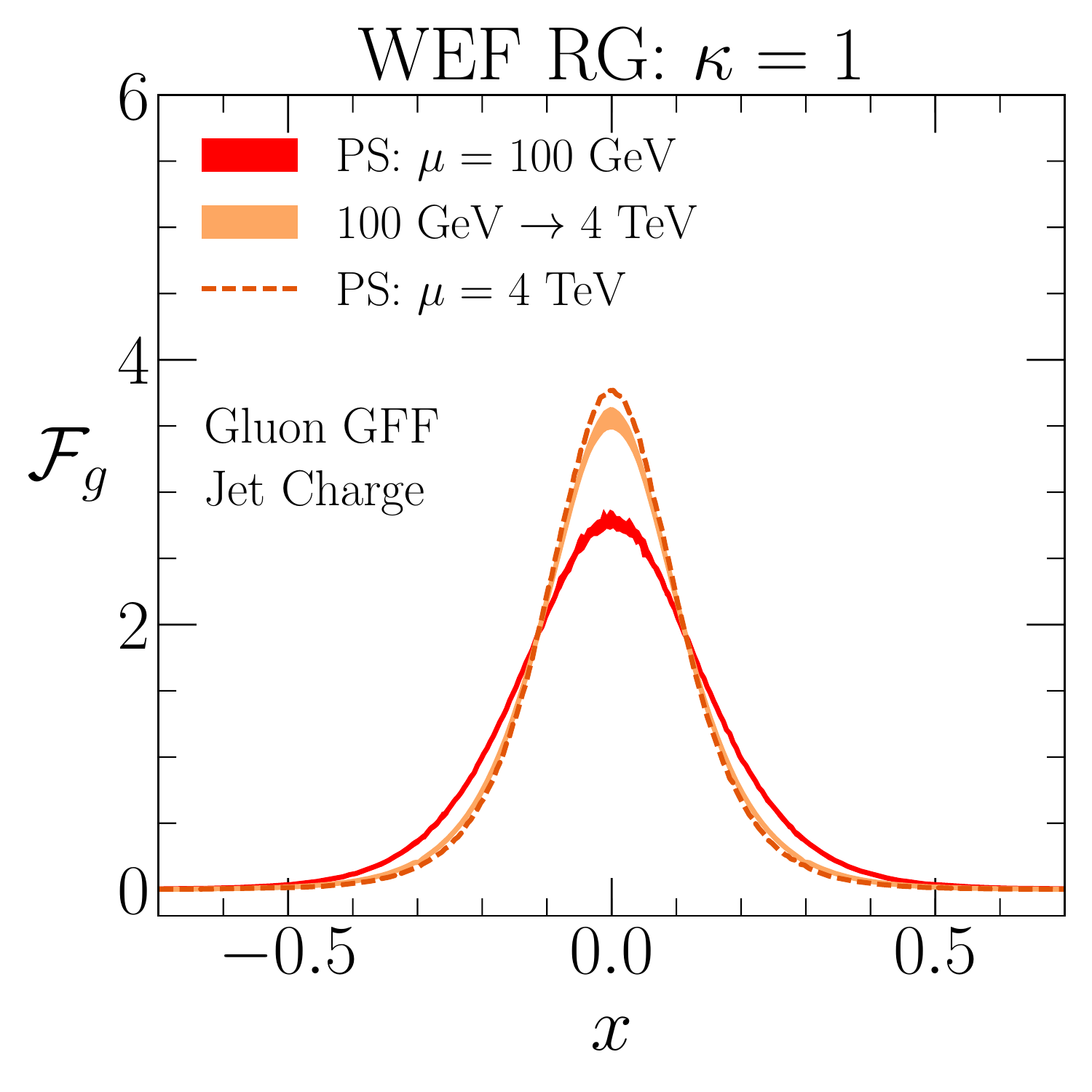}
		\label{fig:wef_gluonf}
}

\subfloat[]{
		\includegraphics[width=0.32\textwidth,trim=0 0 0 0,clip]{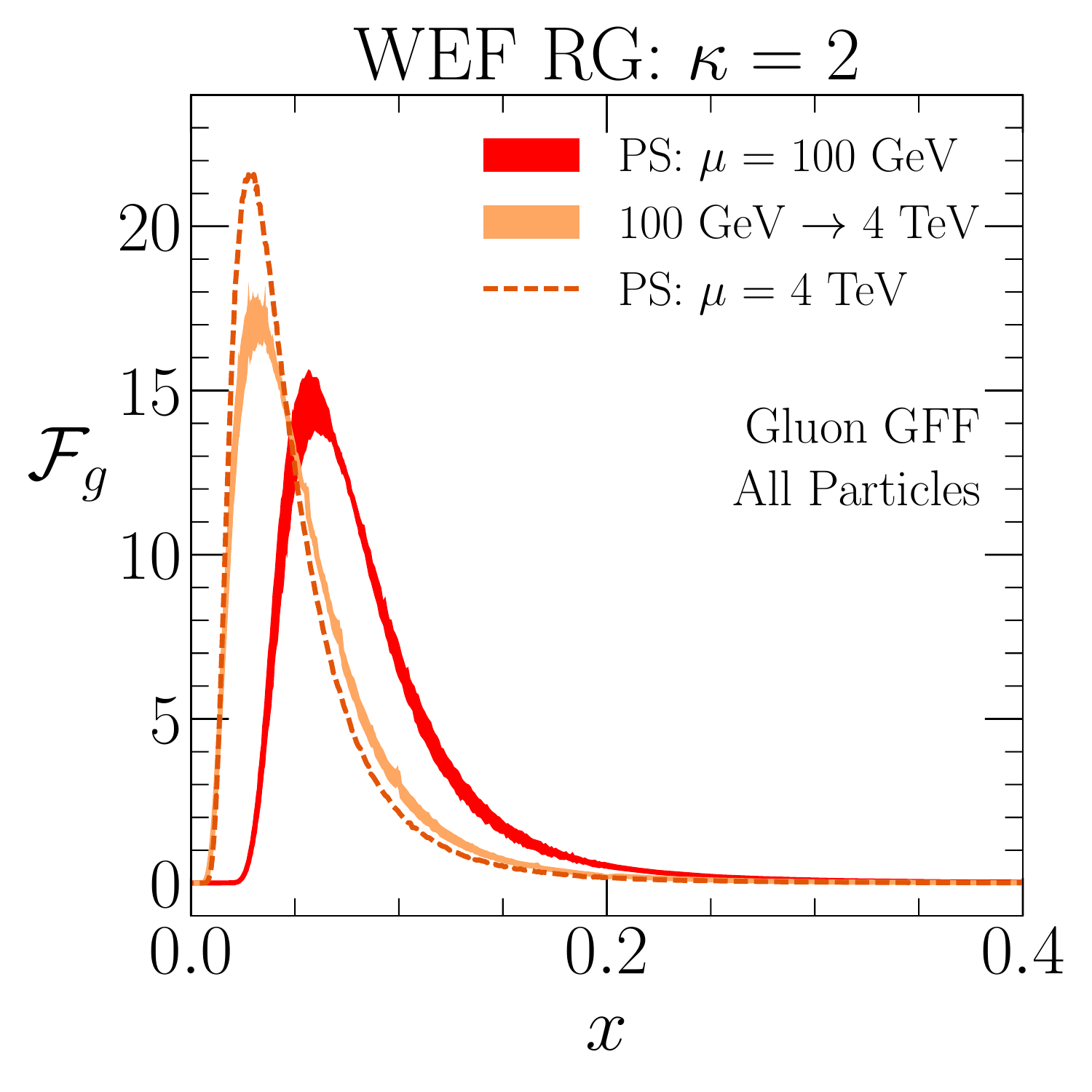}
		\label{fig:wef_gluong}
}
\subfloat[]{
		\includegraphics[width=0.32\textwidth,trim=0 0 0 0,clip]{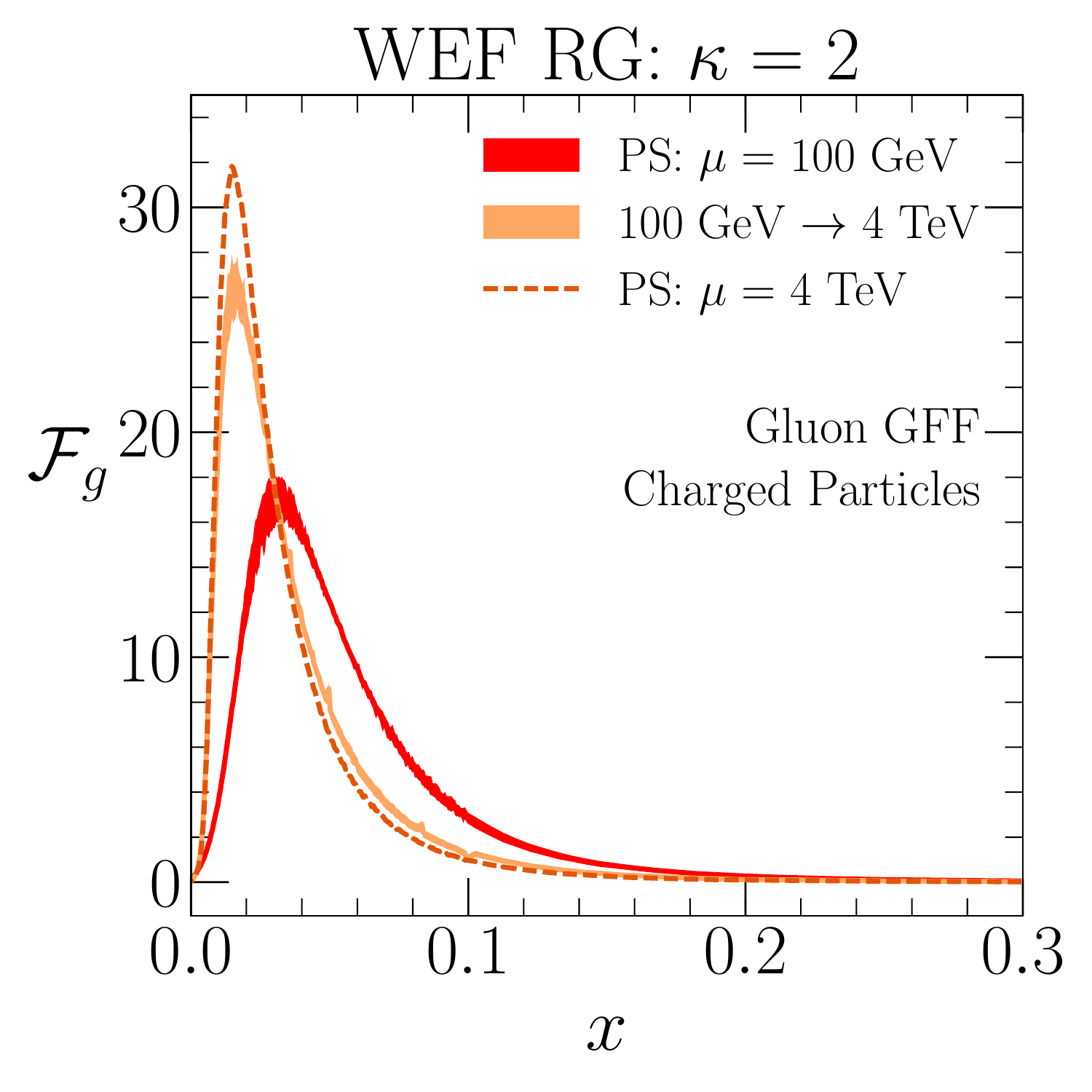}
		\label{fig:wef_gluonh}
}
\subfloat[]{
		\includegraphics[width=0.32\textwidth,trim=0 0 0 0,clip]{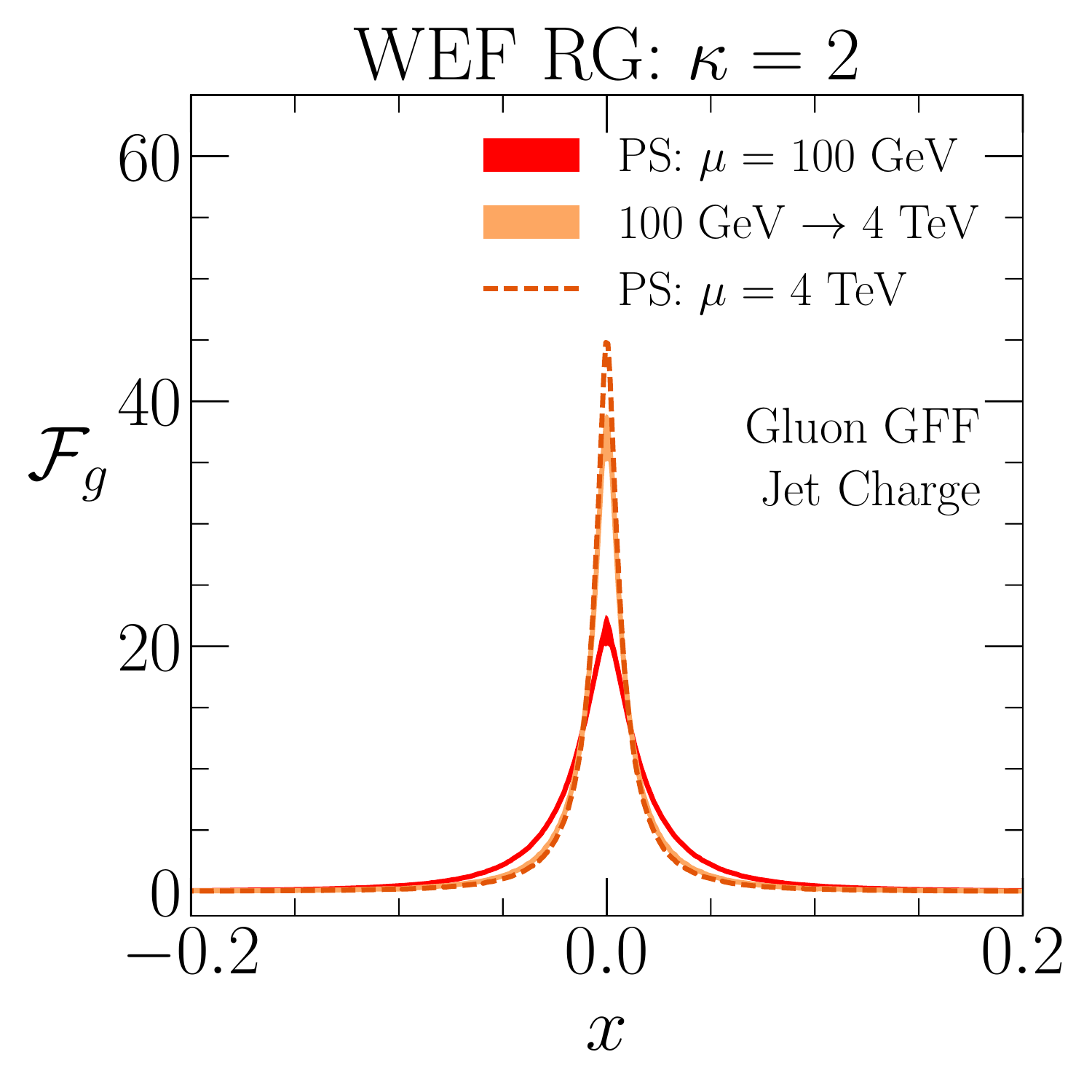}
		\label{fig:wef_gluoni}
}
		\caption{Gluon GFFs of weighted energy fractions with (top row) $\kappa = 0.5$, (middle row) $\kappa = 1$, and (bottom row) $\kappa = 2$.
		Shown are distributions involving (left column) all particles, (middle column) just charged particles, and (right column) charged particles weighted by their charge. 
		The GFFs extracted from parton showers at $\mu = 100$ GeV are shown in solid red.  The result of evolving these initial conditions to $\mu = 4$ TeV are plotted in solid orange, to be compared to the average distribution obtained from parton showers at that value, plotted in dashed orange. The uncertainties come from both varying $R$ and the choice of parton shower (i.e.\ both variations shown in \Fig{fig:extractall}).} 
	\label{fig:wef_gluon}  
\end{figure}

\begin{figure}
\subfloat[]{
	\includegraphics[width=0.32\textwidth,trim=0 0 0 0,clip]{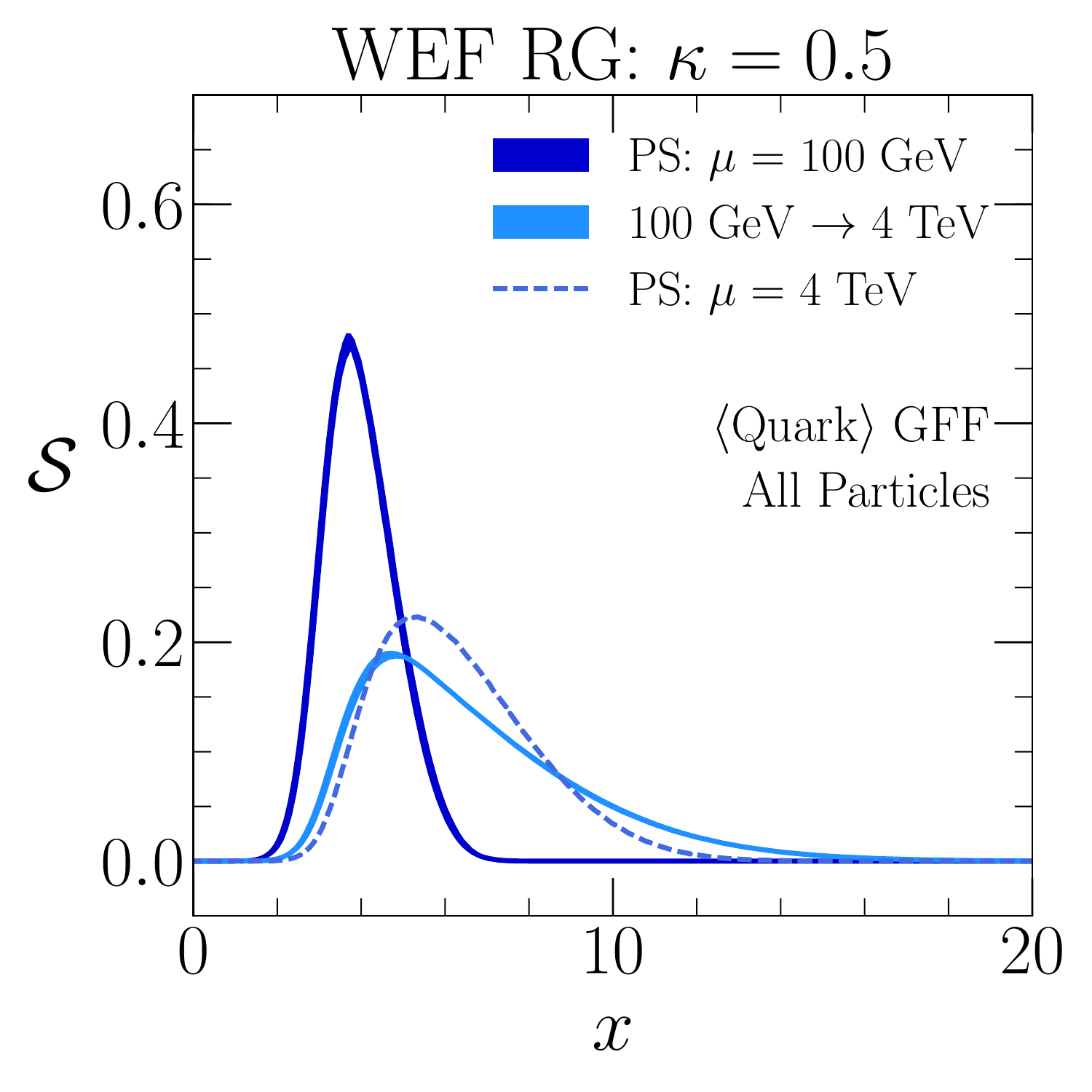}
	\label{fig:wef_quarka}
}
\subfloat[]{
	\includegraphics[width=0.32\textwidth,trim=0 0 0 0,clip]{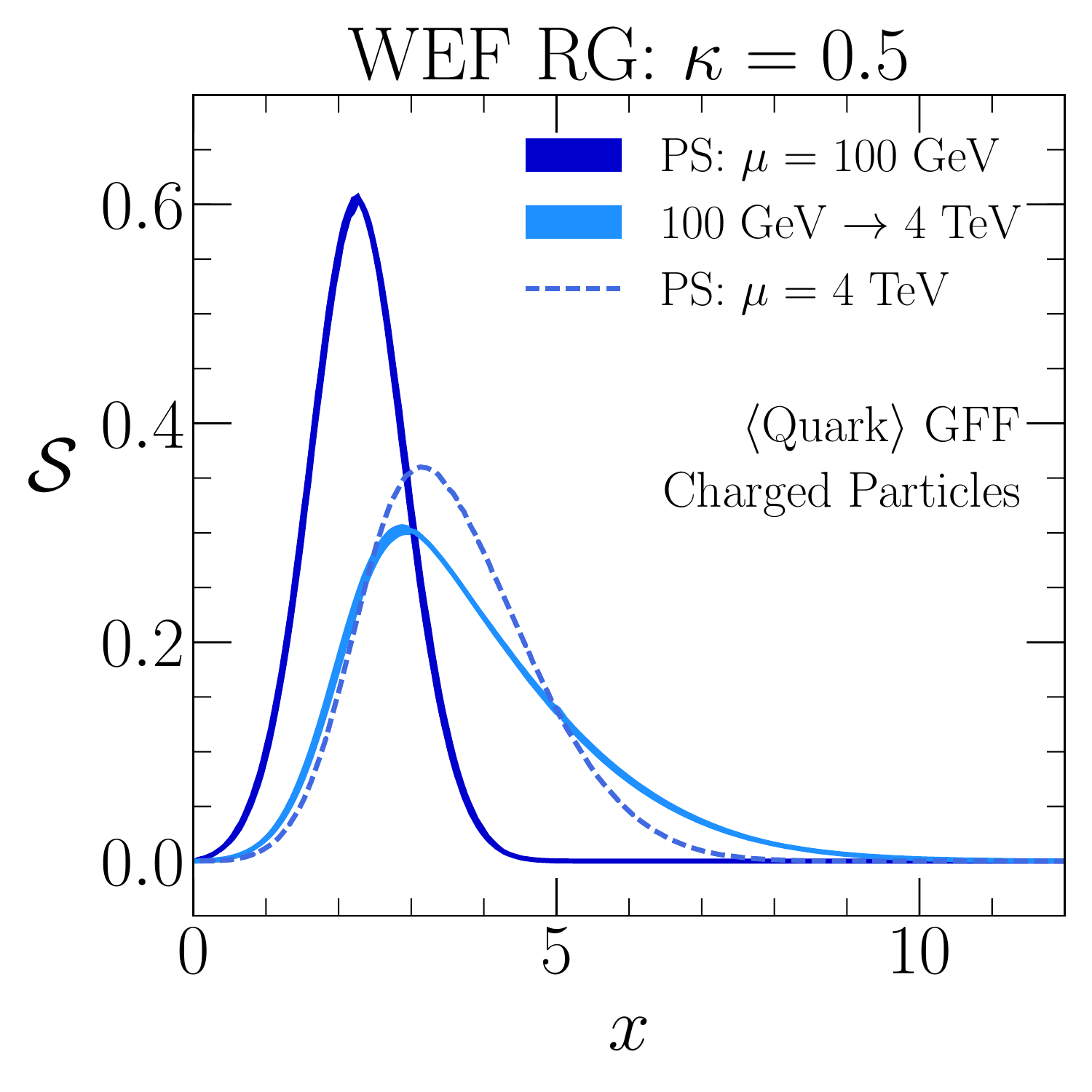}
	\label{fig:wef_quarkb}
}
\subfloat[]{
	\includegraphics[width=0.32\textwidth,trim=0 0 0 0,clip]{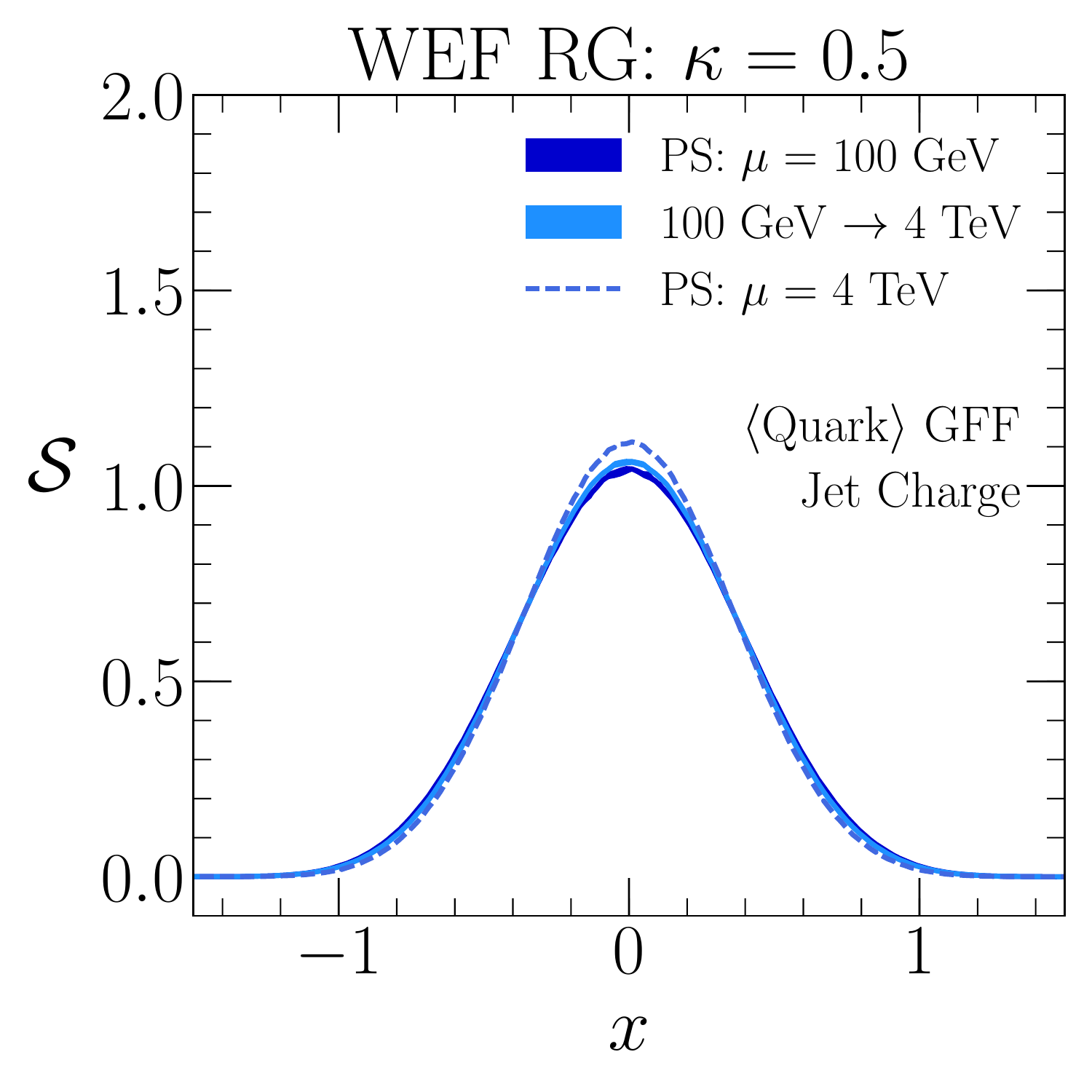}
	\label{fig:wef_quarkc}
}

\subfloat[]{
	\includegraphics[width=0.32\textwidth,trim=0 0 0 0,clip]{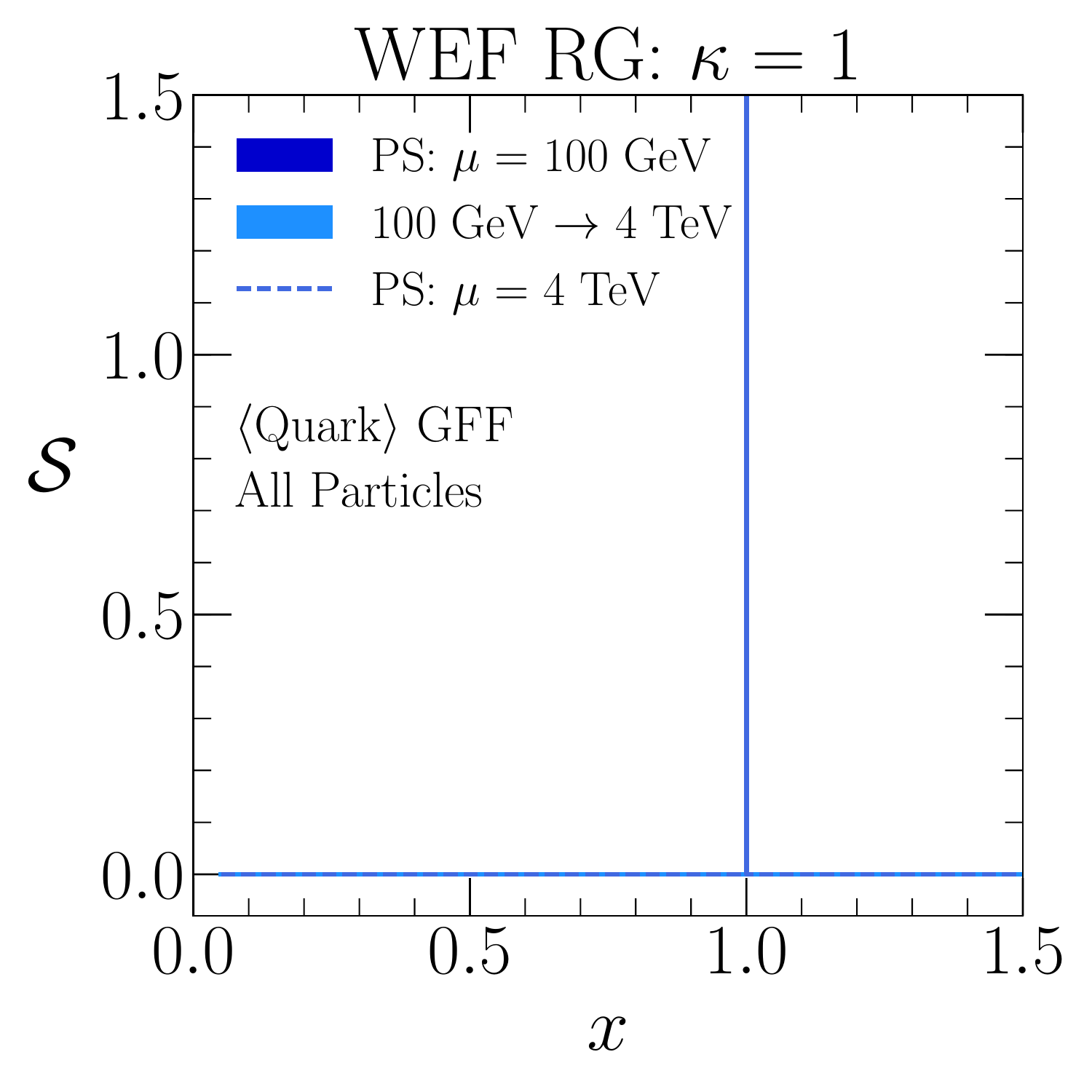}
	\label{fig:wef_quarkd}
}
\subfloat[]{
	\includegraphics[width=0.32\textwidth,trim=0 0 0 0,clip]{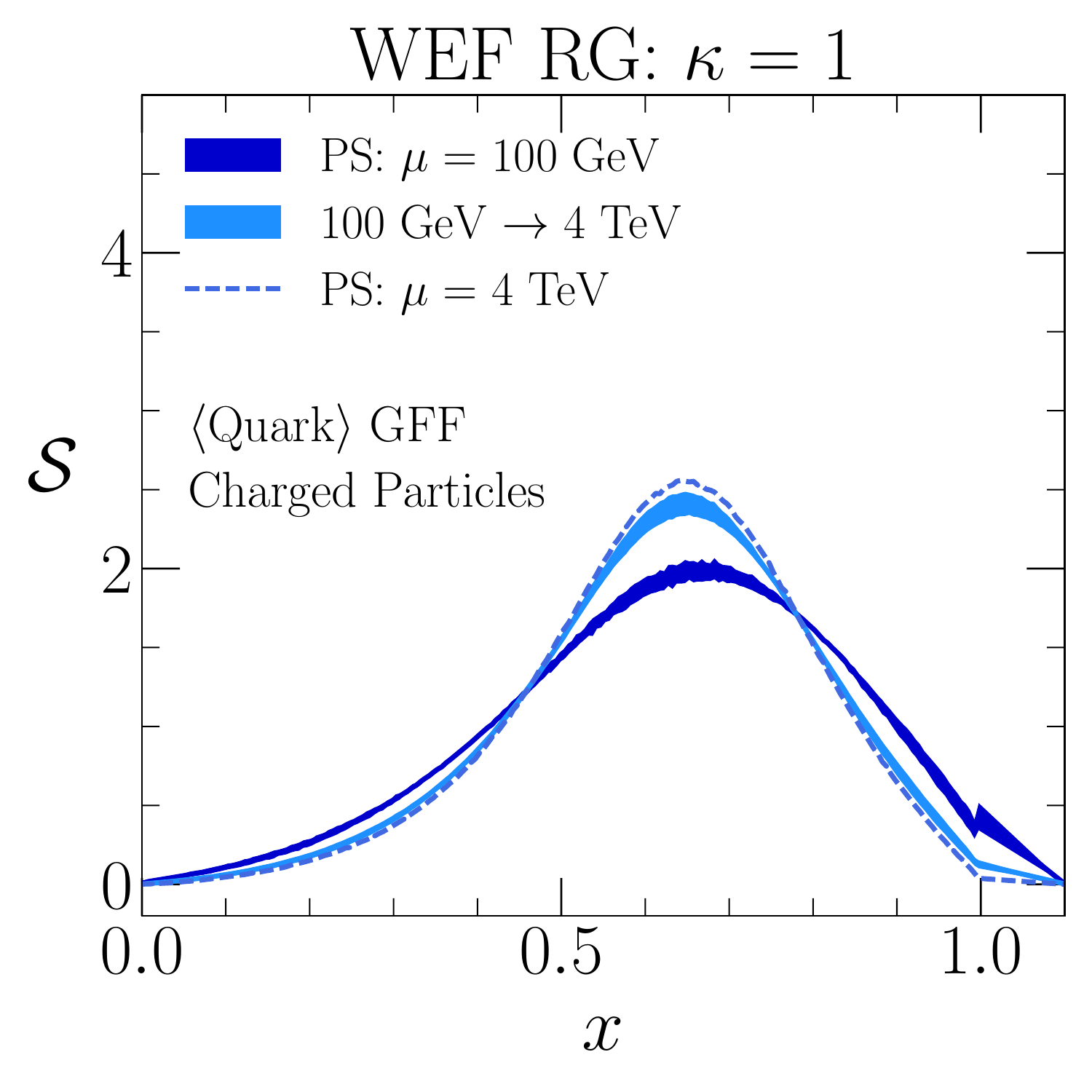}
	\label{fig:wef_quarke}
}
\subfloat[]{
	\includegraphics[width=0.32\textwidth,trim=0 0 0 0,clip]{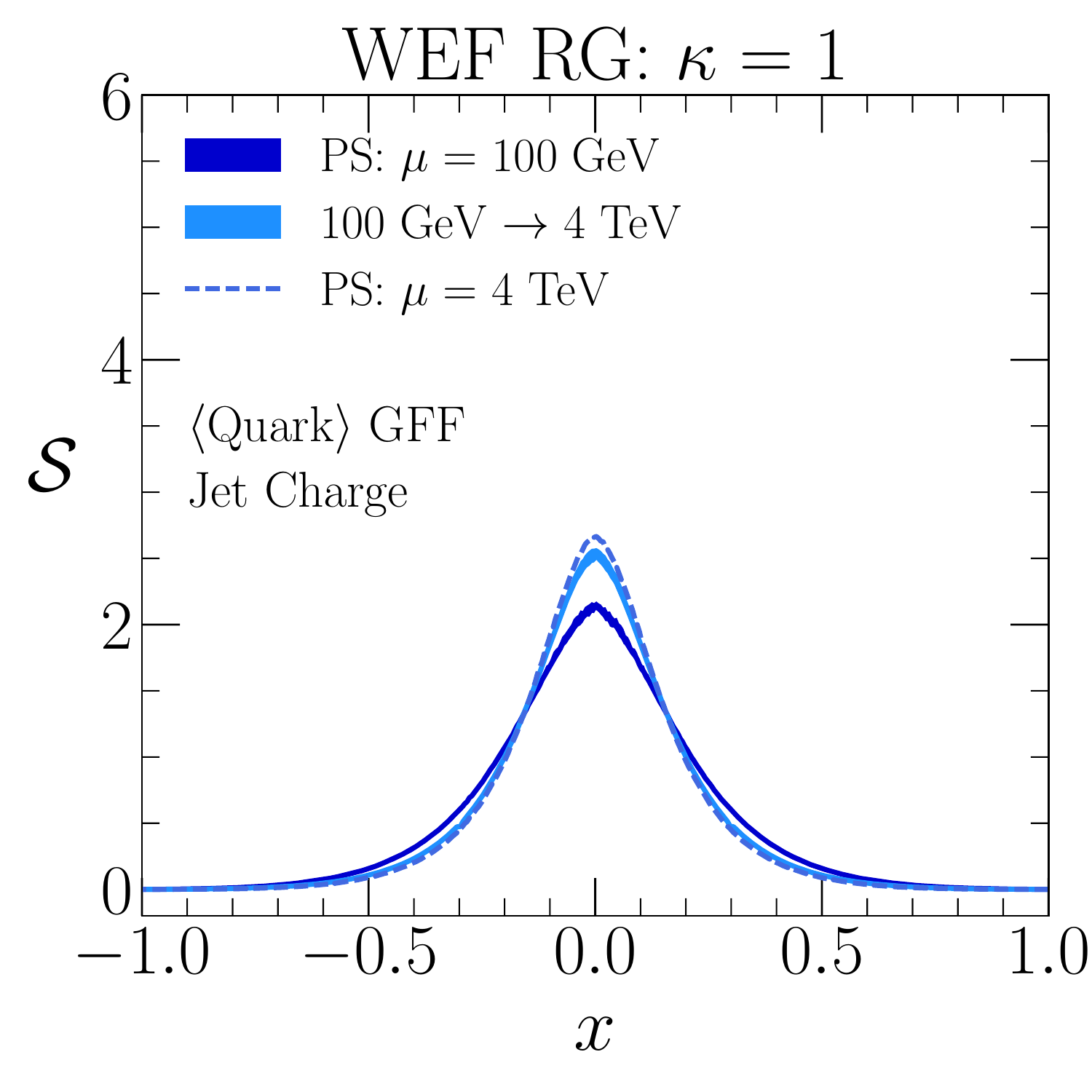}
	\label{fig:wef_quarkf}
}

\subfloat[]{
	\includegraphics[width=0.32\textwidth,trim=0 0 0 0,clip]{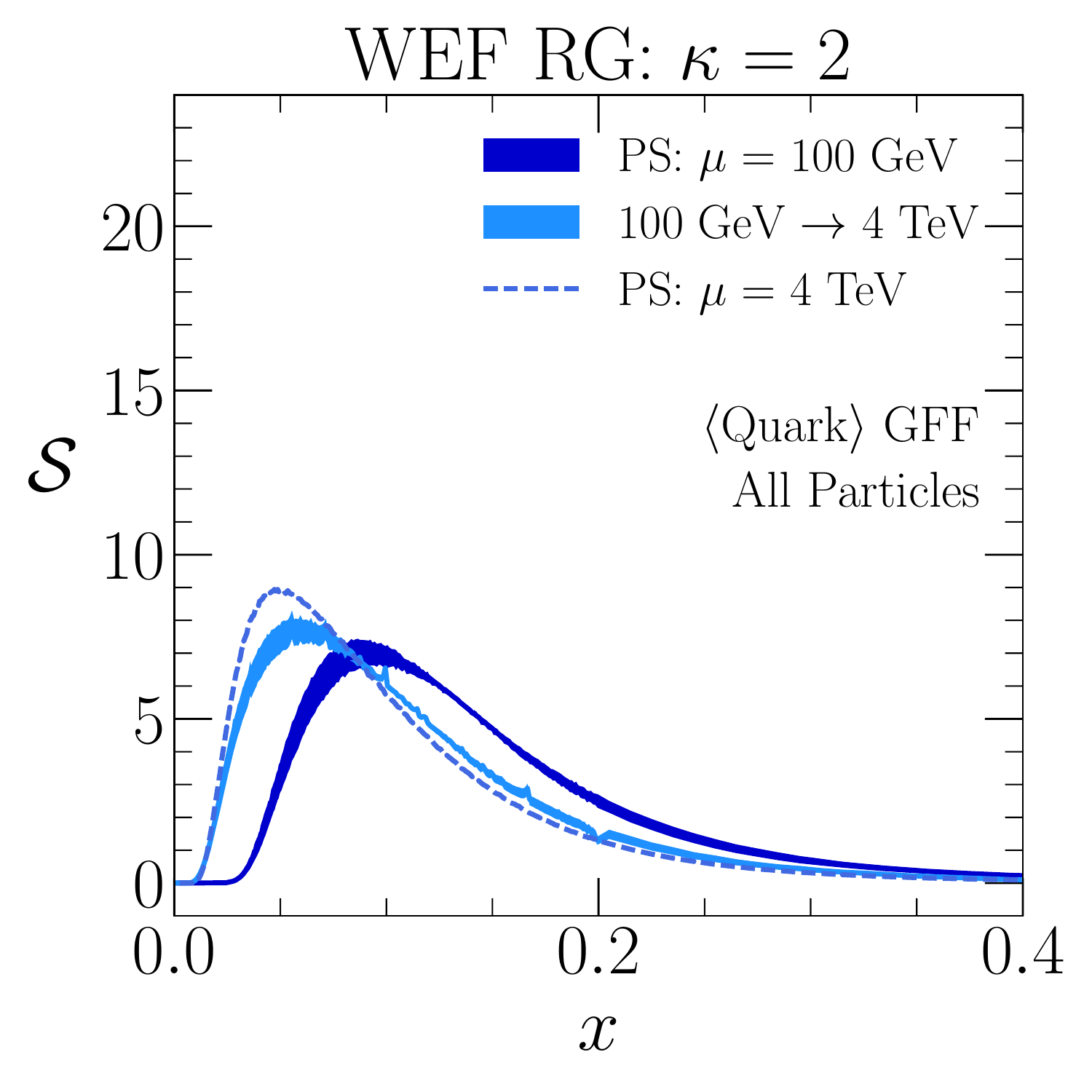}
	\label{fig:wef_quarkg}
}
\subfloat[]{
	\includegraphics[width=0.32\textwidth,trim=0 0 0 0,clip]{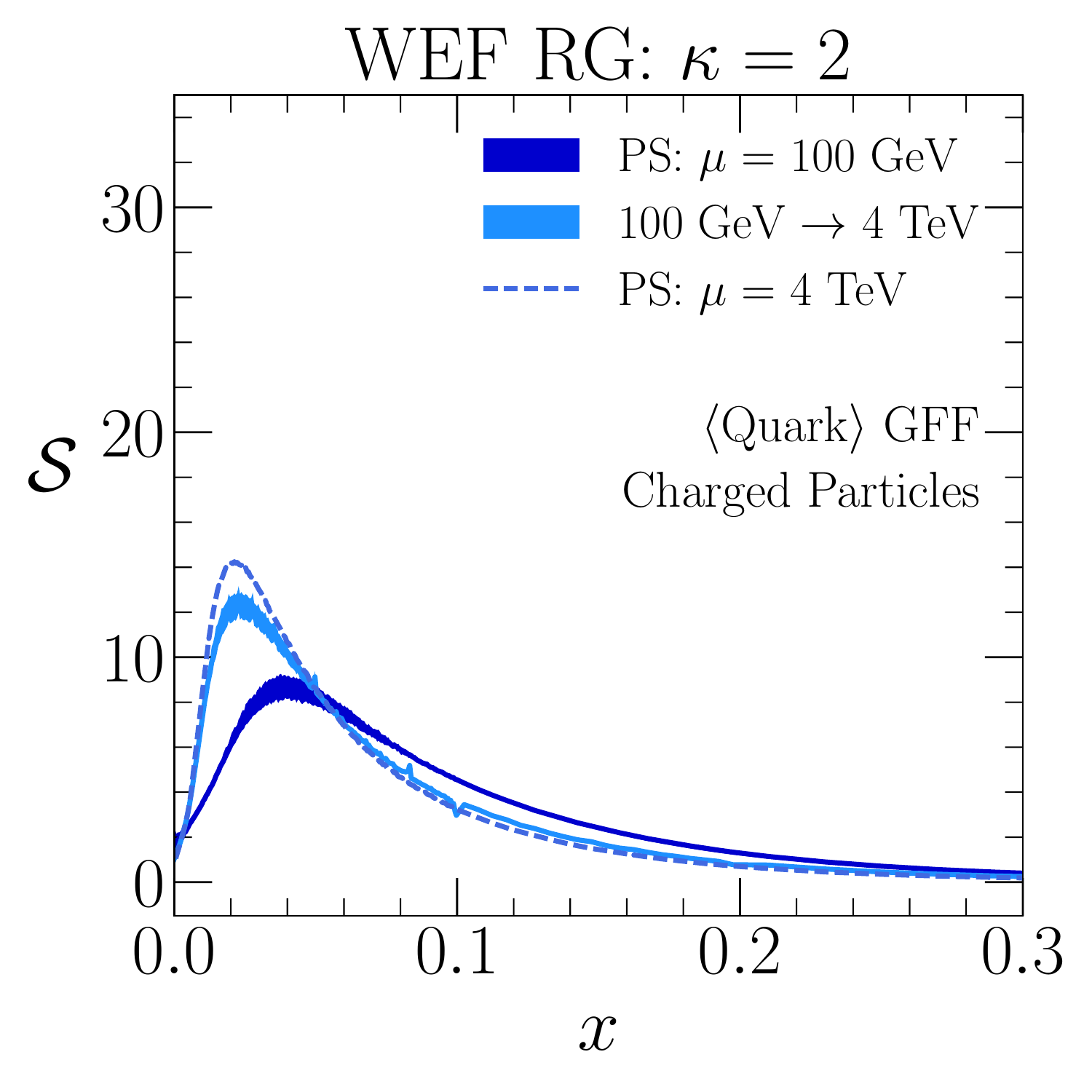}
	\label{fig:wef_quarkh}
}
\subfloat[]{
	\includegraphics[width=0.32\textwidth,trim=0 0 0 0,clip]{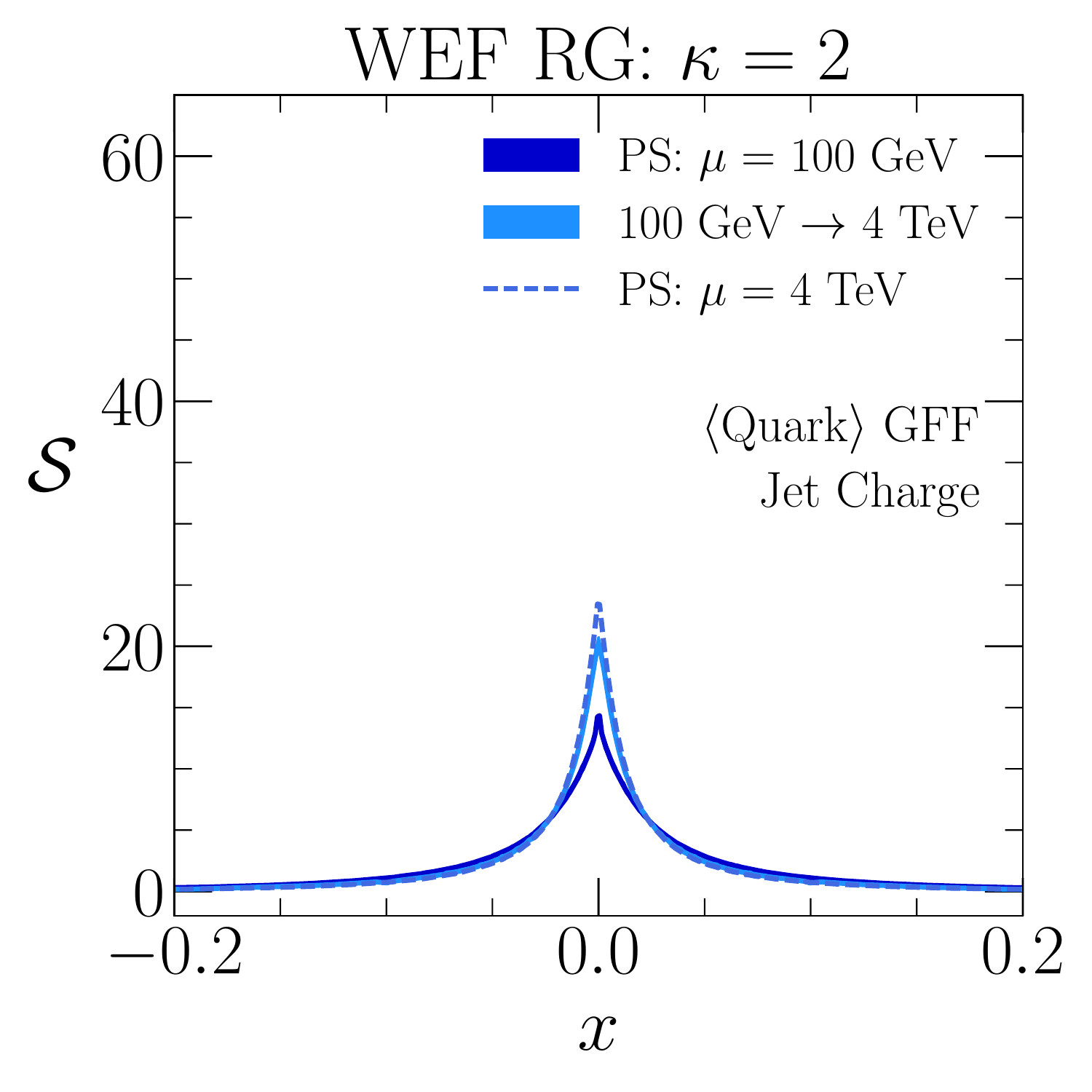}
	\label{fig:wef_quarki}
}
	\caption{Same as \Fig{fig:wef_gluon} but for quark-singlet GFFs, where the distributions extracted from parton showers at $\mu = $ 100 GeV are shown in solid blue, the evolved distribution are shown in solid light blue, and the distributions extracted at $\mu = $ 4 TeV are shown in dashed light blue.}
\label{fig:wef_quark} 
\end{figure}

In \Figs{fig:wef_gluon}{fig:wef_quark}, we present the evolution results for gluon and quark-singlet GFFs respectively, for the weighted energy fractions with $\kappa = \{0.5, 1.0, 2.0\}$.
We test three different choices for the particle weights:  $w_a = 1$ for all particles, $w_a = 1$ ($w_a = 0$) for charged (neutral) particles, and $w_a = Q_a$ with $Q_a$ being the particle's electric charge.
The initial conditions extracted from the parton showers at $\mu = 100$ GeV are the same as those shown in \Fig{fig:extractall}, with the same color scheme of red for gluon GFFs and blue for quark-singlet GFFs.
As described in \Sec{sec:wefextraction}, the uncertainty bands are given by the envelope of values obtained both from varying the jet radius/energy (keeping $\mu$ fixed) and from using different parton showers. 
The evolved distributions to $\mu = 4$ TeV are shown in orange for the gluon GFFs and light blue for the quark-singlet GFFs, where the uncertainty bands show the spread in final values due to the spread in initial conditions.

For comparison, we show in dashed lines the GFFs extracted at $\mu = 4$ TeV, averaged over the three parton showers and three $R$ values.\footnote{The uncertainties from varying the jet radius/energy and changing parton showers at $\mu = 4$ TeV are similar to the ones shown at $\mu = 100$ GeV.}
Overall, our numerical GFF evolution agrees well with parton shower evolution, with both methods giving the same shift in the peak locations.
As previously seen in \Ref{Waalewijn:2012sv}, the two evolution methods agree best for $\kappa \ge 1$, with larger differences seen in the widths of the distributions when $\kappa < 1$.
This is likely because $\kappa < 1$ is more sensitive to collinear fragmentation, with larger expected corrections from higher-order perturbative effects.
Note the expected $\delta$-function when $\kappa = 1$ and $w_a = 1$ for all particles, since the sum of the energy fractions for all particles in the jet equals 1. The  $\kappa \to 1$ limit of weighted energy fractions is discussed  in \Sec{sec:limits} below.

\subsection{Limits}
\label{sec:limits}

There are a few interesting limits of the weighted energy fractions.
For the case of $\kappa = 0$, the energy fractions $z_a$ drop out, so $x$ simply counts the hadrons in the final state, weighted by $w_a$.
Although hadron multiplicity is IR unsafe, it is possible to calculate the evolution of the average hadron multiplicity using fragmentation functions, see e.g.~refs.~\cite{Malaza:1985jd,Lupia:1997in,Bolzoni:2012ii}.
This case requires special care, however, because of the soft gluon singularity of the splitting functions.
IR-safe variants of multiplicity that have only collinear singularities are explored in a forthcoming paper \cite{HarvardInProgress}.

\begin{figure}[t]
\subfloat[]{
		\label{fig:limits-kappa1}
			\includegraphics[width=0.45\textwidth]{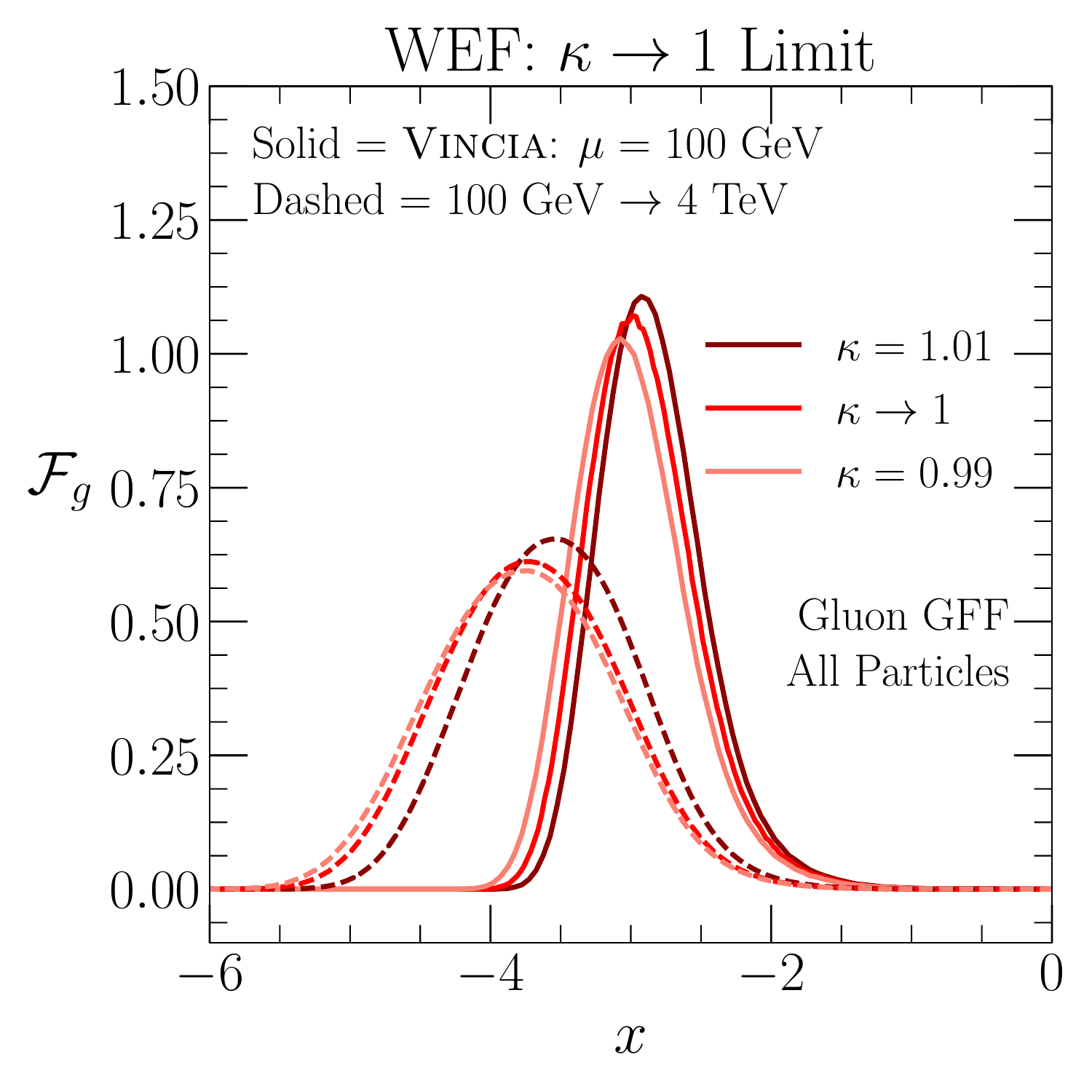}
}
\subfloat[]{
		\label{fig:limits-kappinf}
			\includegraphics[width=0.45\textwidth]{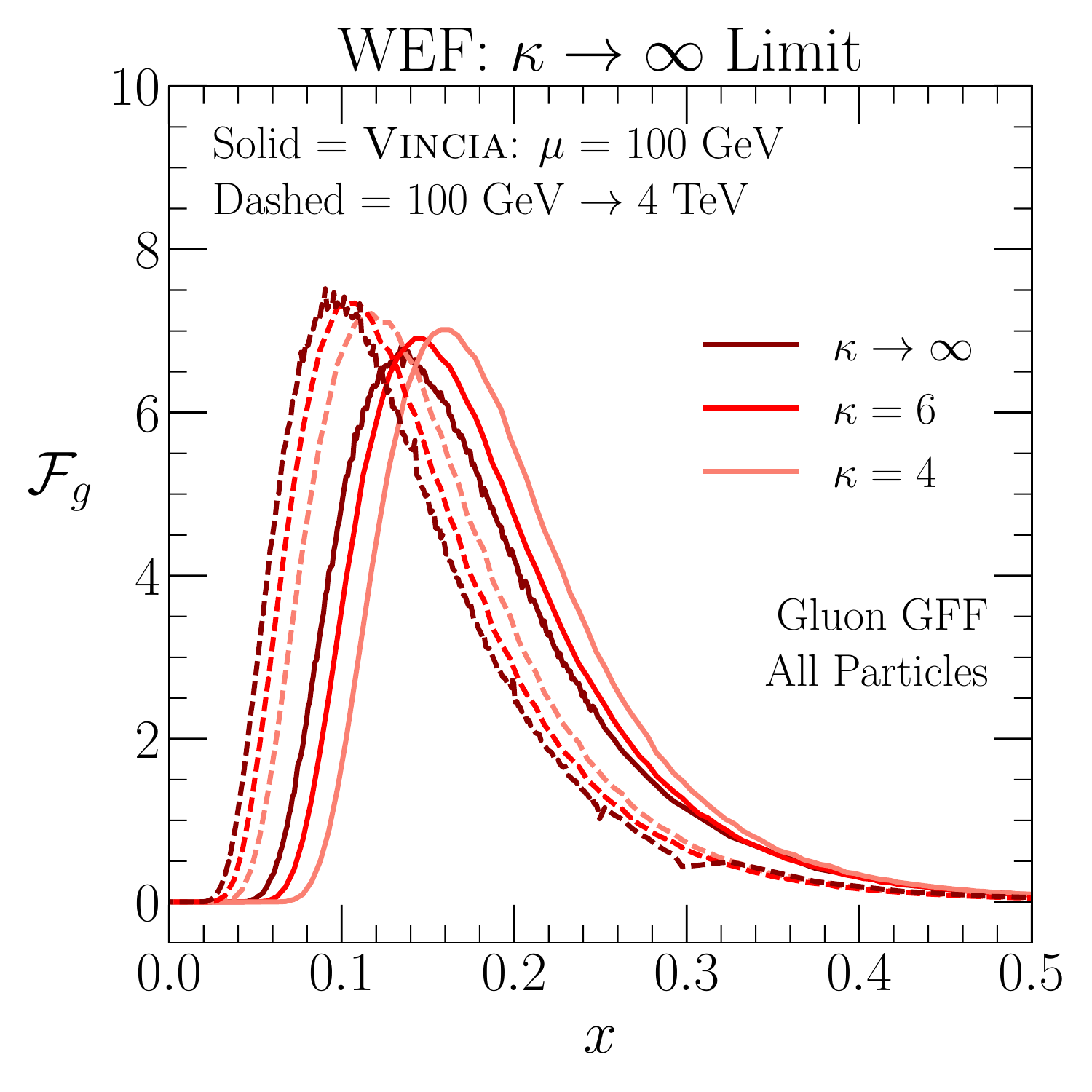}
}
	\caption{Gluon GFFs for (a) the modified weighted energy fractions from \Eq{eq:kappa1} in the $\kappa \to 1$ limit, and (b) the $\kappa$-th root of the weighted energy fractions from \Eq{eq:kappainf} in the $\kappa \to \infty$ limit.  The solid lines show the GFFs extracted from $\textsc{Vincia}$ at $\mu = 100$ GeV, while the dashed lines show the evolution of these GFFs to $\mu = 4$ TeV.  The fact that the limits are smooth is a consistency check on the evolution code.}
	\label{fig:limits}
\end{figure}

For the case of $\kappa =1$ with all hadrons assigned weight 1, the weighted energy fraction simply becomes $x = \sum_a z_a = 1$.  Still, we can expand around the $\kappa \to 1$ limit to find a non-trivial observable \cite{Larkoski:2014pca}.  Consider the modified weighted energy fraction and its limit,
\be
\label{eq:kappa1}
x = \frac{1}{\kappa -1}\bigg[\sum_{a \in \text{jet}} z_a^\kappa -1\bigg], \qquad  \lim_{\kappa \rightarrow 1} x = \sum_{a \in \text{jet}} z_a \ln z_a.
\ee
In the limiting case, the recursion relation becomes
\begin{equation}
\label{eq:kappa1recur}
\hat{x}(z,x_1,x_2) = z\ln z + (1-z)\ln(1-z) +  x_1 \, z +  x_2 \, (1-z),
\end{equation}
with initial hadron weights of $w_a = 0$ (due to the $-1$ in \Eq{eq:kappa1}).  This is easy to verify by testing the three clustering trees in \Fig{fig:trees}.\footnote{Amusingly, the recursion relation in \Eq{eq:kappa1recur} is associative for any choices of initial hadron weights, leading to the fractal observable $x = \sum_{a \in \text{jet}} z_a (w_a + \ln z_a)$.}

The behavior of the evolved GFFs in the $\kappa \rightarrow 1$ limit offers a non-trivial cross check of our evolution code.
Away from the limiting value, the RG evolution can be implemented using the recursion relation in \Eq{eq:wefrecur}.
At the limiting value, we have to use a different RG evolution based on the recursion relation in \Eq{eq:kappa1recur}. 
The smooth convergence of the evolved distributions as $\kappa\rightarrow 1$ is illustrated in \Fig{fig:limits-kappa1}, showing the modified weighted energy fraction from \Eq{eq:kappa1}.
The solid curves show the extraction of the corresponding GFFs at $\mu = 100 $ GeV with $\kappa=0.99$ and $\kappa = 1.01$, which correctly bracket the $\kappa \to 1$ limit.\footnote{In practice, we first extract the $\kappa=0.99$ and $\kappa = 1.01$ distributions for the unmodified weighted energy fraction, and then do a simple change of variables to match the definition in \Eq{eq:kappa1}.}
The dashed curves show the evolution to $\mu = 4$ TeV, where there is again a smooth approach to $\kappa \to 1$.

In the limit that $\kappa \rightarrow \infty$, the most energetic hadron in the jet dominates the sum in \Eq{eq:wefext}.
We can then take the $\kappa$-th root of the weighted energy fraction to have a smooth $\kappa \to \infty$ limit:
\begin{equation}
\label{eq:kappainf}
x = \bigg\vert \sum_{a \in\text{jet}}  w_a z_a^\kappa\bigg\vert^{1/\kappa}, \qquad \lim\limits_{\kappa\rightarrow \infty} x = \max_{w_a \not= 0}{z_a},
\end{equation}
where the maximum is only taken over particles with non-zero weights.
The corresponding recursion relation is
\begin{equation}
\label{eq:kappainfrecur}
\hat{x}(z,x_1,x_2) = \max(|z x_1|,|(1-z)x_2|),
\end{equation}
with modified initial hadron weights of $\tilde{w}_a = |\text{sign}(w_a)| = \{0,1\}$. 
For these modified weights, it is easy to verify that \Eq{eq:kappainfrecur} gives an associative recursion relation using \Fig{fig:trees}.\footnote{It is also possible to keep track of the signs of hadron weights by using an alternative recursion relation $\hat{x}(z,x_1,x_2) = \text{signed-max}(z x_1,(1-z)x_2)$, where the signed-max function takes the term (positive or negative) with the largest absolute value.  Again, this only yields an associative recursion if the hadron weights are a constant multiple of $\{-1,0,1\}$.}

In \Fig{fig:limits-kappinf}, we show the approach to $\kappa\rightarrow \infty$ for the gluon GFFs, considering the case of all particles with equal weight $w_a = 1$.
Here, the finite-$\kappa$ evolution equations use the recursion relation in \Eq{eq:wefrecur} while the $\kappa \to \infty$ limit uses \Eq{eq:kappainfrecur}, and we plot the $\kappa$-th root of the weighted energy fractions as given in \Eq{eq:kappainf}.
Both the extracted distributions at $\mu = 100$ GeV and the evolved distributions to $\mu = 4$ TeV show a smooth transition from $\kappa = 4$ to $\kappa = 6$ to the final $\kappa \to \infty$ limit.
This is again a non-trivial cross check of our evolution code.

\subsection{Moment Space Analysis}
\label{sec:wefmoments}

To gain further insight into the evolution of the GFFs, it is instructive to examine the evolution equations for the first two moments, which are related to averages and widths of the distribution for the fractal observable.
In general, the moments of a GFF are defined as
\begin{equation}
\label{eq:moment}
\overline{\GFF}_i(N,\mu) \equiv \int \text{d}x\, x^N\GFF_i(x,\mu),
\end{equation}
with $N\geq0$.
For the specific case of the weighted energy fractions, it is convenient to introduce a transformed version of the splitting functions
\be
\label{eq:Pmoments}
\overline{P}_{i\rightarrow jk}(\alpha,\beta) \equiv \int \text{d}z\, z^{\alpha}(1-z)^{\beta} P_{i\rightarrow jk}(z), \qquad \overline{P}_{i\rightarrow jk}(\alpha) \equiv \overline{P}_{i\rightarrow jk}(\alpha,0).
\ee
Integrating \Eq{eq:evolution_repeated} against $x^N$, the moment space evolution equation for a weighted energy fraction is
\be
\label{eq:momevolutionwef}
\mu \frac{\text{d}}{\text{d}\mu}\, \overline{\GFF}_i (N,\mu) = \frac{\alpha_s(\mu)}{2\pi} \sum_{j,k} \sum_{M=0}^N \binom{N}{M} \, \overline{P}_{i\rightarrow jk}\big(\kappa(N-M),\kappa M\big) \,  \overline{\GFF}_j (N-M,\mu)\, \overline{\GFF}_k (M,\mu),
\ee
where it is crucial that $N$ is an integer. A derivation of this expression is given in \App{app:momentspace}.

These evolution equations are more compact in the color singlet/non-singlet basis introduced in \Eq{eq:colorsinglet}.
For the quark-non-singlet pieces, the evolution of the first moment (i.e.\ the mean) is given by
\be
\label{eq:1stmomNS}
\mu \frac{\text{d}}{\text{d}\mu}\, \overline{\mathcal{N}}_{ij}(1,\mu) = \frac{\alpha_s(\mu)}{\pi}\, \overline{P}_{q\rightarrow qg}(\kappa)\, \overline{\mathcal{N}}_{ij}(1,\mu).
\ee
Since $\overline{P}_{q\rightarrow qg}(\kappa) < 0$ for all positive $\kappa$, \Eq{eq:1stmomNS} implies that the averages of the different (anti-)quark GFFs functions converge to a common value as $\mu$ evolves upward.
This behavior is expected, since QCD branchings only depend on the parton's color charge, so the low-scale differences between the (anti-)quark flavors, due to e.g.\ electric charge, get washed out at high scales.

\begin{figure}[t]
	\centering
	\includegraphics[width=0.45\textwidth]{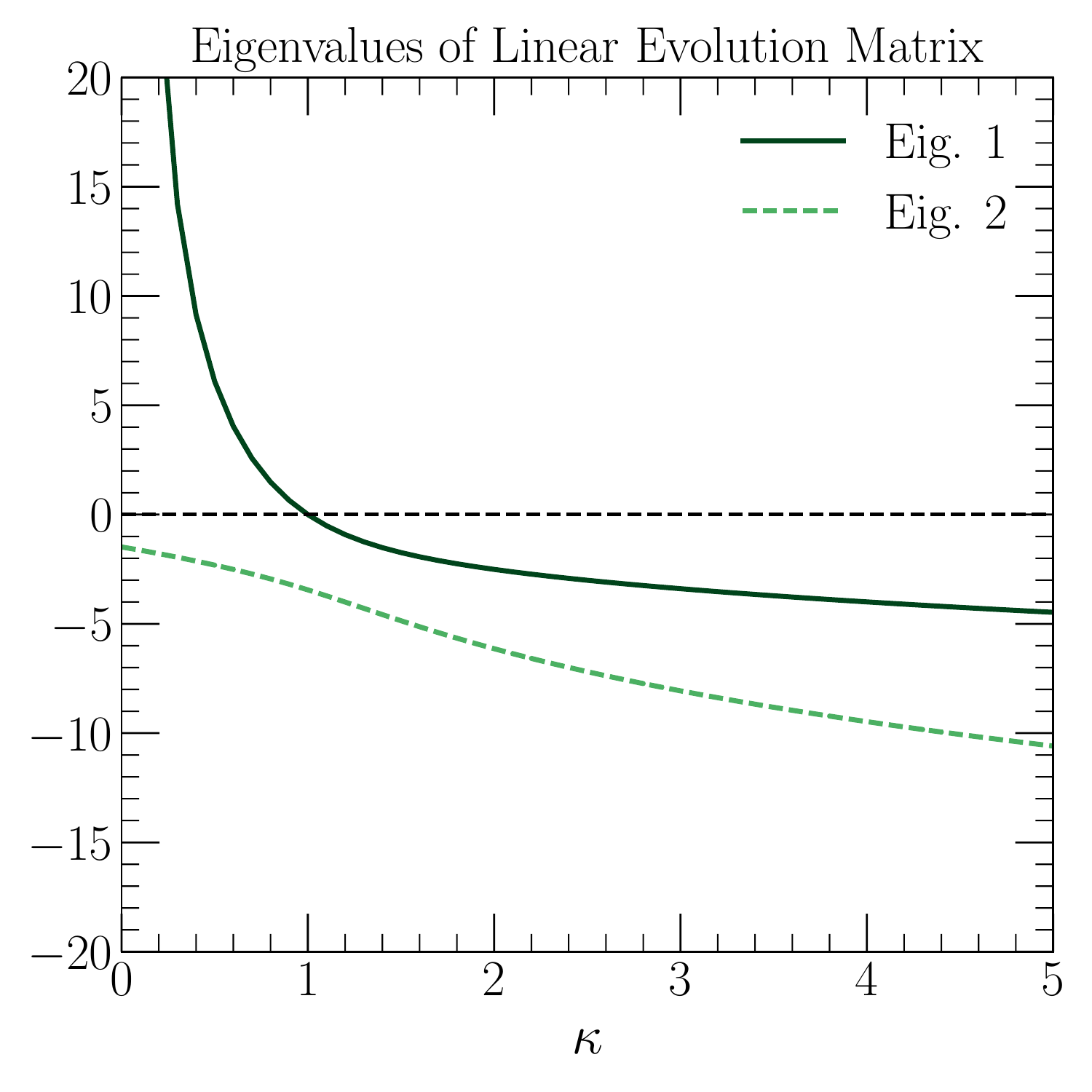}
	\caption{The two eigenvalues of the matrix in \Eq{eq:1stmomS}, as a function of $\kappa$. This matrix governs the evolution of the first moment of weighted energy fraction GFFs.  Only for $\kappa=1$ is there a zero eigenvalue.}\label{fig:eigen}
\end{figure}

The quark-singlet combination mixes with the gluon GFF. For the first moment this is given by
\be
\label{eq:1stmomS}
\mu \frac{\text{d}}{\text{d}\mu} \left(\begin{array}{c}\overline{\mathcal{S}}(1,\mu) \\ \overline{\GFF}_g(1,\mu) \end{array}\right) = \frac{\alpha_s(\mu)}{\pi}\left(\begin{array}{c c} \overline{P}_{q\rightarrow qg}(\kappa) &  \overline{P}_{q\rightarrow gq}(\kappa)\\ 2n_f\overline{P}_{g\rightarrow q\overline{q}}(\kappa) & \overline{P}_{g\rightarrow gg}(\kappa) \end{array}  \right)\left(\begin{array}{c}\overline{\mathcal{S}}(1,\mu) \\ \overline{\GFF}_g(1,\mu) \end{array}\right).
\ee
As shown in \Fig{fig:eigen}, the matrix in \Eq{eq:1stmomS} always has one negative eigenvalue for all $\kappa$, which implies that the first moment of the quark-singlet GFF tries to track the first moment of the gluon GFF.
For example, in the case of $\kappa = 1$, the combination $2C_F\, \overline{\mathcal{S}}(1,\mu)- n_fT_F\, \overline{\GFF}_g(1,\mu)$ asymptotes to zero at high $\mu$.
The second eigenvalue has different signs depending on the value of $\kappa$.
For $\kappa < 1$, it is positive, so the first moments of both the quark-singlet and gluon GFF increase with $\mu$.
For $\kappa > 1$, the second eigenvalue is negative, so the first moments decrease with $\mu$.
For the special case $\kappa = 1$, the second eigenvalue is zero, and the corresponding eigenvector $\overline{\mathcal{S}}(1,\mu) + \overline{\GFF}_g(1,\mu)$ stays constant with $\mu$. 
These broad features agree with the behaviors already seen in \Figs{fig:wef_gluon}{fig:wef_quark}.

\begin{figure}[t]
\subfloat[]{
		\includegraphics[width=0.32\textwidth]{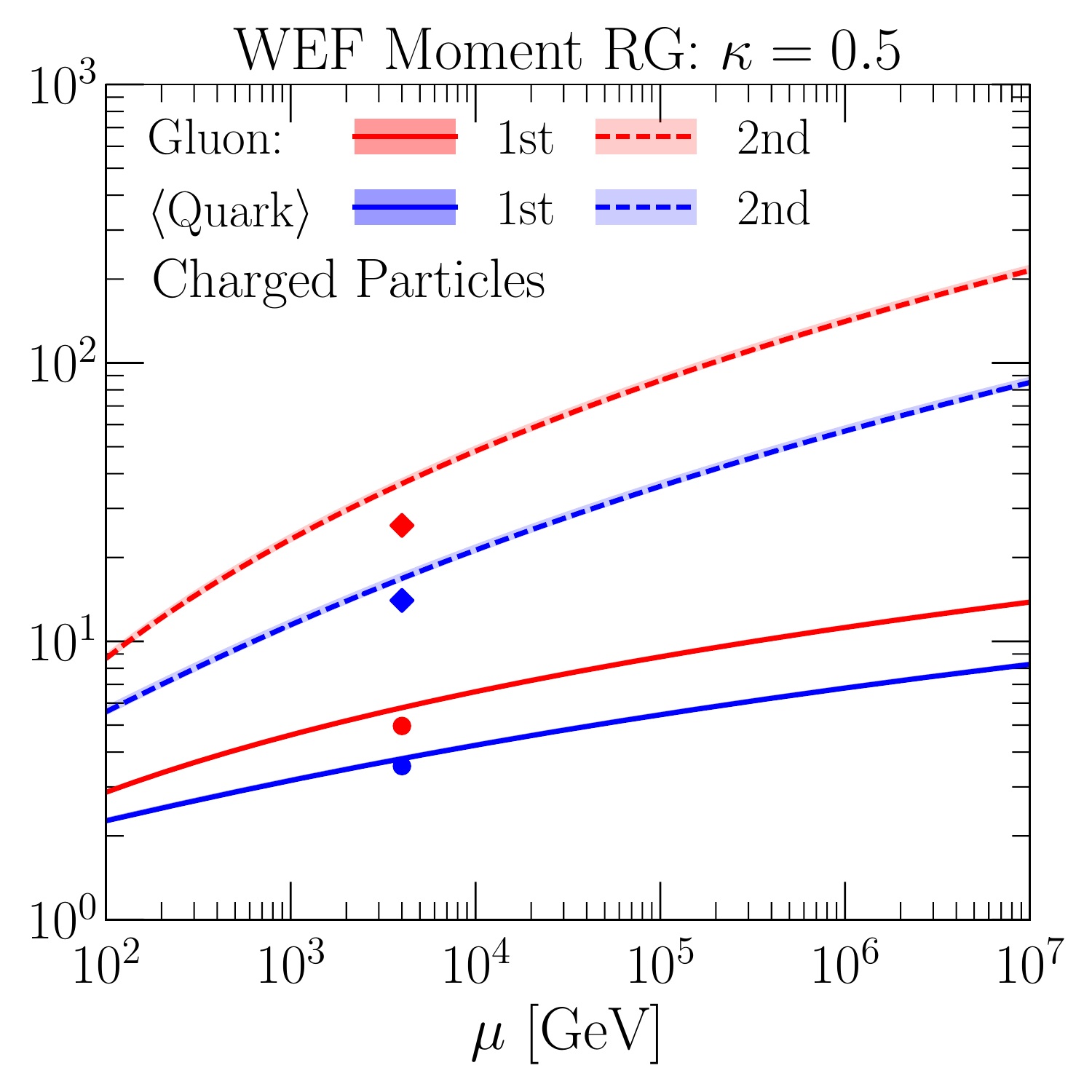}
		 \label{fig:moment-space-wef-a}
}
\subfloat[]{
		\includegraphics[width=0.32\textwidth]{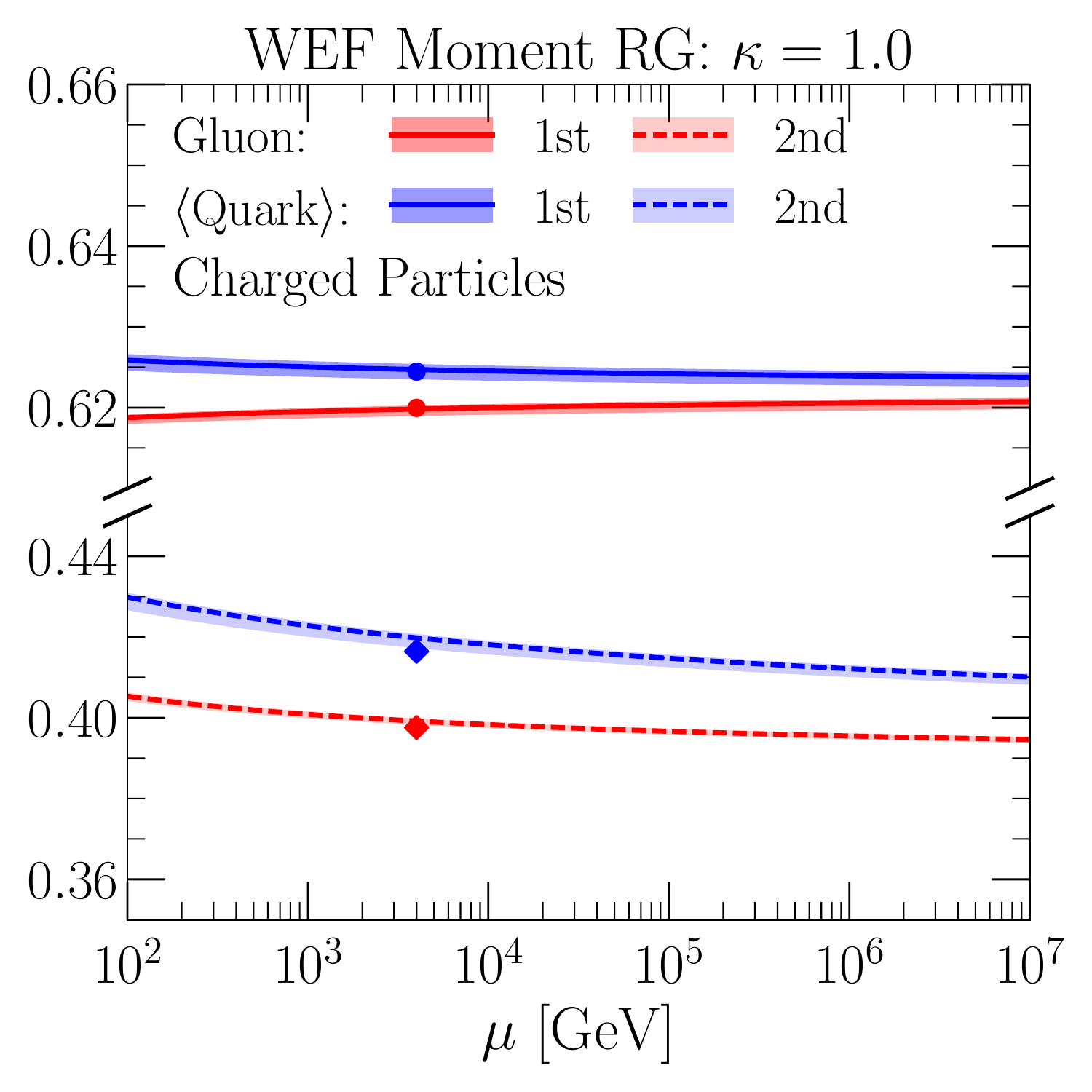}
		 \label{fig:moment-space-wef-c} 
}
\subfloat[]{
		\includegraphics[width=0.32\textwidth]{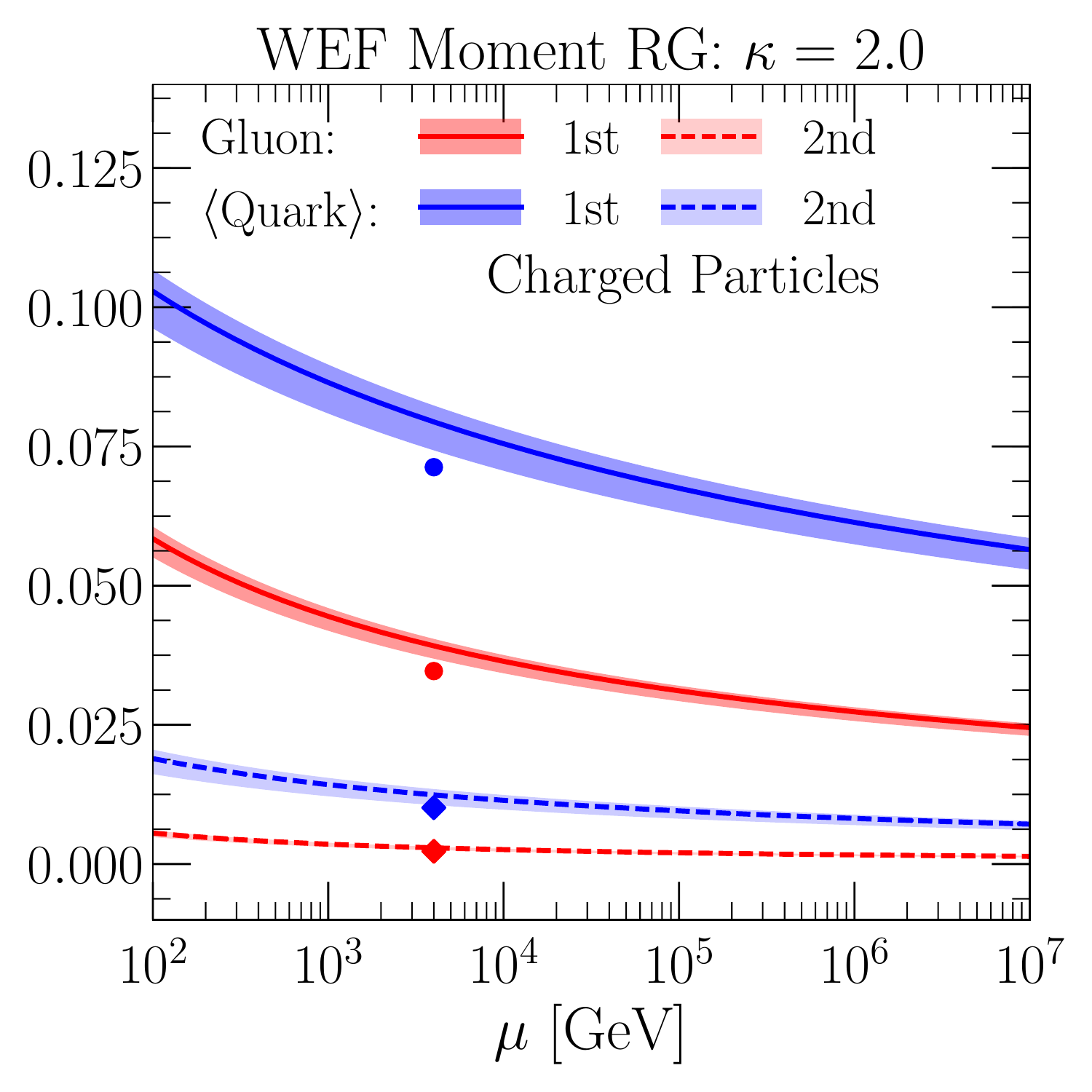}
		 \label{fig:moment-space-wef-e} 
}

\subfloat[]{
		\includegraphics[width=0.32\textwidth]{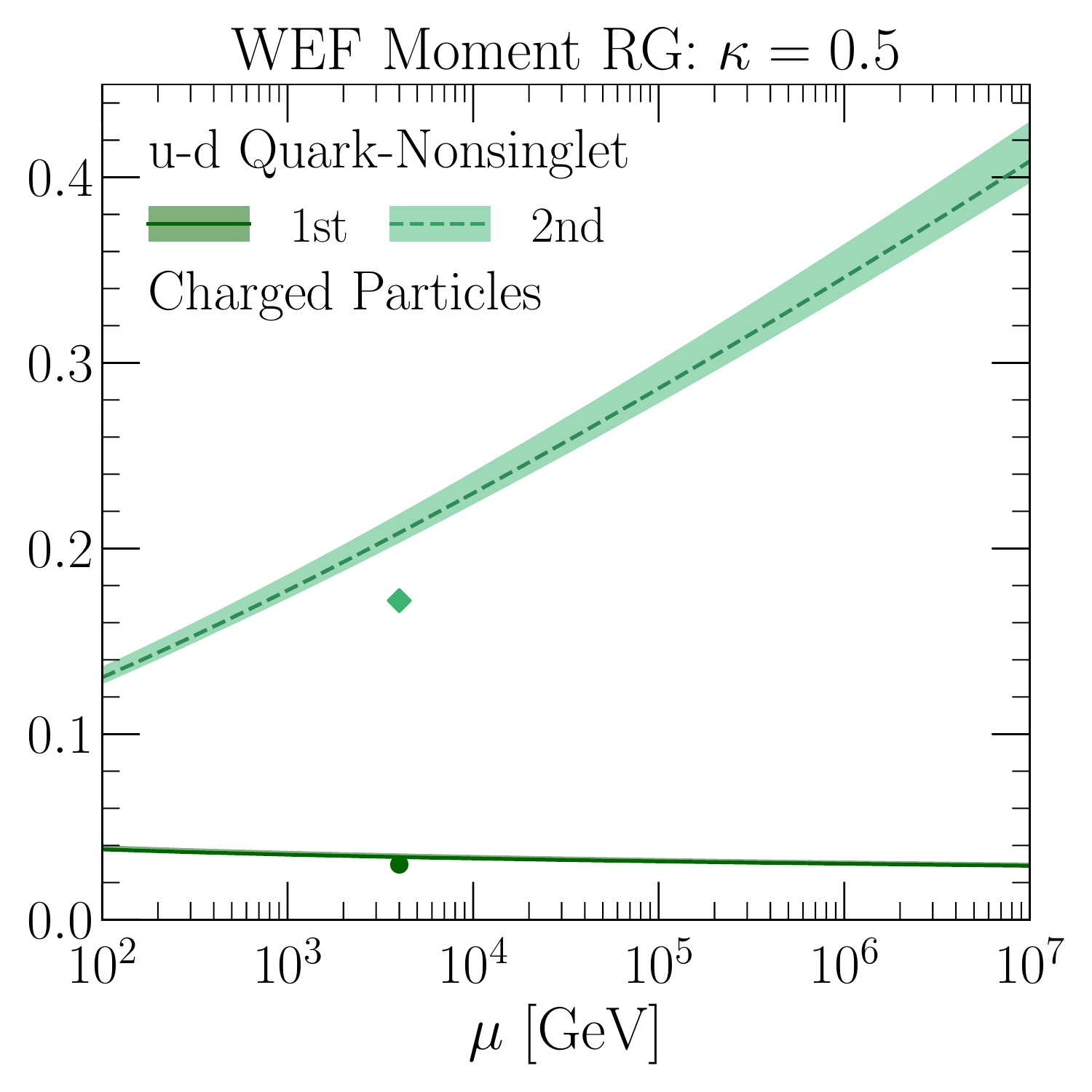}
		 \label{fig:moment-space-wef-b}
}
\subfloat[]{
		\includegraphics[width=0.32\textwidth]{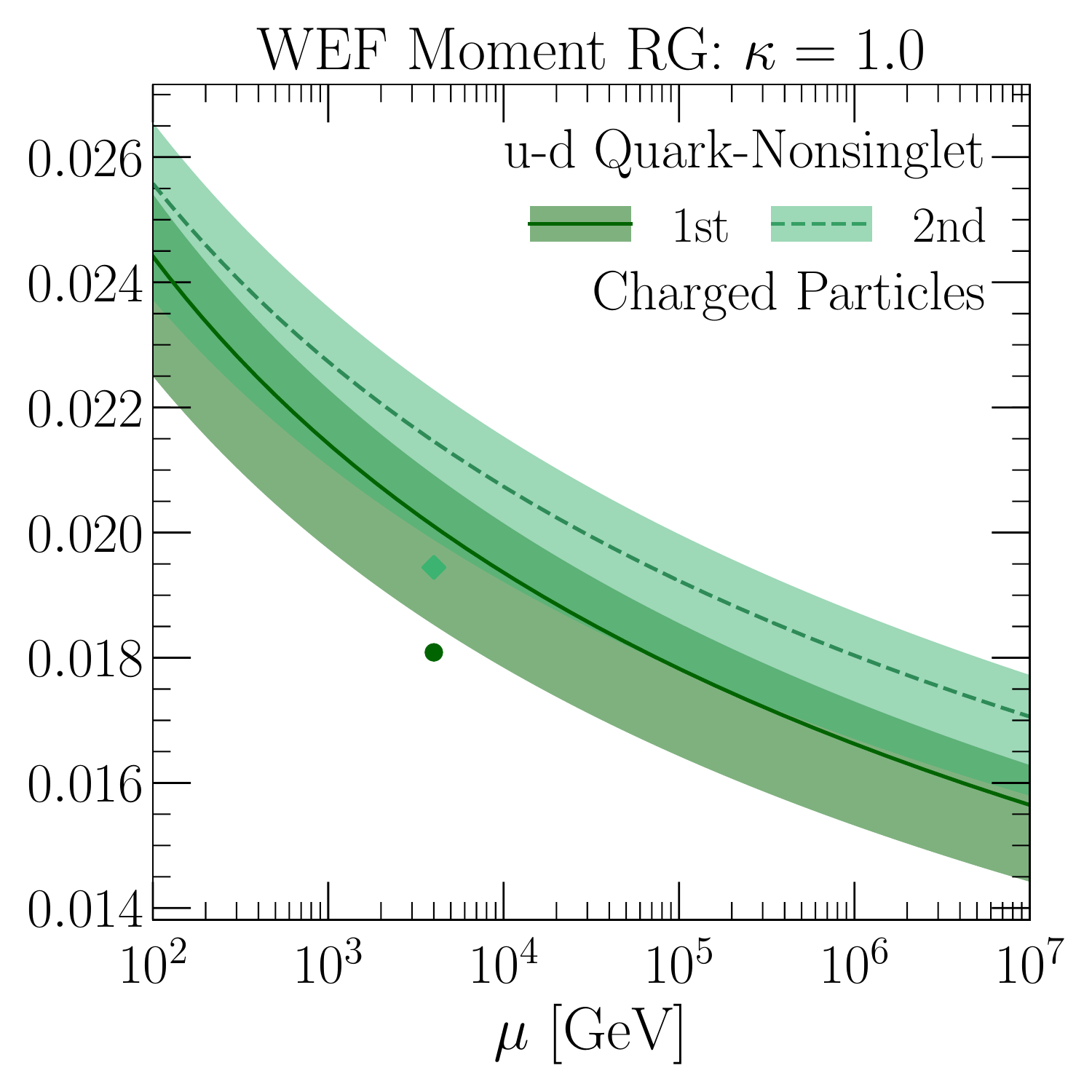}
		 \label{fig:moment-space-wef-d} 
}
\subfloat[]{
		\includegraphics[width=0.32\textwidth]{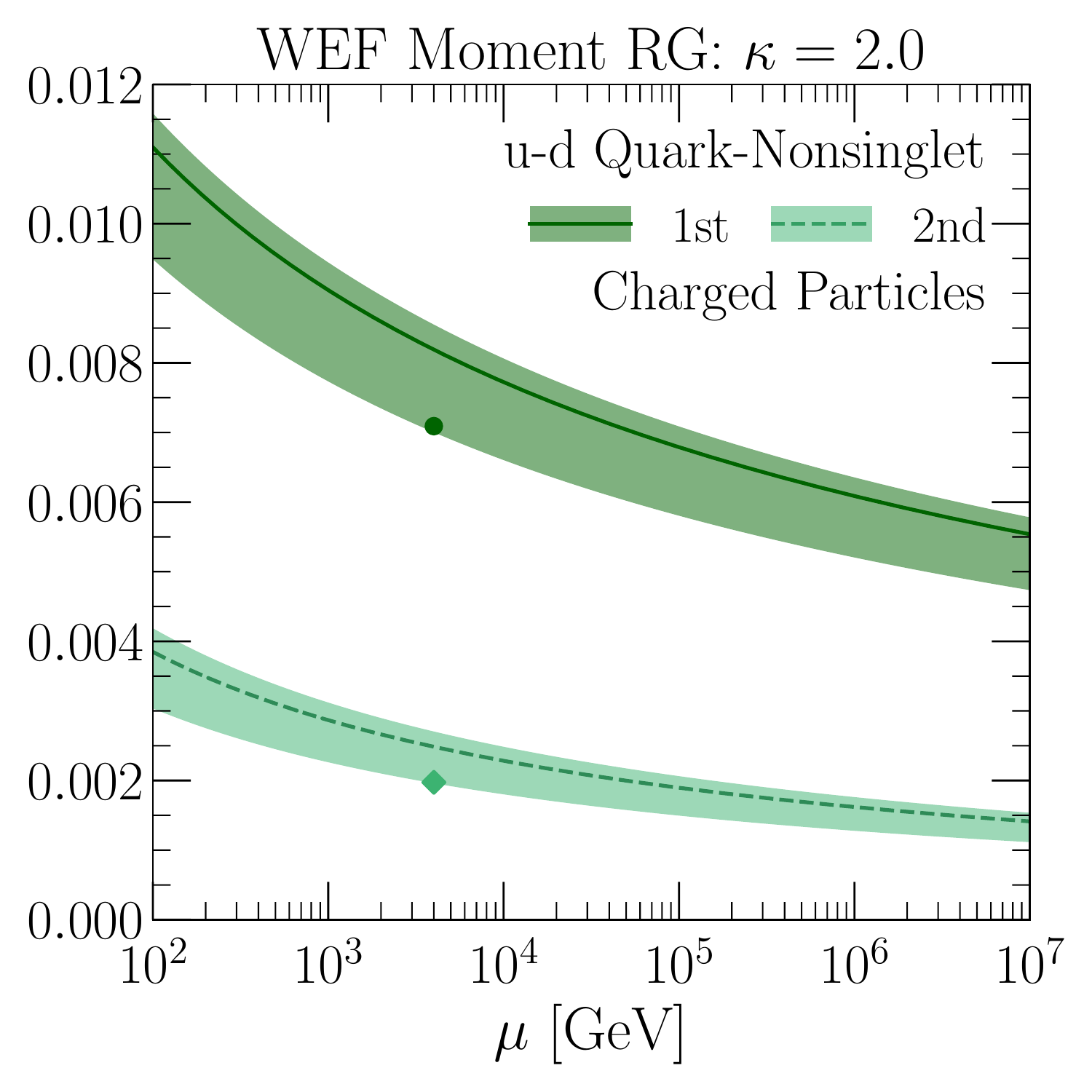}
		 \label{fig:moment-space-wef-f} 
}
	\caption{Evolution of the first and second moments of (top row) the gluon GFFs and quark-singlet GFFs and (bottom row) the $u$-$d$ quark-non-singlet GFFs.   Shown are the first and second GFF moments for weighted energy fractions of charged particles with (left column) $\kappa=0.5$, (middle column) $\kappa=1$, and (right column) $\kappa=2$.   The initial conditions at $\mu = 100$ GeV are obtained from parton showers as described in \Sec{sec:wefextraction}, with uncertainty bands from varying $R$ and changing the parton shower.  The values from the parton shower average at $\mu=4$ TeV are shown as dots (diamonds) for the first (second) moments.}
	\label{fig:moment-space-wef}
\end{figure}

Turning to the second moments, the non-singlet evolution is
\be
\label{eq:2ndmomNS}
\mu \frac{\text{d}}{\text{d}\mu} \overline{\mathcal{N}}_{ij}(2,\mu) = \frac{\alpha_s(\mu)}{\pi}\, \Big [\overline{P}_{q\rightarrow qg}(2\kappa)\,\overline{\mathcal{N}}_{ij}(2,\mu) + 2\overline{P}_{q\rightarrow qg}(\kappa,\kappa)\,\overline{\mathcal{N}}_{ij}(1,\mu)\overline{\GFF}_g(1,\mu) \Big ].
\ee
Since the splitting function in the first term is negative for all values of $\kappa$, this term pushes the second moment of the non-singlet GFFs towards zero as well.   Note, however, that the splitting function in the second term has the opposite sign.  For the weighted energy fractions with $\kappa > 1$, which have $\overline{\GFF}_g(1,\mu) \rightarrow 0$ as $\mu \rightarrow \infty$, this second term is not important, so the different quark GFFs asymptote to the same second moment.  For the weighted energy fractions with $\kappa \le 1$, however, this is not the case. As shown below in \Fig{fig:moment-space-wef-b} for $\kappa = 0.5$, the growth of $\overline{\GFF}_g(1,\mu)$ outpaces the decrease in $\overline{\mathcal{N}}_{ij}(1,\mu)$ from the first term, which leads to differences in the widths (but not the means) of the different quark GFFs.

Assuming the asymptotic behavior $\overline{\mathcal{N}}_{ij}(1,\mu) \to 0$ for simplicity, the evolution of the second moments of the quark-singlet and gluon GFF can be written as
\begin{align}
\label{eq:2ndmomS} \mu \frac{\text{d}}{\text{d}\mu} \left(\begin{array}{c} \overline{\mathcal{S}}(2,\mu) \\ \overline{\GFF}_g(2,\mu) \end{array}\right) &= \frac{\alpha_s(\mu)}{\pi}\left(\begin{array}{c c}  \overline{P}_{q\rightarrow qg}(2\kappa) & \overline{P}_{q\rightarrow gq}(2\kappa)\\ 2n_f\overline{P}_{g\rightarrow q\overline{q}}(2\kappa)  & \overline{P}_{g\rightarrow gg}(2\kappa) \end{array}  \right)\left(\begin{array}{c}\overline{\mathcal{S}}(2,\mu) \\ \overline{\GFF}_g(2,\mu) \end{array}\right) 
\\ & \quad
+\frac{\alpha_s(\mu)}{\pi} \left( \begin{array}{c}2\overline{P}_{q\rightarrow gq}(\kappa,\kappa)\overline{\mathcal{S}}(1,\mu) \overline{\GFF}_g(1,\mu)  \\    2n_f \overline{P}_{g\rightarrow q\overline{q}}(\kappa,\kappa) \left[ \overline{\mathcal{S}}(1,\mu) \right]^2 + \overline{P}_{g\rightarrow gg}(\kappa,\kappa) \left[ \overline{\GFF}_g(1,\mu) \right]^2 \end{array}\right),
\nonumber 
\end{align}
where the assumption allows us to write the nonlinear term as a function of $\overline{S}(1,\mu)$ instead of individual (anti-)quark contributions.\footnote{For the weighted energy fractions in this study, this approximation is very accurate, giving corrections at the per-mille level.}
Due to this nonlinear behavior, we now resort to a numerical analysis.\footnote{Alternatively, if one assumes that $\overline{\GFF}_g(1,\mu)$ and $\overline{S}(1,\mu)$ have also reached their asymptotic behavior, the equation becomes linear again. This approximation turns out to be even more accurate than the assumption $\overline{\mathcal{N}}_{ij}(1,\mu)\rightarrow 0$, though for our numerical studies, we make no simplifications.}

In \Fig{fig:moment-space-wef}, we show an example of the RG evolution of the first and second moments of the gluon GFFs, quark-singlet GFFs, and $u$-$d$ quark-non-singlet GFFs.
Here, we consider weighted energy fractions where charged particles have weight 1 and neutral particles have weight zero, comparing $\kappa = 0.5$, $1$, and $2$.
The evolution starts from GFFs extracted at $\mu=100$ GeV, as described in \Sec{sec:wefextraction}.
The GFF moments are then evolved up to $\mu=10^7$ GeV using the equations above.\footnote{We checked that this agrees with first evolving the full binned distributions and then calculating the first and second moments.}
To connect with the plots in \Figs{fig:wef_gluon}{fig:wef_quark}, we also indicate the first (second) moments extracted from the parton shower average at $\mu =4$ TeV with dots (diamonds).

As expected, the first moments evolve in the direction predicted by the eigenvalues in \Fig{fig:eigen}, with the $\kappa < 1$ first moment moving to larger values as $\mu$ increases, and the $\kappa > 1$ first moment moving to smaller values.
For the boundary case of $\kappa = 1$, the first moment of the gluon and quark singlet GFFs move toward each other, leaving their sum fixed.
The second moments roughly evolve in the same direction as first moments, though with different rates.
The exception is the $\kappa=1$ second moment, where both the gluon and quark singlet values decrease (very slowly), as seen already in \Figs{fig:wef_gluone}{fig:wef_quarke}. 
The first moment of the non-singlet GFFs approaches zero, as indicated by $\overline{P}_{q\rightarrow qg}(\kappa) < 0$. The second moments behave as discussed above, decreasing for $\kappa = 1$ and $\kappa = 2$, and increasing for $\kappa = 0.5$ since $\overline{\GFF}_g(1,\mu)$ grows very large.

We could continue our analysis to third and higher moments, which is a standard way to efficiently solve the DGLAP equations.
An interesting difference with the evolution of the ordinary FFs is that we only get the simple expression in \Eq{eq:momevolutionwef} for integer moments.
In addition, the simple form of \Eq{eq:momevolutionwef} does not hold for general fractal observables with more complicated recursion relations.
For these reasons, we only show the evolution of the first two moments here.
Brief moment-space analyses for the non-associative observables in \Sec{sec:NA} are given in \App{app:momentspace}.

\section{Tree-Dependent Observables}
\label{sec:NA}

We now study fractal jet observables that do depend on the choice of clustering tree.
These are also called non-associative observables, since $x_{\rm A} \not= x_{\rm B} \not= x_{\rm C}$ in the notation of \Eq{eq:ABCtrees}.
We start in \Sec{sec:nodes} with node-product observables, where the recursion relation simplifies to a sum over internal nodes of the tree.
We then turn to a more general family of non-associative observables in \Sec{sec:fulltree}.
%

\subsection{Node Products}\label{sec:nodes}

Node-product observables are based on the recursion relation
\begin{equation}\label{eq:noderecur}
\hat{x} =  x_1 \, z^\kappa + x_2 \, (1-z)^\kappa + (4z(1-z))^{\kappa/2}.
\end{equation}
Note that the last term in \Eq{eq:noderecur} is independent of $x_1$ and $x_2$, and the factor of 4 is added for convenience, to normalize the contribution of a balanced splitting with $z=1/2$ to be 1.
It is straightforward to check that this recursion relation is not associative for generic values of $\kappa$, by considering the three-particle trees in \Fig{fig:trees}.
For the special case of $\kappa = 2$, the recursion relation is associative, yielding an observable closely related to $p_T^D$ (i.e.\ the weighted energy fraction with $\kappa = 2$), 
\be \label{eq:node-product-ptd}
\kappa = 2:  \qquad x = 2 + \sum_{a \in \text{jet}} (w_a - 2) z_a^2.
\ee

For generic values of $\kappa$, this recursion relation simplifies to a sum over the leaves and nodes in the binary tree,
\begin{equation}
\label{eq:nodeext}
x =  \sum_{a \in {\rm jet}} w_a z_a^\kappa + \sum_{\text{nodes}} \left(4 z_L z_R\right)^{\kappa/2}, \qquad z_{L,R} = \frac{E_{L,R}}{E_{\rm jet}},
\end{equation}
where $z_{L,R}$ are the momentum fractions carried by the two branches at this node, relative to the whole jet (i.e.~$z_L + z_R \not= 1$).\footnote{If $z_L$ and $z_R$ had been relative to the node instead, this observable would not be IR safe, as the contribution from an arbitrary soft gluon that subsequently splits collinearly would not be suppressed; see \Eq{eq:IRsafety}.}
To see how this simplification arises, note that the $(4z(1-z))^{\kappa/2}$ term in \Eq{eq:noderecur} adds the product of branch energy fractions to the observable; the $x_1 \, z^\kappa$ and $x_2 \, (1-z)^\kappa$ terms then rescale the energy product to the whole jet momentum.
In this way, node products have intermediate complexity between the weighted energy fractions (with no tree dependence) and more general observables (where the full tree recursion is required).

\begin{figure}[t]
\subfloat[]{
	\includegraphics[width=0.32\textwidth]{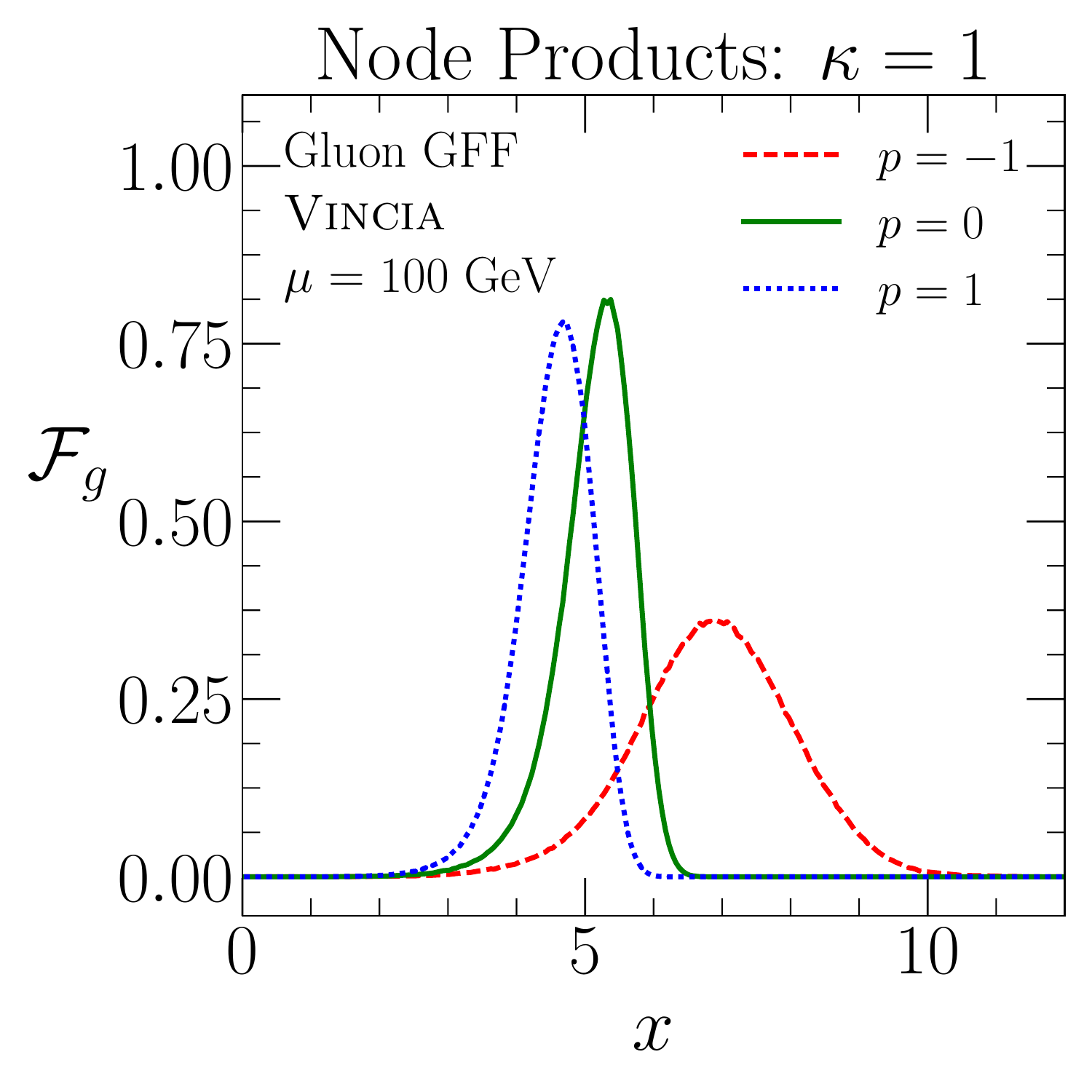}
	\label{fig:node-product-a}
}
\subfloat[]{
	\includegraphics[width=0.32\textwidth]{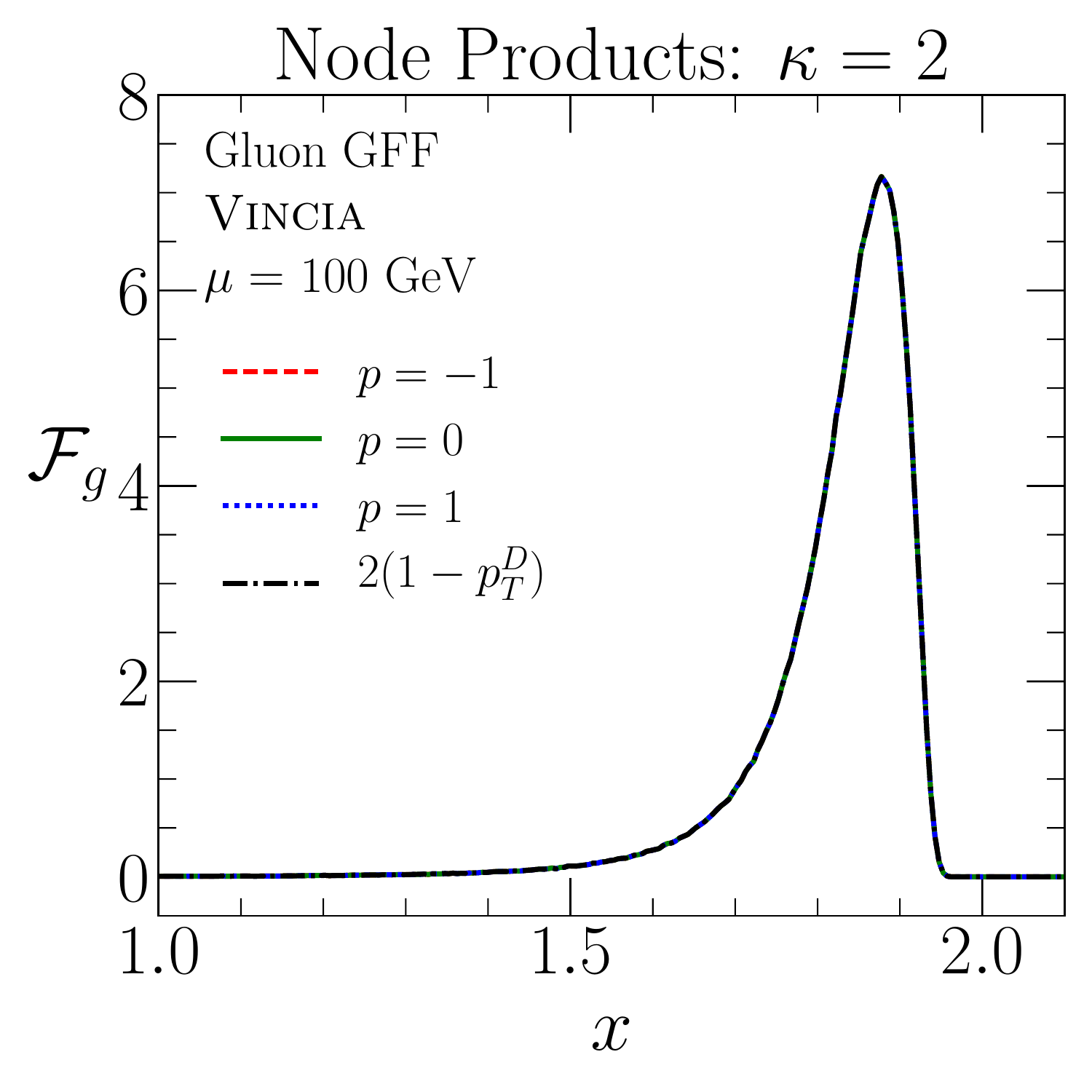}
	\label{fig:node-product-b}
}
\subfloat[]{
	\includegraphics[width=0.32\textwidth]{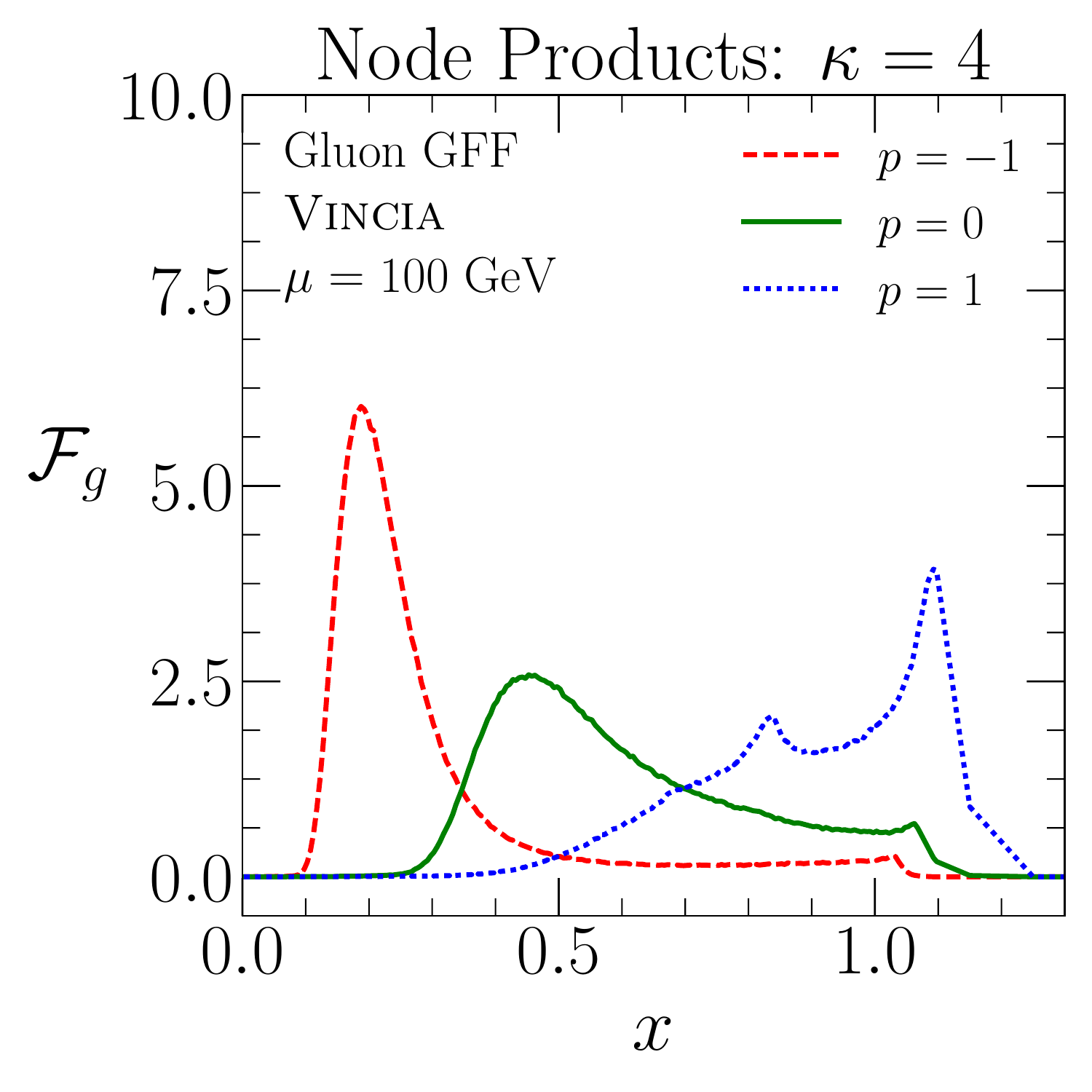}
	\label{fig:node-product-c}
}
	\caption{Gluon GFFs for the node-product observables with $w_a = 0$, taking (a) $\kappa = 1$, (b) $\kappa = 2$, and (c) $\kappa = 4$.  These are extracted from \textsc{Vincia} at $\mu = 100$ GeV.  The tree dependence of these observables is parametrized by the generalized-$k_t$ exponent in \Eq{eq:metric}, with $p=-1$ (anti-$k_t$, red dashed), $p=0$ (C/A, green), and $p=1$ ($k_t$, blue dotted).  For $\kappa = 2$ in (b), there is no tree dependence, as this observable is identical to $2(1-p_T^D)$ (black dot-dashed).}
	\label{fig:node-product}  
\end{figure}

For simplicity, we focus on the case with starting weights of $w_a = 0$, such that the node-product observable only depends on non-leaf nodes, as advertised in \Eq{eq:intro_nodes}.  In \Fig{fig:node-product}, we show the distributions for the gluon GFFs for the node products extracted from \textsc{Vincia} at a jet scale of $\mu = 100$ GeV.  Here, we take $\kappa = \{1, 2, 4\}$, testing three different values of the generalized-$k_t$ clustering exponent $p=\{-1,0,1\}$. 
The tree dependence of this observable for $\kappa=1$ and $\kappa=4$ is evident.  This is particularly true for $\kappa = 4$, where the spikes near $x = 1.1$ (and $x = 0.8$) come from balanced splittings that are more prevalent in $k_t$ trees than C/A or anti-$k_t$ trees.
For $\kappa=2$, the node-product observable is independent of $p$, since it is identical to the associative observable $2(1-p_T^D)$, as shown in \Eq{eq:node-product-ptd}.

\begin{figure}[t]
\subfloat[]{
		\includegraphics[width=0.32\textwidth]{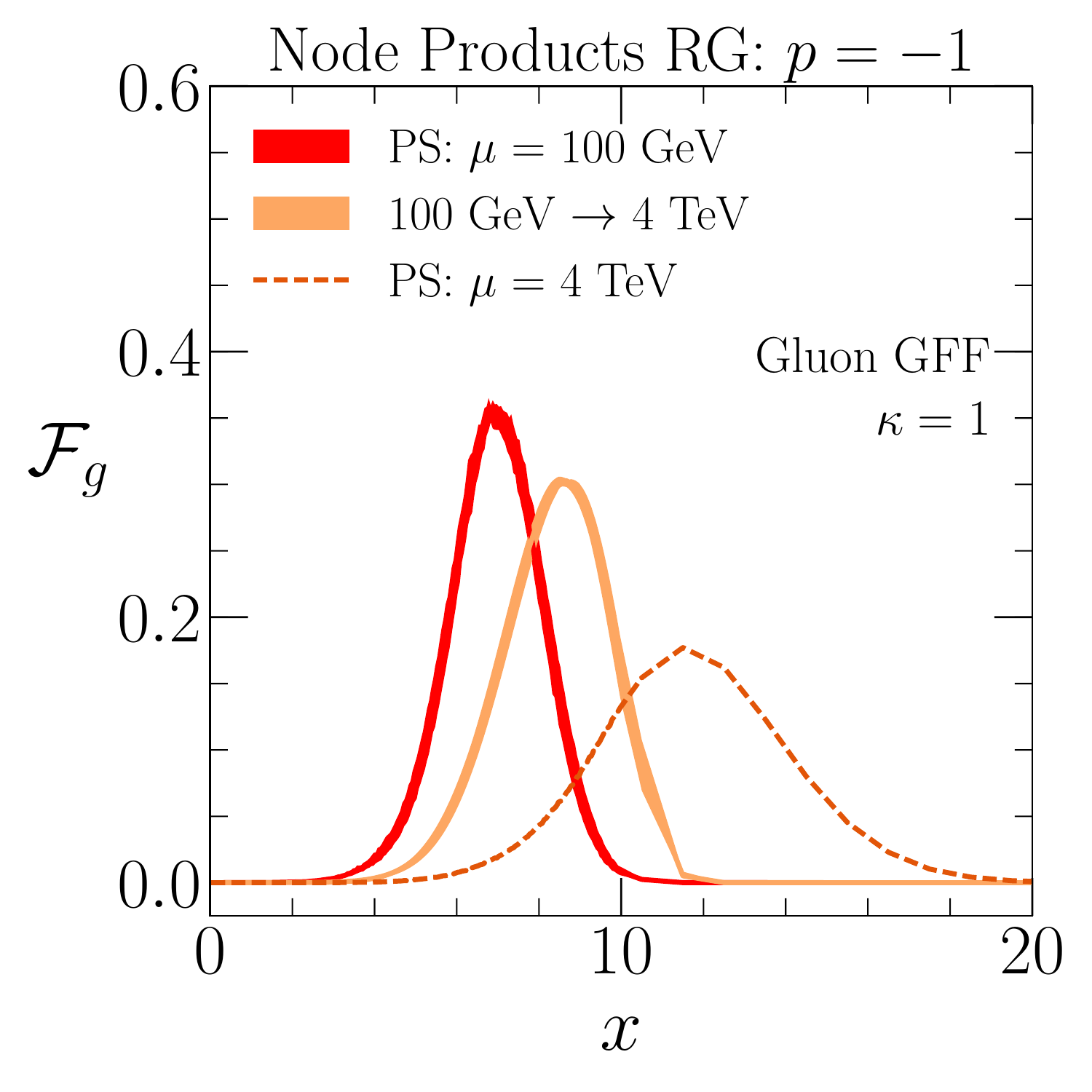}
		 \label{fig:node-evolve-g-TF-1}
}
\subfloat[]{
	\includegraphics[width=0.32\textwidth]{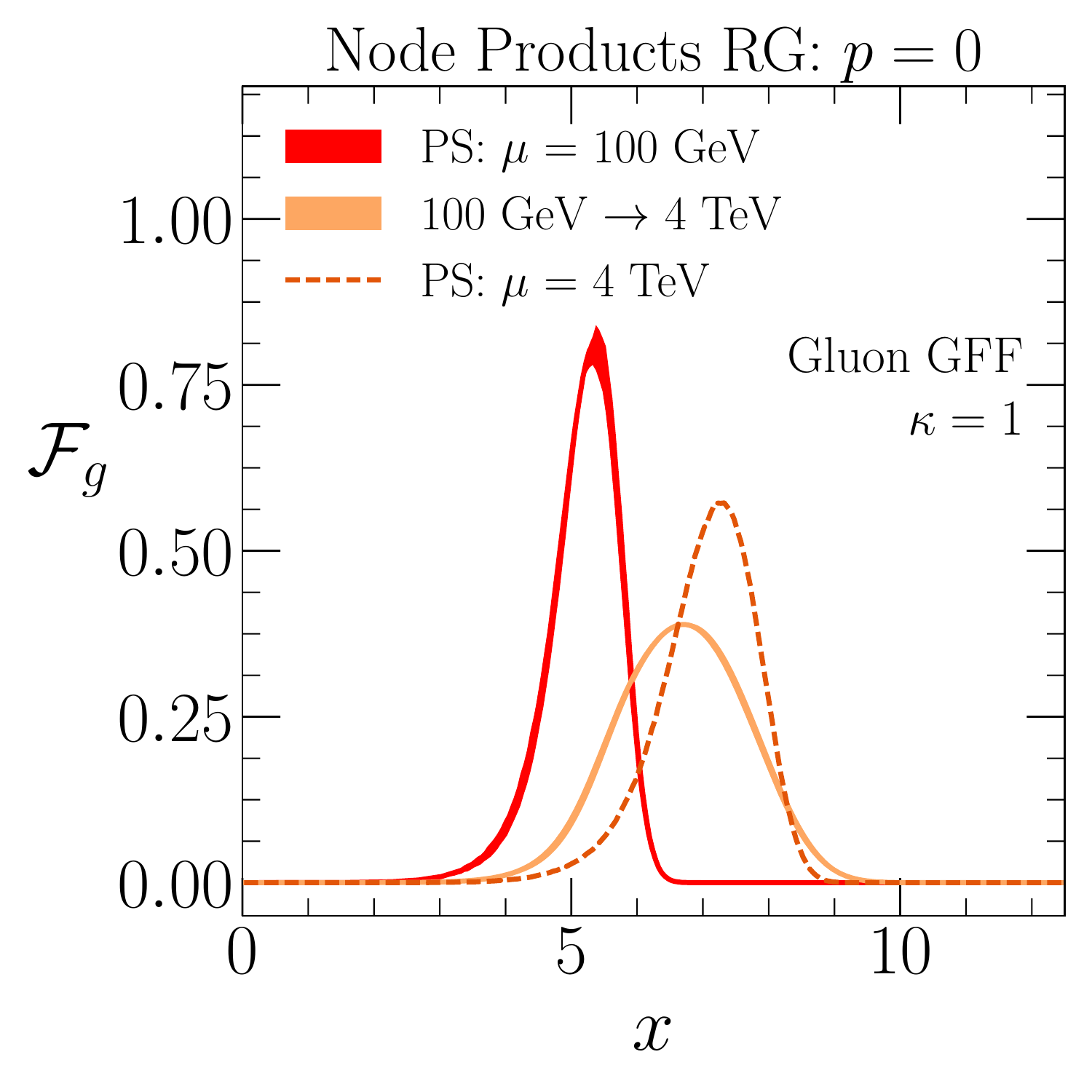}
	\label{fig:node-evolve-g-TF-3}
}
\subfloat[]{
	\includegraphics[width=0.32\textwidth]{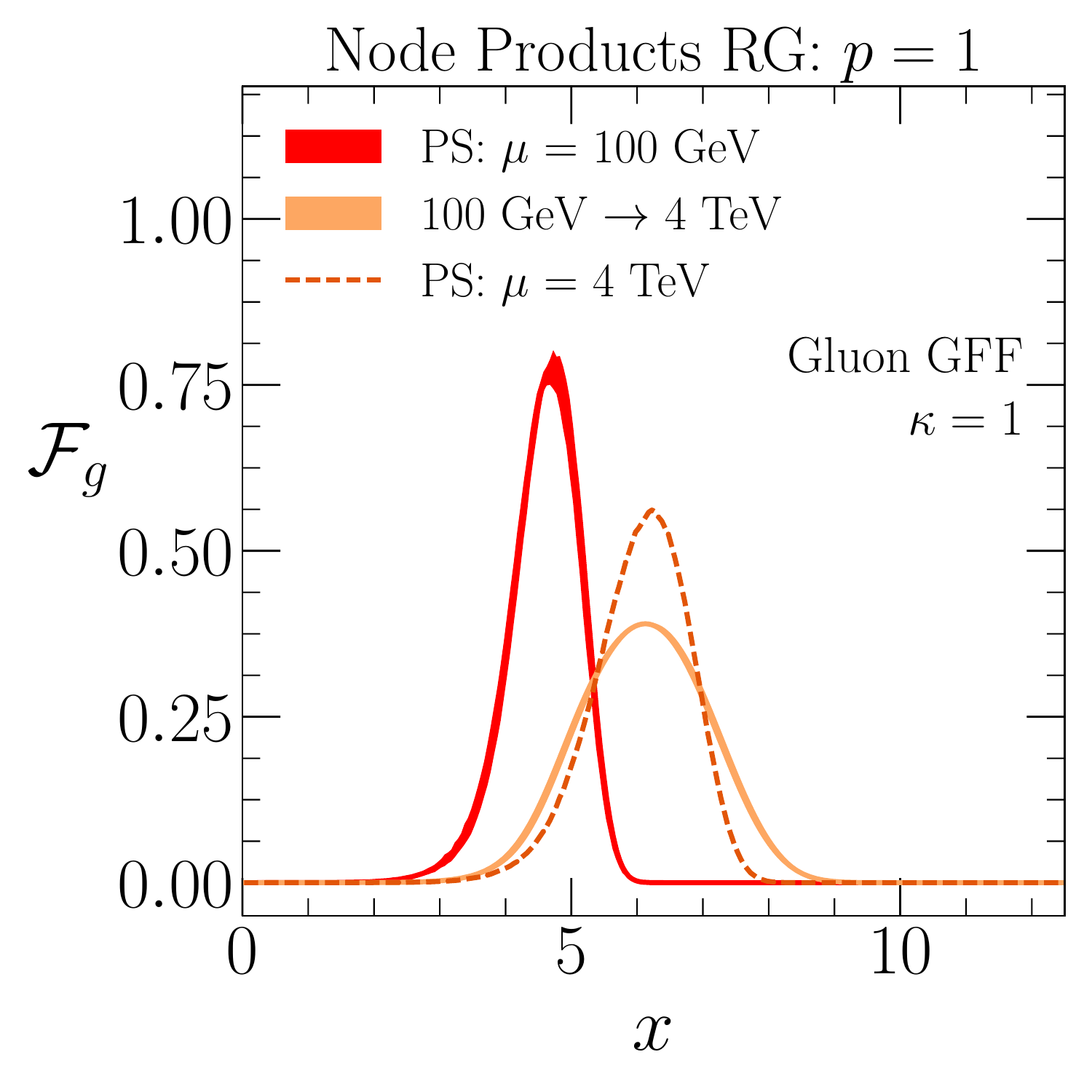}
	\label{fig:node-evolve-g-TF-5}
}

\subfloat[]{
		\includegraphics[width=0.32\textwidth]{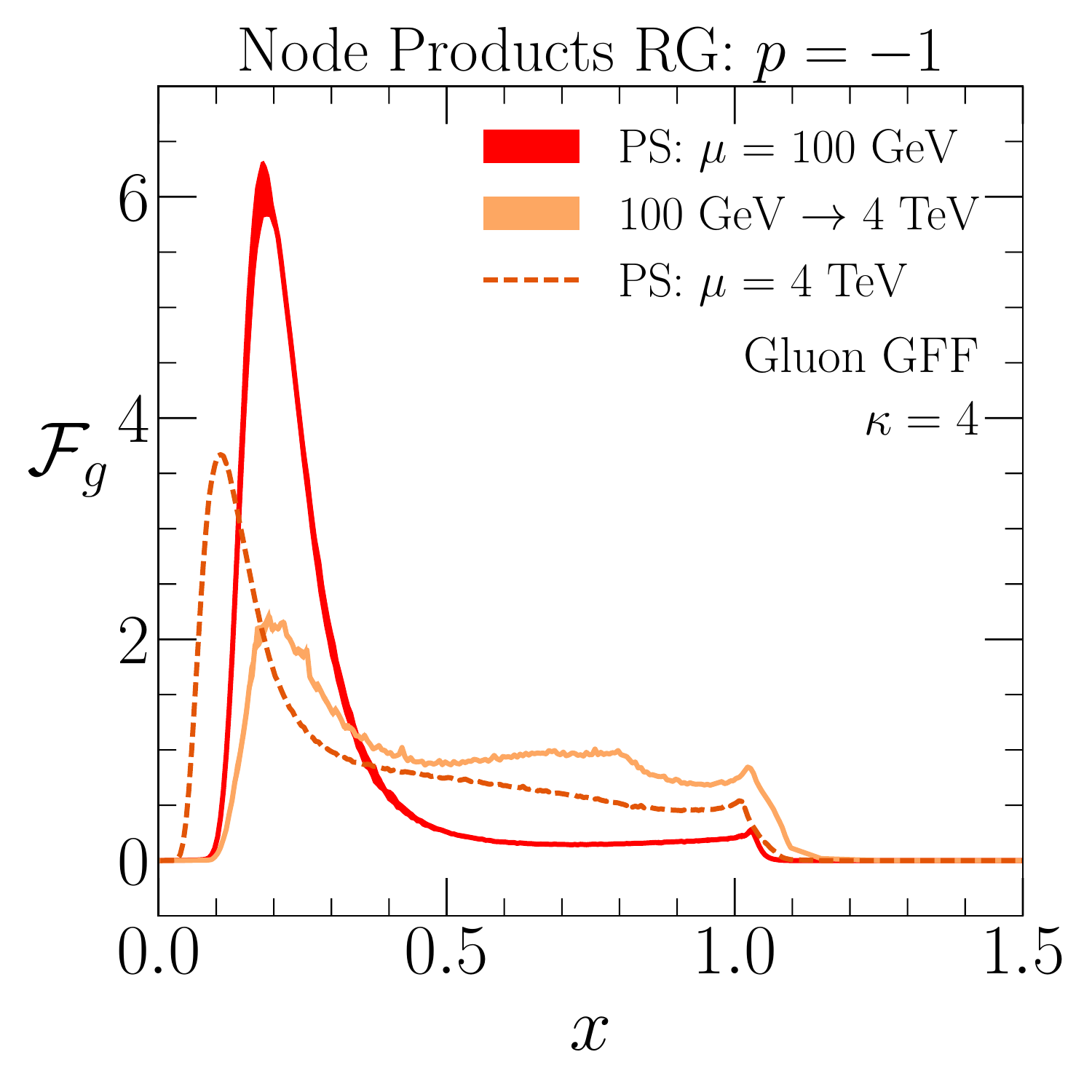}
		 \label{fig:node-evolve-g-TF-2}
}
\subfloat[]{
		\includegraphics[width=0.32\textwidth]{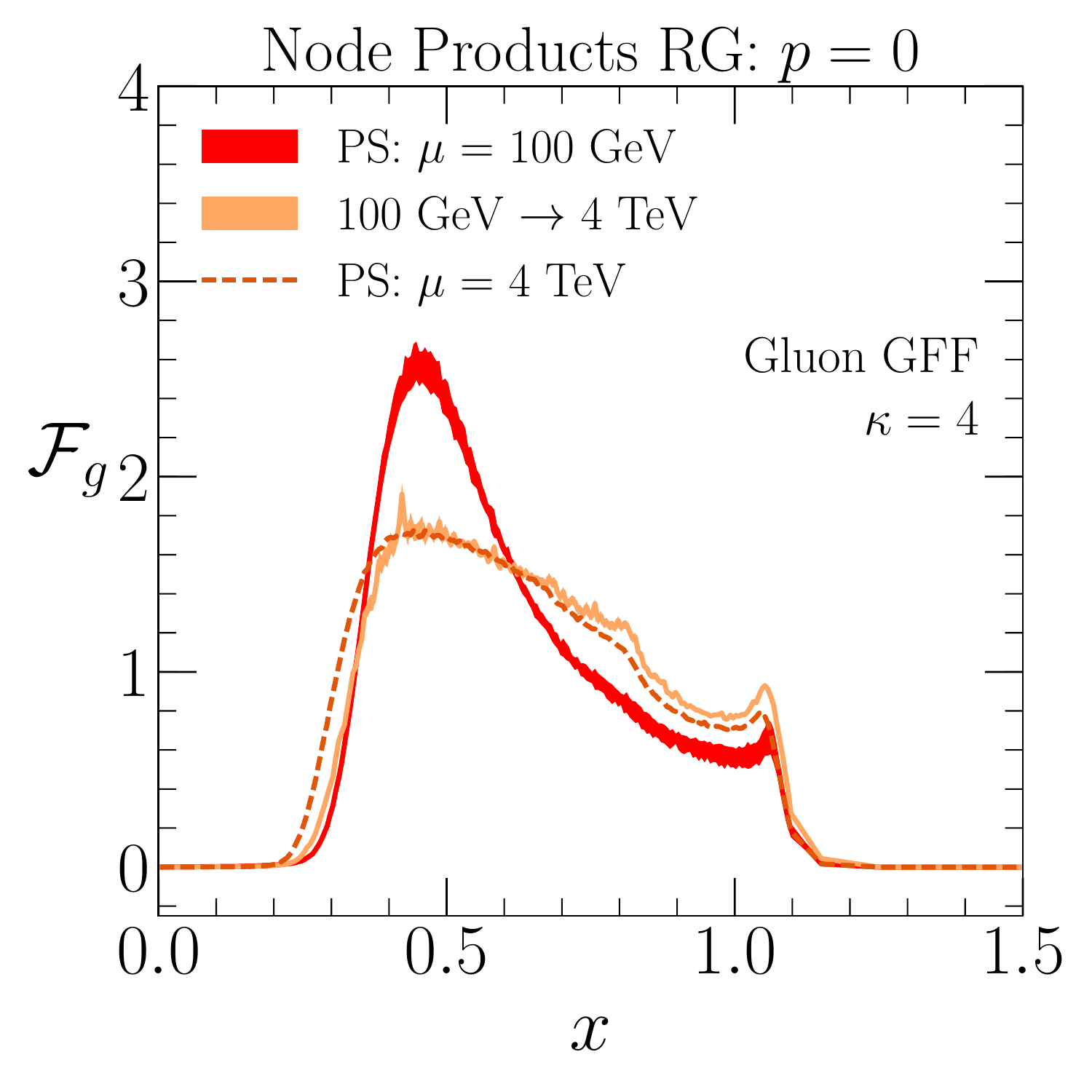}
		 \label{fig:node-evolve-g-TF-4}
}
\subfloat[]{
		\includegraphics[width=0.32\textwidth]{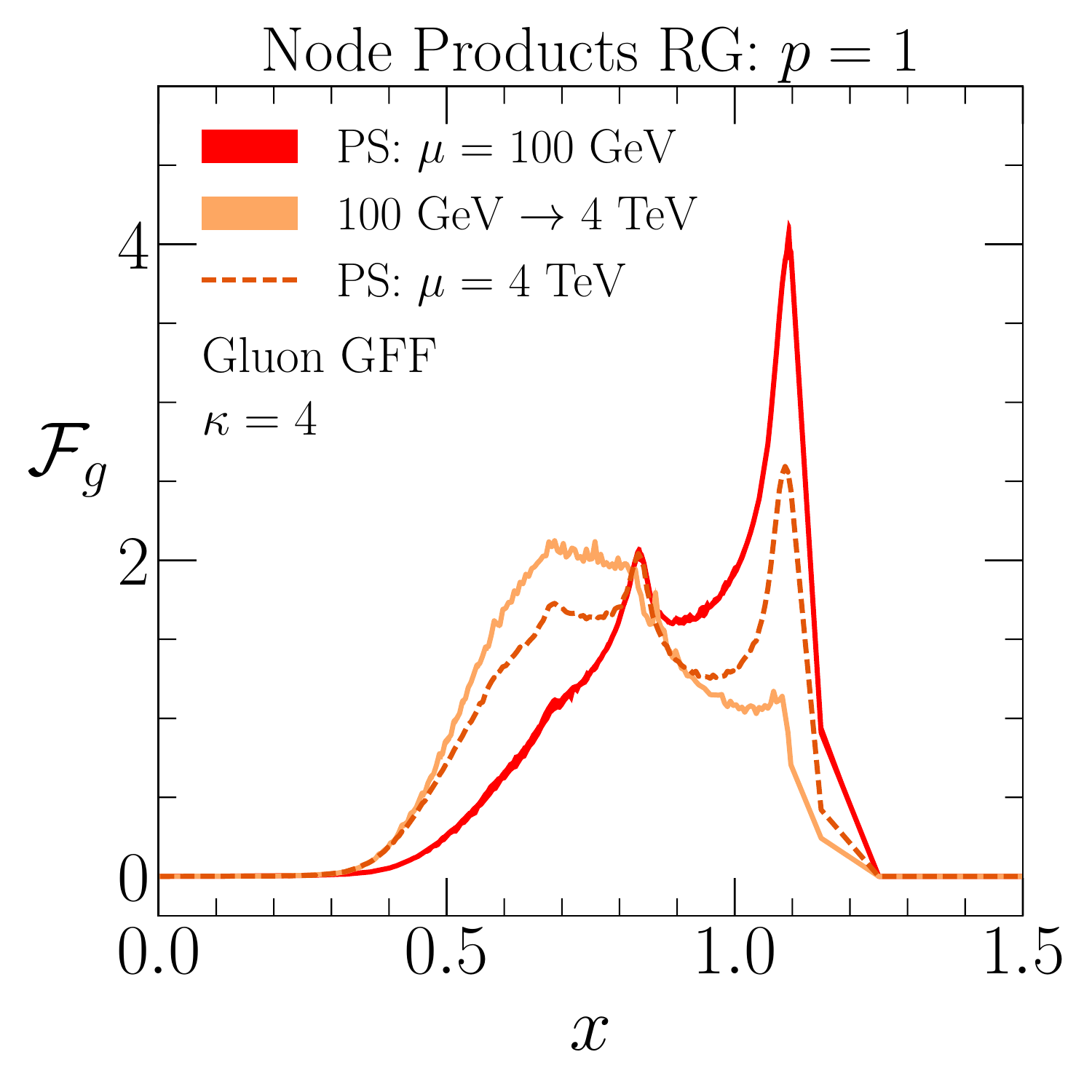}
		 \label{fig:node-evolve-g-TF-6}
}
	\caption{Evolution of the gluon GFFs for node products with (top row) $\kappa=1$ and (bottom row) $\kappa=4$, comparing (left column) $p=-1$, (center column) $p=0$, and (right column) $p=1$. Shown are the gluon GFFs extracted from parton showers at $\mu = 100$ GeV (red solid), the GFFs evolved to $\mu$ = 4 TeV (orange solid), and the GFFs extracted from parton showers at $\mu$ = 4 TeV (orange dashed).  The evolution agrees qualitatively with parton shower predictions, though the agreement is somewhat worse for $p = -1$.} \label{fig:node-evolve-gluon}
\end{figure}

Observables measured on anti-$k_t$ clustering trees tend to be qualitatively distinct from observables measured on $p\ge 0$ trees. This is expected, because C/A and $k_t$ trees are constructed according to angular and $k_t$ ordering, respectively, so these observables more directly mirror the singularity structure of QCD and the expected dynamics of the parton shower. By contrast, anti-$k_t$ trees have a hybrid ordering where angles tend to go from small to large, but energies tend to go from large to small.  Indeed, this reversal in the energy ordering is reflected in \Fig{fig:node-product}, where the product $z_L z_R$ tends to be smaller for anti-$k_t$ trees, leading to larger (smaller) values of node-product observable for $\kappa =1$ ($\kappa = 4$).  Because of this hybrid anti-$k_t$ ordering, one might expect higher-order perturbative corrections to be more important for $p < 0$ when evolving the GFFs, but this can only be confirmed by doing an explicit calculation, which is beyond the scope of the present work.

Despite the fact that different values of $p$ lead to different observables, the leading-order evolution equations are independent of $p$. To check whether this is a sensible feature, we evolve the gluon GFFs in \Fig{fig:node-evolve-gluon} for node products with $\kappa = \{1,4\}$ and $p=\{-1,0,1\}$.
The uncertainty bands in \Fig{fig:node-evolve-gluon} are obtained from the variation of jet radius $R=\{0.3,0.6,0.9\}$ and parton shower PS $=\{\textsc{Vincia},\textsc{Pythia},\textsc{Dire}\}$, as described in \Sec{sec:wefextraction}. 
If the evolution from 100 GeV to 4 TeV would perfectly agree with the extraction at 4 TeV, this would confirm that the evolution is independent of $p$ and all $p$ dependence resides in the initial conditions.  Although the agreement is not perfect, the amount of agreement between the evolution from 100 GeV to 4 TeV and the extraction at 4 TeV seems to be fairly independent of $p$, suggesting that this is a reasonable first approximation. Given the interesting features in the node-product observables as a function of scale, this motivates both higher-order calculations of their RG evolution, as well as measurements in data.

For completeness, we show the evolution of the first and second moments for the node-product observables in \App{app:moment_nodeprod}.

\subsection{Full-Tree Observables}
\label{sec:fulltree}

\begin{figure}[t]
\subfloat[]{
	\includegraphics[width=0.32\textwidth]{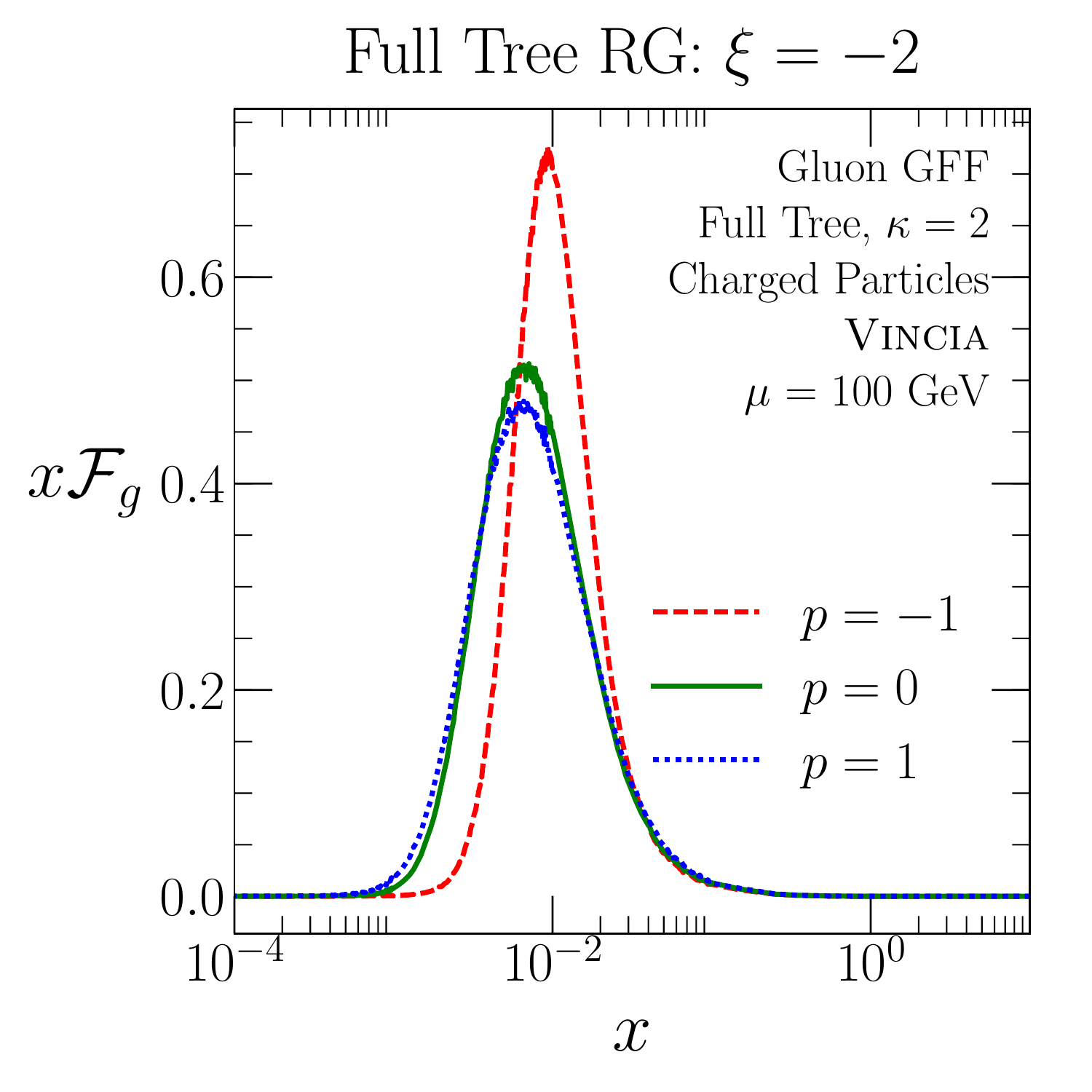}
	 \label{fig:full-tree-extract-a}
}
\subfloat[]{
	\includegraphics[width=0.32\textwidth]{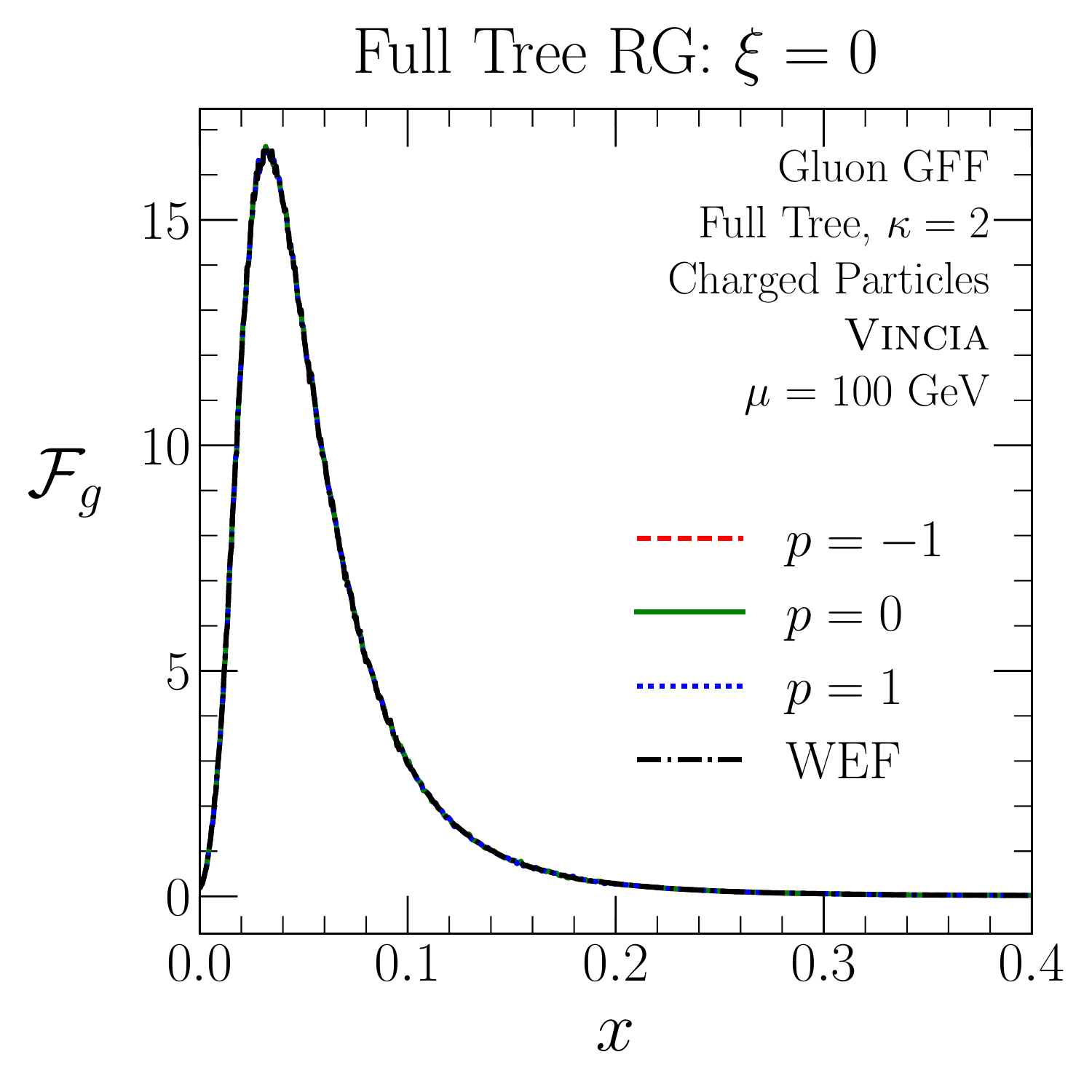}
	\label{fig:full-tree-extract-b}
}
\subfloat[]{
	\includegraphics[width=0.32\textwidth]{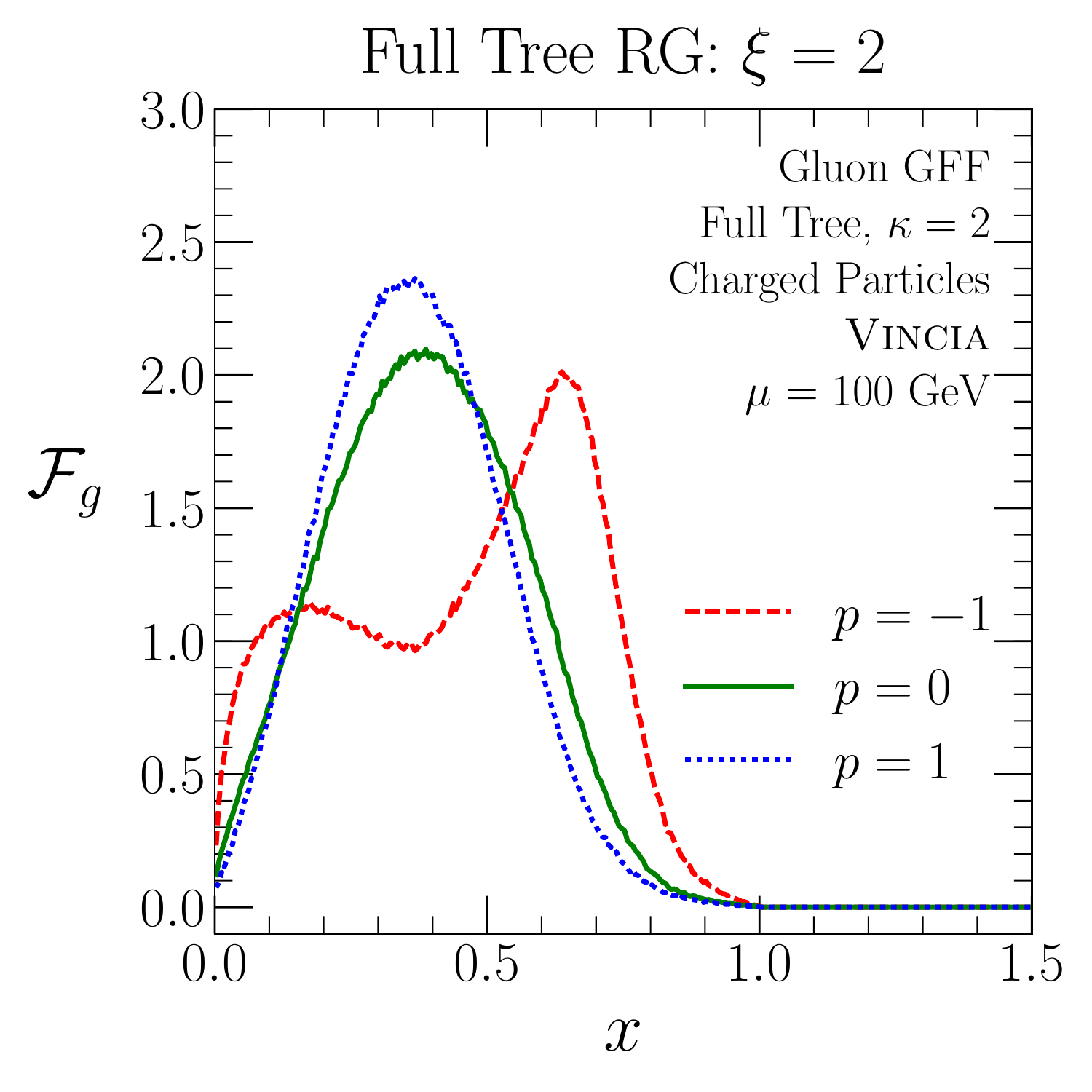}
	 \label{fig:full-tree-extract-c}
}
\caption{Same as \Fig{fig:node-product}, but for the full-tree fractal observable in \Eq{eq:narecursion} defined with $\kappa=2$ on only charged particles, for (a) $\xi=-2$, (b) $\xi=0$, and (c) $\xi=2$. Recall that full-tree observables with $\xi=0$ are the same as weighted energy fractions, so panel (b) is the same as the 100 GeV curve in \Fig{fig:wef_gluonh}, which is plotted as a dash-dotted black line for comparison.} \label{fig:full-tree-extract}
\end{figure}

As our final example of a fractal observable, we present a recursion relation that depends on the full structure of the clustering tree, 
\begin{equation}\label{eq:narecursion}
\hat{x} = \big(z^\kappa x_1 + (1-z)^\kappa x_2 \big) \, e^{\xi z(1-z)}.
\end{equation}
This recursion relation satisfies the requirements in \Eq{eq:IRsafety}, making this observable  IR (but not collinear) safe.
Eq.~\eqref{eq:narecursion} defines a family of fractal observables which depend on the initial particle weights $w_a$, the generalized-$k_t$ clustering exponent $p$, and the parameters $\kappa$ and $\xi$. 
We know of no alternative way to calculate this observable apart from performing the full leaf-to-root recursive traversal of the clustering tree.
Of course, for the special value of $\xi  = 0$, these observables become weighted energy fractions.

\begin{figure}[t]
\subfloat[]{
		\includegraphics[width=0.32\textwidth]{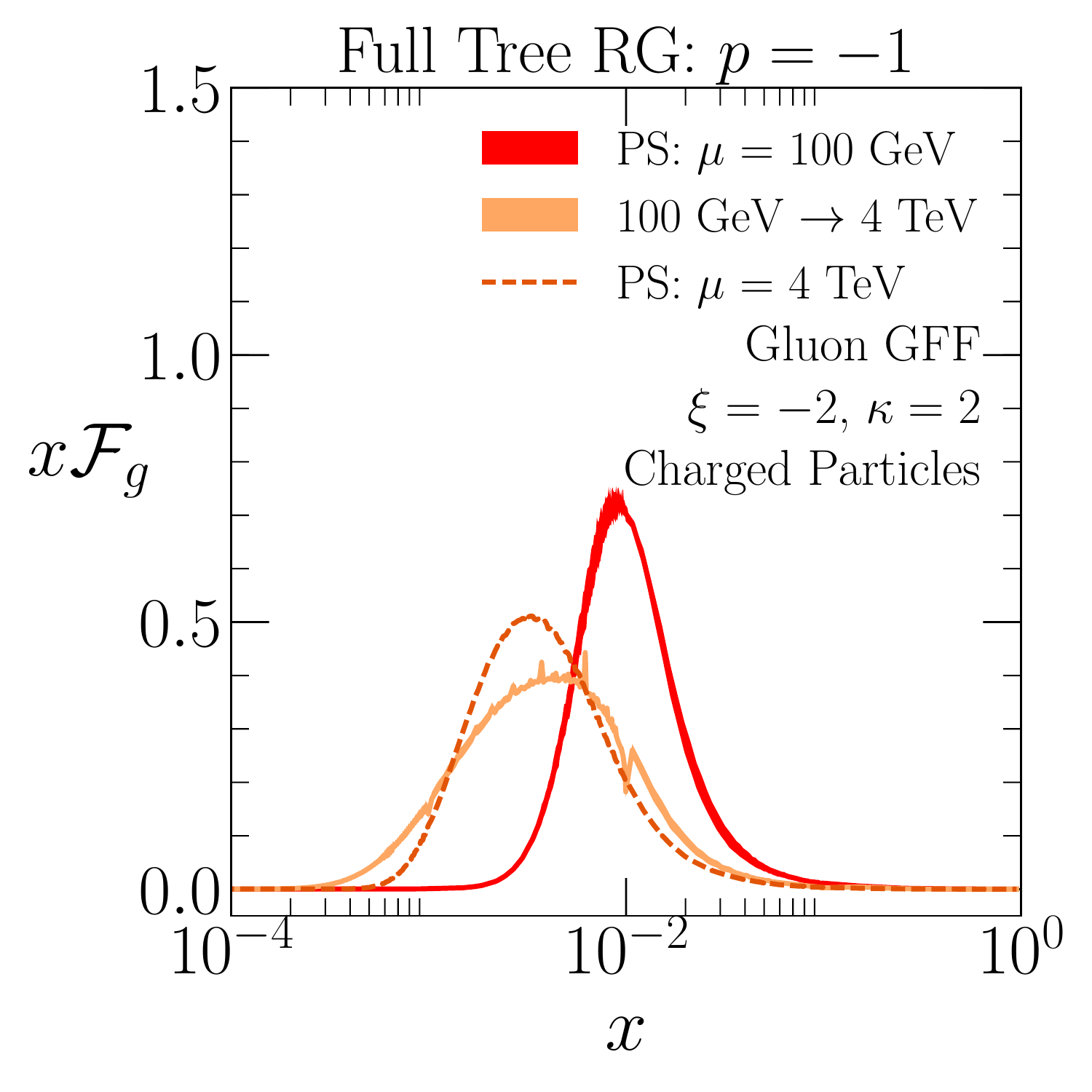}
		 \label{fig:fulltree-evolve-g-TF-1}
}
\subfloat[]{
	\includegraphics[width=0.32\textwidth]{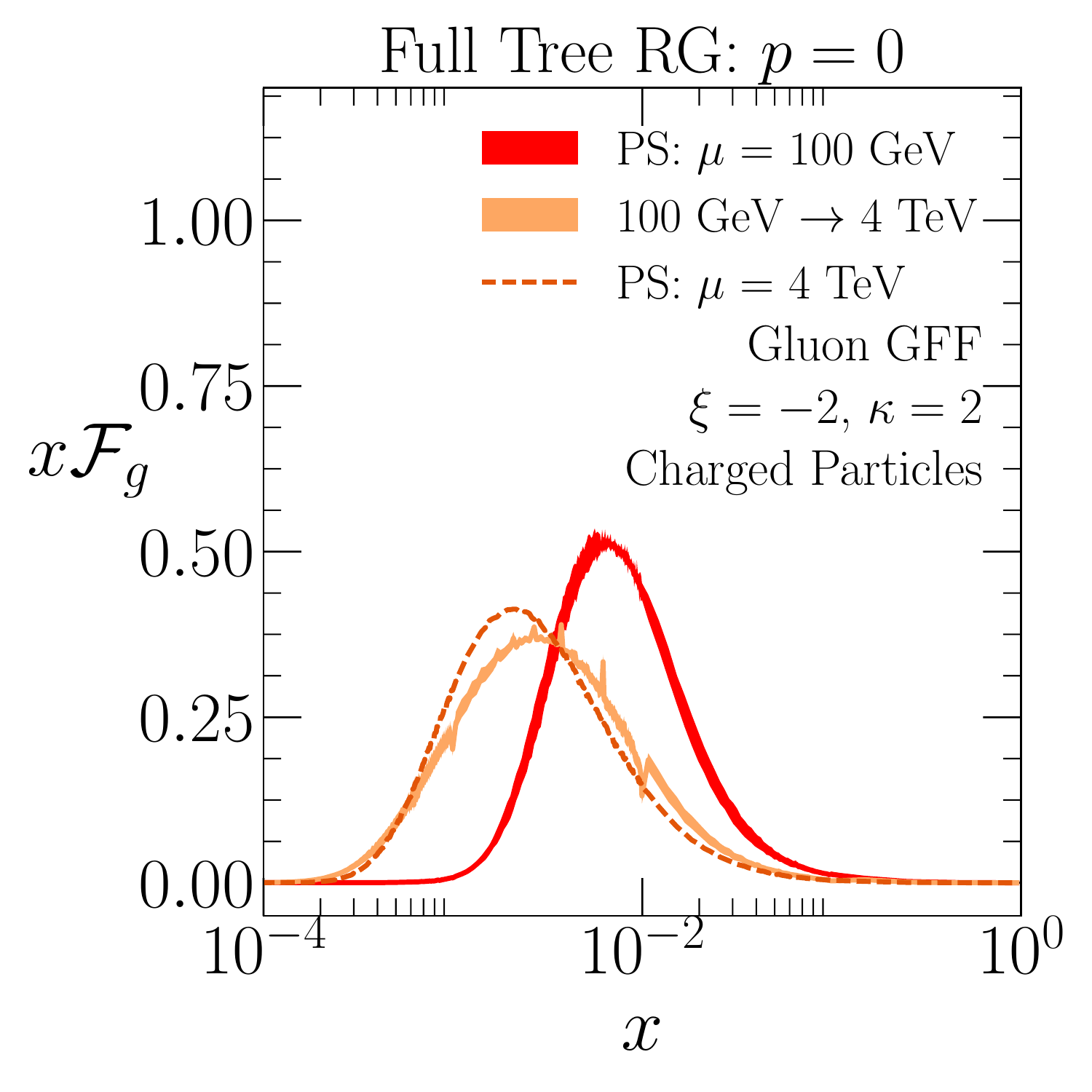}
	\label{fig:fulltree-evolve-g-TF-2}
}
\subfloat[]{
	\includegraphics[width=0.32\textwidth]{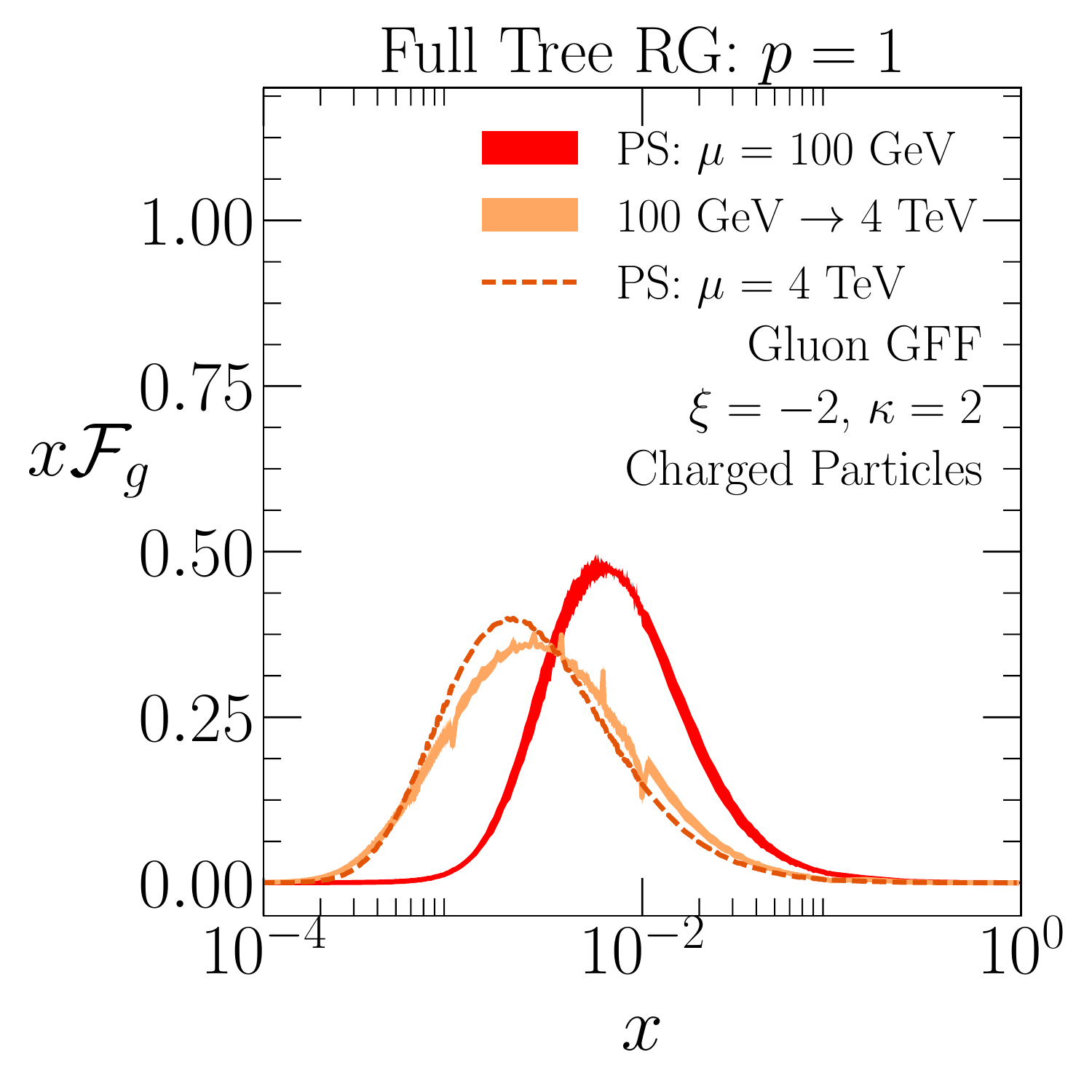}
	\label{fig:fulltree-evolve-g-TF-3}
}

\subfloat[]{
		\includegraphics[width=0.32\textwidth]{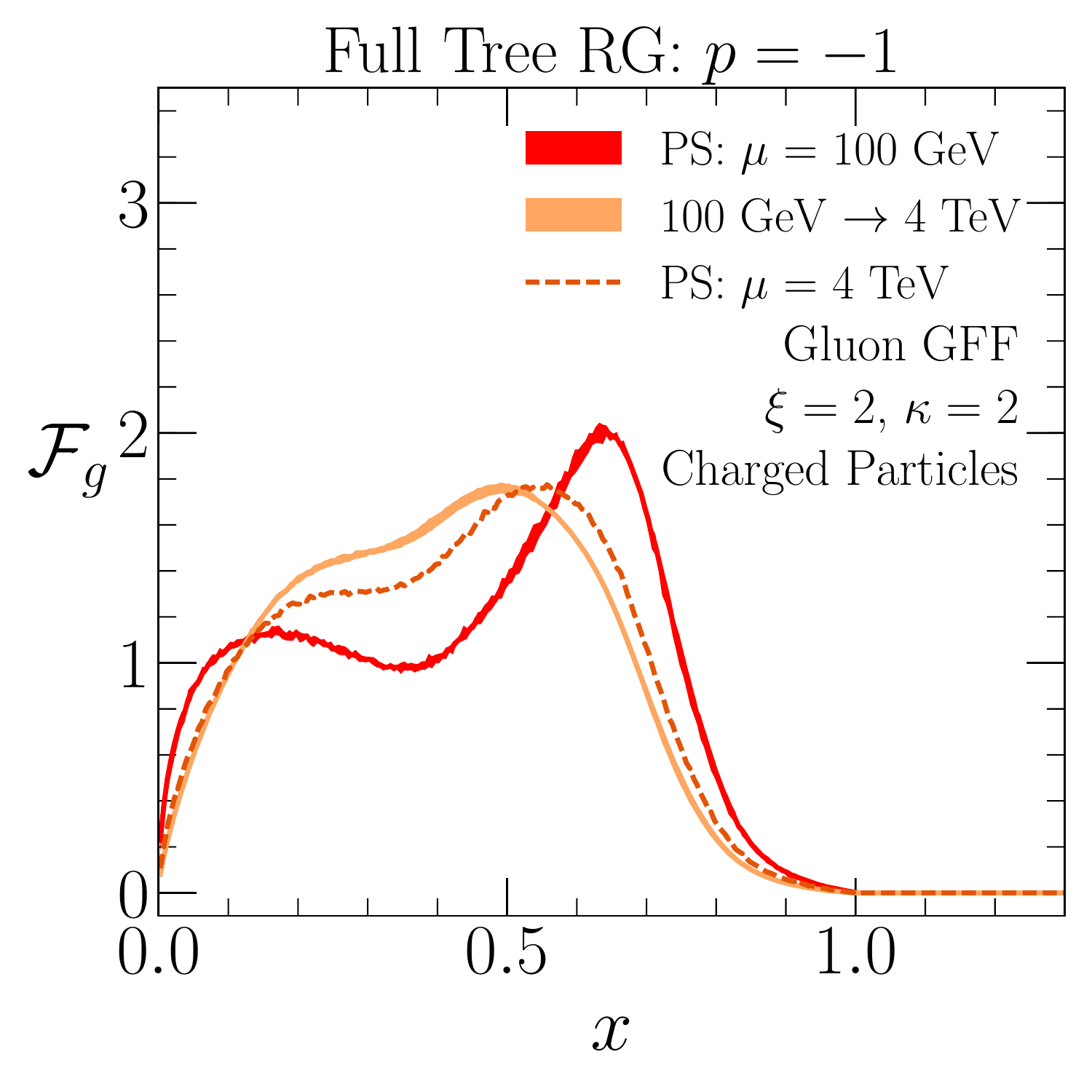}
		 \label{fig:fulltree-evolve-g-TF-4}
}
\subfloat[]{
		\includegraphics[width=0.32\textwidth]{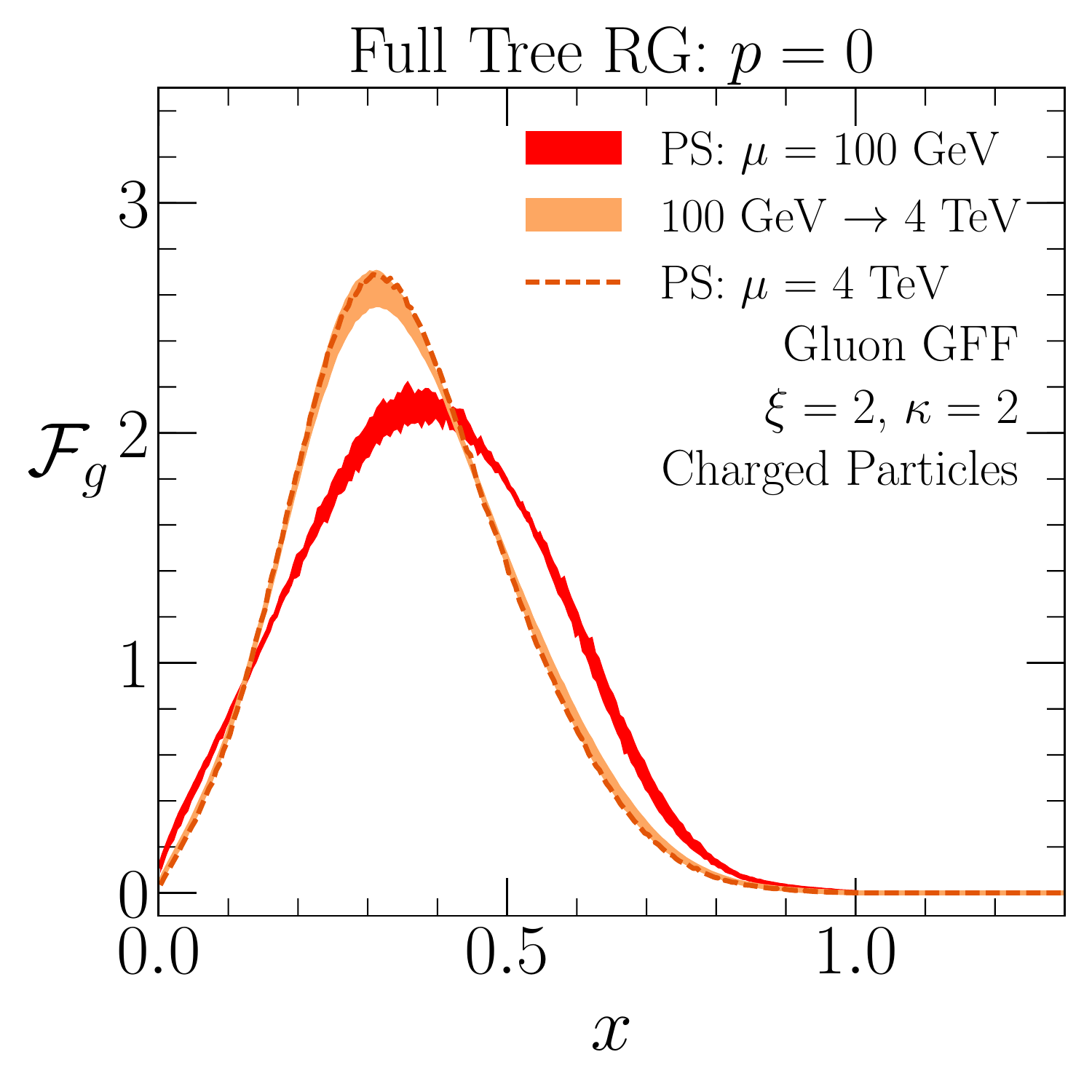}
		 \label{fig:fulltree-evolve-g-TF-5}
}
\subfloat[]{
		\includegraphics[width=0.32\textwidth]{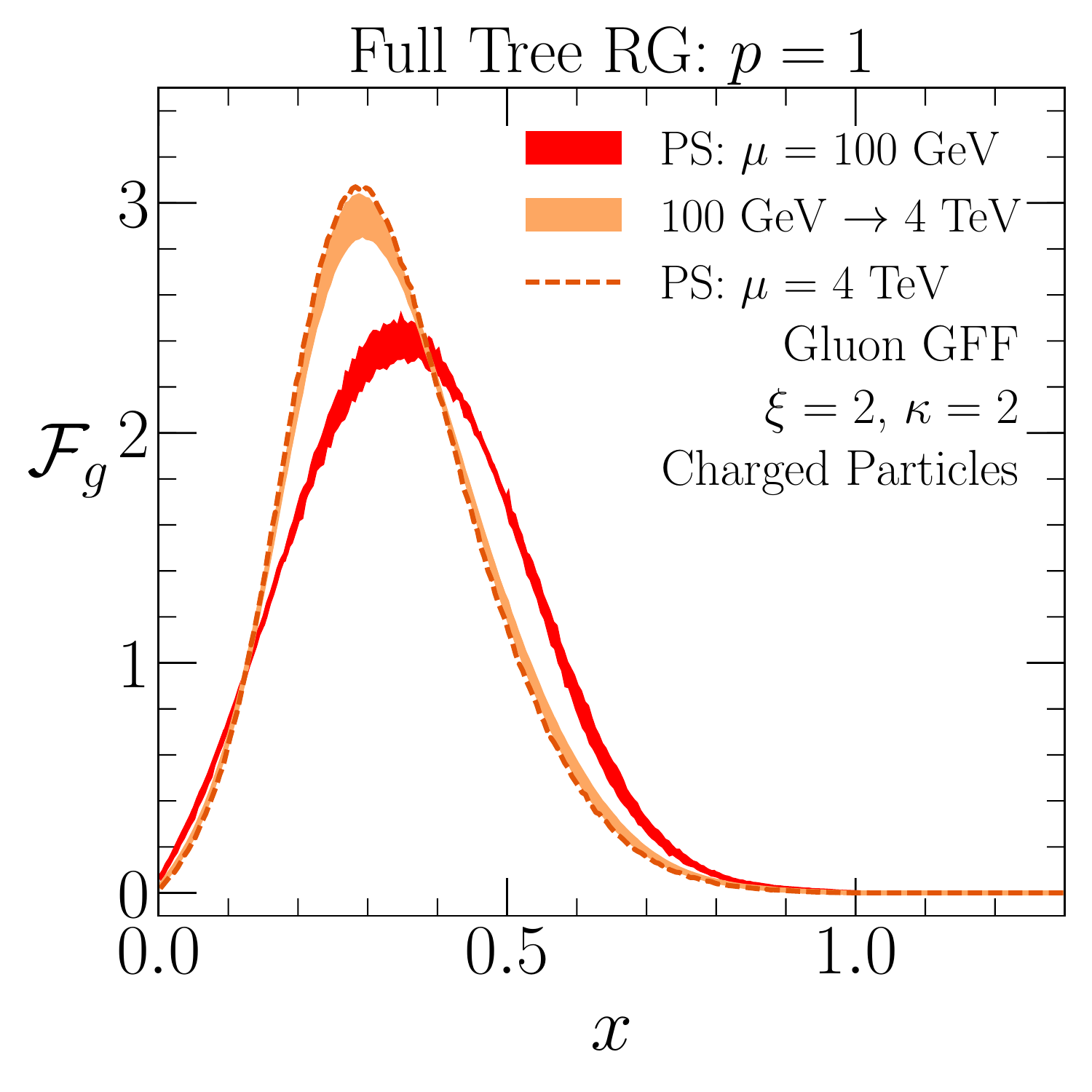}
		 \label{fig:fulltree-evolve-g-TF-6}
}
	\caption{Same as \Fig{fig:node-evolve-gluon}, but for the full-tree fractal observable in \Eq{eq:narecursion} defined with $\kappa=2$ on only charged particles, for (top row) $\xi=-2$ and (bottom row) $\xi=2$.} \label{fig:full-tree-evolve-gluon}
\end{figure}

The tree dependence of this observable is illustrated in \Fig{fig:full-tree-extract} for $\kappa = 2$ and $\xi = \{-2,0, 2\}$, where charged particles are given weight 1 and neutral particles weights 0. 
For nonzero $\xi$, we see that the GFFs depend on the choice of $p$, with rather different behaviors for anti-$k_t$ compared to $k_t$ and C/A. The (associative) observables plotted in \Fig{fig:full-tree-extract-b} are equivalent to the weighted energy fraction with the same weights and $\kappa=2$, shown on this plot for comparison.
Corresponding results for the evolution of the gluon GFFs are shown in \Fig{fig:full-tree-evolve-gluon}.
In this case, it is much clearer that the amount of agreement between the evolution from 100 GeV to 4 TeV and the extraction at 4 TeV is independent of $p$.
Thus, the fact that the leading-order RG evolution is independent of $p$ seems reasonable, even though the GFFs themselves are tree dependent.
This is highlighted by \Fig{fig:fulltree-evolve-g-TF-4}, where the double hump structure at 100 GeV is smoothed out both by the RG evolution equations and the parton shower.

Again for completeness, we discuss the evolution of the first two GFF moments for these full-tree observables in \App{app:momentFT}.

\section{Application in Quark/Gluon Discrimination}
\label{sec:quarkgluon}

Robust and efficient discrimination between quark- and gluon-initiated jets is a key goal of the jet substructure community~\cite{Abdesselam:2010pt,Altheimer:2012mn,Altheimer:2013yza,Adams:2015hiv}, with applications both in searches for physics beyond the SM and precision tests of QCD (see further discussions in \cite{Gallicchio:2011xq,Gallicchio:2012ez,Larkoski:2013eya,Larkoski:2014pca,Bhattacherjee:2015psa,Badger:2016bpw,FerreiradeLima:2016gcz,Komiske:2016rsd,Davighi:2017hok,Gras:2017jty}).
Weighted energy fractions are already used for quark/gluon discrimination, specifically the $p_T^D$ observable~\cite{Pandolfi:1480598,Chatrchyan:2012sn} used by CMS in its quark-gluon likelihood analysis \cite{CMS:2013kfa}.
Here, we explore the potential discrimination power of non-associative fractal jet observables, corresponding to non-associative variants of $p_T^D$.
An alternative application of the GFF formalism to quark/gluon discrimination will be presented in \Ref{HarvardInProgress}.

It is not immediately obvious that non-associativity should be a valuable feature to help distinguish quark- from gluon-initiated jets.  Compared to $p_T^D$, non-associative observables are of course sensitive to the angular structure of the jet through the clustering tree.  Then again, discriminants like the (generalized) angularities~\cite{Berger:2003iw,Almeida:2008yp,Ellis:2010rwa,Larkoski:2014pca} and energy correlation functions \cite{Larkoski:2013eya} also encode angular information about particles in the jet, either their angular distance to the jet axis or their pairwise angular distance to each other.  As we will see, there are non-associative observables that do exhibit better performance than $p_T^D$, at least in the context of a parton shower study, but we do not (yet) understand the origin of that improvement from first principles.

\begin{figure}[t]
\subfloat[]{
		\includegraphics[width=0.45\textwidth]{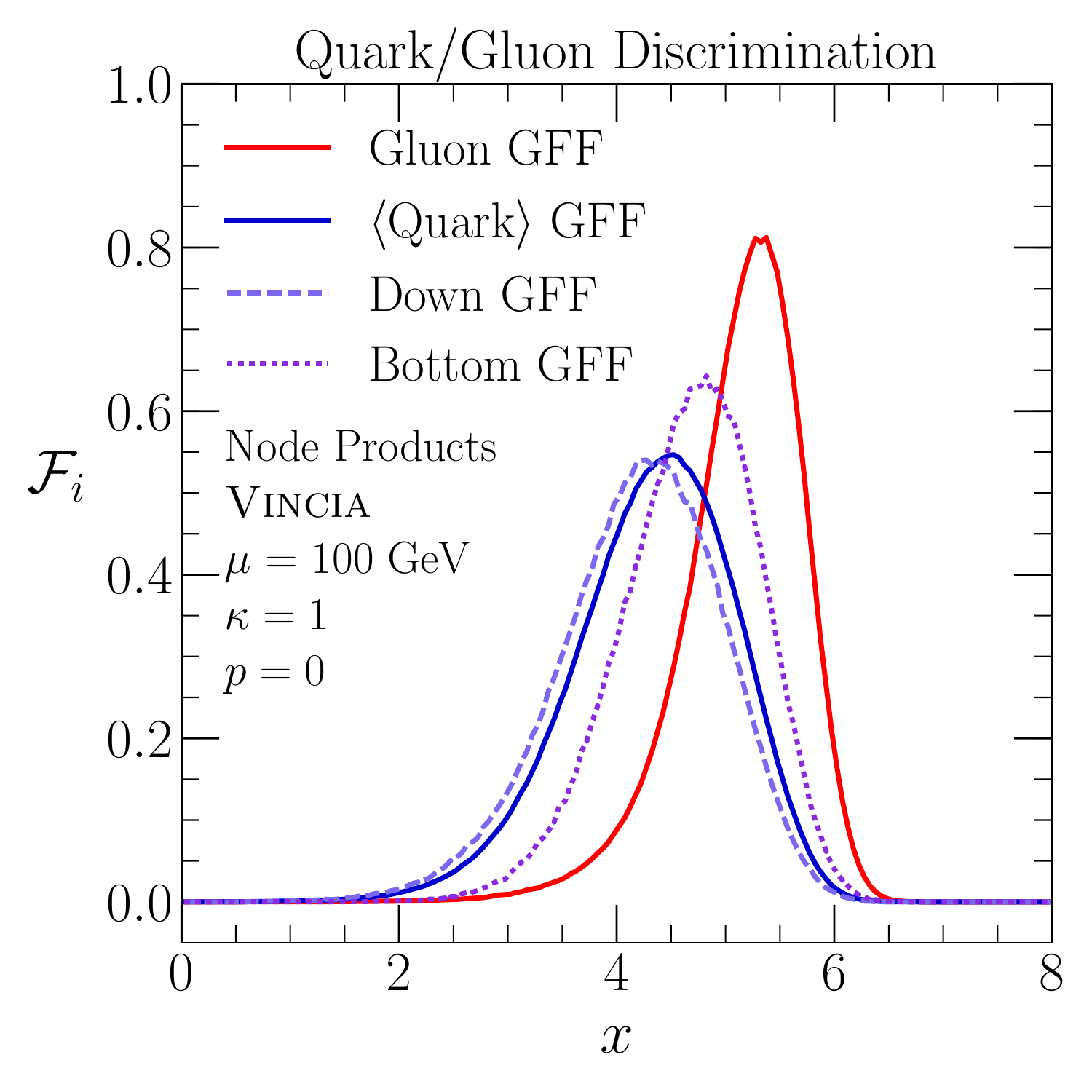}
		 \label{fig:qgdiscrimination-a}
}
\subfloat[]{
		\includegraphics[width=0.45\textwidth]{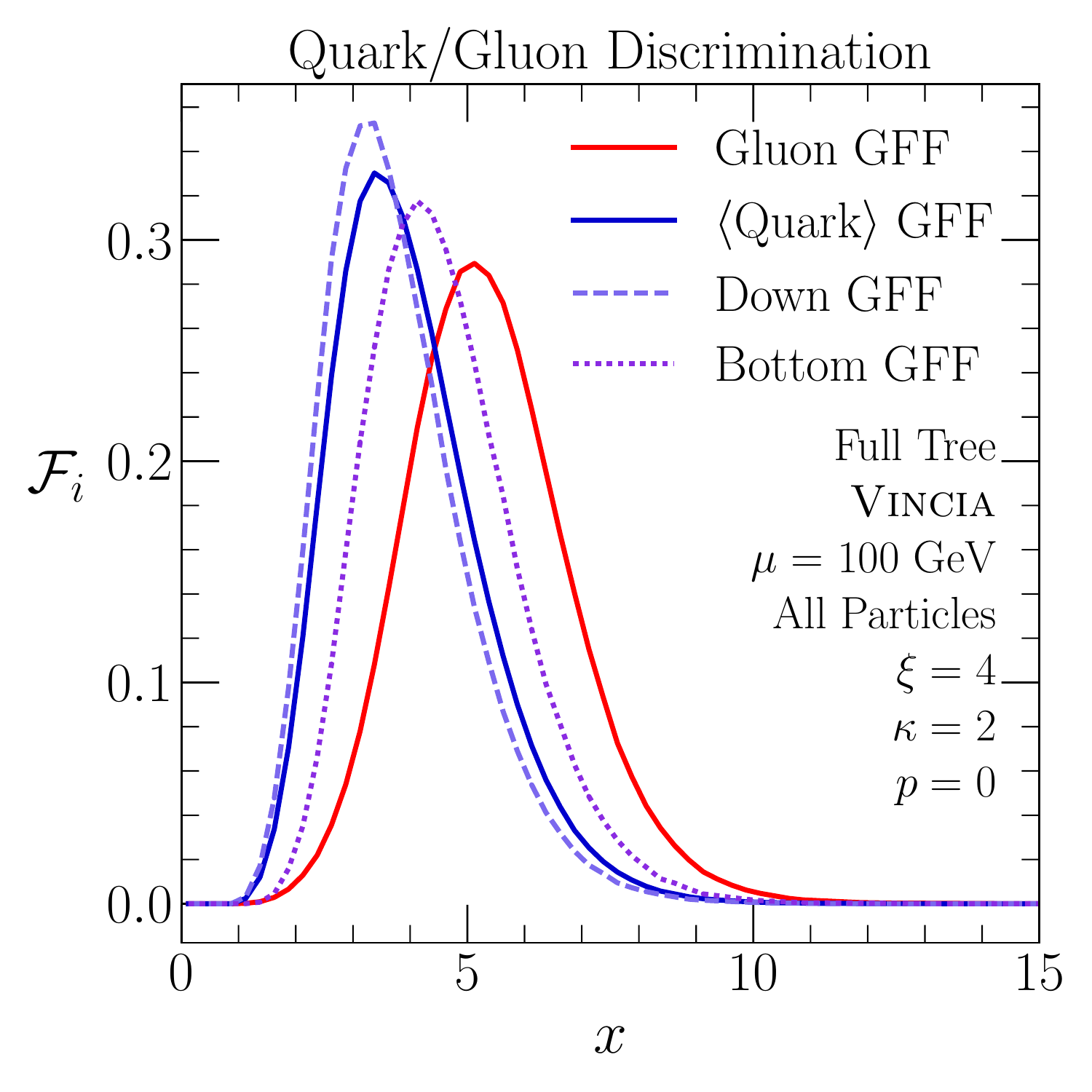}
		 \label{fig:qgdiscrimination-b}
}
	\caption{GFFs for two strong quark/gluon discriminants based on C/A trees:  (a) the node-product observable with $\kappa = 1$, and (b) the full-tree observable with $\kappa = 2$ and $\xi = 4$ with all particle weights one.  Shown are the gluon GFF (red solid), quark-singlet GFF (blue solid), down-quark GFF (light-blue dashed), and bottom-quark GFF (violet dotted) as extracted from $\textsc{Vincia}$ at $\mu = 100$ GeV. }\label{fig:qgdiscrimination}
\end{figure}

Here, our primary interest in non-associative observables is for testing the evolution of quark/gluon discrimination power as a function of RG scale $\mu$.  As recently studied in \Refs{Badger:2016bpw,Gras:2017jty}, different parton showers exhibit different quark/gluon discrimination trends as a function of jet energy.  Therefore, the study of fractal jet observables might help identify which higher-order effects in the parton shower are most important for correctly modeling the radiation patterns of quarks and gluons.

As an initial investigation into non-associative fractal observables for quark/gluon discrimination, we consider some examples of the node-product and full-tree observables from \Sec{sec:NA}.
In \Fig{fig:qgdiscrimination}, we show two good quark/gluon discriminants, comparing the gluon GFF distribution to the quark-singlet GFF distribution.
We also show the down-quark and bottom-quark GFFs as a cross check.
An example of a node-product observable from \Eq{eq:nodeext} is shown in \Fig{fig:qgdiscrimination-a}, where we take $\kappa = 1$ and $w_a = 0$ on a C/A tree.
An example of a full-tree observable from \Eq{eq:narecursion} is shown in \Fig{fig:qgdiscrimination-b}, where we take $\kappa=2$ and $\xi=4$ on a C/A tree with all particles given weight 1.
There are noticeable differences between the gluon and quark-singlet GFFs which can be exploited for the purposes of discrimination.
Among the observables we tested, these two performed among the best, outperforming, for example, variants using only charged particles. 

\begin{figure}[t]
\subfloat[]{
		\includegraphics[width=0.45\textwidth]{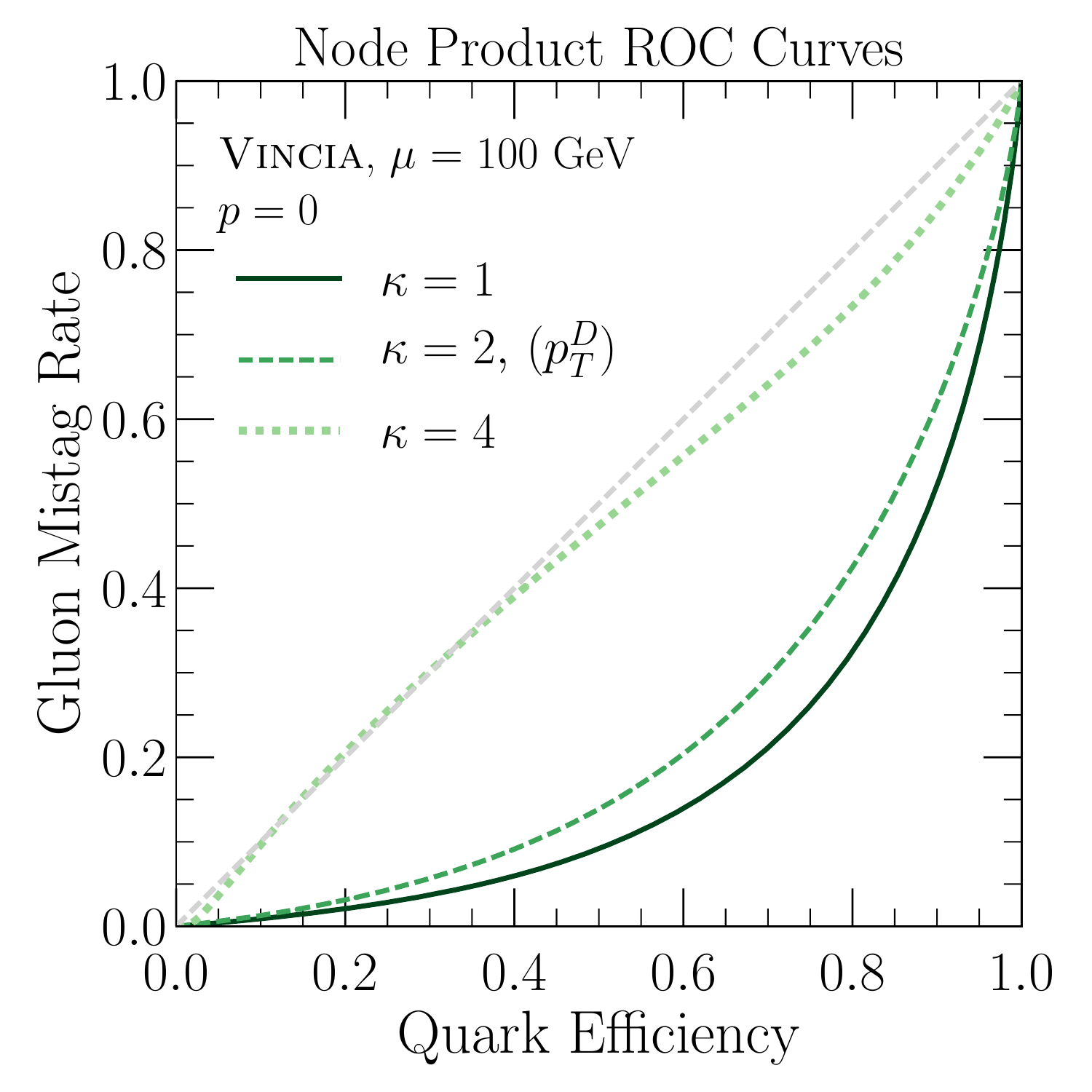}
		 \label{fig:ROC_nodes-a}
}
\subfloat[]{
		\includegraphics[width=0.45\textwidth]{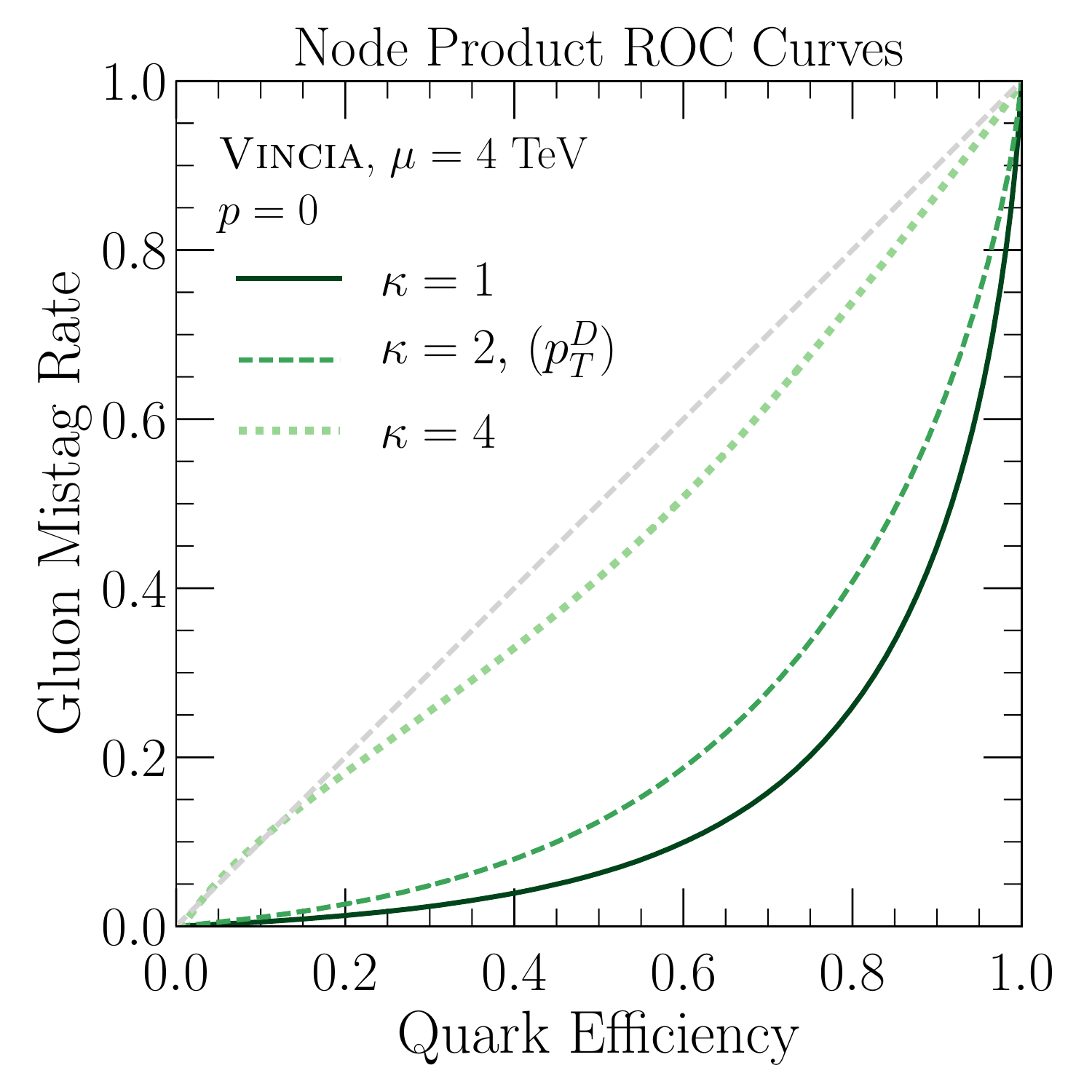}
		 \label{fig:ROC_nodes-b}
}
\caption{Quark/gluon ROC curves from \textsc{Vincia} for the node-product observables at (a) $\mu = 100$ GeV and (b) $\mu = 4$ TeV. The curves correspond to $\kappa=1$ (dark green solid), $\kappa=2$ (green dashed), and $\kappa=4$ (light green dotted).  Note that the $\kappa =2$ case has the same ROC curve as $p_T^D$, and the gray dashed line represents an observable with no discrimination power.}
\label{fig:ROC_nodes}
\end{figure}

\begin{figure}[t]
\subfloat[]{
		\includegraphics[width=0.32\textwidth]{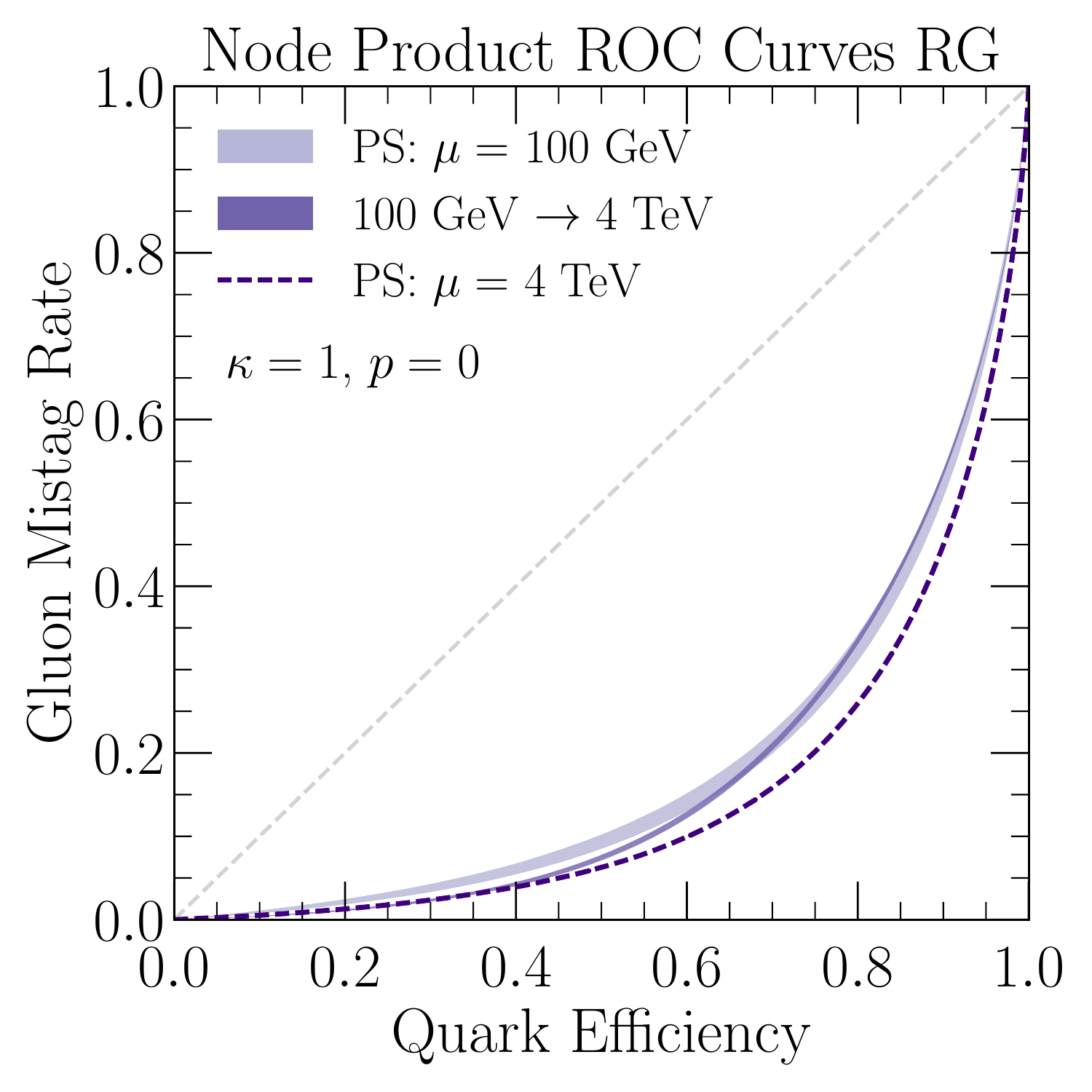}
		 \label{fig:ROC-evolution-nodes-a}
}
\subfloat[]{
		\includegraphics[width=0.32\textwidth]{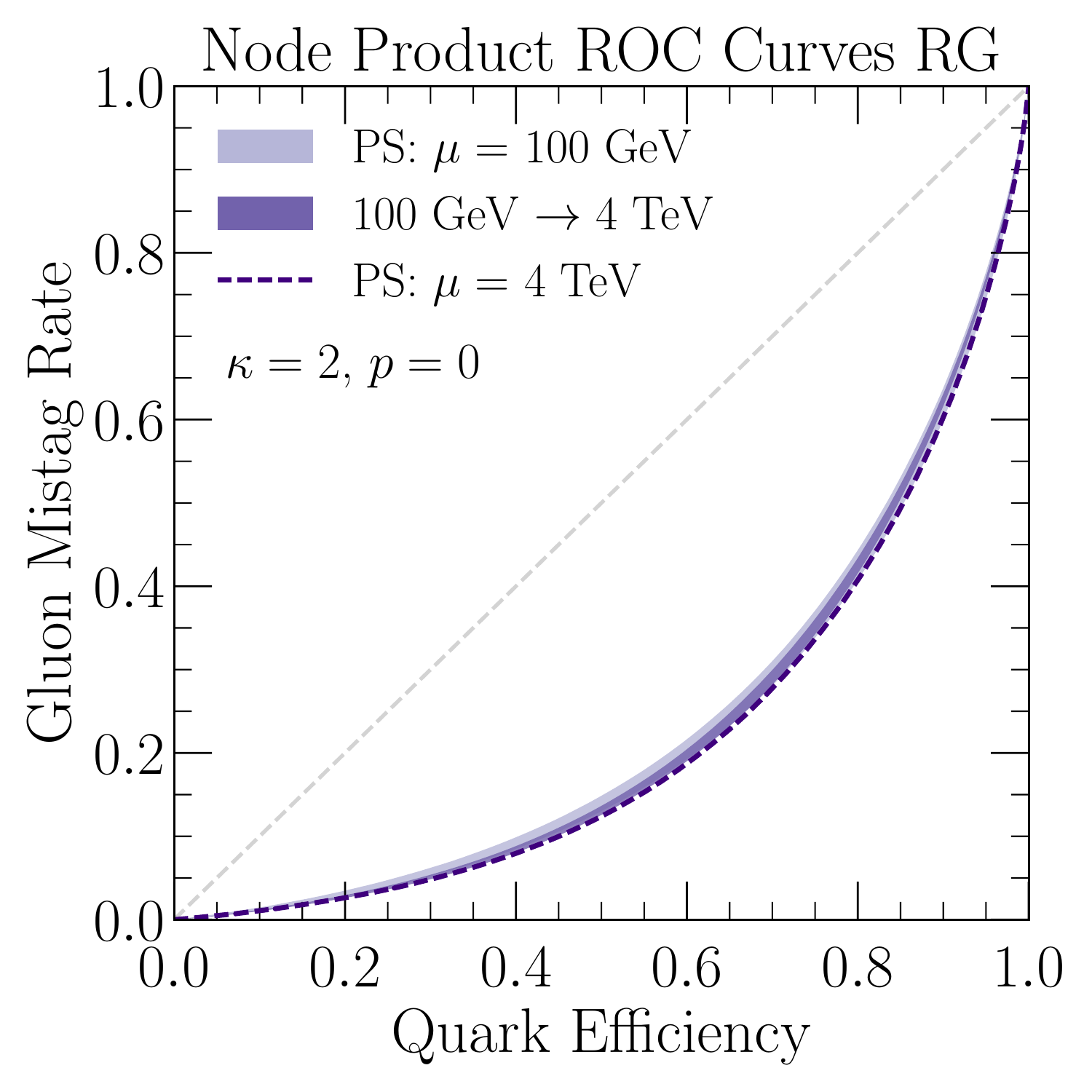}
		 \label{fig:ROC-evolution-nodes-b}
}
\subfloat[]{
		\includegraphics[width=0.32\textwidth]{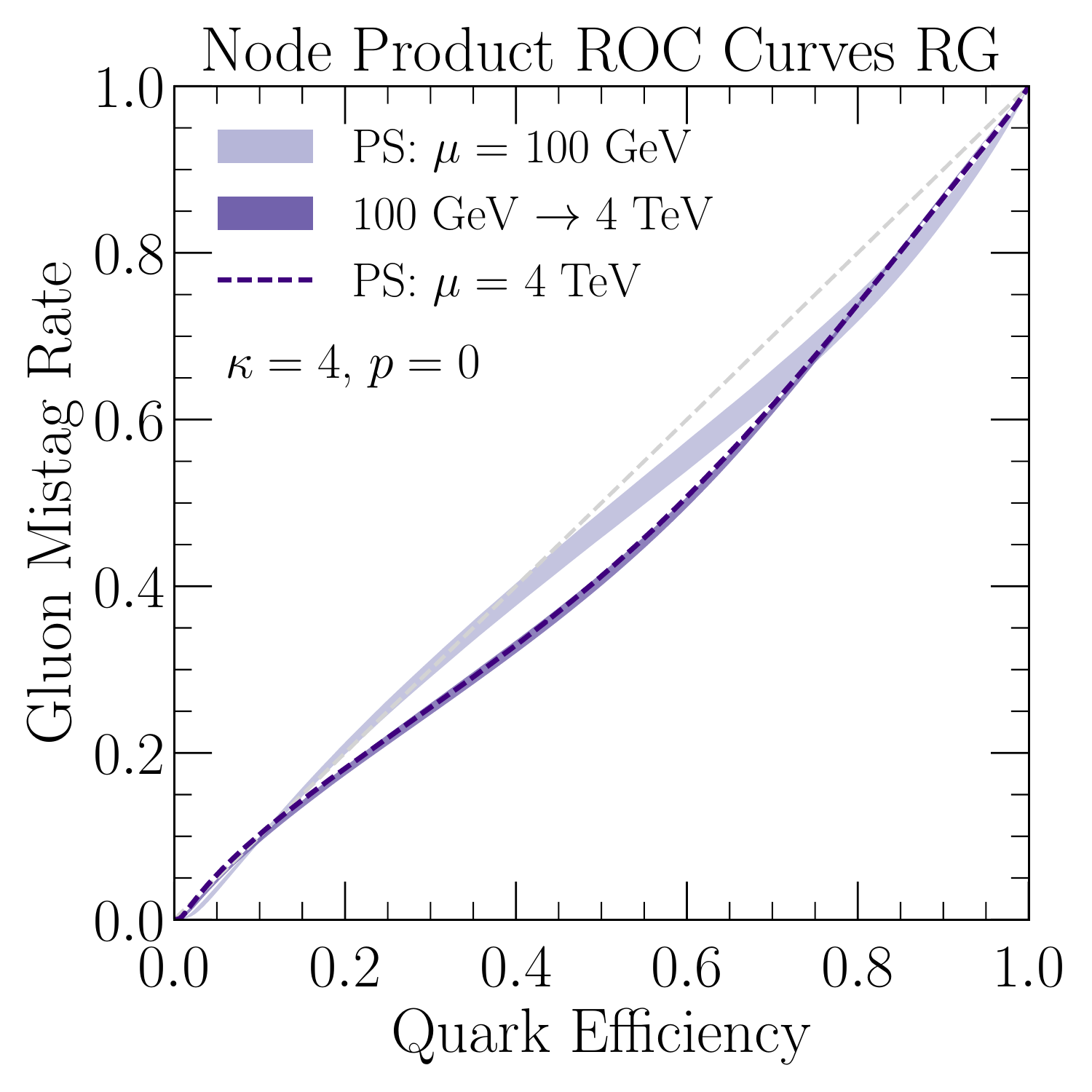}
		 \label{fig:ROC-evolution-nodes-c}
}
	\caption{Evolution of the ROC curves for node-product observables with (a) $\kappa = 1$, (b) $\kappa = 2$ (equivalent to $p_T^D$), and (c) $\kappa = 4$.  Shown are the ROC curves extracted from parton showers at 100 GeV (light purple band) and 4 TeV (dark purple, dashed), as well as the ROC curve obtained from evolving the GFF from $\mu = 100$ GeV to 4 TeV (medium purple band). The spread of these curves is obtained from calculating the ROC curves from the spread of distributions, as described in \Sec{sec:wefextraction}.}
	\label{fig:ROC-evolve-nodes} 
\end{figure}

To evaluate the potential quark/gluon discrimination power more quantitatively, we show ROC (receiver operating characteristic) curves showing the efficiency of identifying quark jets against the mistag rate for gluon jets.
These plots are obtained from \textsc{Vincia}, comparing the discrimination performance at $\mu = 100$ GeV to $\mu = 4$ \TeV.
In \Fig{fig:ROC_nodes}, we show variants of the node-product observables defined on C/A trees for $\kappa = \{1, 2, 4\}$, recalling that $\kappa = 2$ is the same as $2 (1 - p_T^D)$.
The node product with $\kappa = 1$ exhibits much better discrimination power than $\kappa=2$, especially at $\mu = 4$ \TeV. The discrimination power does continue increasing (slowly) with lower $\kappa$, but approaching the $\kappa\rightarrow 0$ limit, the observable becomes IR unsafe and the GFF formalism no longer applies. 

We can check whether this jet-energy dependence is reasonable using the RG evolution equations, as shown in \Fig{fig:ROC-evolve-nodes}.
For $\kappa = 1$, the discrimination power does indeed increase with increasing $\mu$, but not as much as predicted by the parton showers. This could have already been anticipated from the results in \Fig{fig:node-evolve-g-TF-3}, where the RG-evolved gluon GFF does not shift as dramatically as predicted in the parton showers.
This could either be a sign that the parton showers are too aggressive in their evolution, or that higher-order terms in the evolution equation are important for getting the proper shape of the $\kappa = 1$ distribution.
For $\kappa = 2$, the evolution of the ROC curves according to \Eq{eq:evolution_repeated} does match the evolution in the parton shower, but this evolution is very slight, less than the spread in the ROC curves at either scale from varying $R$ and the parton shower. 
For $\kappa = 4$, the discrimination power is poor at all scales, but the evolution matches well between \Eq{eq:evolution_repeated} and the parton showers.

\begin{figure}[t]
\subfloat[]{
		\includegraphics[width=0.45\textwidth]{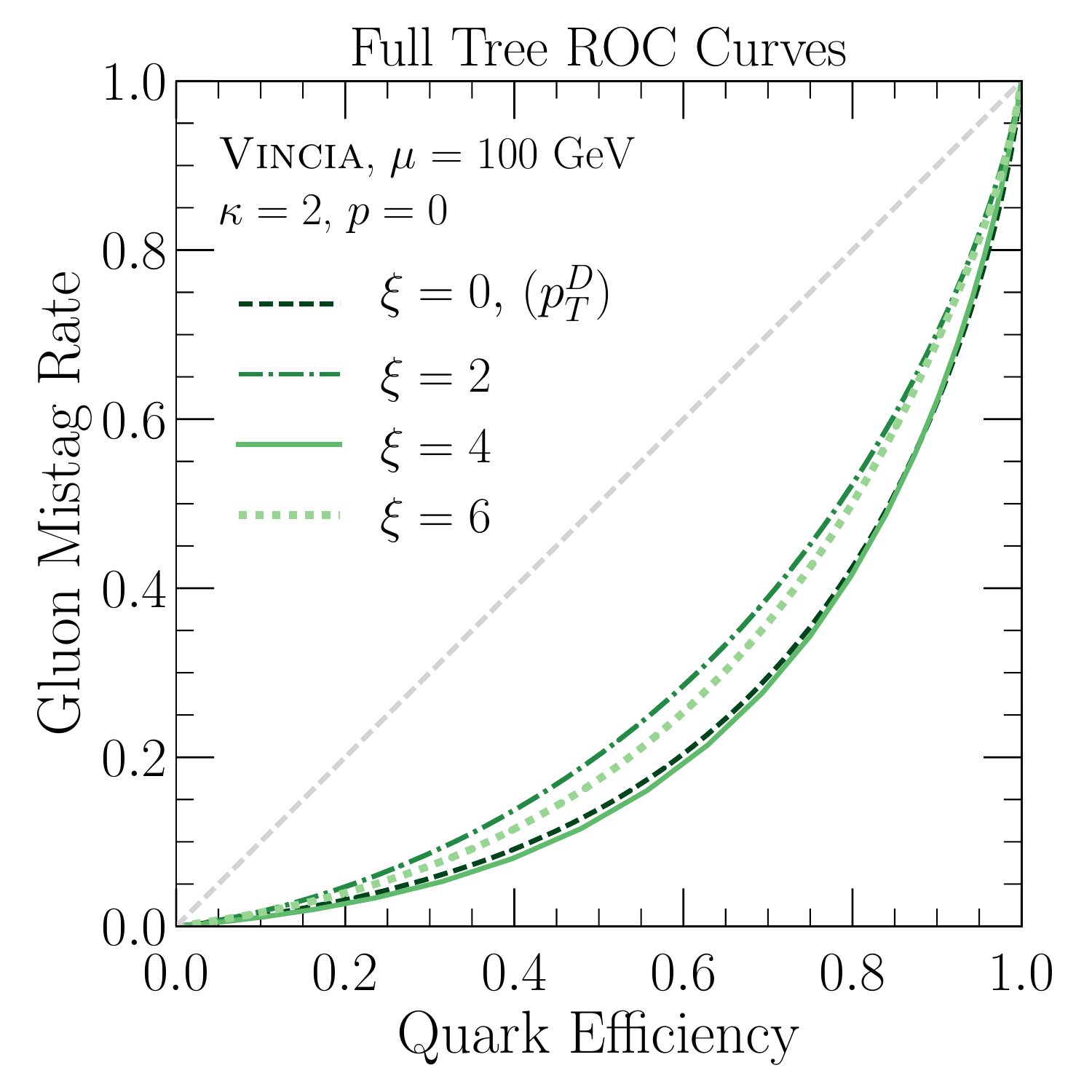}
		 \label{fig:ROC_fulltree-a}
}
\subfloat[]{
		\includegraphics[width=0.45\textwidth]{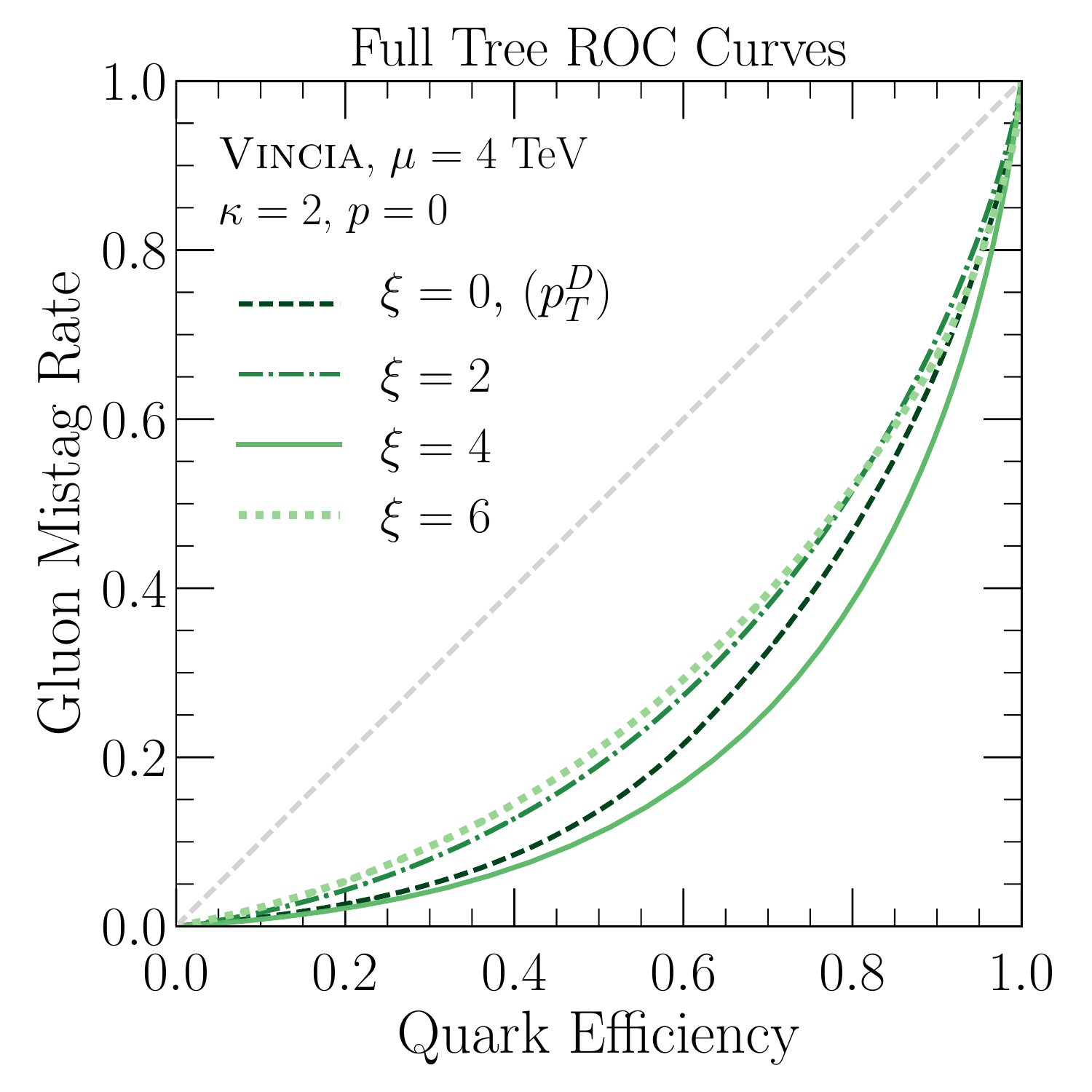}
		 \label{fig:ROC_fulltree-b}
}
\caption{Same as \Fig{fig:ROC_nodes} but for the full-tree observables with $\kappa=2$ and $\xi=\{0,2,4,6\}$.  Note that the $\xi = 0$ case is identical to $p_T^D$.}\label{fig:ROC_fulltree} 
\end{figure}

\begin{figure}[t]
\subfloat[]{
		\includegraphics[width=0.32\textwidth]{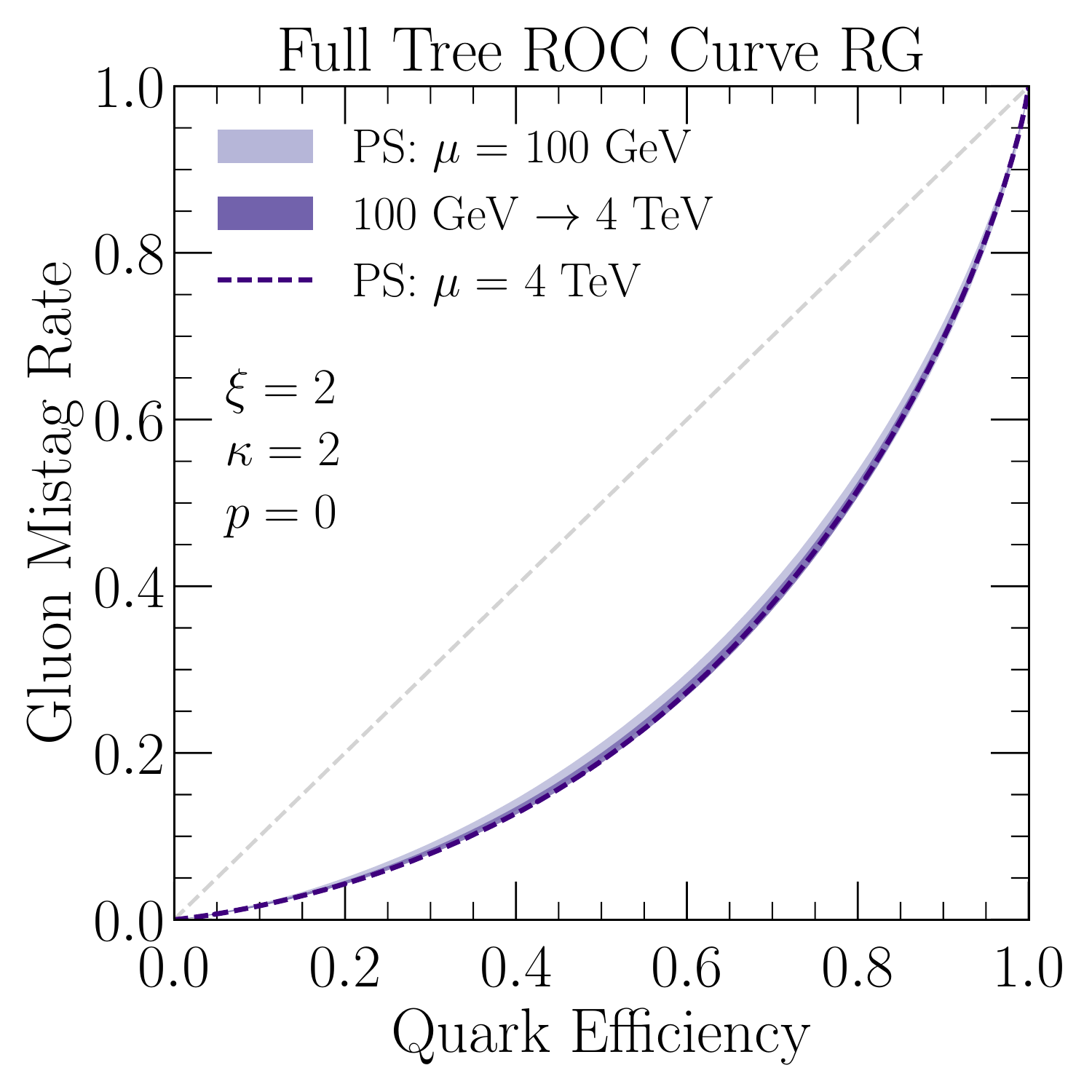}
		 \label{fig:ROC-evolve-fulltree-a}
}
\subfloat[]{
		\includegraphics[width=0.32\textwidth]{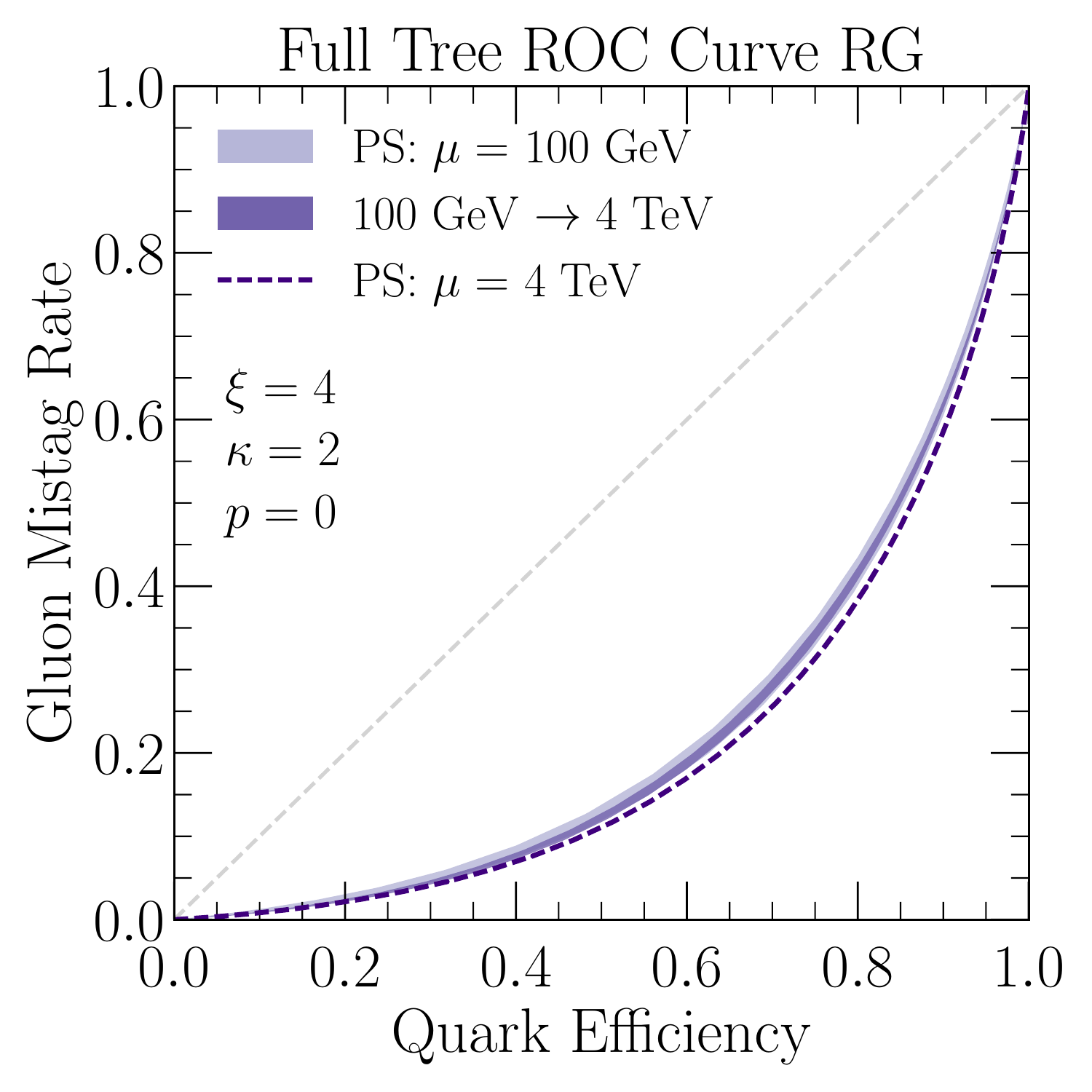}
		 \label{fig:ROC-evolve-fulltree-b}
}
\subfloat[]{
		\includegraphics[width=0.32\textwidth]{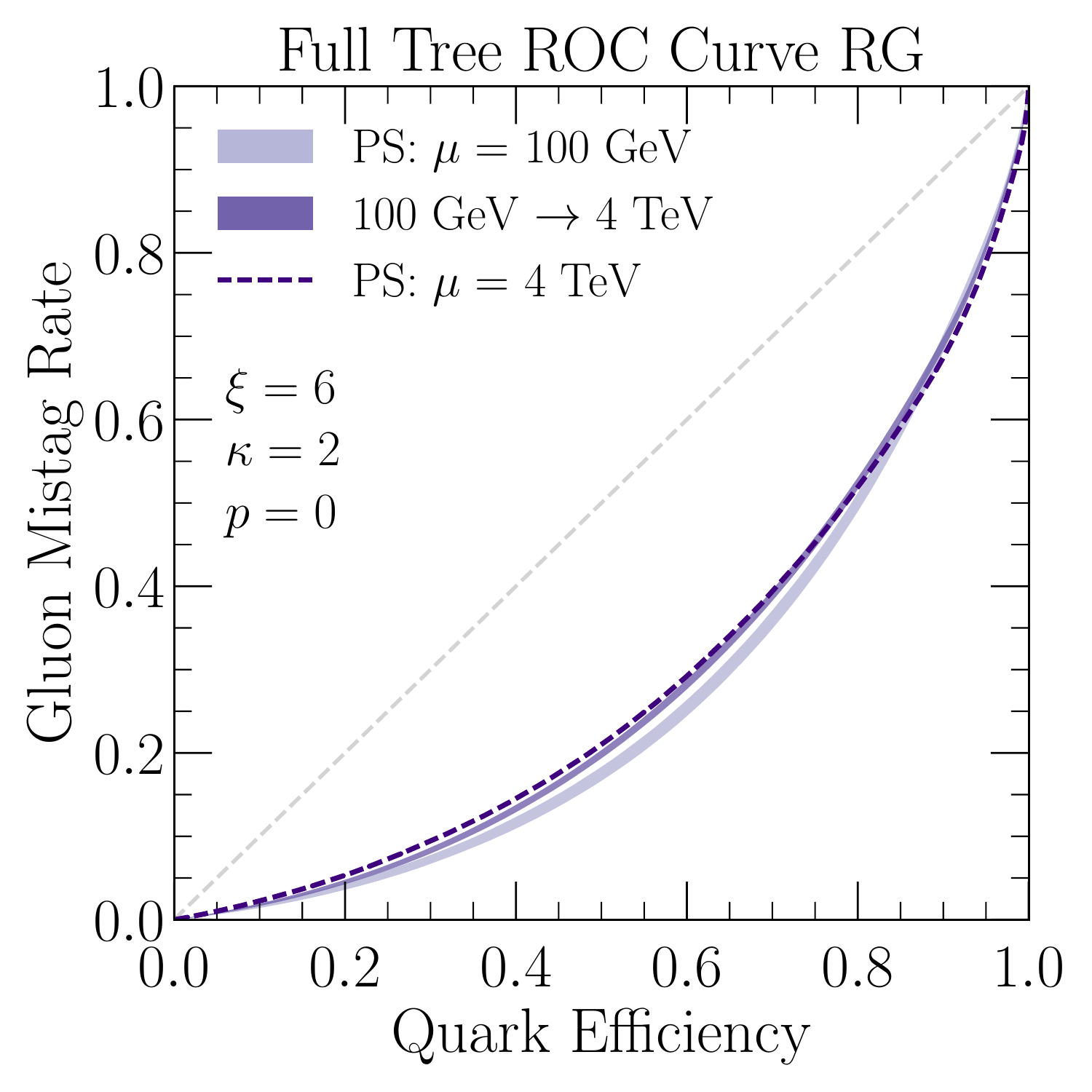}
		 \label{fig:ROC-evolve-fulltree-c}
}
	\caption{Same as \Fig{fig:ROC-evolve-nodes} but for the full-tree observables with $\kappa=2$ and (a) $\xi=2$, (b) $\xi = 4$, and (c) $\xi = 6$.  The $\xi = 0$ case is identical to $p_T^D$, shown in \Fig{fig:ROC-evolution-nodes-b}.}
	\label{fig:ROC-evolve-fulltree} 
\end{figure}

We next turn to the full-tree observables in \Fig{fig:ROC_fulltree}, using a C/A tree with $\kappa = 2$ on all particles.  We compare $\xi=\{0,2,4,6\}$, where $\xi = 0$ is identical to $p_T^D$.
The $\xi = 4$ observable yields comparable performance to $p_T^D$ at $\mu = 100$ GeV, but performs somewhat better than $p_T^D$ at $\mu = 4$ TeV. 
Note that the quark/gluon discrimination power is not monotonic as a function of $\xi$.
We can again check whether this evolution is reasonable using the RG equations, as shown in \Fig{fig:ROC-evolve-fulltree}. For all three $\xi$ values, the evolution of the ROC curves in \Eq{eq:evolution_repeated} matches the parton shower, but the evolution is extremely slow.

As emphasized in \Ref{Larkoski:2014pca}, predicting the quark/gluon discrimination power from first principles is a much more challenging task than  predicting the distributions themselves.
Because the ROC curve shapes depend sensitively on the overlap between the quark and gluon distributions, small changes in the distribution shapes can lead to large changes in the predicted discrimination power. This is especially evident in \Fig{fig:ROC-evolve-fulltree}, where the uncertainties in the ROC curves at the same scale are generally larger than the evolution between scales.
This highlights the importance of precision calculations for correctly predicting quark/gluon discrimination behavior.

\section{Fractal Observables from Subjets}
\label{sec:angular}

\begin{figure}[t]
	\centering
	\begin{tikzpicture}
	\node[circle,minimum size=1cm,draw] at (4,3) (x) {$\hat{x}_2$};
	\node[circle,minimum size=1cm,draw] at (2,1) (x12) {$\hat{x}_2$};
	\node[circle,minimum size=1cm,draw] at (6,1) (x34) {$\hat{x}_2$};
	\draw[->,ultra thick] (x12) -- (x);
	\draw[->, ultra thick] (x34) -- (x);
	\node[diamond,draw,inner sep=7pt,aspect=0.75] at (1,-0.5) (w1) {};
	\node[diamond,draw,inner sep=7pt,aspect=0.75] at (3,-0.5) (w2) {};
	\node[diamond,draw,inner sep=7pt,aspect=0.75] at (5,-0.5) (w3) {};
	\node[diamond,draw,inner sep=3pt,aspect=0.75] at (7,-0.5) (w4) {$x_1$};
	\draw[->,ultra thick] (w1) -- (x12);
	\draw[->,ultra thick] (w2) -- (x12);
	\draw[->,ultra thick] (w3) -- (x34);
	\draw[->,ultra thick] (w4) -- (x34);
	\node[circle,minimum size=0.3cm,inner sep=2pt,draw] at (11,-0.5) (y) {$\hat{x}_1$};
	\node[circle,minimum size=0.3cm,inner sep=2pt,draw] at (10,-1.5) (y12) {$\hat{x}_1$};
	\node[circle,minimum size=0.3cm,inner sep=2pt,draw] at (12,-1.5) (y34) {$\hat{x}_1$};
	\draw[->,ultra thick] (y12) -- (y);
	\draw[->, ultra thick] (y34) -- (y);
	\node[diamond,draw,inner sep=4pt,aspect=0.75] at (9.5,-2.8) (z1) {};
	\node[diamond,draw,inner sep=4pt,aspect=0.75] at (10.5,-2.8) (z2) {};
	\node[diamond,draw,inner sep=4pt,aspect=0.75] at (11.5,-2.8) (z3) {};
	\node[diamond,draw,inner sep=4pt,aspect=0.75] at (12.5,-2.8) (z4) {};
	\draw[->,ultra thick] (z1) -- (y12);
	\draw[->,ultra thick] (z2) -- (y12);
	\draw[->,ultra thick] (z3) -- (y34);
	\draw[->,ultra thick] (z4) -- (y34);
	\draw[->,ultra thick,dashed] (y) -- (w4)
	node[draw=none,fill=none,midway,sloped,anchor=center,above] {$\theta > R_{\rm sub}$};
	\end{tikzpicture}
	\caption{Modified fractal jet observables where the recursion relation changes at a characteristic scale $R_{\rm sub}$.  When using a C/A tree, it is possible to switch the recursion relation from $\hat x_1$ to $\hat x_2$ for angular scales $\theta > R_{\rm sub}$. This is equivalent to determining the observable $\hat x_1$ on all subjets of radius $R_{\rm sub}$ and then using these as initial weights for the tree with $\hat x_2$.
	} \label{fig:fractalsubjets}
\end{figure}
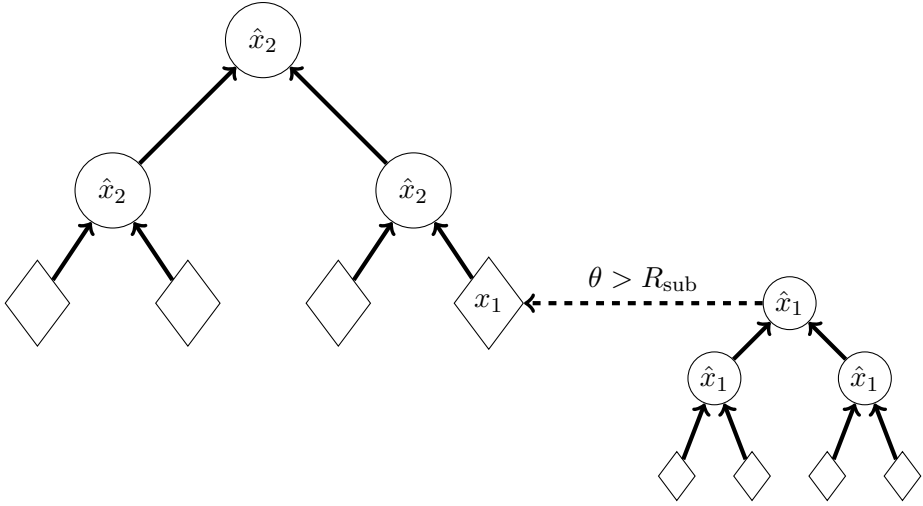

As our final investigation into the structure of fractal jet observables, we now consider the possibility that the recursion relation in \Eq{eq:recurse} is modified to depend on the angular scale of the clustering.
For simplicity, we only consider observables defined on angular-ordered C/A clustering trees, since in that case the depth in the C/A tree is directly associated with an angular scale $\theta$.
This opens up the possibility to define a modified recursion relation with $\theta$ dependence, for example,
\begin{align}\label{eq:subjet}
  \hat{x}(z,x_1,x_2) = \bigg\{ 
  \begin{tabular}{ll} $\hat{x}_1(z,x_1,x_2)$ & if $\theta < R_{\rm sub}$, \\
  $\hat{x}_2(z,x_1,x_2)$ & if $\theta > R_{\rm sub}$.
  \end{tabular} 
\end{align}
As shown in \Fig{fig:fractalsubjets}, the nodes as defined by $\hat{x}_1$ become the starting weights for the subsequent nodes defined by $\hat{x}_2$. 

It is straightforward to implement the leading-logarithmic resummation of an observable defined by \Eq{eq:subjet}.  Starting from a low-energy boundary condition, this involves an initial evolution to the scale 
\begin{equation}
 \mu_{\rm sub} = E_{\rm jet} R_{\rm sub}
\end{equation}
using \Eq{eq:evolution_repeated} with the recursion relation $\hat{x}_1$, followed by an evolution to $\mu = E_{\rm jet} R$ using $\hat{x}_2$ instead.
The discontinuity in anomalous dimensions of the evolution equations across the threshold $\mu_{\rm sub}$ will be compensated by a fixed-order correction at that scale, but this only enters at next-to-leading-logarithmic order. 

\begin{figure}[t]
	\includegraphics[width=0.45\textwidth]{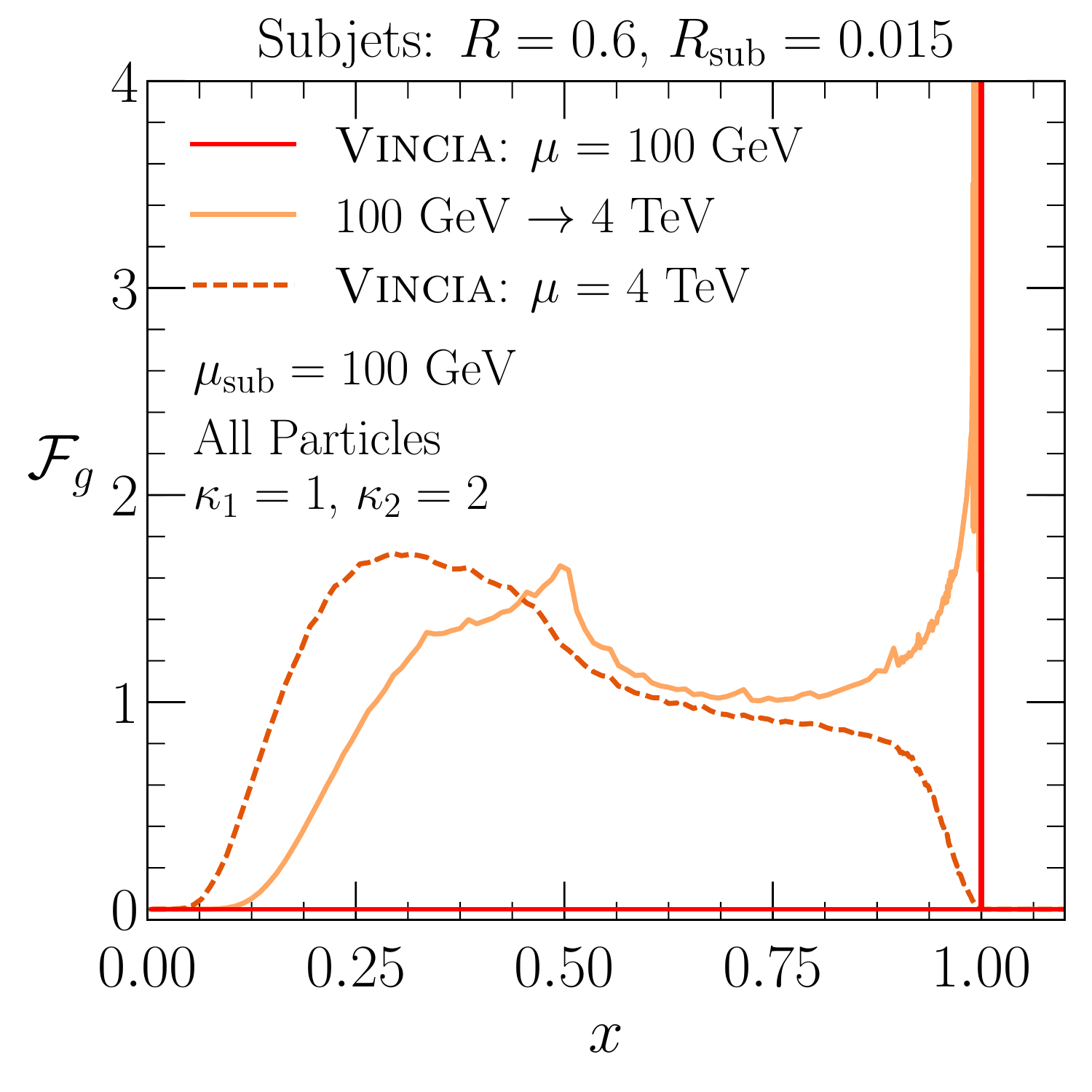}
	\caption{Evolution of the fractal observable defined by equations \Eqs{eq:subjet}{eq:subjetdelta}, where $\hat{x}_1$ and $\hat{x}_2$ are given by weighted energy fractions measured on all particles with $\kappa_1 = 1$ and $\kappa_2 = 2$, respectively.}
	\label{fig:subjet} 
\end{figure}

One interesting case is when the observable defined at small angular scales $\theta < R_{\rm sub}$ is the weighted energy fraction of all particles with $\kappa=1$.  This observable is simply 1 for each of the branches, so the GFFs at the scale $\mu_{\rm sub}$ are 
\begin{equation}\label{eq:subjetdelta}
\GFF_i(x,\mu_{\rm sub}) = \delta(1-x)
\,,\end{equation}
which are then the input for the fractal observable $\hat x_2$ for $\theta>R_{\rm sub}$. 
This effectively removes the sensitivity to nonperturbative physics, allowing us to calculate fractal observables analytically, as long as the scale $\mu_{\rm sub}$ is perturbative. An example of this kind of observable is shown in \Fig{fig:subjet}, where the observable is clustered using the recursion relation \Eq{eq:wefrecur} with $\kappa=1$ for angles $\theta < R_{\rm sub}$ and $\kappa=2$ for $\theta > R_{\rm sub}$. 
The spike at $x=1$ persists in the numerical evolution, even with very fine bins and a large amount of computing time.\footnote{The generating functional approach (see e.g.~ref.~\cite{Ellis:1991qj}) provides an alternative implementation of the evolution in \Eq{eq:evolution} that can be used to resum (sub)jet radius logarithms~\cite{Dasgupta:2014yra}. This approach may be more amenable to an initial condition with a delta function.}
This feature is not seen in the $\textsc{Vincia}$ evolution, which at every stage in the parton shower uses a scale closer to $\mu \simeq  z \, E_{\rm jet} \, \theta$, where $z$ and $\theta$ are the momentum fraction and opening angle of the splitting. Compared to our choice of $\mu = E_{\rm jet}R$ for the shower as a whole, we would expect the $\textsc{Vincia}$ scale, which corresponds to a larger coupling, to accelerate the depletion of the $\delta$ function in the evolution. It will be interesting to see if this behavior persists with higher-order evolution equations. 

An alternative way of viewing the above prescription is that we can build fractal jet observables not just out of hadrons but also out of subjets of radius $R_{\rm sub}$, thus enlarging the range of applicability of the GFF framework.
By taking $R_{\rm sub}$ not too small, the observable becomes perturbative.
On the other hand, we still want $R_{\rm sub} \ll R$, such that the leading logarithms of $R/R_{\rm sub}$ dominate the observable and \Eq{eq:evolution_repeated} gives a reliable description of its behavior.

\section{Conclusions}
\label{sec:conclusion}

To date, the bulk of analytic jet physics studies are based on either single-hadron fragmentation functions or IRC-safe jet shapes.
In this paper, we emphasized the intermediate possibility of IR-safe but collinear-unsafe jet observables defined on a \emph{subset} of hadrons.
We started by introducing the framework of Generalized Fragmentations Functions (GFFs), which are applicable to general collinear-unsafe jet observables.
The GFFs are universal functions that absorb collinear singularities order by order in $\alpha_s$, which not only restores calculational control, but also implies that the GFFs evolve under a nonlinear version of the DGLAP equations.
We then discussed fractal jet observables, defined recursively on an IRC-safe clustering tree with certain initial hadron weights, which satisfy a self-similar RG evolution at leading order given by \Eq{eq:evolution_repeated}.
The higher order evolution is no longer universal, but still self-similar, and has the schematic form in \Eq{eq:higherorder}.

The simplest fractal jet observables are those with associative recursion relations, whose value does not depend on the choice of clustering tree.
This is indeed the case for the weighted energy fractions, studied in \Sec{sec:WEF}, which include several observables already in use at colliders, including $p_T^D$, weighted jet charge, and track fractions.
More exotic fractal jet observables depend on the clustering sequence, including the node-product and full-tree observables studied in \Sec{sec:NA}.
Remarkably, the structure of the RG evolution for these observables is independent of the clustering tree at leading order.

As one potential application of fractal observables, we studied whether non-associative observables could be useful for quark/gluon discrimination.
Indeed, we found examples in \Sec{sec:quarkgluon} which do perform better than the weighted energy fraction $p_T^D$ currently used by CMS.
Though the GFF formalism does not allow us to predict the absolute discrimination power of collinear-unsafe observables, it does allow us to predict the RG evolution of the discrimination power, a feature that will be further exploited in \Ref{HarvardInProgress}.
To gain more perturbative control, one can work with fractal observables defined on subjets (instead of hadrons), as briefly discussed in \Sec{sec:angular}.

Looking to the future, the next step for fractal jet observables is pushing beyond the leading-order evolution equations.
This will require computing the bare GFFs to higher orders in $\alpha_s$, as well as extracting GFFs using the matching scheme sketched in \Eq{eq:cross_section}, and presented in detail at next-to-leading order for $e^+e^-$ collisions in \App{app:matching}.
More ambitiously, one would like to study correlations between two or more fractal jet observables, which would require multivariate GFFs.
Such correlations are known to be important for improved quark/gluon discrimination \cite{Gallicchio:2012ez,Larkoski:2014pca,Komiske:2016rsd}, though even for IRC-safe jet shapes, there are relatively few multivariate studies \cite{Larkoski:2013paa,Larkoski:2014tva,Procura:2014cba}.
Together with the work in this paper, higher-order and correlation studies would facilitate a deeper understanding of jet fragmentation, with important consequences for analyses at the LHC and future collider experiments.

\begin{acknowledgments}
We thank Christopher Frye and Andrew Larkoski for helpful comments on quark/gluon discrimination.
The work of B.E., J.T., and K.Z. is supported by the DOE under grant contract numbers DE-SC-00012567 and DE-SC-00015476.
M.P.\ is supported by a Marie Curie Intra-European Fellowship of the
European Community's 7th Framework Programme under contract number
PIEF-GA-2013-622527.
W.W.~is supported by the European Research Council under grant ERC-STG-2015-677323, and the D-ITP consortium, a program of the Netherlands Organization for Scientific Research (NWO) that is funded by the Dutch Ministry of Education, Culture and Science (OCW). 
\end{acknowledgments}

\appendix

\section{Generalized Fragmentation in Inclusive Jet Production}
\label{app:matching}

In this appendix, we explicitly verify \Eq{eq:cross_section} at $\mathcal{O}(\alpha_s)$. We first calculate the left-hand side of this equation for the measurement of the fractal variable $x$ together with the fraction of the center-of-mass energy carried by the jet, $z_J \equiv 2E_{\rm jet}/E_{\rm cm}$. Assuming that $R$ is not so large that all final-state partons get clustered into one jet, we get
\begin{align} \label{eq:calc}
\frac{1}{\sigma^{(0)}}\frac{\text{d}\sigma}{\text{d}z_J\, \text{d}x}
&= \frac{1}{\sigma^{(0)}}\int\! {\rm d}y_1\, {\rm d}y_2\, 
\frac{\text{d}\sigma}{\text{d}y_1 \text{d}y_2}\,
\bigg\{\sum_{i<j}
\theta(R - \phi_{ij})\bigg( 
  \delta(z_J - y_k) \GFF^{(0)}_k(x,\mu)
 \nonumber \\ & \quad 
 + \delta(z_J - y_i - y_j)
 \int \text{d}x_1\, \text{d}x_2\, \GFF^{(0)}_i(x_1,\mu)\, \GFF^{(0)}_j(x_2,\mu)\,
 \delta\bigg[x-\hat{x}\Big(\frac{y_i}{y_i+y_j},x_1,x_2\Big)\bigg]
 \bigg)
 \nonumber \\ & \quad +
 \theta(\phi_{12}-R)\,  \theta(\phi_{13}-R)\, \theta(\phi_{23}-R)
 \sum_i \delta(z_J - y_i)\, \GFF^{(0)}_i(x,\mu) \bigg\}
.\end{align}
Here, ${i,j}=1,2,3$ and $y_i$ is the parton momentum fraction normalized such that $y_1 + y_2 + y_3=2$. In the following calculations, we  identify parton 1 with the quark, 2 with the antiquark, and 3 with the gluon. The angle $\phi_{ij}$ between partons $i$ and $j$ is given by
\begin{equation} \label{eq:ang}
  \phi_{ij} = \arccos\bigg[1 - \frac{2(1-y_k)}{y_i\,y_j}\bigg]
\,,\end{equation}
and $k$ denotes the parton different from $i$ and $j$. Although the angle $\phi_{ij}$ becomes ambiguous when $y_i$ or $y_j$ is zero, IR safety ensures that the measurement is not.
The term in \Eq{eq:calc} with $\phi_{ij}< R$ describes the situation where partons $i$ and $j$ are clustered in a jet but parton $k$ is in a separate jet. The final term, where all $\phi_{ij}>R$, corresponds to the situation where all partons are in separate jets. Each of the three partons has a leading-order GFF attached to it.
The squared matrix element that enters in \Eq{eq:calc} is given up to $\mathcal{O}(\al_s)$  by
\begin{align}\label{eq:matrixelement}
\frac{1}{\sigma^{(0)}}\frac{\text{d}\sigma}{\text{d}y_1 \text{d}y_2} &= 
\delta(1\!-\!y_1) \delta(1\!-\!y_2) +
\frac{\alpha_sC_F}{2\pi} \bigg\{ \frac{\theta(1-y_3) (y_1^2 + y_2^2)}{2(1\!-\!y_1)_+ (1\!-\!y_2)_+} + \Big(\frac{\pi^2}{2} - 4\Big) \delta(1\!-\!y_1)\delta(1\!-\!y_2)
\nonumber \\ & \quad
+ \delta(1-y_2)\bigg[ 
\frac{P_{q \to q g}(y_1)}{C_F}
\Big(-\frac{1}{\epsilon_\text{IR}} +  \ln \frac{y_1E_{\rm cm}^2}{\mu^2} \Big)
 + (1+y_1^2)\Big(\frac{\ln(1-y_1)}{1-y_1}\Big)_+ + 1-y_1 \bigg]  
\nonumber \\ & \quad
+ (y_1\leftrightarrow y_2)\bigg\}
,\end{align}
where 
\begin{equation}
P_{q \to q g}(y) = C_F \Big(\frac{1+y^2}{1-y}\Big)_+
.\end{equation}

Let us now focus on the right-hand side of \Eq{eq:cross_section}. In our case, the coefficients $C_i$ are the standard ones for inclusive fragmentation in $e^+e^-$ collisions \cite{Curci:1980uw,Floratos:1981hs,Altarelli:1979kv}
 since the only kinematic variable appearing on the left-hand side of \Eq{eq:calc} is the jet energy fraction $z_J$:
\begin{align} \label{eq:C_coeff}
C_q(z, E_{\rm cm}, \mu) &= \delta(1-z) + \frac{\al_s}{2\pi} \bigg\{ P_{q \to q g}(z) \ln \frac{E_{\rm cm}^2}{\mu^2} + C_F \bigg[ (1+z^2) \bigg(\Big(\frac{\ln(1-z)}{1-z}\Big)_+ 
\nonumber \\ & \quad
+ \frac{2\ln z}{1-z}\bigg)  - \frac32 \frac{1}{(1-z)}_+  + \delta(1-z) \Big(\frac{2\pi^2}{3}-\frac92\Big) - \frac32 z+ \frac52 \bigg] \bigg\}
\,, \nonumber \\
C_g(z, E_{\rm cm}, \mu) &= \frac{\al_s}{2\pi} P_{q \to q g}(1-z)  \Big( \ln \frac{E_{\rm cm}^2}{\mu^2} + \ln(1-z) + 2\ln z \Big).
\end{align}

The coefficients $  \mathcal{J}^{(1)}_{q\to qg}$ and $\mathcal{J}^{(1)}_{q\to gq}$ for an $e^+ e^-$ $k_T$-like jet algorithm were calculated using the $\overline{\rm MS}$ scheme in ref.~\cite{Waalewijn:2012sv}, 
\begin{align} \label{eq:Jqqgcoeff}
  \mathcal{J}^{(1)}_{q\to qg}(z, E_{\rm jet} R, \mu) &= \frac{\alpha_s}{2\pi} \bigg\{2C_F L^2\, \delta(1\!-\!z)  
   + \big[2P_{q \to q g}(z) \!-\! 3 C_F\, \delta(1\!-\!z)\big] L 
   + C_F \bigg[  4 z \Big(\frac{\ln(1\!-\!z)}{1\!-\!z}\Big)_{\!+}  
   \nonumber \\ & \quad   
   + 2(1-z) \ln (1-z)
  + 2 \Big(\frac{1+z^2}{1-z}\Big) \ln z + 1- z
   - \frac{\pi^2}{12}\, \delta(1-z)   
    \bigg] \bigg\}
  \,, \nonumber \\
  \mathcal{J}^{(1)}_{q\to gq}(z, E_{\rm jet} R, \mu) &=   \mathcal{J}^{(1)}_{q\to qg}(1-z, E_{\rm jet} R, \mu)
\,, \end{align}
while $\mathcal{J}^{(1)}_{q\to q}$ and $\mathcal{J}^{(1)}_{q\to g}$ are given by the finite terms of eq.~(2.34) and eq.~(2.35) in~ref.~\cite{Kang:2016mcy}
\begin{align} \label{eq:Jqqcoeff}
 \mathcal{J}^{(1)}_{q \to q}(z, E_{\rm jet} R,\mu) 
  &= \frac{\alpha_s}{2\pi} \bigg[C_F \delta(1-z)\Big(-2 L^2 + 3 L  +\frac{\pi^2}{12} \Big) 
\nonumber \\ & \quad
-2L P_{q \to q g}(z) -2C_F(1+z^2)\Big(\frac{\ln(1-z)}{1-z}\Big)_+ -C_F(1-z)  \bigg] 
, \nonumber \\
 \mathcal{J}_{q\to g}^{(1)} (z,E_{\rm jet} R,\mu) 
 &=  \mathcal{J}_{q\to q}^{(1)} \Big(1-z,\frac{1-z}{z}E_{\rm jet} R,\mu\Big) 
 \,, 
 \end{align}
where 
\begin{equation}
L \equiv \ln \Big(\frac{E_{\rm jet} R}{\mu}\Big)
.\end{equation}
The coefficients for anti-quarks are identical.
Note that the relation between $\mathcal{J}^{(1)}_{q \to q}$ and  $\mathcal{J}^{(1)}_{q \to g}$ is not simply $z \leftrightarrow 1-z$, because the jet energy $E_{\rm jet}$ rather than the energy of the initiating parton is held fixed. Since $\mathcal{J}^{(1)}_{q \to q}$ and $\mathcal{J}^{(1)}_{q \to qg}$ describe the same splitting in complementary regions of phase space (in-jet versus out-of-jet), their sum vanishes in dimensional regularization,
\begin{align} \label{eq:J_rel}
\mathcal{J}_{q\to qg}^{(1)} (z,E_{\rm jet} R,\mu) + \mathcal{J}^{(1)}_{q\to q}(z, z\,E_{\rm jet} R, \mu) = 0\,.
\end{align}
The final ingredient we need is the renormalized one-loop expression for the GFF (see \Eq{eq:GFFNLO}),
\begin{align}
\GFF_i (x) &= \GFF^{(0)}_i (x) - \frac{1}{2\,\epsilon_\text{IR}} \sum_{j,k} \int \text{d}z\, \frac{\alpha_s(\mu)}{2\pi} P_{i\rightarrow jk}(z) 
\nonumber \\ & \quad \times
 \int \text{d}x_1\, \text{d}x_2\, \GFF_j^{(0)}(x_1,\mu)\, \GFF_k^{(0)}(x_2,\mu)\, \delta[x-\hat{x}(z,x_1,x_2)]
\,.\end{align}

Let us first verify the cancellation of IR divergences between left- and right-hand sides in \Eq{eq:cross_section}. On the latter, these solely come from $C_q^{(0)}(z_J, E_{\rm cm},\mu) [\GFF_q^{(1)}(x,\mu)+\GFF_{\bar q}^{(1)}(x,\mu)]$. On the left-hand side, we find
\begin{align} \label{eq:calc_IR}
\left. \frac{1}{\sigma^{(0)}}\frac{\text{d}\sigma}{\text{d}z_J\, \text{d}x}\right \vert_{\rm{IR\,div}}
&= \int\! {\rm d}y_1\, {\rm d}y_2\, 
\frac{\alpha_s}{2\pi} \bigg[-\frac{1}{\epsilon_\text{IR}}\, \delta(1-y_1) 
P_{q \to q g}(y_2)
\bigg]\,
\delta(z_J - 1)
\bigg[ 
 \GFF^{(0)}_q(x,\mu)
 \nonumber \\ & \quad 
 +  \int \text{d}x_1\, \text{d}x_2\, \GFF^{(0)}_{\bar q}(x_1,\mu)\, \GFF^{(0)}_g(x_2,\mu)\,
 \delta\Big(x-\hat{x}(y_2,x_1,x_2)\Big)
 \bigg] + (q \leftrightarrow \bar q)
\nonumber \\
& = \delta(z_J -1) [\GFF_q^{(1)}(x,\mu)+\GFF_{\bar q}^{(1)}(x,\mu)]
,\end{align}
which demonstrate the cancellation of the IR divergences.
Note that the term on the first line of \Eq{eq:calc_IR} proportional to $\GFF_i^{(0)}$ does not contribute here because it is $y_2$-independent and
\begin{align}
 \int\! \text{d} y_2\, P_{q \to q g}(y_2) = 0
\,.\end{align}

To verify that also the finite terms match in \Eq{eq:cross_section}, we expand the angular constraint in the small $R$ limit as
\begin{align}
 \theta(R-\phi_{ij}) \approx \theta\Big(\frac{R^2}{4} - \frac{1-y_k}{y_i\,y_j}\Big) 
,\end{align}
which implies $y_k \approx 1$ and $y_j \approx 1 - y_i$.
We first consider the $\theta(R - \phi_{13})$ term in \Eq{eq:calc}, which gives
\begin{align} \label{eq:calc_2}
\left. \frac{1}{\sigma^{(0)}}\frac{\text{d}\sigma}{\text{d}z_J\, \text{d}x} \right \vert_{13}
&= \frac{\alpha_sC_F}{2\pi} \int\! {\rm d}y_1\, {\rm d}y_2\, 
 \bigg\{ \frac{1+ y_1^2}{(1-y_1)}_+ \frac{1}{(1-y_2)}_+ + (\pi^2 - 8) \delta(1-y_1)\delta(1-y_2)
\nonumber \\ & \quad
+ \delta(1-y_2)\bigg[ 
\frac{P_{q \to q g}(y_1)}{C_F}
 \ln \frac{y_1E_{\rm cm}^2}{\mu^2} 
 + (1+y_1^2)\Big(\frac{\ln(1-y_1)}{1-y_1}\Big)_+ + 1-y_1 \bigg]  
\bigg\}
\nonumber \\ & \quad \times
\theta\Big(\frac{R^2}{4} - \frac{1-y_2}{y_1\,(1-y_1)}\Big) 
\bigg[ 
  \delta(z_J - 1) \GFF^{(0)}_{\bar q}(x,\mu)
 \nonumber \\ & \quad 
 + \delta(z_J - 1)
 \int \text{d}x_1\, \text{d}x_2\, \GFF^{(0)}_q(x_1,\mu)\, \GFF^{(0)}_g(x_2,\mu)\,
 \delta\big(x-\hat{x}(y_1,x_1,x_2)\big)
 \bigg]
 \nonumber \\
 &= \frac{\alpha_s}{2\pi} \int\! \text{d}z\,\bigg\{
P_{q \to q g}(z)
 \ln \frac{z^2 E_{\rm jet}^2 R^2}{\mu^2} 
 + C_F \bigg[ 2(1+z^2)\Big(\frac{\ln(1-z)}{1-z}\Big)_+ 
\nonumber \\ & \quad 
 + 1-z 
  +(\dots) \delta(1-z)
 \bigg] \bigg\}
\bigg[ 
  \delta(z_J - 1) \GFF^{(0)}_{\bar q}(x,\mu)
 \nonumber \\ & \quad 
 + \delta(z_J - 1)
 \int \text{d}x_1\, \text{d}x_2\, \GFF^{(0)}_q(x_1,\mu)\, \GFF^{(0)}_g(x_2,\mu)\,
 \delta\big(x-\hat{x}(z,x_1,x_2)\big)
 \bigg]
 \nonumber \\
 &=
 \delta(1 - z_J)\, \int \text{d}z\, \text{d}x_1\, \text{d}x_2\, \mathcal{J}^{(1)}_{q\rightarrow qg}(z, E_{\rm jet} R,\mu)\, \GFF_q (x_1,\mu)\, \GFF_g (x_2,\mu)
  \nonumber \\ & \quad \times
  \delta[x-\hat{x}(z,x_1,x_2)]  + (\dots)_{13}
.\end{align}
As the integral over $y_2$ yields a $\ln(1-y_1)$, the resulting $\ln(1-y_1)/(1-y)_+$ is not properly regularized, leaving the coefficient of $\delta(1-z)$ undetermined. As we will see, however, this ambiguity cancels exactly against the one arising from $\mathcal{J}_{q \to q}^{(1)}$, due to \Eq{eq:J_rel}. The $\theta(R - \phi_{23})$ term gives the corresponding contribution with quark and anti-quark interchanged, whereas the $\theta(R - \phi_{12})$ term is $\mathcal{O}(R^2)$ suppressed due to the $e^+ e^- \to q \bar{q} g$ squared matrix element.

For the last contribution in \Eq{eq:calc}, we rewrite 
\begin{align}
 \theta(\phi_{12}-R)\,  \theta(\phi_{13}-R)\, \theta(\phi_{23}-R) 
 = 1 -  \theta(R-\phi_{12}) - \theta(R - \phi_{13}) - \theta(R - \phi_{23}) \,.
\end{align}
where the first term in the sum corresponds to the calculation of the matching coefficients for inclusive fragmentation, thus yielding the $C_i(z_J, E_{\rm cm},\mu) \GFF_i(x,\mu)$ contribution on the right-hand side of \Eq{eq:cross_section}. For the remaining terms, we can follow the same strategy as in \Eq{eq:calc_2}. For example, the $- \theta(R - \phi_{13})$ term gives
\begin{align} \label{eq:calc_3}
\left. \frac{1}{\sigma^{(0)}}\frac{\text{d}\sigma}{\text{d}z_J\, \text{d}x} \right \vert_{-13}
&= - \frac{\alpha_s C_F}{2\pi} \int\! {\rm d}y_1\, {\rm d}y_2\, 
 \bigg\{ \frac{1+ y_1^2}{(1\!-\!y_1)}_+ \frac{1}{(1\!-\!y_2)}_+ + (\pi^2 - 8) \delta(1\!-\!y_1)\delta(1\!-\!y_2)
\nonumber \\ & \quad
+ \delta(1-y_2)\bigg[ 
\frac{P_{q \to q g}(y_1)}{C_F}
 \ln \frac{y_1E_{\rm cm}^2}{\mu^2} 
 + (1+y_1^2)\Big(\frac{\ln(1-y_1)}{1-y_1}\Big)_+ + 1-y_1 \bigg]  
\bigg\}
\nonumber \\ & \quad \times
\theta\Big(\frac{R^2}{4} - \frac{1-y_2}{y_1\,(1-y_1)}\Big) 
\bigg[ 
  \delta(z_J - y_1) \GFF^{(0)}_q(x,\mu) 
 \nonumber \\ & \quad +  
  \delta(z_J - 1) \GFF^{(0)}_{\bar q}(x,\mu) +
  \delta(z_J - 1 + y_1) \GFF^{(0)}_g(x,\mu)
 \bigg]
 \nonumber \\
 &= -\frac{\alpha_s}{2\pi} \int\! \text{d}z\,\bigg\{
P_{q \to q g}(z)
 \ln \frac{z^2 E_{\rm cm}^2 R^2}{4\mu^2} 
 + C_F \bigg[ 2(1+z^2)\Big(\frac{\ln(1-z)}{1-z}\Big)_+ 
\nonumber \\ & \quad 
 + 1-z 
  +(\dots) \delta(1-z)
 \bigg] \bigg\}
\bigg[ 
  \delta(z_J - z) \GFF^{(0)}_q(x,\mu) 
 \nonumber \\ & \quad +  
  \delta(z_J - 1) \GFF^{(0)}_{\bar q}(x,\mu) +
  \delta(z_J - 1 + z) \GFF^{(0)}_g(x,\mu)
 \bigg]
 \nonumber \\
 &= \mathcal{J}^{(1)}_{q \to q} (z_J, E_{\rm jet} R,\mu)\, \GFF_q(x,\mu) 
 + \mathcal{J}^{(1)}_{q \to g} (z_J, E_{\rm jet} R,\mu)\, \GFF_g(x,\mu)- (\dots)_{13}
.\end{align}
The similarity with the calculation in \Eq{eq:calc_2} and 
the relationship between $\mathcal{J}^{(1)}_{q \to q}$, $\mathcal{J}^{(1)}_{q \to g}$ in
\Eq{eq:Jqqcoeff} together with \Eq{eq:J_rel} make this straightforward to verify. The $(\dots)_{13}$ term cancels in the sum with \Eq{eq:calc_2}. The $- \theta(R - \phi_{23})$ term corresponds to the term with quark and anti-quark interchanged and the $- \theta(R - \phi_{12})$ contribution is again suppressed by $\mathcal{O}(R^2)$. This completes the check of \Eq{eq:cross_section} at $\mathcal{O}(\alpha_s)$.

\section{A Non-Fractal Example: Sums of Weighted Energy Fractions}
\label{app:sums_wef}

While \Eq{eq:recurse} is rather general, there are of course many collinear-unsafe observables that are not fractal jet observables.
In this appendix, we give an explicit example of an observable that does not satisfy the requirements in \Sec{sec:fractal_preamble}.

Consider two weighted energy fractions
\be
x = \sum_{i\in \text{jet}} w_i \, z_i^\kappa, \qquad y = \sum_{i\in \text{jet}} v_i \, z_i^\lambda,
\ee
for particle weights $w_i$ and $v_i$, and energy exponents $\kappa$ and $\lambda$.  Individually, $x$ and $y$ are described by the evolution equation in \Eq{eq:evolution_repeated}.  On the other hand, their sum
\be
t = x + y
\ee
is not a fractal jet observable, though it still can be described by a GFF.

To see this, consider the GFF for $t$, $\GFF_i(t)$, which can be written in terms of a joint GFF for $x$ and $y$ as
\be
\label{eg:GFFfromjoint}
\GFF_i(t) =  \int \text{d} x \,  \text{d} y \, \GFF_i(x, y) \, \delta[t - x - y].
\ee
The evolution equation for the joint GFF follows from the analysis in \Eq{eq:GFFNLO}, leading to 
\begin{align}
\mu \frac{\text{d}}{\text{d}\mu}  \GFF_i(x,y; \mu) =  \frac{\alpha_s(\mu)}{2\pi} \sum_{j,k} & \int\text{d}z\, \text{d}x_1\, \text{d}x_2\, \text{d}y_1\, \text{d}y_2\,  P_{i\rightarrow jk}(z)\, \GFF_j (x_1,y_1;\mu)\, \GFF_k(x_2,y_2;\mu) \nonumber \\
& \quad \times \delta\big[x - z^\kappa x_1 - (1-z)^\kappa x_2 \big] \, \delta \big[ y - z^\lambda y_1 - (1-z)^\lambda y_2 \big].
\label{eg:GFFjointevolution}
\end{align}
Plugging \Eq{eg:GFFjointevolution} into \Eq{eg:GFFfromjoint}, we can insert a factor of
\be
1 \equiv \int \text{d} t_1\, \text{d}t_2\,  \delta[t_1-x_1-y_1] \, \delta[t_2-x_2-y_2]
\ee
to perform the integrals over $y_1$ and $y_2$.  The resulting equation is
\begin{align}
\mu \frac{\text{d}}{\text{d}\mu} \GFF_i(t; \mu) &= \frac{\alpha_s(\mu)}{2\pi} \sum_{j,k} \int \text{d}z\, \text{d}t_1\, \text{d}t_2\, \text{d}x_1\, \text{d}x_2\, P_{i\rightarrow jk}(z)\, \GFF_j(x_1,t_1-x_1)\, \GFF_k(x_2,t_2-x_2) 
\nonumber \\ &\quad \times 
\delta\big[t - z^\lambda t_1 - (1-z)^\lambda t_2 - (z^\kappa-z^\lambda)x_1 - \big((1-z)^\kappa -(1-z)^\lambda \big)x_2\big].
\label{eq:frustratingequation}
\end{align}
As written, this is a valid GFF evolution equation, but the GFF for $t$ explicitly involves the joint GFF for $x$ and $y$, so we do not get an evolution equation of the form of \Eq{eq:evolution_repeated}.

If and only if $\kappa = \lambda$, can we cancel the $x_1$ and $x_2$ terms inside of the $\delta$ function in \Eq{eq:frustratingequation}.  In that case, we can rewrite the joint probabilities as probability densities for the sums $t_1 = x_1 + y_1$ and $t_2 = x_2 + y_2$, so that the evolution equation is of the desired fractal form.  Of course, $\kappa = \lambda$ just corresponds to a regular weighted energy fraction with weights $w_i + v_i$, so this is not a new fractal observable.

\section{Software Implementation}
\label{app:implementation}

The software to perform the RG evolution in this paper is available from the authors upon request.  In this paper, we discuss some of the specifics of its implementation.  A public version of the code is planned for a release some time in the future.

\subsection{Running Coupling}

Because we only perform leading-order evolution, the running of $\alpha_s$ is strictly speaking only required at leading-logarithmic accuracy.
In our implementation, though, the running of the strong coupling is included using the $\beta$ function at $\mathcal{O}(\alpha_s^3)$,
\begin{equation}\label{eq:rg-two-loop}
\mu \frac{\text{d}\alpha_s(\mu)}{\text{d}\mu} = -2\alpha_s \bigg(\beta_0\left(\frac{\alpha_s}{4\pi}\right) + \beta_1\left(\frac{\alpha_s}{4\pi}\right)^2 \bigg),
\end{equation}
\begin{equation}\label{eq:beta-function}
\beta_0 = \frac{11}{3}C_A-\frac{4}{3}T_F n_f, \qquad \beta_1 = \frac{34}{3}C_A^2 - \frac{20}{3} C_A T_F n_f - 4C_F T_F n_f.
\end{equation}
The running coupling at the scale $\mu$ is given by solving \Eq{eq:rg-two-loop} iteratively to order $\mathcal{O}(\alpha_s^3)$,
\begin{equation}\label{eq:running-alpha}
\alpha_s(\mu) = \frac{4\pi}{\beta_0}\bigg( \frac{1}{L} - \frac{\beta_1}{\beta_0^2 L^2}\ln L\bigg),
\end{equation}
where $L = \ln \frac{\mu^2}{\Lambda_{\rm QCD}^2}$. Using the PDG value $\alpha_s(M_Z) = 0.1181$ gives the boundary condition $\Lambda_{\rm QCD} = 0.2275$ GeV. The group theory factors for QCD are $C_F = \frac{4}{3}$, $T_F = \frac{1}{2}$, and $C_A = 3$.  For applications to the LHC running at 13 TeV, the number of quark flavors is $n_f = 5$.

\subsection{Discretization}

The evolution equation in \Eq{eq:evolution_repeated} can be solved by binning the values of the GFFs in the $x$ variable. If the GFF domain is partitioned into $N$ bins, \Eq{eq:evolution_repeated} becomes a set of $(2n_f + 1)N$ coupled ordinary differential equations. The evolution equation for the binned GFF for bin $n$, $\widetilde{\GFF}_i (n,\mu)$, is given by\footnote{This equation is written for $N$ equal-width bins for simplicity of notation. The generalization to unequal bins is straightforward, and the software implementation is set up to handle variable bin widths if desired.}
\begin{align}\label{eq:discretization}
\frac{\text{d}}{\text{d}\ln \mu}\, \widetilde{\GFF}_i (n,\mu) &\equiv  \frac{\text{d}}{\text{d}\ln \mu} N \int_{\frac{(n-1)}{N}}^{n/N} \text{d}x\, \GFF_i(x,\mu) \\
&= \frac{N}{2}\sum_{j,k} \int_{\frac{(n-1)}{N}}^{n/N} \text{d}x\, \sum_{n_1,n_2} \int_{\frac{(n_1-1)}{N}}^{n_1/N} \text{d}x_1\, \int_{\frac{(n_2-1)}{N}}^{n_2/N} \text{d}x_2\, \int_0^1 \text{d}z\, P_{i\rightarrow jk}(z) \nonumber \\
&\quad \times \GFF_j(x_1,\mu)\, \GFF_k(x_2,\mu)\, \delta\big[x-\hat{x}(z,x_1,x_2)\big] \nonumber\\
&= \frac{N}{2}\sum_{j,k} \int_{\frac{(n-1)}{N}}^{n/N} \text{d}x\, \sum_{n_1,n_2} \int_0^1 \text{d}z \, P_{i\rightarrow jk}(z) \widetilde{\GFF}_j(n_1,\mu)\, \widetilde{\GFF}_k(n_2,\mu)\, \delta\big[x-\hat{x}(z,x_{n_1},x_{n_2})\big],
\nonumber\end{align}
where $x_{n_1}$ and $x_{n_2}$ are the positions of the midpoints of the $n_1$-th and $n_2$-th bins. Note that \Eq{eq:discretization} is written in terms of $\ln \mu$ instead of $\mu$, since this is how the evolution was implemented numerically to make the step size and numerical errors more consistent.
In principle, the $\delta$ function could be used to carry out the $z$ integral exactly. In practice, it is easier to discretize the $z$ integral and use the $\delta$ function to choose the $x$-bin corresponding to each triplet $(z,x_1,x_2)$. This is because inverting $\hat{x}$ to solve for $z$ analytically for general $x_1$ and $x_2$ is not possible. Doing so in advance separately for each value of $x$, $x_1$ and $x_2$ can be prohibitively memory intensive for large numbers of bins.

The splitting functions are approximated by the analytic value of their integral over the width of the bin.  For our analysis, we need the following splitting functions:
\begin{align}
P_{q\rightarrow gq}(z) &= P_{q\rightarrow qg}(1-z) = C_F \left( \frac{1+(1-z)^2}{z_+} + \frac{3}{2}\delta[z]\right), \nonumber \\
P_{g\rightarrow q\bar{q}}(z) &= T_F \left(z^2 + (1-z)^2\right), \nonumber \\
P_{g\rightarrow gg}(z) &= 2C_A\left(\frac{1-z}{z_+} + \frac{z}{(1-z)}_+ \!+ z(1-z)\right) + \frac{\beta_0}{2}\big(\delta[1-z]+\delta[z]\big),
\end{align}
where $P_{q\rightarrow gq}(z)$ is the splitting function for a quark radiating a gluon with momentum fraction $z$, the integration constant for integrals of the plus distributions are fixed by 
\be
\int_0^1\,\frac{\text{d}z}{z_+} = 0\,, \qquad
\int_0^1\,\frac{\text{d}z}{(1-z)}_+ = 0\,,
\ee
and $\beta_0$ is given in \Eq{eq:beta-function}.\footnote{The $1/z_+$ and $\delta(z)$ terms in $P_{q\rightarrow gq}(z)$ and $P_{g\rightarrow gg}(z)$ are necessary because the evolution in \Eq{eq:evolution} requires distributions that are also regulated at $z=0$.} 
When performing the integration, terms with a plus-function regulator must be handled correctly for the endpoint bins. 
If the regulated functions have the following primitives
\begin{equation}
\frac{\text{d}F(z)}{\text{d}z} = \frac{f(z)}{z}, \qquad \frac{\text{d}G(z)}{\text{d}z} = \frac{g(z)}{1-z},
\end{equation}
then their integrals over the $n$-th bin are implemented by
\begin{align}
\int_{z-0.5\Delta z}^{z+0.5\Delta z} \text{d}z' \frac{f(z')}{z'_+} &= 
\begin{cases} 
F(z+0.5\,\Delta z)-F(z-0.5\,\Delta z) &n \ne 0,\\
F(z+0.5\,\Delta z) & n = 0,
\end{cases}
\nonumber \\ 
\int_{z-0.5\Delta z}^{z+0.5\Delta z} \text{d}z' \frac{g(z')}{(1-z')_+} &= 
\begin{cases} 
G(z+0.5\,\Delta z)-G(z-0.5\,\Delta z)& n \ne n_{\text{final}}, \\
G(z-0.5\,\Delta z) & n = n_{\text{final}}.
\end{cases}
\end{align}

\begin{figure}[t]
\subfloat[]{
		\includegraphics[width=0.45\textwidth]{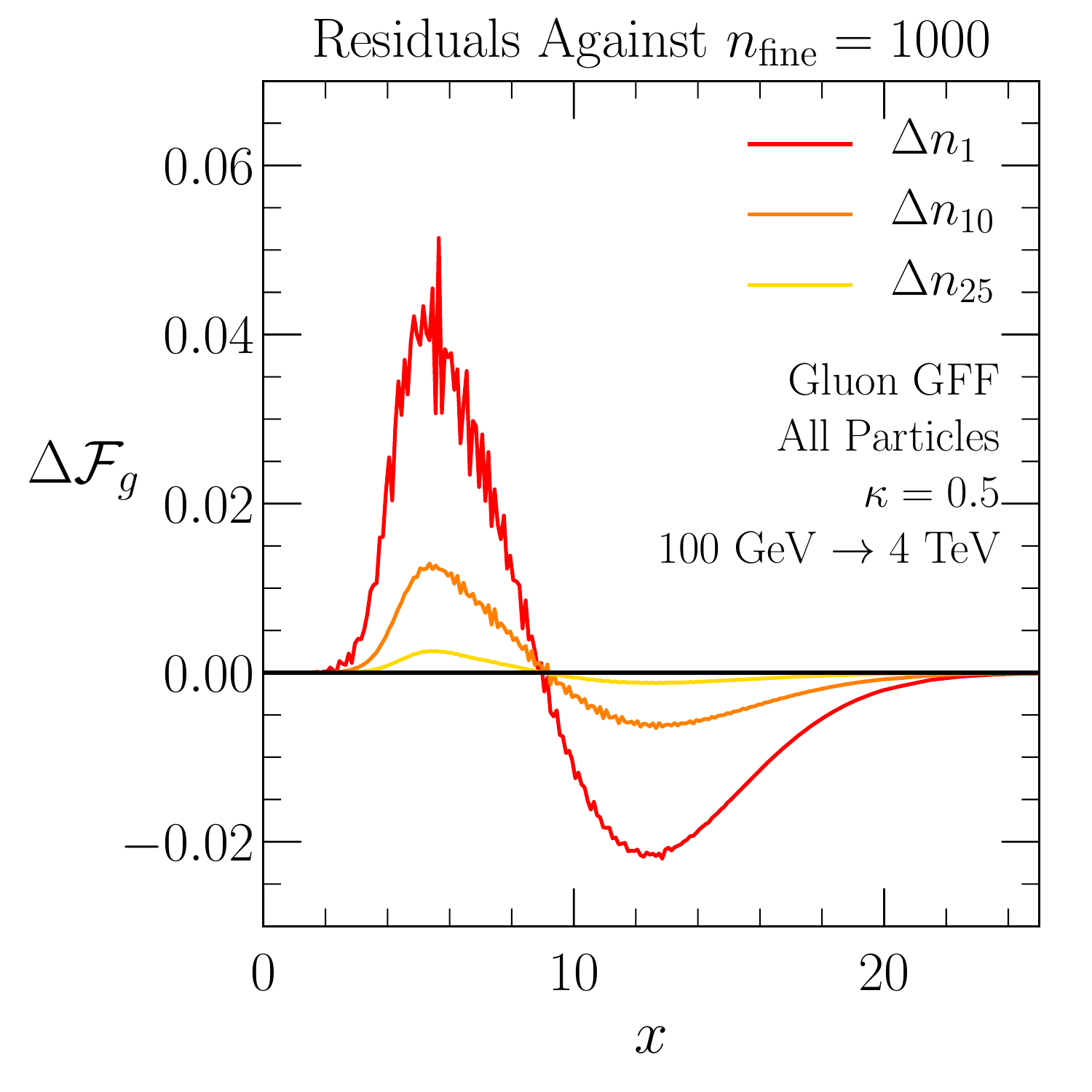}
		 \label{fig:subbins-a}
}
\subfloat[]{
		\includegraphics[width=0.45\textwidth]{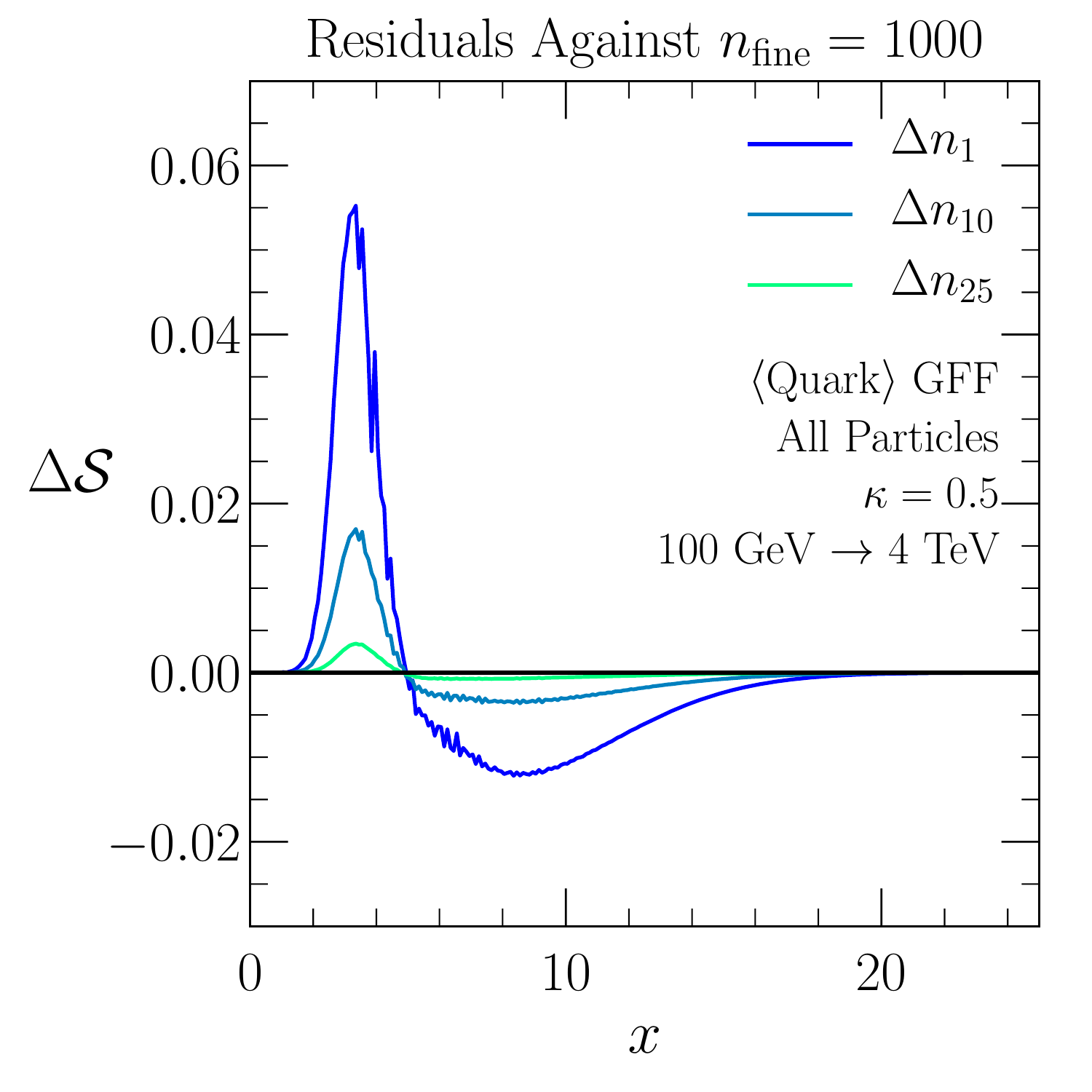}
		 \label{fig:subbins-b}
}
	\caption{\label{fig:subbins}Sensitivity of the evolution from $\mu =100$ GeV to 4 TeV on the choice of fine bin width.  Shown are the (left) gluon GFF and (right) quark-singlet GFF for the weighted energy fraction with $\kappa=0.5$.  The  curves labeled $\Delta n_X$ are the difference between the result using $n_{\rm fine} = X$ and the result using $n_{\rm fine} = 1000$.  For the default value of $n_{\rm fine}=100$ used in this paper, the results are indistinguishable by eye. 
}
\end{figure}

In our implementation, the integration range $z\in [0,1]$ is divided into $n_{\text{rough}}$ bins, and the first and last bin are then further subdivided by a factor of $n_{\text{fine}}$. The user can specify these two parameters. For the results presented in this paper, the values used were $n_{\text{rough}} = 1000$ and $n_{\text{fine}} = 100$. The finer division of the endpoint bins is necessary to accurately capture the singular behavior of the splitting functions near $z=0$ and $z=1$. For many GFFs, this is not necessary, but consider the weighted energy fractions, whose recursion relation satisfies
\begin{equation}
\hat{x}(z,x_1,x_2) = z^\kappa x_1 + (1-z)^\kappa x_2 \implies \frac{\partial \hat{x}}{\partial z} = \kappa \, (z^{\kappa-1}x_1 - (1-z)^{\kappa-1}x_2).
\end{equation}
For $\kappa < 1$, there are poles in the derivative of $\hat{x}$ at $z=0$ and $z=1$, resulting in a noticeable dependence on $n_{\text{fine}}$. This is shown in \Fig{fig:subbins} for the case of $\kappa = 0.5$, with all particle weights one.  Once we increase $n_{\rm fine}=100\rightarrow 1000$, the maximum change in the value of the evolved GFFs in a single $x$-bin is less than 0.06$\%$.

\subsection{Runge-Kutta Algorithm}

After the discretization in \Eq{eq:discretization}, the RG evolution is performed with an embedded fifth-order Runge-Kutta method adapted from \Ref{numrecipes}. This method requires six evaluations of the right side of \Eq{eq:evolution_repeated}, which on the $k$th step can be combined to give a fifth-order estimate $y_{k+1}$ of the desired function after a step of size $h_k$. These computations can be recombined with different coefficients to give a fourth-order Runge-Kutta estimate $y_{k+1}^*$. The difference between these two methods then gives an estimate of the local truncation error. 
The error estimated this way applies to the fourth-order value $y_{k+1}^*$, but we take the (more accurate) fifth-order value. This ensures that our solution is actually slightly more accurate than our error indicates. Estimating the error on this fifth-order solution would require calculating a still-higher order step. 

Once a step $h_k$ is taken, with an error $\mathcal{E}_k$, we would like to choose an appropriate trial value for our next step. This fourth-order error estimate scales as $\mathcal{O}(h^5)$, so we choose the next step, $h_{k+1}$, to be
\begin{equation}
h_{k+1} = \begin{cases} S \, h_k \vert \frac{\mathcal{E}_{k+1}}{\mathcal{E}_k}\vert^{0.20} & \mathcal{E}_{k+1} > \mathcal{E}_k, \\
S \, h_k \vert \frac{\mathcal{E}_{k+1}}{\mathcal{E}_k}\vert^{0.25} & \mathcal{E}_{k+1} < \mathcal{E}_n.
\end{cases}
\end{equation}
Here,  $\mathcal{E}_{k+1}$ is the projected error in the $(k+1)$th step, and $S$ is a safety factor taken to be $0.9$. This formula allows the step size to grow if the error is much smaller than our tolerance. If the error is larger than the tolerance, the step fails, and is retried with a smaller step.

It is important that the algorithm be able to dynamically change step size in order to evolve a solution efficiently while keeping errors within desired limits. At low scales, the strong coupling grows large, and the solution changes rapidly. Numerical precision therefore requires small step sizes in this region. At high scales, asymptotic freedom ensures that the solutions change slowly, so much larger step sizes result in the same level of accuracy. This procedure requires a prescription for the maximal acceptable error. For a system of $M\equiv (2n_f+1)n$ coupled ODEs, there is a separate $\mathcal{E}_k^m$ for each $m \in M$. The step is considered a failure unless every equation is within its error tolerance. The error $\mathcal{E}^m_k$ for the $m$th equation on the $k$th step is required to satisfy
\begin{equation}\label{eq:error-bound}
\Bigg\vert \frac{\mathcal{E}_k^m}{\vert y_k^m\vert + \vert h_k\big(\text{d}y^m_k/\text{d}\ln \mu\big)\vert +10^{-6}}\Bigg\vert < \epsilon .
\end{equation}
The value $\epsilon$ is an overall upper limit which was set to $10^{-9}$ for the GFF evolution. The last numerical term in the denominator is required to avoid artificially large errors when the domain of the GFFs input into the program exceeds the actual support of the GFF.  As an additional constraint, our algorithm sets a maximum step size of ${\rm d}\ln \mu \le 0.4$.  Note that the same step size is used for every equation in the system. 

\begin{figure}[p]
\subfloat[]{\includegraphics[width=0.45\textwidth]{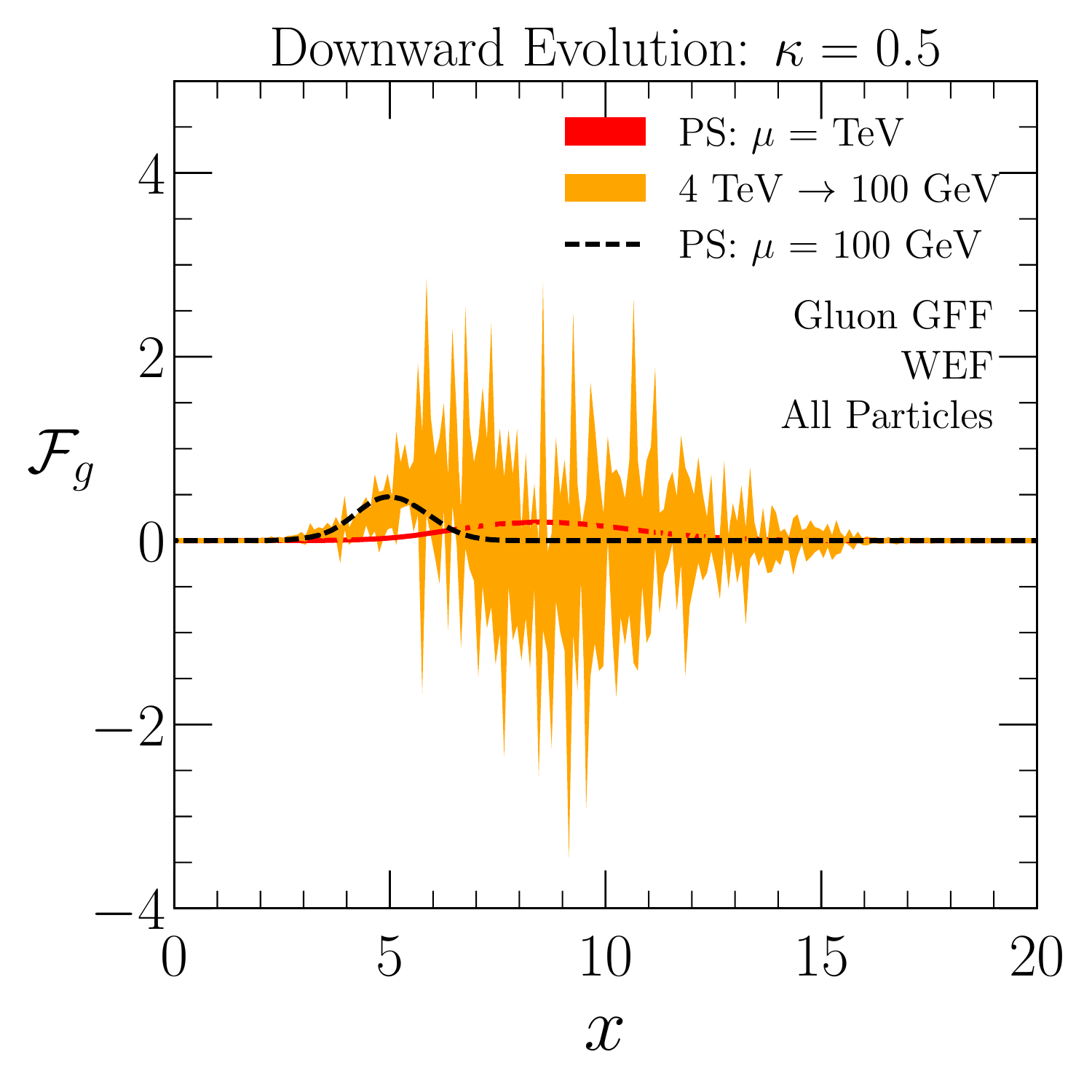}
\label{fig:numerical-stability-g}}
\subfloat[]{\includegraphics[width=0.45\textwidth]{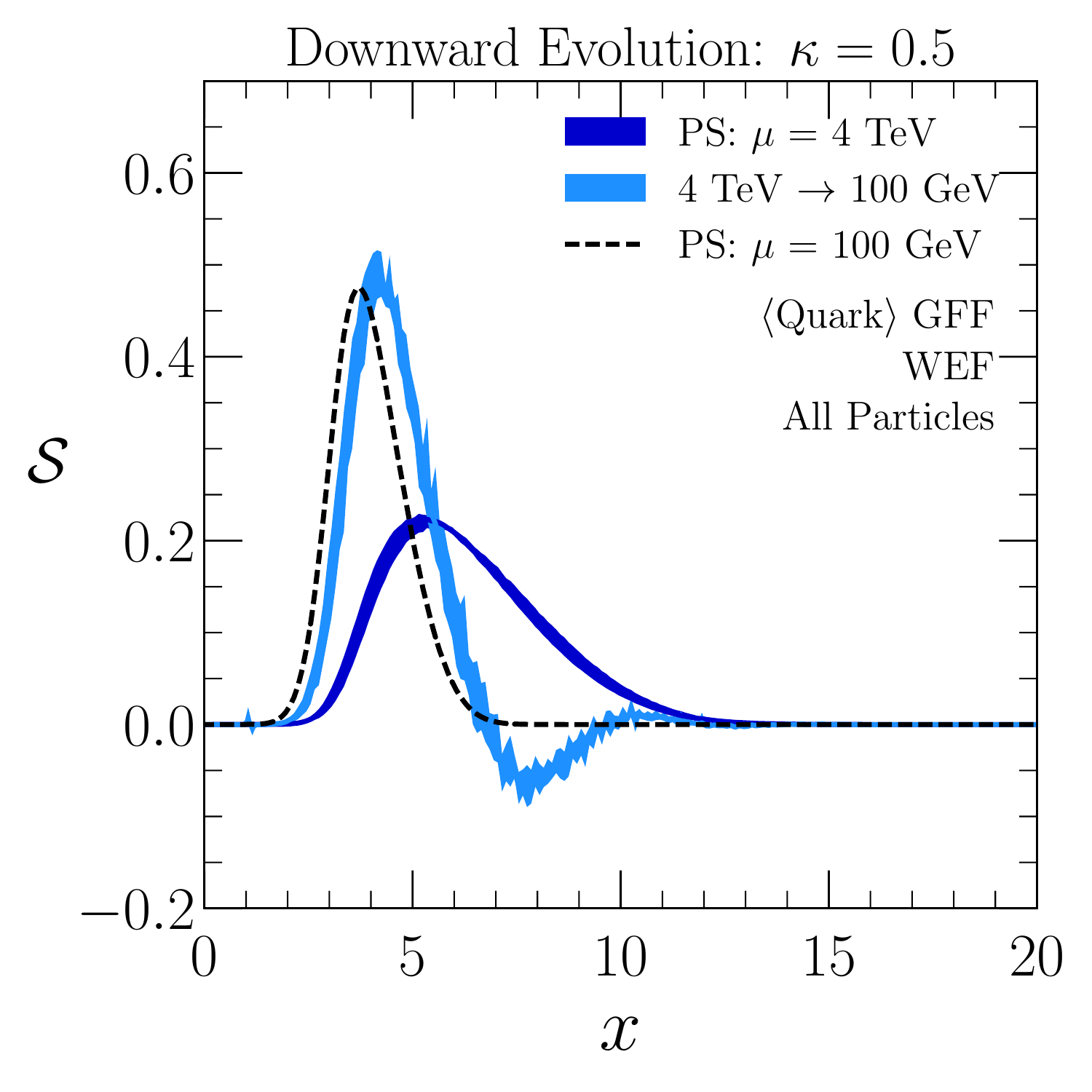}
\label{fig:numerical-stability-q}
}

\subfloat[]{\includegraphics[width=0.45\textwidth]{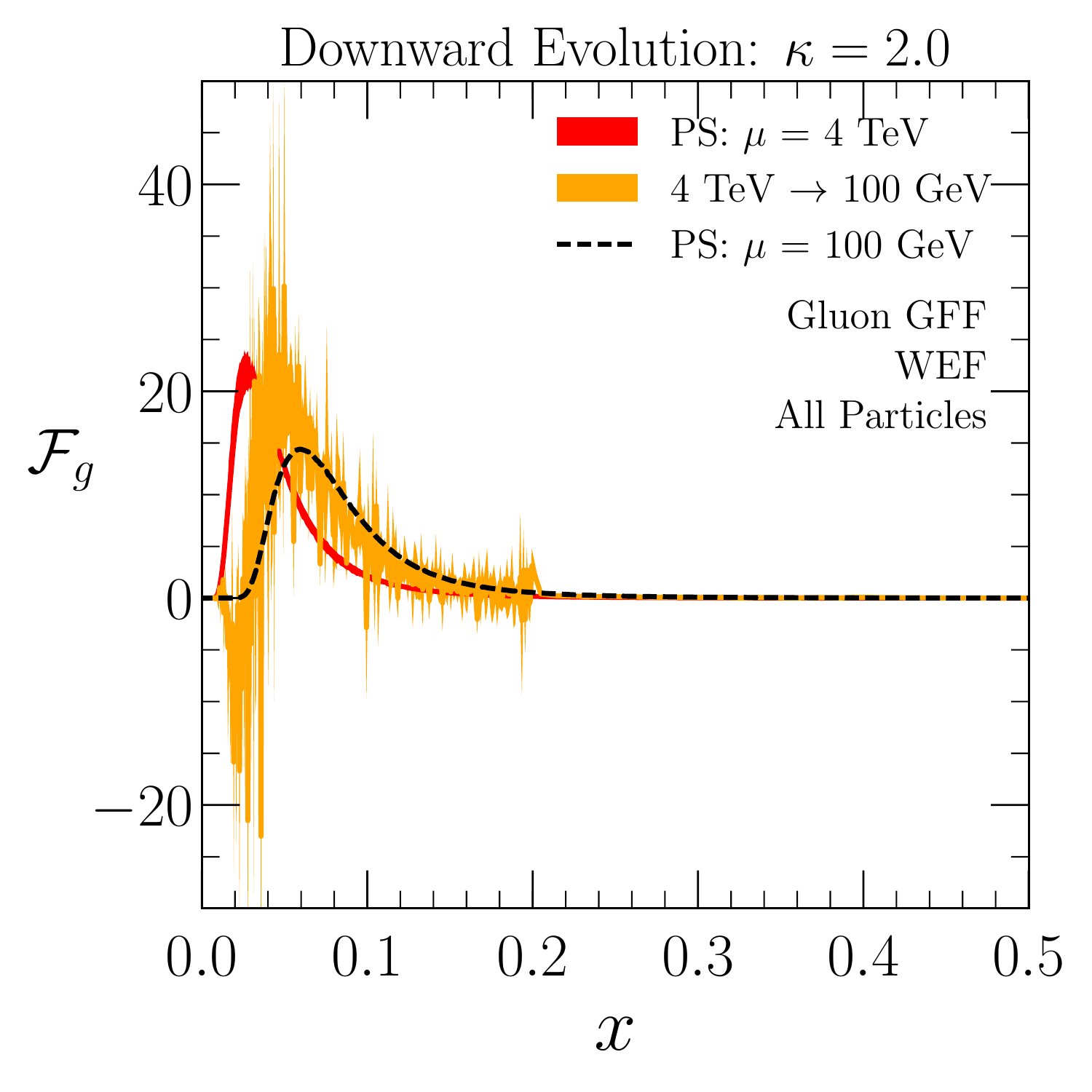}
\label{fig:numerical-stability-g-2}
}
\subfloat[]{\includegraphics[width=0.45\textwidth]{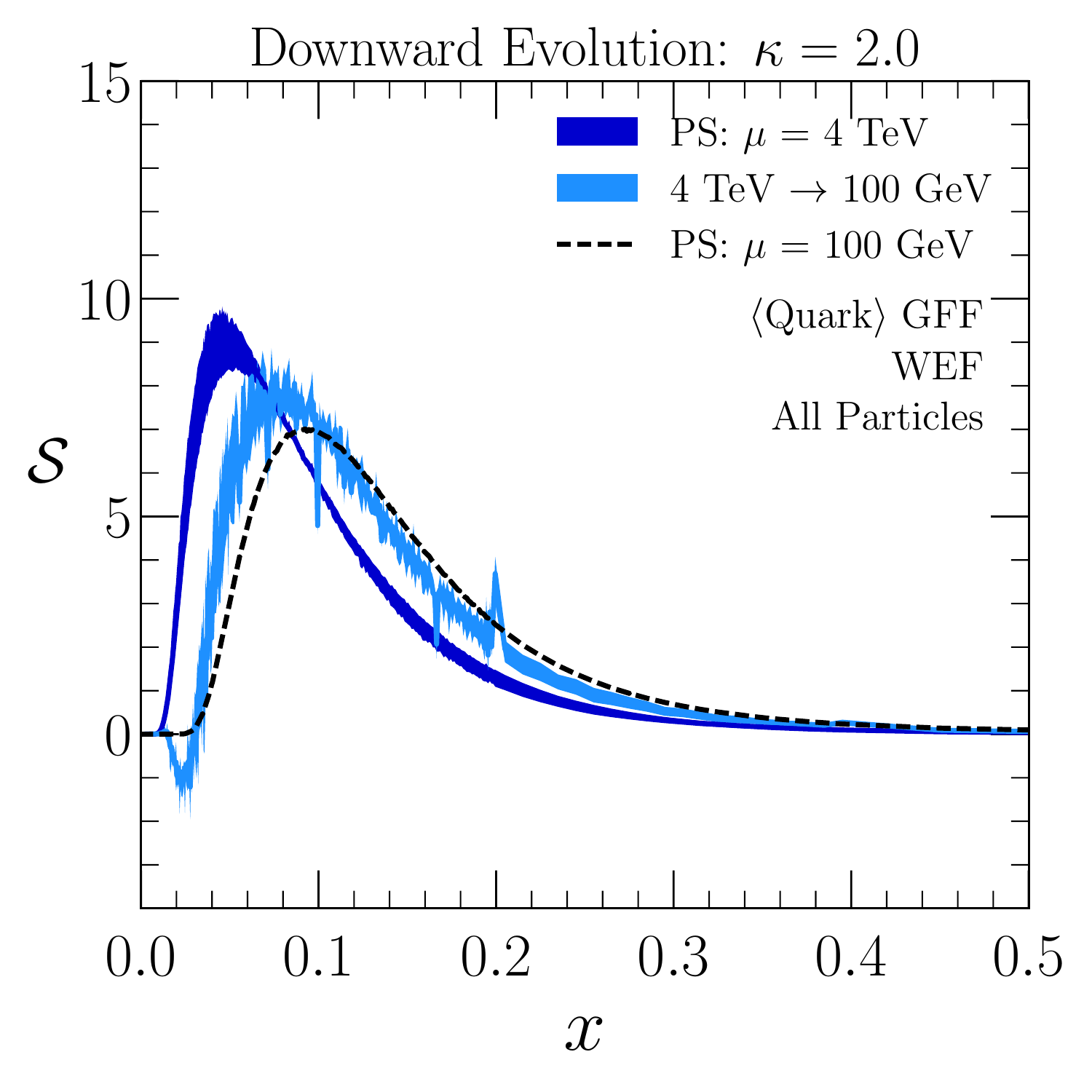}
\label{fig:numerical-stability-q-2}
}
\caption{Downward evolution from $\mu = 4$ TeV to $\mu = 100$ GeV of the (left column) gluon GFF and (right column) quark-singlet GFF with (top row) $\kappa=0.5$ and (bottom row) $\kappa=2.0$.  The envelopes of the evolved distributions are constructed as in \Sec{sec:wefextraction} by varying the jet radius $R$ and the choice of parton shower, which highlight the numerical instability of downward evolution.} \label{fig:numerical-stability-1}
\end{figure}

\section{Numerical Stability}
\label{app:numerical_stability}

All of the RG results in this paper are based on the numerical solution of \Eq{eq:evolution_repeated} for upwards evolution in the scale $\mu$.
The reason is because downward evolution is numerically unstable, in the sense that small irregularities in the initial conditions amplify into large fluctuations, especially for the gluon GFFs.  This behavior is illustrated in \Fig{fig:numerical-stability-1}, where gluon and quark-singlet GFFs are evolved downward from 4 TeV to 100 GeV. 

Heuristically, if evolution upwards in scale is analogous to convolution of the GFFs, evolution downwards is akin to deconvolution, a problem known to be ill-posed. To verify that the instability is inherent to the differential equation, and not merely a numerical artifact, we checked that the envelope shown in \Fig{fig:numerical-stability-1} is not affected by choosing a smaller step size or more stringent error bound in the Runge-Kutta algorithm.
To get a sensible result, one could use a numerical regularization method such as Tikhonov regularization \cite{tikhonov1977solutions}, though we do not do so here.  Note that in general, if the evolution in one direction is stable, such that small fluctuations get washed out, the evolution is expected to be unstable in the reverse direction.

\section{Moment Space Details}
\label{app:momentspace}

In this appendix, we give details of the moment space analysis from \Sec{sec:wefmoments}, as well as perform similar analyses for the non-associative observables from \Sec{sec:NA}.
The moments of the GFFs are defined by
\begin{equation}\label{eq:mom_def}
\overline{\GFF}_i (N,\mu) = \int \text{d}x\, x^N \GFF_i(x,\mu)
\,,\end{equation}
where the zeroth moment is just the normalization,
\begin{equation} \label{eq:GFF_norm}
\overline{\GFF}_i (0,\mu) = \int \text{d}x\, \GFF_i(x,\mu) = 1
.\end{equation}
This convention follows the standard nomenclature of probability theory.  Applying $\int_{-\infty}^{+\infty} \text{d}x\, x^N$ to both sides of the evolution equation in \Eq{eq:evolution_repeated} gives the moment space evolution equation,
\begin{equation}
\label{eq:moment-space-general}
\mu \frac{\text{d}}{\text{d}\mu} \overline{\GFF}_i (N,\mu) = \frac{1}{2}\sum_{j,k} \int \text{d}z\, \text{d}x_1\, \text{d}x_2\, \big(\hat{x}(z,x_1,x_2)\big)^N\, \frac{\alpha_s(\mu)}{\pi} P_{i\rightarrow jk}(z)\, \GFF_j (x_1,\mu)\, \GFF_k (x_2,\mu).
\end{equation}
In order to proceed further, we need the specific form of the recursion relation, $\hat{x}$. We now discuss the details for each of the sets of observables studied in this paper.  

\subsection{Weighted Energy Fractions}
\label{app:moment_wef}

Inserting the weighted energy fraction recursion relation \Eq{eq:wefrecur} into \Eq{eq:moment-space-general} leads to
\begin{align}
\mu \frac{\text{d}}{\text{d}\mu}\, \overline{\GFF}_i (N,\mu) &= \frac{\alpha_s(\mu)}{2\pi} \sum_{j,k} \sum_{M=0}^N \binom{N}{M} \, \int_0^1 \text{d}z\, z^{\kappa(N-M)}(1-z)^{\kappa M} P_{i\rightarrow jk}(z) 
\nonumber \\
&\quad \times 
\int \text{d}x_1\, x_1^{N-M} \GFF_j (x_1,\mu)\, \int \text{d}x_2\, x_2^{M} \GFF_k (x_2,\mu),
\end{align}
assuming that $N$ is integer and using the binomial theorem.
As in \Eq{eq:Pmoments}, the moments of the splitting functions are defined as
\begin{equation}
\overline{P}_{i\rightarrow j,k} (N,M) = \int_0^1 \text{d}z\, z^N(1-z)^M P_{i\rightarrow j,k}(z)
\,,\end{equation}
with the convention that $\overline{P}_{i\rightarrow j,k}(N)\equiv \overline{P}_{i\rightarrow j,k}(N,0)$. 
For any real $N > 0$, they can be expressed in terms of the digamma function $\psi_0(N)$ and the Euler-Mascheroni constant $\gamma_E$,
\begin{align}\label{eq:wef-splitting-kernels}
	\overline{P}_{q\rightarrow qg}(N) &= C_F\left(\frac{3}{2} + \frac{1}{N+1} + \frac{1}{N+2} -2\gamma_E - 2\psi_0(N+3)\right)\,, \nonumber \\
	\overline{P}_{q\rightarrow gq}(N) &= C_F\left(\frac{N^2 + 3N + 4}{N(N+1)(N+2)} \right)\,, \nonumber \\
	\overline{P}_{g\rightarrow q\bar{q}}(N) &= T_F\left(\frac{N^2 + 3N + 4}{(N+1)(N+2)(N+3)}\right)\,, \nonumber \\
	\overline{P}_{g\rightarrow gg}(N) &= 2C_A\left(\frac{11}{12}  + \frac{2(N^2 + 3N + 3)}{N(N+1)(N+2)(N+3)} - \gamma_E - \psi_0(N+2) \right) - \frac{2}{3}T_Fn_f
\,.\end{align}
Alternatively, one can use the harmonic number function, $H_N=\gamma_E + \psi_0(N+1)$.
These expressions for all positive real numbers are necessary to evaluate the moment space evolution equation in \Eq{eq:momevolutionwef} for non-integer $\kappa$. Note that $N$ is shifted up by one from the expression usually seen in the literature, because our convention for moments in \Eq{eq:mom_def} is shifted by one as well compared to Mellin moments.

\subsection{Node Products}
\label{app:moment_nodeprod}

\begin{figure}[t]
\subfloat[]{
		\includegraphics[width=0.32\textwidth]{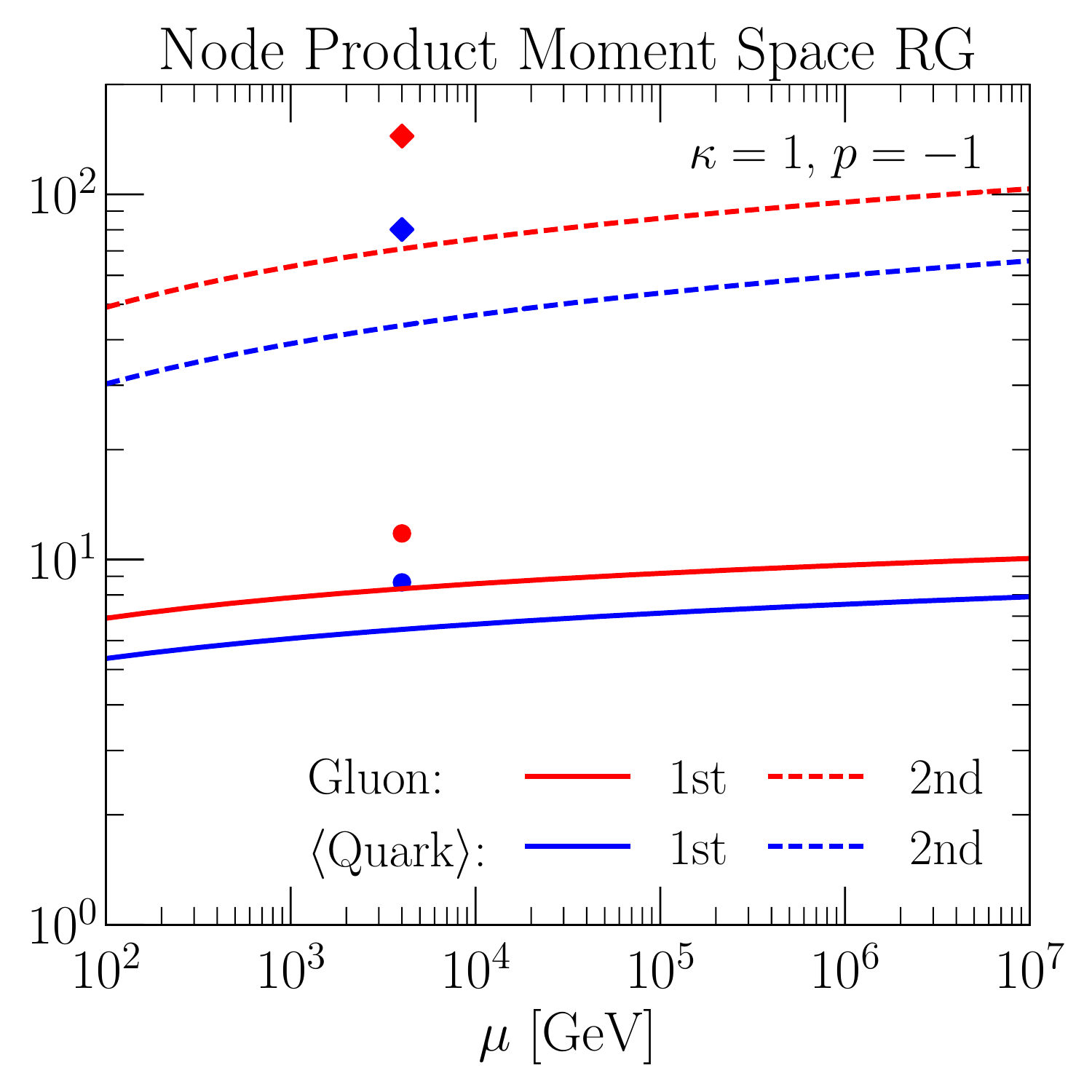}
		 \label{fig:moment-space-node-a} 
}
\subfloat[]{
		\includegraphics[width=0.32\textwidth]{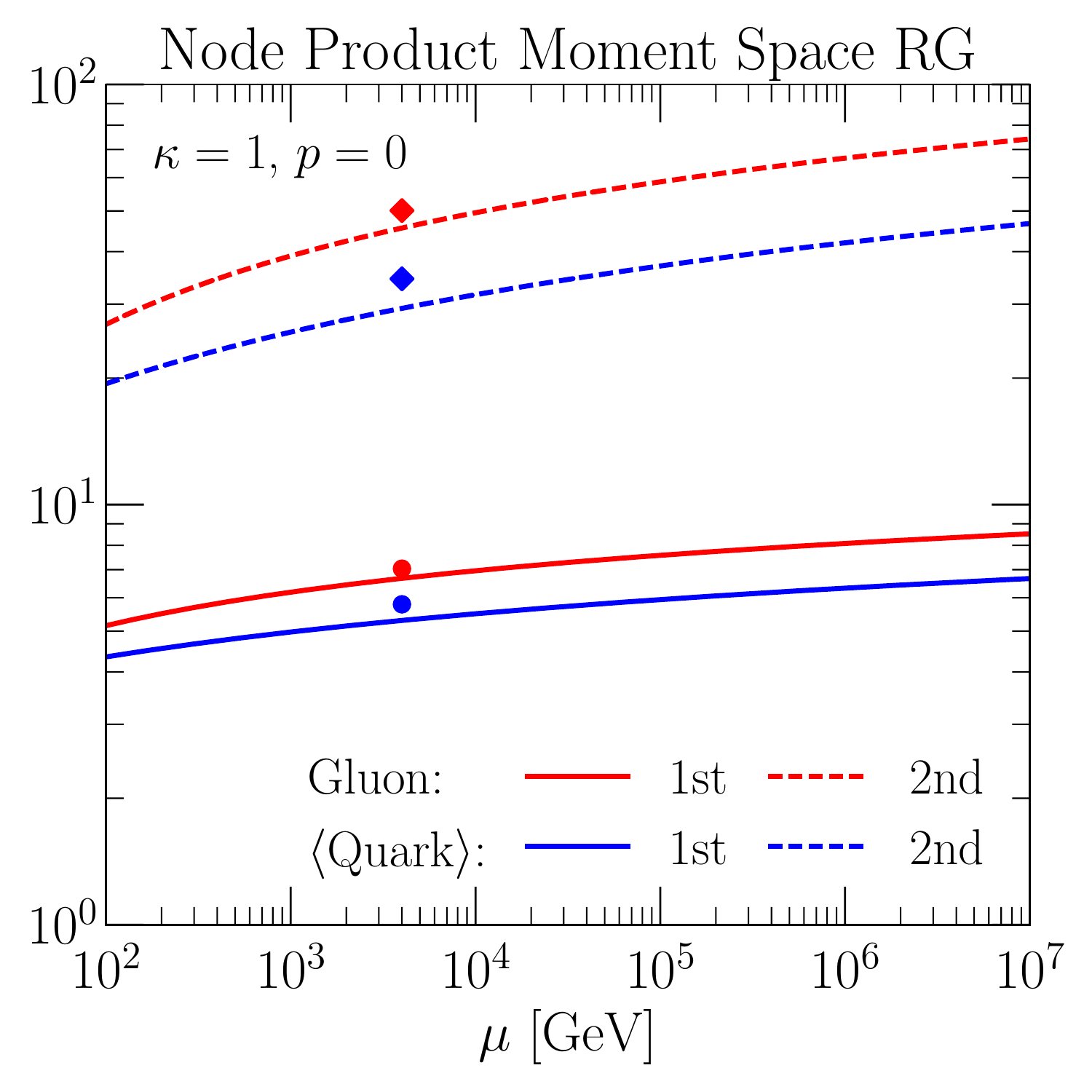}
		 \label{fig:moment-space-node-c}
}
\subfloat[]{
		\includegraphics[width=0.32\textwidth]{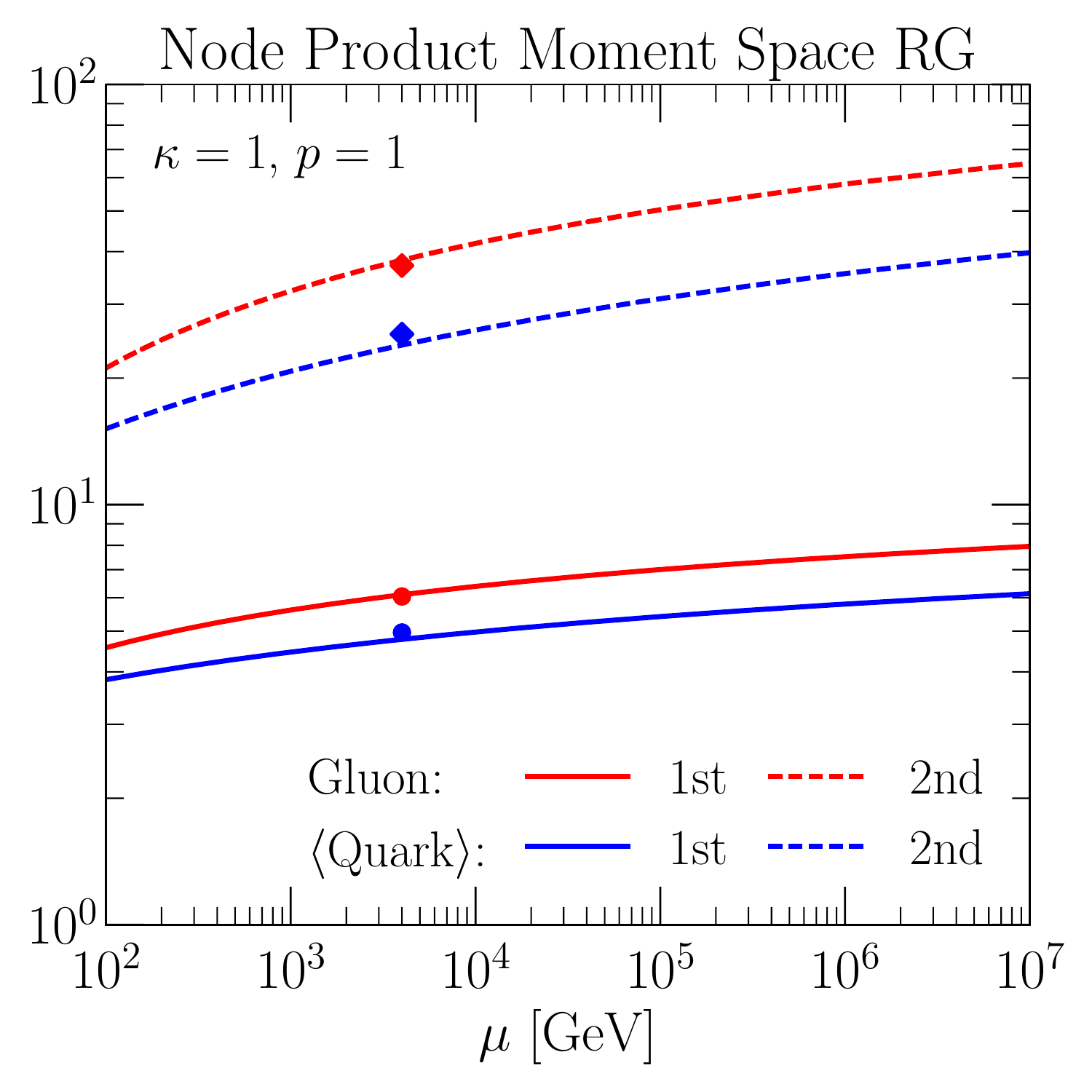}
		 \label{fig:moment-space-node-e}
}

\subfloat[]{
		\includegraphics[width=0.32\textwidth]{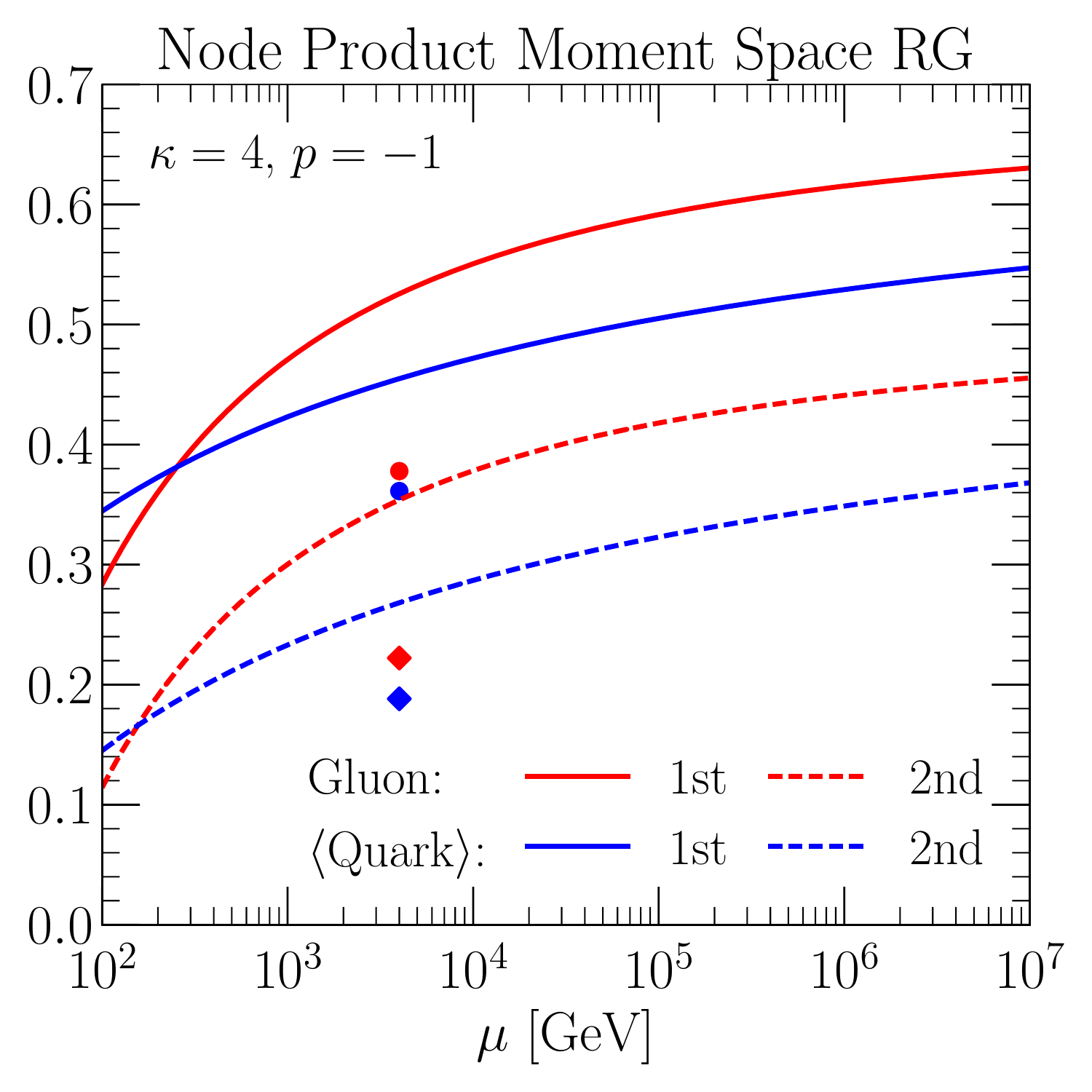}
		 \label{fig:moment-space-node-b}
}
\subfloat[]{
		\includegraphics[width=0.32\textwidth]{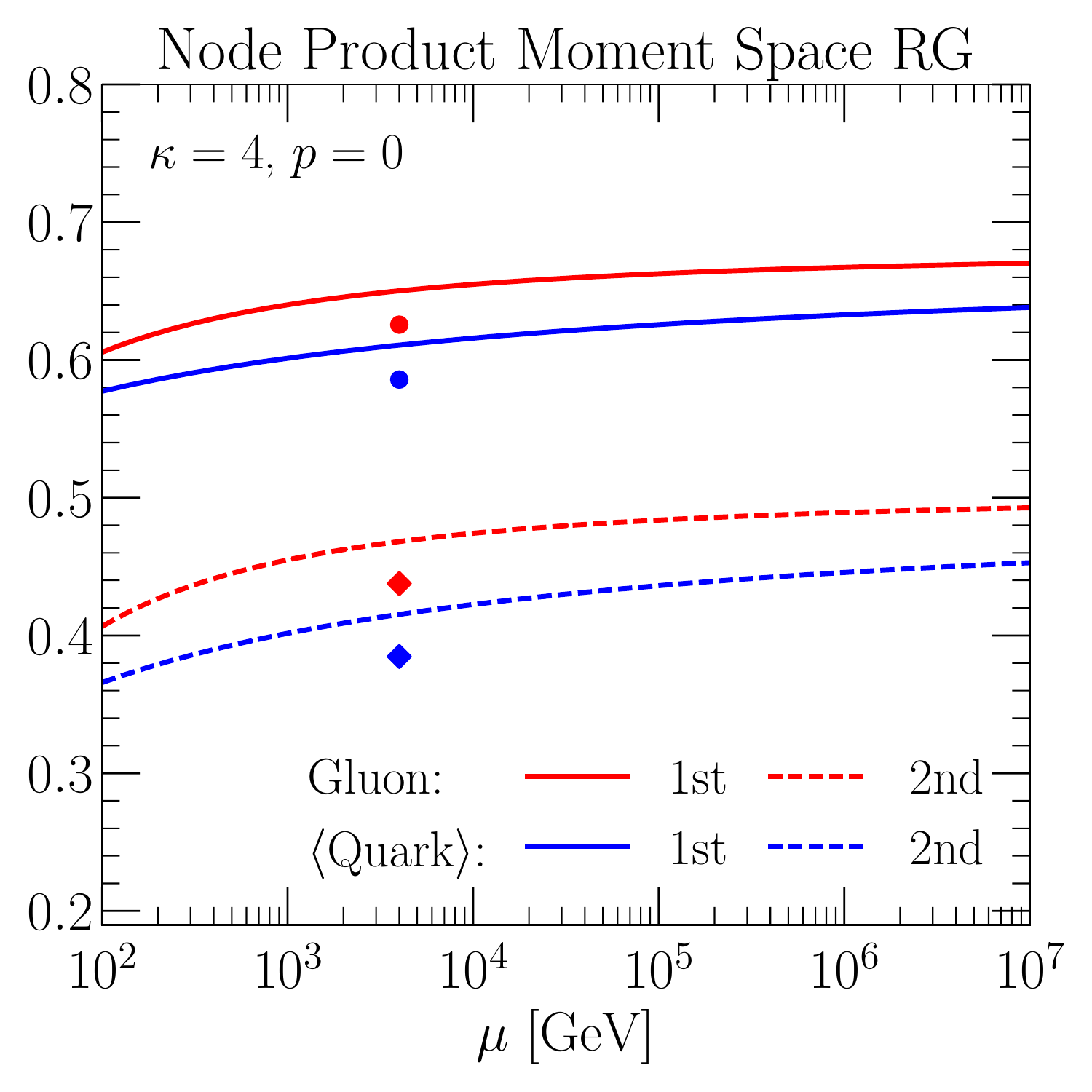}
		 \label{fig:moment-space-node-d}
}
\subfloat[]{
		\includegraphics[width=0.32\textwidth]{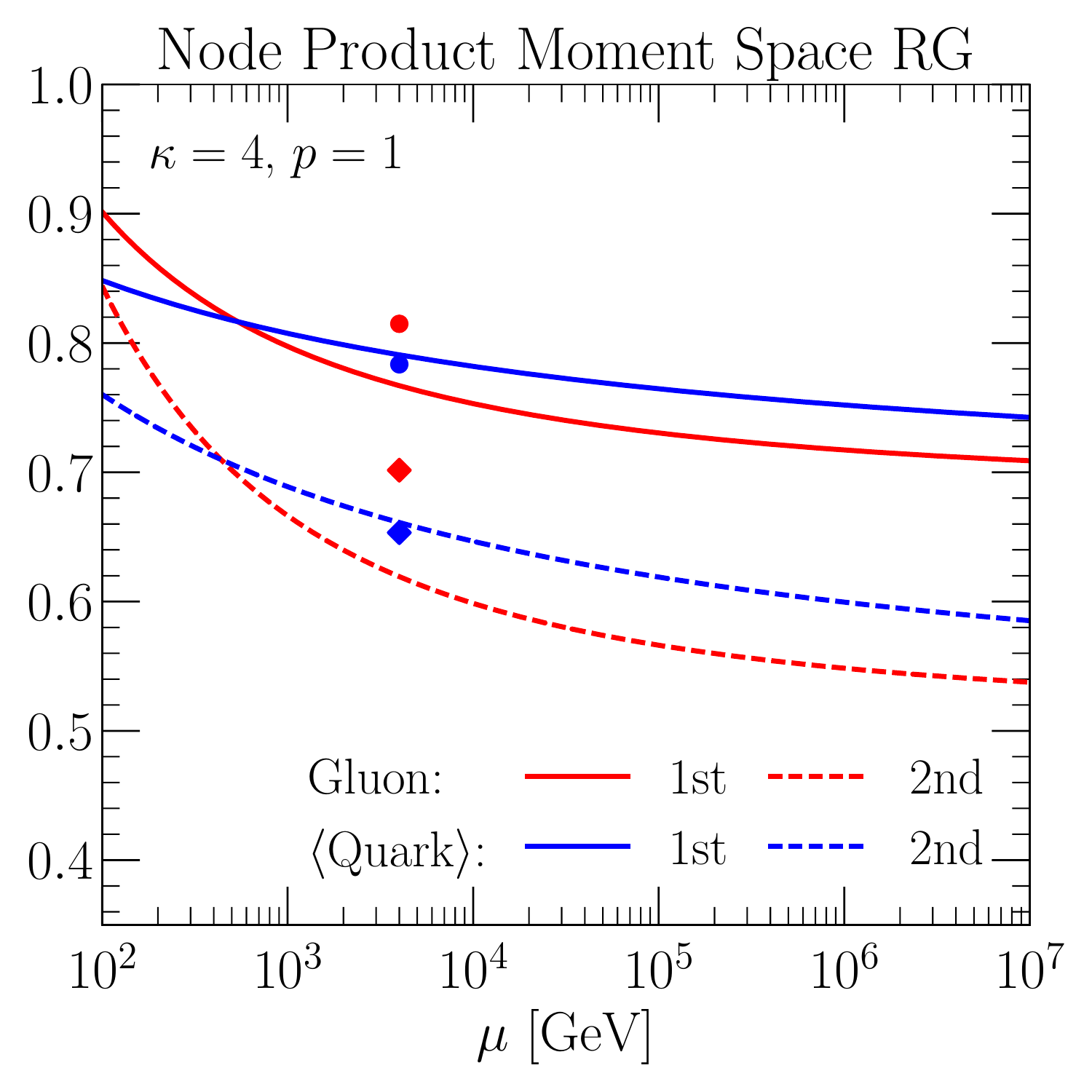}
		 \label{fig:moment-space-node-f}
}
	\caption{\label{fig:moment-space-node}Moment space evolution of the node-product observables with (top row) $\kappa=1$ and (bottom row) $\kappa=4$ for the generalized-$k-t$ clustering trees with (left column) $p=-1$, (middle column) $p=0$, and (right column) $p=1$.  Shown are the first (solid curves) and second (dashed curves) moments of gluon (red) and quark-singlet (blue) GFFs.  The first (second) moments extracted from the parton shower average at  $\mu =$ 4 TeV are shown as points (diamonds).}
\end{figure}

We now insert the recursion relation for the node products from \Eq{eq:noderecur} into \Eq{eq:moment-space-general}.  This leads to evolution equations with additional terms compared to those for the weighted energy fractions. These terms have splitting kernels of the form
\begin{equation}
\int_0^1 \text{d}z\, \big(4z(1-z)\big)^a z^b(1-z)^cP_{i\rightarrow j,k}(z)
\end{equation}
for $a>0$ and $b,c\ge 0$. These integrals are convergent, so no plus function regulators are required. They can also be performed analytically for general $a$, $b$, and $c$. Explicitly, the first moments of the quark-singlet and gluon GFFs evolve as
\begin{align}
\label{eq:node-1st-mom}
\mu \frac{\text{d}}{\text{d}\mu} \left(\begin{array}{c}\overline{\mathcal{S}}(1,\mu)  \\ \overline{\GFF}_g(1,\mu) \end{array}\right) = \frac{\alpha_s(\mu)}{\pi} &\left(\begin{array}{c c} \overline{P}_{q\rightarrow qg}(\kappa) &  \overline{P}_{q\rightarrow gq}(\kappa)\\ 2n_f\overline{P}_{g\rightarrow q\overline{q}}(\kappa) & \overline{P}_{g\rightarrow gg}(\kappa) \end{array}  \right)\left(\begin{array}{c}\overline{\mathcal{S}}(1,\mu) \\ \overline{\GFF}_g(1,\mu) \end{array}\right)
+ \frac{\alpha_s(\mu)}{\pi}\left(\begin{array}{c}\overline{P}_{q1}^\text{Node}(\kappa) \\ \overline{P}_{g1}^\text{Node}(\kappa) \end{array}\right)
.\end{align}
The additional constant terms are defined as
\begin{align}
\overline{P}_{q1}^\text{Node}(\kappa) &\equiv \frac{1}{2} \int_0^1 \text{d}z\, \big(P_{q\rightarrow qg}(z) + P_{q\rightarrow gq}(z)\big)\big(4z(1-z)\big)^{\kappa/2}\,, 
\nonumber \\
\overline{P}_{g1}^\text{Node}(\kappa) &\equiv \frac{1}{2} \int_0^1 \text{d}z\, \big(2n_f P_{g\rightarrow q\bar{q}}(z) + P_{q\rightarrow gg}(z)\big)\big(4z(1-z)\big)^{\kappa/2}
\,,\end{align}
which can be evaluated in terms of $\Gamma$ functions.
The additional terms drop out of the equation for the first moments of the non-singlet GFFs, so these still evolve according to \Eq{eq:1stmomNS}. The third term in \Eq{eq:noderecur} leads to several more terms in the evolution equations for higher moments.

In \Fig{fig:moment-space-node}, we plot the $\mu$ evolution of the gluon and quark-singlet GFF moments for node products with $\kappa =\{1,4\}$ and $p=\{-1,0,1\}$.  The first and second moments were computed at the scale $\mu=100$ GeV from the GFFs in \Fig{fig:node-product}, averaged over the different parton showers and $R$ values (as described in \Sec{sec:wefextraction}). These average moments were evolved to the scale $\mu=10^7$ GeV using \Eq{eq:node-1st-mom} and the corresponding second moment equation.  For comparison, the first and second moments of the GFFs extracted from the parton shower average at the scale $\mu =$ 4 TeV are shown as dots and diamonds, respectively.

\subsection{Full-Tree Observables}
\label{app:momentFT}

\begin{figure}[t]
\subfloat[]{
		\includegraphics[width=0.32\textwidth]{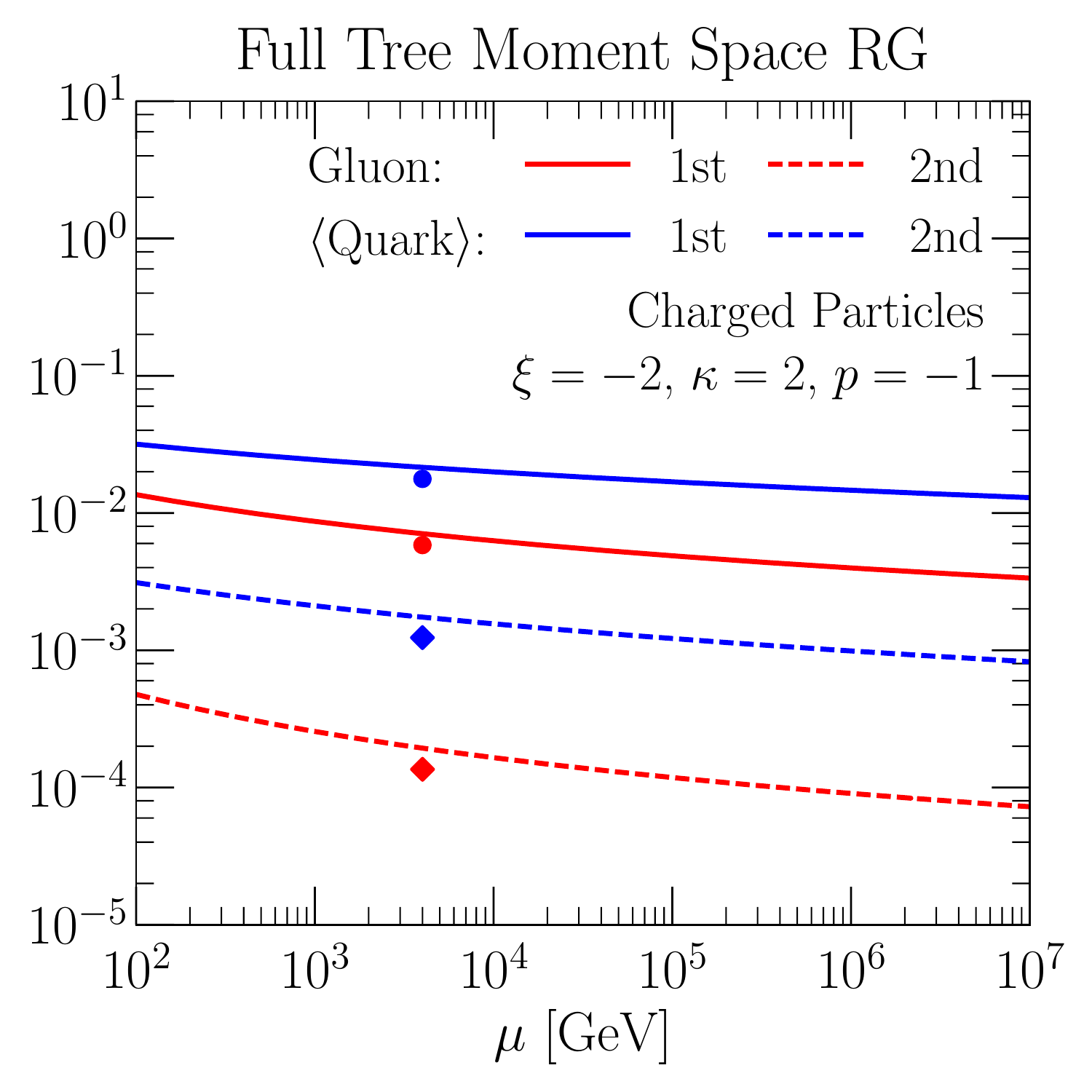}
		 \label{fig:moment-space-fulltree-a} 
}
\subfloat[]{
		\includegraphics[width=0.32\textwidth]{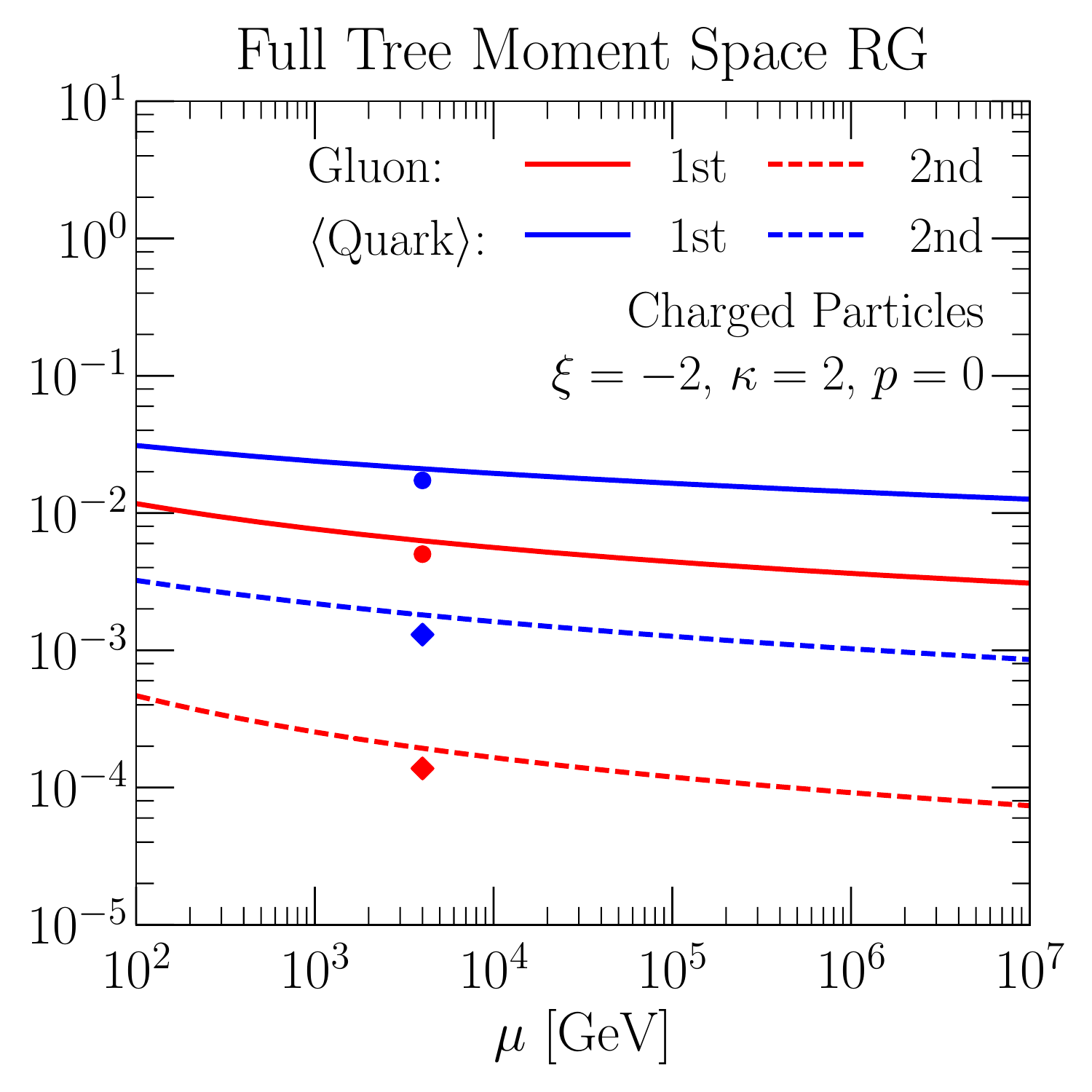}
		 \label{fig:moment-space-fulltree-c}
}
\subfloat[]{
		\includegraphics[width=0.32\textwidth]{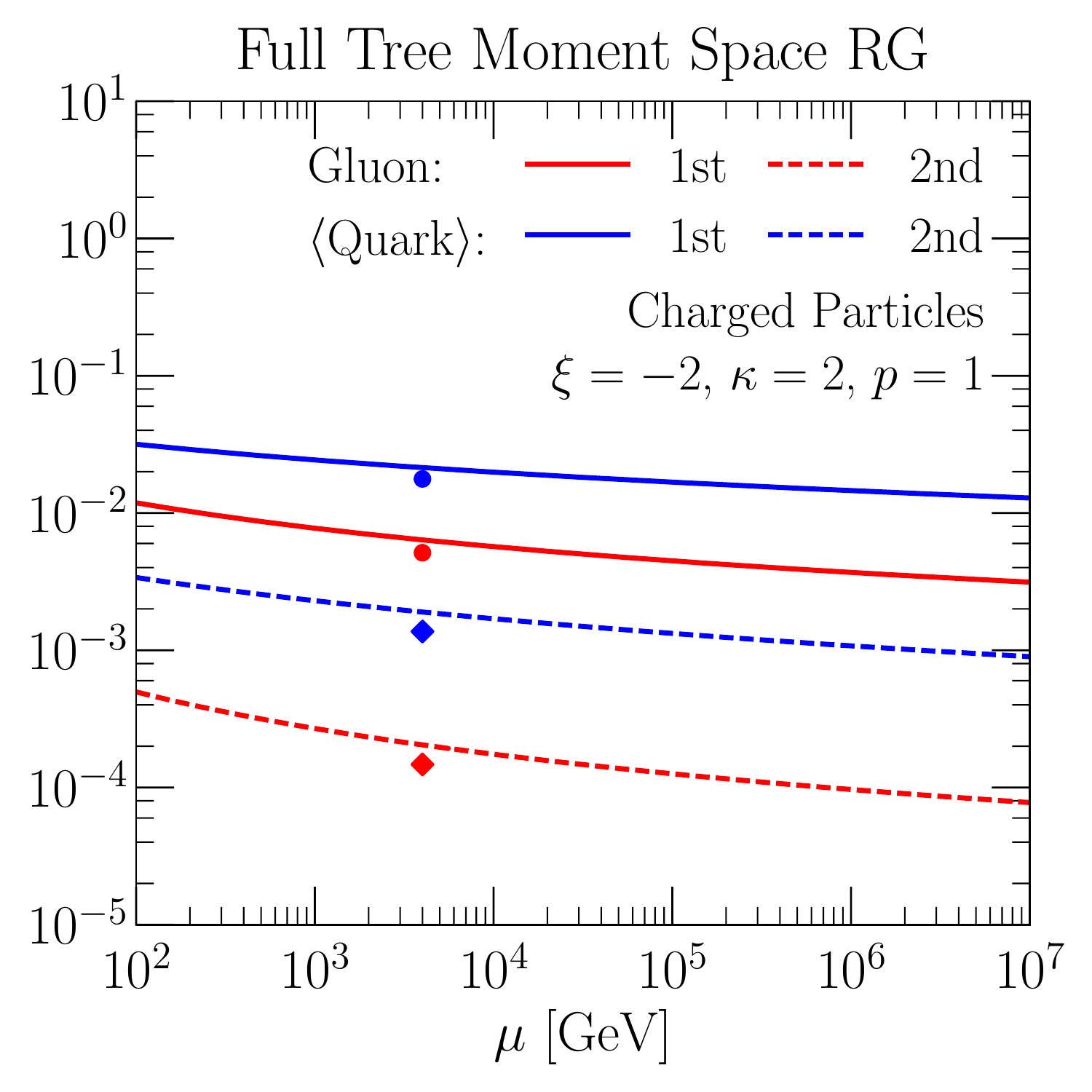}
		 \label{fig:moment-space-fulltree-e}
}

\subfloat[]{
		\includegraphics[width=0.32\textwidth]{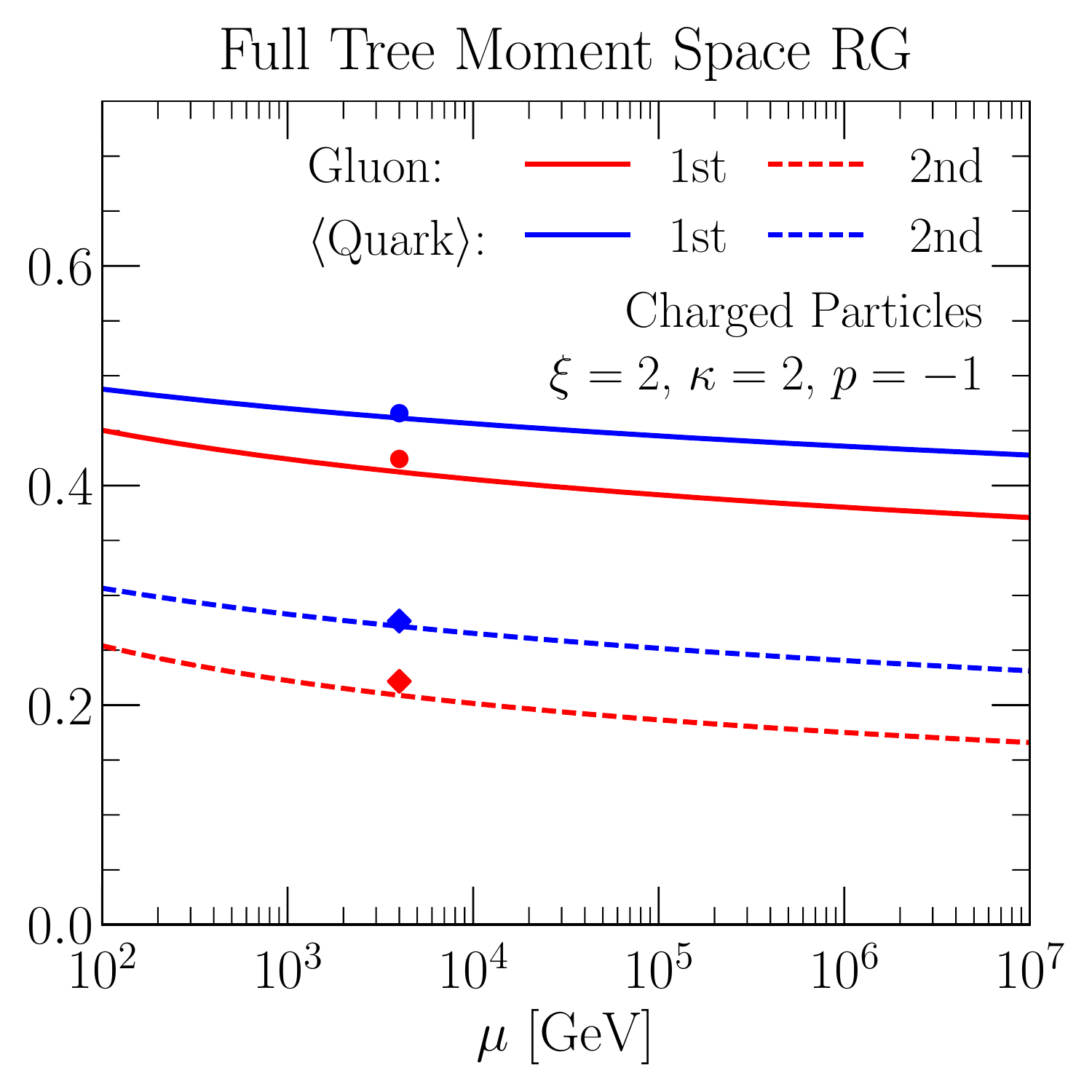}
		 \label{fig:moment-space-fulltree-b}
}
\subfloat[]{
		\includegraphics[width=0.32\textwidth]{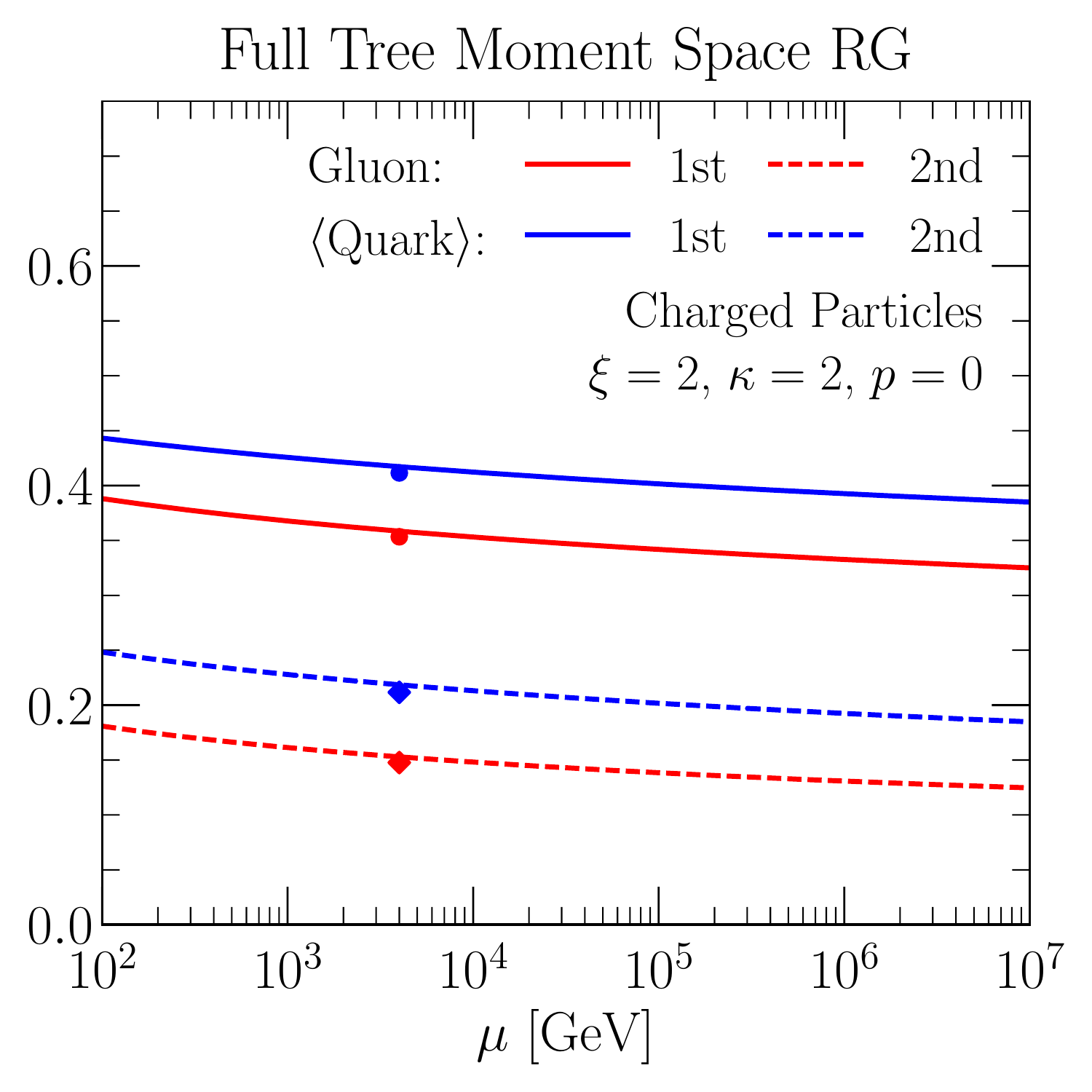}
		 \label{fig:moment-space-fulltree-d}
}
\subfloat[]{
		\includegraphics[width=0.32\textwidth]{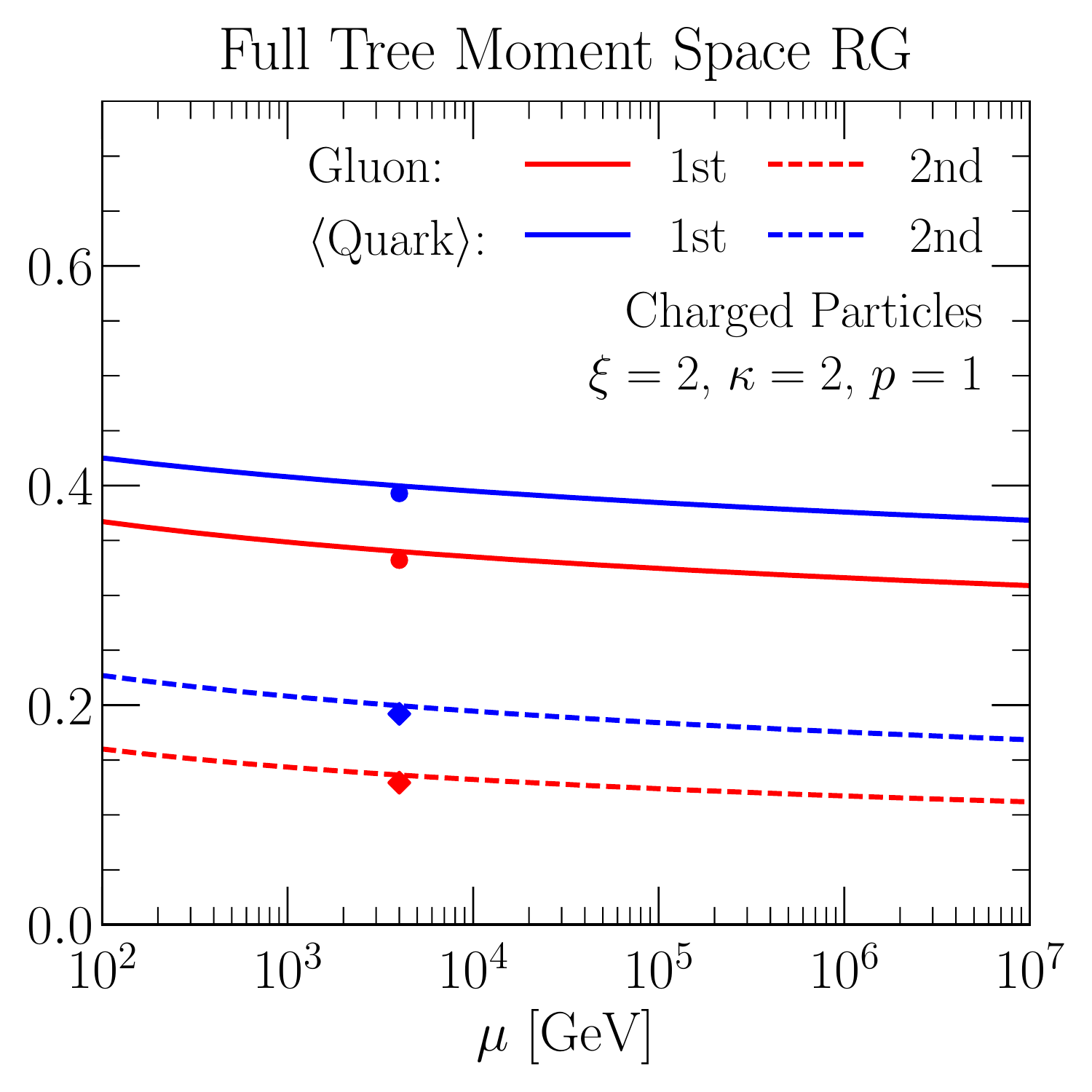}
		 \label{fig:moment-space-fulltree-f}
}
	\caption{\label{fig:fulltreemom}The same as \Fig{fig:moment-space-node}, except now for the full-tree observables with $\kappa = 2$ measured on charged particles, with (top row) $\xi=-2$ and (bottom row) $\xi=2$.}
\end{figure}

For full-tree observables with recursion relation given in \Eq{eq:narecursion}, the moment space evolution equations are of the same general form as for the weighted energy fractions,
\begin{align}
 \mu \frac{\text{d}}{\text{d}\mu} \overline{\GFF}_i(N,\mu) = \frac{\alpha_s(\mu)}{2\pi} \sum_{j,k} \sum_{M=0}^N \binom{N}{M} \, \overline{P}_{i\rightarrow j,k}^{\rm FT}(N,M) \, \overline{\GFF}_j(N-M,\mu) \, \overline{\GFF}_k(M,\mu),
\end{align}
but with different splitting kernels,
\begin{equation}
\overline{P}_{i\rightarrow j,k}^{\rm FT}(N,M) \equiv \int_0^1 \text{d}z\, e^{N\xi z(1-z)} z^{\kappa (N-M)}(1-z)^{\kappa M} P_{i\rightarrow j,k} (z).
\end{equation}
To our knowledge, these integrals do not have a closed form solution for general values of the parameters $\kappa$ and $\xi$, but it is straightforward to evaluate them numerically.  If $M=0$ or $M=N$, these integrals are sensitive to the plus-prescription in the splitting functions.  Explicitly, for the first moment in the quark-singlet basis, 
\begin{align}
\mu \frac{\text{d}}{\text{d}\mu} \overline{\mathcal{S}}(1,\mu) &= \frac{\alpha_s(\mu)}{\pi}  \bigg[ C_F\bigg(\frac{3}{2} + \int_0^1 \text{d}z\, \frac{e^{\xi z(1-z)}z^\kappa (1+z^2) - 2}{1-z} \bigg) \,  \overline{\mathcal{S}}(1,\mu) \\
& \qquad \qquad \quad  + C_F \int_0^1 \text{d}z\, \left( e^{\xi z (1-z)} z^{\kappa-1} (1+(1-z)^2) \right) \, \overline{\GFF}_g(1,\mu)\bigg],
\nonumber \\
\mu \frac{\text{d}}{\text{d}\mu} \overline{\GFF}_g(1,\mu) &= \frac{\alpha_s(\mu)}{\pi}  \bigg[ 2n_f T_F\int_0^1 \text{d}z\, \left(e^{\xi z(1-z)}z^\kappa (z^2+(1-z)^2) \right) \, \overline{\mathcal{S}}(1,\mu)
\nonumber \\
& \qquad \qquad \quad  + 2C_A \int_0^1 \text{d}z\, \bigg( e^{\xi z (1-z)} (z^{\kappa-1}(1-z) + z^{\kappa+1} (1-z)) 
\nonumber \\
& \qquad \qquad \qquad \qquad \qquad \qquad
 + \frac{e^{\xi z(1-z)} z^{\kappa+1}-1}{1-z}  + \frac{11}{6}-\frac{2}{3}\frac{T_F n_f}{C_A} \bigg) \, \overline{\GFF}_g(1,\mu)\bigg]. \nonumber
\end{align}
In \Fig{fig:fulltreemom}, we show the evolution of the first two moments of the GFFs for $\kappa=2$, $\xi =\{-2,2\}$, and $p=\{-1,0,1\}$. In this case, the evolution agrees well with the value extracted from the parton shower average at $\mu=4$ TeV.

\bibliography{GFFbiblio}

\providecommand{\href}[2]{#2}\begingroup\raggedright\begin{thebibliography}{10}

\bibitem{Georgi:1977mg}
H.~Georgi and H.~D. Politzer, \emph{{Quark Decay Functions and Heavy Hadron
  Production in QCD}},
  \href{http://dx.doi.org/10.1016/0550-3213(78)90269-9}{\emph{Nucl. Phys.}
  {\bfseries B136} (1978) 445--460}.

\bibitem{Mueller:1978xu}
A.~H. Mueller, \emph{{Cut Vertices and their Renormalization: A Generalization
  of the Wilson Expansion}},
  \href{http://dx.doi.org/10.1103/PhysRevD.18.3705}{\emph{Phys. Rev.}
  {\bfseries D18} (1978) 3705}.

\bibitem{Ellis:1978ty}
R.~K. Ellis, H.~Georgi, M.~Machacek, H.~D. Politzer and G.~G. Ross,
  \emph{{Perturbation Theory and the Parton Model in QCD}},
  \href{http://dx.doi.org/10.1016/0550-3213(79)90105-6}{\emph{Nucl. Phys.}
  {\bfseries B152} (1979) 285--329}.

\bibitem{Curci:1980uw}
G.~Curci, W.~Furmanski and R.~Petronzio, \emph{{Evolution of Parton Densities
  Beyond Leading Order: The Nonsinglet Case}},
  \href{http://dx.doi.org/10.1016/0550-3213(80)90003-6}{\emph{Nucl. Phys.}
  {\bfseries B175} (1980) 27--92}.

\bibitem{Altarelli:1981ax}
G.~Altarelli, \emph{{Partons in Quantum Chromodynamics}},
  \href{http://dx.doi.org/10.1016/0370-1573(82)90127-2}{\emph{Phys. Rept.}
  {\bfseries 81} (1982) 1}.

\bibitem{Collins:1981uk}
J.~C. Collins and D.~E. Soper, \emph{{Back-To-Back Jets in QCD}},
  \href{http://dx.doi.org/10.1016/0550-3213(81)90339-4}{\emph{Nucl. Phys.}
  {\bfseries B193} (1981) 381}.

\bibitem{Collins:1981uw}
J.~C. Collins and D.~E. Soper, \emph{{Parton Distribution and Decay
  Functions}},
  \href{http://dx.doi.org/10.1016/0550-3213(82)90021-9}{\emph{Nucl. Phys.}
  {\bfseries B194} (1982) 445--492}.

\bibitem{Lipatov:1974qm}
L.~N. Lipatov, \emph{{The parton model and perturbation theory}}, {\emph{Sov.
  J. Nucl. Phys.} {\bfseries 20} (1975) 94--102}.

\bibitem{Gribov:1972ri}
V.~N. Gribov and L.~N. Lipatov, \emph{{Deep inelastic e p scattering in
  perturbation theory}}, {\emph{Sov. J. Nucl. Phys.} {\bfseries 15} (1972)
  438--450}.

\bibitem{Altarelli:1977zs}
G.~Altarelli and G.~Parisi, \emph{{Asymptotic Freedom in Parton Language}},
  \href{http://dx.doi.org/10.1016/0550-3213(77)90384-4}{\emph{Nucl. Phys.}
  {\bfseries B126} (1977) 298--318}.

\bibitem{Dokshitzer:1977sg}
Y.~L. Dokshitzer, \emph{{Calculation of the Structure Functions for Deep
  Inelastic Scattering and $e^+ e^-$ Annihilation by Perturbation Theory in
  Quantum Chromodynamics.}}, {\emph{Sov. Phys. JETP} {\bfseries 46} (1977)
  641--653}.

\bibitem{Davighi:2017hok}
J.~Davighi and P.~Harris, \emph{{Fractal based observables to probe jet
  substructure of quarks and gluons}},
  \href{https://arxiv.org/abs/1703.00914}{{\ttfamily 1703.00914}}.

\bibitem{FIELD19781}
R.~Field and R.~Feynman, \emph{A parametrization of the properties of quark
  jets},
  \href{http://dx.doi.org/http://dx.doi.org/10.1016/0550-3213(78)90015-9}{\emph{Nuclear
  Physics B} {\bfseries 136} (1978) 1 -- 76}.

\bibitem{Waalewijn:2012sv}
W.~J. Waalewijn, \emph{{Calculating the Charge of a Jet}},
  \href{http://dx.doi.org/10.1103/PhysRevD.86.094030}{\emph{Phys. Rev.}
  {\bfseries D86} (2012) 094030},
  [\href{https://arxiv.org/abs/1209.3019}{{\ttfamily 1209.3019}}].

\bibitem{Chang:2013rca}
H.-M. Chang, M.~Procura, J.~Thaler and W.~J. Waalewijn, \emph{{Calculating
  Track-Based Observables for the LHC}},
  \href{http://dx.doi.org/10.1103/PhysRevLett.111.102002}{\emph{Phys. Rev.
  Lett.} {\bfseries 111} (2013) 102002},
  [\href{https://arxiv.org/abs/1303.6637}{{\ttfamily 1303.6637}}].

\bibitem{Chang:2013iba}
H.-M. Chang, M.~Procura, J.~Thaler and W.~J. Waalewijn, \emph{{Calculating
  Track Thrust with Track Functions}},
  \href{http://dx.doi.org/10.1103/PhysRevD.88.034030}{\emph{Phys. Rev.}
  {\bfseries D88} (2013) 034030},
  [\href{https://arxiv.org/abs/1306.6630}{{\ttfamily 1306.6630}}].

\bibitem{Pandolfi:1480598}
F.~Pandolfi and D.~Del~Re, \emph{Search for the Standard Model Higgs Boson in
  the $H\to ZZ\to llqq$ Decay Channel at CMS}.
\newblock PhD thesis, Zurich, ETH, 2012.

\bibitem{Chatrchyan:2012sn}
{\scshape CMS} collaboration, S.~Chatrchyan et~al., \emph{{Search for a Higgs
  boson in the decay channel $H \to ZZ^{(*)} \to q \bar q \ell^- \ell^+$ in
  $pp$ collisions at $\sqrt{s}=7$ TeV}},
  \href{http://dx.doi.org/10.1007/JHEP04(2012)036}{\emph{JHEP} {\bfseries 04}
  (2012) 036}, [\href{https://arxiv.org/abs/1202.1416}{{\ttfamily 1202.1416}}].

\bibitem{Larkoski:2014pca}
A.~J. Larkoski, J.~Thaler and W.~J. Waalewijn, \emph{{Gaining (Mutual)
  Information about Quark/Gluon Discrimination}},
  \href{http://dx.doi.org/10.1007/JHEP11(2014)129}{\emph{JHEP} {\bfseries 11}
  (2014) 129}, [\href{https://arxiv.org/abs/1408.3122}{{\ttfamily 1408.3122}}].

\bibitem{Gallicchio:2011xq}
J.~Gallicchio and M.~D. Schwartz, \emph{{Quark and Gluon Tagging at the LHC}},
  \href{http://dx.doi.org/10.1103/PhysRevLett.107.172001}{\emph{Phys. Rev.
  Lett.} {\bfseries 107} (2011) 172001},
  [\href{https://arxiv.org/abs/1106.3076}{{\ttfamily 1106.3076}}].

\bibitem{Gallicchio:2012ez}
J.~Gallicchio and M.~D. Schwartz, \emph{{Quark and Gluon Jet Substructure}},
  \href{http://dx.doi.org/10.1007/JHEP04(2013)090}{\emph{JHEP} {\bfseries 04}
  (2013) 090}, [\href{https://arxiv.org/abs/1211.7038}{{\ttfamily 1211.7038}}].

\bibitem{Larkoski:2013eya}
A.~J. Larkoski, G.~P. Salam and J.~Thaler, \emph{{Energy Correlation Functions
  for Jet Substructure}},
  \href{http://dx.doi.org/10.1007/JHEP06(2013)108}{\emph{JHEP} {\bfseries 06}
  (2013) 108}, [\href{https://arxiv.org/abs/1305.0007}{{\ttfamily 1305.0007}}].

\bibitem{Bhattacherjee:2015psa}
B.~Bhattacherjee, S.~Mukhopadhyay, M.~M. Nojiri, Y.~Sakaki and B.~R. Webber,
  \emph{{Associated jet and subjet rates in light-quark and gluon jet
  discrimination}},
  \href{http://dx.doi.org/10.1007/JHEP04(2015)131}{\emph{JHEP} {\bfseries 04}
  (2015) 131}, [\href{https://arxiv.org/abs/1501.04794}{{\ttfamily
  1501.04794}}].

\bibitem{Badger:2016bpw}
J.~R. Andersen et~al., \emph{{Les Houches 2015: Physics at TeV Colliders
  Standard Model Working Group Report}},  in \emph{{9th Les Houches Workshop on
  Physics at TeV Colliders (PhysTeV 2015) Les Houches, France, June 1-19,
  2015}}, 2016.
\newblock \href{https://arxiv.org/abs/1605.04692}{{\ttfamily 1605.04692}}.

\bibitem{FerreiradeLima:2016gcz}
D.~Ferreira~de Lima, P.~Petrov, D.~Soper and M.~Spannowsky, \emph{{Quark-Gluon
  tagging with Shower Deconstruction: Unearthing dark matter and Higgs
  couplings}}, \href{http://dx.doi.org/10.1103/PhysRevD.95.034001}{\emph{Phys.
  Rev.} {\bfseries D95} (2017) 034001},
  [\href{https://arxiv.org/abs/1607.06031}{{\ttfamily 1607.06031}}].

\bibitem{Komiske:2016rsd}
P.~T. Komiske, E.~M. Metodiev and M.~D. Schwartz, \emph{{Deep learning in
  color: towards automated quark/gluon jet discrimination}},
  \href{http://dx.doi.org/10.1007/JHEP01(2017)110}{\emph{JHEP} {\bfseries 01}
  (2017) 110}, [\href{https://arxiv.org/abs/1612.01551}{{\ttfamily
  1612.01551}}].

\bibitem{Gras:2017jty}
P.~Gras, S.~Hoeche, D.~Kar, A.~Larkoski, L.~Lönnblad, S.~Plätzer et~al.,
  \emph{{Systematics of quark/gluon tagging}},
  \href{https://arxiv.org/abs/1704.03878}{{\ttfamily 1704.03878}}.

\bibitem{Dokshitzer:1997in}
Y.~L. Dokshitzer, G.~D. Leder, S.~Moretti and B.~R. Webber, \emph{{Better jet
  clustering algorithms}},
  \href{http://dx.doi.org/10.1088/1126-6708/1997/08/001}{\emph{JHEP} {\bfseries
  08} (1997) 001}, [\href{https://arxiv.org/abs/hep-ph/9707323}{{\ttfamily
  hep-ph/9707323}}].

\bibitem{Wobisch:1998wt}
M.~Wobisch and T.~Wengler, \emph{{Hadronization corrections to jet
  cross-sections in deep inelastic scattering}},  in \emph{{Monte Carlo
  generators for HERA physics. Proceedings, Workshop, Hamburg, Germany,
  1998-1999}}, pp.~270--279, 1998.
\newblock \href{https://arxiv.org/abs/hep-ph/9907280}{{\ttfamily
  hep-ph/9907280}}.

\bibitem{Dasgupta:2014yra}
M.~Dasgupta, F.~Dreyer, G.~P. Salam and G.~Soyez, \emph{{Small-radius jets to
  all orders in QCD}},
  \href{http://dx.doi.org/10.1007/JHEP04(2015)039}{\emph{JHEP} {\bfseries 04}
  (2015) 039}, [\href{https://arxiv.org/abs/1411.5182}{{\ttfamily 1411.5182}}].

\bibitem{Kang:2016mcy}
Z.-B. Kang, F.~Ringer and I.~Vitev, \emph{{The semi-inclusive jet function in
  SCET and small radius resummation for inclusive jet production}},
  \href{http://dx.doi.org/10.1007/JHEP10(2016)125}{\emph{JHEP} {\bfseries 10}
  (2016) 125}, [\href{https://arxiv.org/abs/1606.06732}{{\ttfamily
  1606.06732}}].

\bibitem{Dai:2016hzf}
L.~Dai, C.~Kim and A.~K. Leibovich, \emph{{Fragmentation of a Jet with Small
  Radius}}, \href{http://dx.doi.org/10.1103/PhysRevD.94.114023}{\emph{Phys.
  Rev.} {\bfseries D94} (2016) 114023},
  [\href{https://arxiv.org/abs/1606.07411}{{\ttfamily 1606.07411}}].

\bibitem{Sukhatme:1980vs}
U.~Sukhatme and K.~Lassila, \emph{{$Q^2$ Evolution of Multi-Hadron
  Fragmentation Functions}},
  \href{http://dx.doi.org/10.1103/PhysRevD.22.1184}{\emph{Phys. Rev.}
  {\bfseries D22} (1980) 1184}.

\bibitem{Vendramin:1981te}
I.~Vendramin, \emph{{Two-Hadron Fragmentation Functions: A Study of Their $Q^2$
  Evolution}}, \href{http://dx.doi.org/10.1007/BF02731692}{\emph{Nuovo Cim.}
  {\bfseries A66} (1981) 339}.

\bibitem{Trentadue:1993ka}
L.~Trentadue and G.~Veneziano, \emph{{Fracture functions: An Improved
  description of inclusive hard processes in QCD}},
  \href{http://dx.doi.org/10.1016/0370-2693(94)90292-5}{\emph{Phys. Lett.}
  {\bfseries B323} (1994) 201--211}.

\bibitem{Graudenz:1994dq}
D.~Graudenz, \emph{{One particle inclusive processes in deeply inelastic lepton
  - nucleon scattering}},
  \href{http://dx.doi.org/10.1016/0550-3213(94)90606-8}{\emph{Nucl. Phys.}
  {\bfseries B432} (1994) 351--376},
  [\href{https://arxiv.org/abs/hep-ph/9406274}{{\ttfamily hep-ph/9406274}}].

\bibitem{Furmanski:1981cw}
W.~Furmanski and R.~Petronzio, \emph{{Lepton - Hadron Processes Beyond Leading
  Order in Quantum Chromodynamics}},
  \href{http://dx.doi.org/10.1007/BF01578280}{\emph{Z. Phys.} {\bfseries C11}
  (1982) 293}.

\bibitem{Collins:1989gx}
J.~C. Collins, D.~E. Soper and G.~F. Sterman, \emph{{Factorization of Hard
  Processes in QCD}},
  \href{http://dx.doi.org/10.1142/9789814503266_0001}{\emph{Adv. Ser. Direct.
  High Energy Phys.} {\bfseries 5} (1989) 1--91},
  [\href{https://arxiv.org/abs/hep-ph/0409313}{{\ttfamily hep-ph/0409313}}].

\bibitem{Olive:2016xmw}
{\scshape Particle Data Group} collaboration, C.~Patrignani et~al.,
  \emph{{Review of Particle Physics}},
  \href{http://dx.doi.org/10.1088/1674-1137/40/10/100001}{\emph{Chin. Phys.}
  {\bfseries C40} (2016) 100001}.

\bibitem{Kang:2016ehg}
Z.-B. Kang, F.~Ringer and I.~Vitev, \emph{{Jet substructure using
  semi-inclusive jet functions in SCET}},
  \href{http://dx.doi.org/10.1007/JHEP11(2016)155}{\emph{JHEP} {\bfseries 11}
  (2016) 155}, [\href{https://arxiv.org/abs/1606.07063}{{\ttfamily
  1606.07063}}].

\bibitem{Procura:2009vm}
M.~Procura and I.~W. Stewart, \emph{{Quark Fragmentation within an Identified
  Jet}}, \href{http://dx.doi.org/10.1103/PhysRevD.81.074009,
  10.1103/PhysRevD.83.039902}{\emph{Phys. Rev.} {\bfseries D81} (2010) 074009},
  [\href{https://arxiv.org/abs/0911.4980}{{\ttfamily 0911.4980}}].

\bibitem{Jain:2011xz}
A.~Jain, M.~Procura and W.~J. Waalewijn, \emph{{Parton Fragmentation within an
  Identified Jet at NNLL}},
  \href{http://dx.doi.org/10.1007/JHEP05(2011)035}{\emph{JHEP} {\bfseries 05}
  (2011) 035}, [\href{https://arxiv.org/abs/1101.4953}{{\ttfamily 1101.4953}}].

\bibitem{Liu:2010ng}
X.~Liu, \emph{{SCET approach to top quark decay}},
  \href{http://dx.doi.org/10.1016/j.physletb.2011.03.055}{\emph{Phys. Lett.}
  {\bfseries B699} (2011) 87--92},
  [\href{https://arxiv.org/abs/1011.3872}{{\ttfamily 1011.3872}}].

\bibitem{Jain:2011iu}
A.~Jain, M.~Procura and W.~J. Waalewijn, \emph{{Fully-Unintegrated Parton
  Distribution and Fragmentation Functions at Perturbative $k_T$}},
  \href{http://dx.doi.org/10.1007/JHEP04(2012)132}{\emph{JHEP} {\bfseries 04}
  (2012) 132}, [\href{https://arxiv.org/abs/1110.0839}{{\ttfamily 1110.0839}}].

\bibitem{Cacciari:2008gp}
M.~Cacciari, G.~P. Salam and G.~Soyez, \emph{{The Anti-$k_t$ jet clustering
  algorithm}},
  \href{http://dx.doi.org/10.1088/1126-6708/2008/04/063}{\emph{JHEP} {\bfseries
  04} (2008) 063}, [\href{https://arxiv.org/abs/0802.1189}{{\ttfamily
  0802.1189}}].

\bibitem{Catani:1993hr}
S.~Catani, Y.~L. Dokshitzer, M.~H. Seymour and B.~R. Webber,
  \emph{{Longitudinally invariant $k_\perp$ clustering algorithms for hadron
  hadron collisions}},
  \href{http://dx.doi.org/10.1016/0550-3213(93)90166-M}{\emph{Nucl. Phys.}
  {\bfseries B406} (1993) 187--224}.

\bibitem{Ellis:1993tq}
S.~D. Ellis and D.~E. Soper, \emph{{Successive combination jet algorithm for
  hadron collisions}},
  \href{http://dx.doi.org/10.1103/PhysRevD.48.3160}{\emph{Phys. Rev.}
  {\bfseries D48} (1993) 3160--3166},
  [\href{https://arxiv.org/abs/hep-ph/9305266}{{\ttfamily hep-ph/9305266}}].

\bibitem{Ilten:2017rbd}
P.~Ilten, N.~L. Rodd, J.~Thaler and M.~Williams, \emph{{Disentangling Heavy
  Flavor at Colliders}},  \href{https://arxiv.org/abs/1702.02947}{{\ttfamily
  1702.02947}}.

\bibitem{HarvardInProgress}
C.~Frye, A.~J. Larkoski, J.~Thaler and K.~Zhou, \emph{{Casimir meets Poisson:
  Improved quark/gluon discrimination with counting observables}}, {\emph{in
  progress\!\!} }.

\bibitem{Krohn:2012fg}
D.~Krohn, M.~D. Schwartz, T.~Lin and W.~J. Waalewijn, \emph{{Jet Charge at the
  LHC}}, \href{http://dx.doi.org/10.1103/PhysRevLett.110.212001}{\emph{Phys.
  Rev. Lett.} {\bfseries 110} (2013) 212001},
  [\href{https://arxiv.org/abs/1209.2421}{{\ttfamily 1209.2421}}].

\bibitem{Stuart:1989db}
{\scshape AMY} collaboration, D.~Stuart et~al., \emph{{Forward - backward
  charge asymmetry in $e^+e^- \to$ hadron jets}},
  \href{http://dx.doi.org/10.1103/PhysRevLett.64.983}{\emph{Phys. Rev. Lett.}
  {\bfseries 64} (1990) 983}.

\bibitem{Decamp:1991se}
{\scshape ALEPH} collaboration, D.~Decamp et~al., \emph{{Measurement of charge
  asymmetry in hadronic Z decays}},
  \href{http://dx.doi.org/10.1016/0370-2693(91)90844-G}{\emph{Phys. Lett.}
  {\bfseries B259} (1991) 377--388}.

\bibitem{Berge:1979qg}
{\scshape Fermilab-Serpukhov-Moscow-Michigan} collaboration, J.~P. Berge
  et~al., \emph{{Net Charge in Deep Inelastic anti-neutrino - Nucleon
  Scattering}},
  \href{http://dx.doi.org/10.1016/0370-2693(80)90456-6}{\emph{Phys. Lett.}
  {\bfseries B91} (1980) 311--313}.

\bibitem{Buskulic:1992sq}
{\scshape ALEPH} collaboration, D.~Buskulic et~al., \emph{{Measurement of B -
  anti-B mixing at the Z using a jet charge method}},
  \href{http://dx.doi.org/10.1016/0370-2693(92)91943-4}{\emph{Phys. Lett.}
  {\bfseries B284} (1992) 177--190}.

\bibitem{Abazov:2006vd}
{\scshape D0} collaboration, V.~M. Abazov et~al., \emph{{Experimental
  discrimination between charge 2e/3 top quark and charge 4e/3 exotic quark
  production scenarios}},
  \href{http://dx.doi.org/10.1103/PhysRevLett.98.041801}{\emph{Phys. Rev.
  Lett.} {\bfseries 98} (2007) 041801},
  [\href{https://arxiv.org/abs/hep-ex/0608044}{{\ttfamily hep-ex/0608044}}].

\bibitem{Aad:2015cua}
{\scshape ATLAS} collaboration, G.~Aad et~al., \emph{{Measurement of jet charge
  in dijet events from $\sqrt{s}$=8 TeV pp collisions with the ATLAS
  detector}}, \href{http://dx.doi.org/10.1103/PhysRevD.93.052003}{\emph{Phys.
  Rev.} {\bfseries D93} (2016) 052003},
  [\href{https://arxiv.org/abs/1509.05190}{{\ttfamily 1509.05190}}].

\bibitem{Sjostrand:2014zea}
T.~Sj{\"o}strand, S.~Ask, J.~R. Christiansen, R.~Corke, N.~Desai, P.~Ilten
  et~al., \emph{{An Introduction to PYTHIA 8.2}},
  \href{http://dx.doi.org/10.1016/j.cpc.2015.01.024}{\emph{Comput. Phys.
  Commun.} {\bfseries 191} (2015) 159--177},
  [\href{https://arxiv.org/abs/1410.3012}{{\ttfamily 1410.3012}}].

\bibitem{Cacciari:2011ma}
M.~Cacciari, G.~P. Salam and G.~Soyez, \emph{{FastJet User Manual}},
  \href{http://dx.doi.org/10.1140/epjc/s10052-012-1896-2}{\emph{Eur. Phys. J.}
  {\bfseries C72} (2012) 1896},
  [\href{https://arxiv.org/abs/1111.6097}{{\ttfamily 1111.6097}}].

\bibitem{Giele:2007di}
W.~T. Giele, D.~A. Kosower and P.~Z. Skands, \emph{{A simple shower and
  matching algorithm}},
  \href{http://dx.doi.org/10.1103/PhysRevD.78.014026}{\emph{Phys. Rev.}
  {\bfseries D78} (2008) 014026},
  [\href{https://arxiv.org/abs/0707.3652}{{\ttfamily 0707.3652}}].

\bibitem{Hoche:2015sya}
S.~H{\"o}che and S.~Prestel, \emph{{The midpoint between dipole and parton
  showers}}, \href{http://dx.doi.org/10.1140/epjc/s10052-015-3684-2}{\emph{Eur.
  Phys. J.} {\bfseries C75} (2015) 461},
  [\href{https://arxiv.org/abs/1506.05057}{{\ttfamily 1506.05057}}].

\bibitem{Malaza:1985jd}
E.~D. Malaza and B.~R. Webber, \emph{{Multiplicity Distributions in Quark and
  Gluon Jets}},
  \href{http://dx.doi.org/10.1016/0550-3213(86)90138-0}{\emph{Nucl. Phys.}
  {\bfseries B267} (1986) 702--713}.

\bibitem{Lupia:1997in}
S.~Lupia and W.~Ochs, \emph{{Unified QCD description of hadron and jet
  multiplicities}},
  \href{http://dx.doi.org/10.1016/S0370-2693(97)01388-9}{\emph{Phys. Lett.}
  {\bfseries B418} (1998) 214--222},
  [\href{https://arxiv.org/abs/hep-ph/9707393}{{\ttfamily hep-ph/9707393}}].

\bibitem{Bolzoni:2012ii}
P.~Bolzoni, B.~A. Kniehl and A.~V. Kotikov, \emph{{Gluon and quark jet
  multiplicities at N$^3$LO+NNLL}},
  \href{http://dx.doi.org/10.1103/PhysRevLett.109.242002}{\emph{Phys. Rev.
  Lett.} {\bfseries 109} (2012) 242002},
  [\href{https://arxiv.org/abs/1209.5914}{{\ttfamily 1209.5914}}].

\bibitem{Abdesselam:2010pt}
A.~Abdesselam, E.~B. Kuutmann, U.~Bitenc, G.~Brooijmans, J.~Butterworth et~al.,
  \emph{{Boosted objects: A Probe of beyond the Standard Model physics}},
  \href{http://dx.doi.org/10.1140/epjc/s10052-011-1661-y}{\emph{Eur. Phys. J.}
  {\bfseries C71} (2011) 1661},
  [\href{https://arxiv.org/abs/1012.5412}{{\ttfamily 1012.5412}}].

\bibitem{Altheimer:2012mn}
A.~Altheimer, S.~Arora, L.~Asquith, G.~Brooijmans, J.~Butterworth et~al.,
  \emph{{Jet Substructure at the Tevatron and LHC: New results, new tools, new
  benchmarks}}, \href{http://dx.doi.org/10.1088/0954-3899/39/6/063001}{\emph{J.
  Phys.} {\bfseries G39} (2012) 063001},
  [\href{https://arxiv.org/abs/1201.0008}{{\ttfamily 1201.0008}}].

\bibitem{Altheimer:2013yza}
A.~Altheimer et~al., \emph{{Boosted objects and jet substructure at the LHC.
  Report of BOOST2012, held at IFIC Valencia, 23rd-27th of July 2012}},
  \href{http://dx.doi.org/10.1140/epjc/s10052-014-2792-8}{\emph{Eur. Phys. J.}
  {\bfseries C74} (2014) 2792},
  [\href{https://arxiv.org/abs/1311.2708}{{\ttfamily 1311.2708}}].

\bibitem{Adams:2015hiv}
D.~Adams et~al., \emph{{Towards an Understanding of the Correlations in Jet
  Substructure}},
  \href{http://dx.doi.org/10.1140/epjc/s10052-015-3587-2}{\emph{Eur. Phys. J.}
  {\bfseries C75} (2015) 409},
  [\href{https://arxiv.org/abs/1504.00679}{{\ttfamily 1504.00679}}].

\bibitem{CMS:2013kfa}
{\scshape CMS} collaboration, \emph{{Performance of quark/gluon discrimination
  in 8 TeV pp data}},  Tech. Rep. CMS-PAS-JME-13-002, 2013.

\bibitem{Berger:2003iw}
C.~F. Berger, T.~Kucs and G.~F. Sterman, \emph{{Event shape / energy flow
  correlations}},
  \href{http://dx.doi.org/10.1103/PhysRevD.68.014012}{\emph{Phys. Rev.}
  {\bfseries D68} (2003) 014012},
  [\href{https://arxiv.org/abs/hep-ph/0303051}{{\ttfamily hep-ph/0303051}}].

\bibitem{Almeida:2008yp}
L.~G. Almeida, S.~J. Lee, G.~Perez, G.~F. Sterman, I.~Sung and J.~Virzi,
  \emph{{Substructure of high-$p_T$ Jets at the LHC}},
  \href{http://dx.doi.org/10.1103/PhysRevD.79.074017}{\emph{Phys. Rev.}
  {\bfseries D79} (2009) 074017},
  [\href{https://arxiv.org/abs/0807.0234}{{\ttfamily 0807.0234}}].

\bibitem{Ellis:2010rwa}
S.~D. Ellis, C.~K. Vermilion, J.~R. Walsh, A.~Hornig and C.~Lee, \emph{{Jet
  Shapes and Jet Algorithms in SCET}},
  \href{http://dx.doi.org/10.1007/JHEP11(2010)101}{\emph{JHEP} {\bfseries 11}
  (2010) 101}, [\href{https://arxiv.org/abs/1001.0014}{{\ttfamily 1001.0014}}].

\bibitem{Ellis:1991qj}
R.~K. Ellis, W.~J. Stirling and B.~R. Webber, \emph{{QCD and collider
  physics}}, {\emph{Camb. Monogr. Part. Phys. Nucl. Phys. Cosmol.} {\bfseries
  8} (1996) 1--435}.

\bibitem{Larkoski:2013paa}
A.~J. Larkoski and J.~Thaler, \emph{{Unsafe but Calculable: Ratios of
  Angularities in Perturbative QCD}},
  \href{http://dx.doi.org/10.1007/JHEP09(2013)137}{\emph{JHEP} {\bfseries 09}
  (2013) 137}, [\href{https://arxiv.org/abs/1307.1699}{{\ttfamily 1307.1699}}].

\bibitem{Larkoski:2014tva}
A.~J. Larkoski, I.~Moult and D.~Neill, \emph{{Toward Multi-Differential Cross
  Sections: Measuring Two Angularities on a Single Jet}},
  \href{http://dx.doi.org/10.1007/JHEP09(2014)046}{\emph{JHEP} {\bfseries 09}
  (2014) 046}, [\href{https://arxiv.org/abs/1401.4458}{{\ttfamily 1401.4458}}].

\bibitem{Procura:2014cba}
M.~Procura, W.~J. Waalewijn and L.~Zeune, \emph{{Resummation of
  Double-Differential Cross Sections and Fully-Unintegrated Parton Distribution
  Functions}}, \href{http://dx.doi.org/10.1007/JHEP02(2015)117}{\emph{JHEP}
  {\bfseries 02} (2015) 117},
  [\href{https://arxiv.org/abs/1410.6483}{{\ttfamily 1410.6483}}].

\bibitem{Floratos:1981hs}
E.~G. Floratos, C.~Kounnas and R.~Lacaze, \emph{{Higher Order QCD Effects in
  Inclusive Annihilation and Deep Inelastic Scattering}},
  \href{http://dx.doi.org/10.1016/0550-3213(81)90434-X}{\emph{Nucl. Phys.}
  {\bfseries B192} (1981) 417--462}.

\bibitem{Altarelli:1979kv}
G.~Altarelli, R.~K. Ellis, G.~Martinelli and S.-Y. Pi, \emph{{Processes
  Involving Fragmentation Functions Beyond the Leading Order in QCD}},
  \href{http://dx.doi.org/10.1016/0550-3213(79)90062-2}{\emph{Nucl. Phys.}
  {\bfseries B160} (1979) 301--329}.

\bibitem{numrecipes}
V.~Press, Teukolsky and Flannery, \emph{{Numerical Recipes in C}}.
\newblock Cambridge University Press, 2~ed., 1992.

\bibitem{tikhonov1977solutions}
A.~N. Tikhonov and V.~Y. Arsenin, \emph{{Solutions of ill-posed problems}}.
\newblock Winston, Washington DC, 1977.

\end{thebibliography}\endgroup
\bibliographystyle{JHEP}

\end{document}